\documentclass[10pt,oneside,a4paper]{article}

\usepackage{latexsym}

\usepackage[dvips]{graphics,lscape,rotating}
\usepackage{euscript}
\usepackage{psfrag}
\usepackage{bbm} 
\usepackage{amsfonts,amsmath,amssymb}
\usepackage{mathrsfs}
\usepackage{pstricks,array,longtable,pst-all}
\usepackage{longtable}

\numberwithin{equation}{section}

\newcommand{\fcmt}[2]{\fbox{\mbox{\begin{minipage}[t]{#1cm} #2 \end{minipage}}}}
\newcommand{\cmt}[2]{\mbox{\begin{minipage}[t]{#1cm} #2 \end{minipage}}}
\newcommand{\mb}[1]{\mathbbm{#1}}  
\newcommand{\Muserfunction}[1]{A}

\newcommand{\PI}[3]{\big[\pi_{{#2}}(#1)\big]_{{#3}}}
\newcommand{\PIbar}[3]{\overline{\big[\pi_{{#2}}(#1)\big]}_{{#3}}}
\newcommand{\Sbra}[3]{\big<~#1~#2~#3~\big|}
\newcommand{\Sket}[3]{\big|~#1~#2~#3~\big>}
\newcommand{\ket}[1]{\big|~#1~\big>}
\newcommand{\bra}[1]{\big<~#1~\big|}
\newcommand{\scaprod}[2]{\big<~#1~\big|~#2~\big>}
\newcommand{\NN}{\nonumber\\}
\newcommand{\CGC}[3]{\big<~#1~;~#2~\big|~#3~\big>}

\newcommand{\barr}[1]{\begin{eqnarray}\begin{array}{#1}}
\newcommand{\earr}{\end{array}\end{eqnarray}}                       
\newcommand{\SixJ}[6]{\left\{\begin{array}{ccc}
                               #1 & #2 & #3\\
			       #4 & #5 & #6  
			    \end{array} \right\}}

\newcommand{\MC}[1]{\mathcal{#1}}

\DeclareMathOperator{\vol}{vol}
\DeclareMathOperator{\SU}{SU}
\newcommand{\eval}{\lambda_{\hat V}}
\newcommand{\numspinconfigs}{N_{\vec{j}-\mathrm{configs}}}
\newcommand{\numevals}{N_{\mathrm{evals}}}
\newcommand{\nevals}{N_{\mathrm{evals}}}

\newcommand{\jmax}{j_{\mathrm{max}}}
\newcommand{\nbins}{N_{\mathrm{bins}}}

\newcommand{\code}[1]{\texttt{#1}}
\newcommand{\sigconfs}{$\vec\sigma$-configurations}
\newcommand{\sigconf}{$\vec\sigma$-configuration}
\newcommand{\signconfs}{$\vec\epsilon$-configurations}
\newcommand{\mineval}{\lambda_{\hat{V}}^{(\mathrm{min})}}
\newcommand{\minev}{\mineval}
\newcommand{\maxev}{\lambda_{\hat{V}}^{(\mathrm{max})}}
\newcommand{\maxevfit}{\lambda_{\hat{V}}^{(\mathrm{max}),(fit)}}
\newcommand{\maxeval}{\maxev}
\newcommand{\thresh}{T}
\newcommand{\gtsim}{\gtrsim}
\renewcommand{\S}{section\ }

\def\ba{\begin{eqnarray}}
\def\ea{\end{eqnarray}}
\def\be{\begin{equation}}
\def\ee{\end{equation}}

\newtheorem{Theorem}{Theorem}[section]                             
\newtheorem{Definition}{Definition}[section]

\DeclareMathOperator{\tr}{tr}
\DeclareMathOperator{\sgn}{sgn}

\begin{document}

\title{Properties of the Volume Operator\\ in Loop Quantum Gravity II: Detailed Presentation}
\author{Johannes Brunnemann$^1$\thanks{brunnemann@math.uni-hamburg.de}~~,
David Rideout$^2$\thanks{drideout@perimeterinstitute.ca}\\
\\
$^1$Department of Mathematics, University of Hamburg,  20146 Hamburg, Germany\\ $^2$Perimeter Institute for Theoretical Physics, Waterloo, Ontario N2L 2Y5, Canada}

\maketitle

\begin{abstract}
{The properties of the Volume operator in Loop Quantum Gravity, as
constructed by Ashtekar and Lewandowski, are analyzed for the first time at
generic vertices of valence greater than four.  We find that the occurrence
of a smallest non-zero eigenvalue is dependent upon the geometry of the
underlying graph, and is not a property of the Volume operator itself.  The
present analysis benefits from the general simplified formula for matrix
elements of the Volume operator derived in \cite{Volume Paper}, making it
feasible to implement it on a computer as a matrix which is then diagonalized
numerically.  The resulting eigenvalues serve as a database to investigate
the spectral properties of the volume operator.  Analytical results on the
spectrum at 4-valent vertices are included.  This is a companion paper to
\cite{NumVolSpecLetter}, providing details of the analysis presented there.}
\end{abstract}
\tableofcontents
\section{ Introduction}
During the last 15 years major achievements have been obtained towards constructing a quantum theory of gravity in a background independent way. 
Within the framework of Loop Quantum Gravity (LQG) 
\cite{TT:big script,Rovelli: LQG Intro,Ashtekar:2004eh} it has been possible to reformulate General Relativity as a constrained\footnote{
Upon rewriting General Relativity in Hamilton's formalism,
the background independence of the underlying theory is manifested in the occurrence of constraints which 
must be fulfilled by physical configurations in the thus constructed model.
By `background independence' we refer to independence from a choice of fixed background geometry.}
gauge field theory which is then quantized using a refined version\footnote{This quantization program is referred to as Refined Algebraic Quantization (RAQ), see \cite{TT:big script} for a pedagogical introduction.} of canonical quantization which is based on \cite{Dirac constrained system}. In order to complete this program several conceptual and technical problems have to be solved. One major task is to understand and evaluate the action of the operators in the quantum theory which correspond to the classical constraints. Following the program of Refined Algebraic Quantization these operators have to be imposed on the up to now kinematical theory. 
A key ingredient \cite{TT:QSD V,Thiemann:2005zg} for this is the understanding of the quantum operator corresponding to the classical volume of a region in three dimensional Riemannian space.

Note that there are different versions of this operator due to Ashtekar and Lewandowski \cite{AL: Quantized Geometry: Volume Operators} and  due to Rovelli and Smolin \cite{Rovelli Smolin Volume Operator}  resulting from two different regularization schemes.
We will focus here on the first version which considers  
diffeomorphism invariant properties of the graphs, namely the relative orientation of the graph edges underlying the kinematical basis states encoded in sign factors, whereas the latter 
is insensitive to this information. 
Moreover the latter regularization \cite{Rovelli Smolin Volume Operator} seems to lead to difficulties when one attempts to reconstruct the action of the electric flux operators from the volume operator \cite{Consistency Check I,Consistency Check II}.
In what follows we will always imply the operator due to Ashtekar and Lewandowski when using the phrase 'volume operator'.
This paper contains the details on the results presented in the companion paper \cite{NumVolSpecLetter},
on the properties of the volume operator in Loop Quantum Gravity.  
We recommend the reader to have a look at \cite{NumVolSpecLetter} first in order to get an overview before getting into the details. 
Note that the basic notation which is used throughout this paper 
is introduced in \cite{NumVolSpecLetter}. Also an introduction to details of Loop Quantum Gravity essential for our analysis is provided in \cite{NumVolSpecLetter} and will not be reproduced here. 
\\

This paper is organized as follows.
In section 2 we recall the construction (regularization scheme) of the volume operator from the classical volume expression of a spatial region in order to clarify the origin of the combinatorial sign factor
and the recoupling
in the resulting operator expression. Here we closely follow \cite{TT:vopelm}.
Starting from \cite{Volume Paper} we then present in section 3 a more elaborate discussion of the computation of the matrix elements of the volume operator, in particular all cases of special arguments  of the general matrix element formula obtained in \cite{Volume Paper} are worked out in detail.
Subsequently in section 4 we show how considerations of gauge invariance impose conditions on the combinatorial sign factors, 
which reveals
a kind of  `self-regulating' property of this operator with respect to high valent vertices, providing a suppression of the volume contribution of certain high valent vertices. 
Next we present a detailed analysis on the set of diffeomorphism inequivalent embeddings which exist for the edges of an $N$-valent vertex 
into a spatial three dimensional Riemannian manifold. This analysis is accompanied by a numerical Monte Carlo sprinkling computation. 
Section 5 then describes the technical details of the computational implementation.
In the following section 6 the results of the detailed numerical analysis of the volume spectrum is presented.
The eigenvalues are computed via singular value decomposition, utilizing the LAPACK library \cite{lapack}.  
The numerical work employs the 
Cactus computational framework \cite{cactus}, which provides automatic parallelism, facilitates use of the LAPACK library (including managing inter-language procedure calls), and places the code in a modular context from which it can easily be used by others for future investigations.
Finally in section 7 we present new analytic results on properties of the volume operator at four-valent gauge invariant vertices. We show that its spectrum is simple and give an expression of the volume eigenstates in terms of its matrix elements and its eigenvalues. Moreover we show how one can (under mild assumptions on the four spins at the vertex) obtain an analytic lower bound of the smallest non-zero eigenvalue, solving an outstanding problem of \cite{Volume Paper}.
Our results are then briefly summarized in section 8; for a detailed summary we refer the reader to the companion paper \cite{NumVolSpecLetter}.
An appendix contains background material regarding spin networks and angular momentum theory, in order to make the present series of papers self contained and accessible to 
non-specialists.

\section{ Definition and Derivation of the Volume Operator }
Let us briefly introduce our notation first. Note that a basic introduction to LQG is given in the companion paper \cite{NumVolSpecLetter} and will not be reproduced here. For more details the reader is referred to \cite{TT:big script,Rovelli: LQG Intro,Ashtekar:2004eh}.  
 
As General Relativity is treated in Hamiltonian formalism as an $SU(2)$-gauge field theory, four dimensional space time is foliated into three dimensional spatial slices $\Sigma$ with induced metric $q_{ab}(x)$ and orthogonal foliation direction parametrized by a real foliation parameter.  
The canonical variables are then densitized triads $E^a_i(x)$ and a connection $A^j_b(y)$, at $x, y \in \Sigma$. Usually $i,j=1,2,3$ are $su(2)$-directions, and $a,b=1,2,3$ are spatial tensor indices. The set of classical (smooth, continuous) connection configurations is denoted by $\MC{A}$, the set of distributional connections is called $\overline{\MC{A}}$. 
The Poisson brackets of the $E^a_i(x)$ and the connection $A^j_b(y)$ are then given by
\be\label{class P A}
   \big\{E^a_i(x),A_b^j(y) \big\}=\kappa \delta^a_b\delta^j_i\delta(x,y)
\ee
Here $\kappa = 8\pi G_N$, $G_N$ being Newton's constant. Electric fluxes $E_i(S)$ are constructed by integration of the dual electric field over two dimensional orientable surfaces $S\subset\Sigma$ and the integral of the connection along one dimensional oriented piecewise analytic edges $e\subset\Sigma$ giving the holonomy $h_e(A)$ for each $A\in \MC{A}$. In this way (\ref{class P A}) is regularized.
A collection of edges is called a graph $\gamma$. Its set of edges is called $E(\gamma)$. It is constructed such that two edges mutually intersect at most in their beginning/end point, called a vertex $v$ in the vertex set $V(\gamma)$. Continuous maps $f=f_\gamma\circ p_\gamma$ are called  cylindrical functions, the set of cylindrical functions is denoted by CYL. Here 
$p_\gamma:\MC{A}\rightarrow \{h_e(A)\}_{e\in E(\gamma)}\in SU(2)^{|E(\gamma)|}$
and 
$f_\gamma:\{h_e(A)\}_{e\in E(\gamma)}\in SU(2)^{|E(\gamma)|}\rightarrow \mb{C}$. Sometimes we will also refer to $f_\gamma$ as cylindrical functions. Moreover one finds that  spin network functions (SNF)
$T_{\gamma\vec{j}\vec{m}\vec{n}}(A)=\prod_{e\in E(\gamma)}\sqrt{2j_e+1} \PI{h_e(A)}{j_e}{m_en_e}$
serve as a basis for CYL. Here $\PI{h_e(A)}{j_e}{m_en_e}$ is a representation matrix element function of an irreducible $SU(2)$ representation of weight $j_e$ which is associated to every $e\in E(\gamma)$, and $m_e,n_e=-j_e,\ldots j_e$ denote the matrix element.

\subsection{Classical Starting Point}

In this section we will closely follow the construction presented in \cite{TT:vopelm}.

Let us review the most important steps in order to write down a well defined operator $\hat{V}$ acting on the kinematical Hilbert space of LQG. The classical expression for the volume of an open, connected three dimensional spatial region $R$ given by:
\ba\label{Classical Volume Expression}
   V(R)&=&\int_R~d^3x~\sqrt{\det q(x)}
   \nonumber
   \\
   &=&\int_R~d^3x~\sqrt{\Big|\frac{1}{3!} \epsilon^{ijk}\epsilon_{abc}E^a_i(x)E^b_j(x)E^c_k(x) \Big|}
\ea
where we choose the Riemannian signature $\det(q) > 0$ $\forall x$ and
have used the fact that the square root of the determinant of the spatial metric $q_{ab}(x)$ can be re-expressed in terms of the densitized triads $E^a_i(x)$, as introduced in \cite{NumVolSpecLetter}. 

\subsection{Regularization Scheme}
 
The regularization procedure displayed here will be similar to the regularization of (\ref{class P A}) by smearing the dual of electric fields $E^a_i(x)$ over two dimensional surfaces $S$ and the connection $A^j_b(y)$ along one dimensional edges $e$ in order to obtain its holonomy $h_e(A)$, however the idea is here to define the smearing surfaces somewhat intrinsically, adapted to a graph $\gamma$ being the support for a cylindrical function $f_\gamma$.

Let us introduce the characteristic function $\chi_\Delta(p,x)$ of a cube in a coordinate frame $x$ centered at a point $p$ and spanned by the three vectors $\vec{\Delta}_\rho=\Delta_\rho\vec{n}_\rho$, the  $\vec{n}_\rho$ being normal vectors in the frame $x$. The cube $\Delta$ has coordinate volume 
$\textrm{vol}(\Delta)=2^3\Delta_1\Delta_2\Delta_3\det(\vec{n}_1,\vec{n}_2,\vec{n}_3)$. Then $\chi_\Delta(p,x)$ can be defined as 
\be
   \chi_\Delta(p,x)=\displaystyle\prod_{\rho=1}^{3}\Theta\Big(\Delta_\rho-\big|~\big<n^\rho,(x-p)^\rho\big>~ \big| \Big)
\ee
and  $\big<\cdot \big|\cdot\big>$ is the Euclidean inner product, $\Theta(z)$ is the usual unit step function with $\Theta(z)=0$ if $z< 0$, $\Theta(z)=\frac{1}{2}$ if $z=0$ and $\Theta(z)=1$ if $z> 0$. 
\begin{figure}[hbt]
    \center
    \psfrag{x}{$x$}
    \psfrag{p}{$p$}
    \psfrag{xp}{$x-p$}
    \psfrag{D1}{$2\Delta_1$}
    \psfrag{D2}{$2\Delta_2$}
    \psfrag{D3}{$2\Delta_3$}
    \psfrag{n1}{$\blue\vec{n}_1$}
    \psfrag{n2}{$\blue\vec{n}_2$}
    \psfrag{n3}{$\blue\vec{n}_3$}
    \psfrag{O}{$0$}
    \includegraphics[height=5cm]{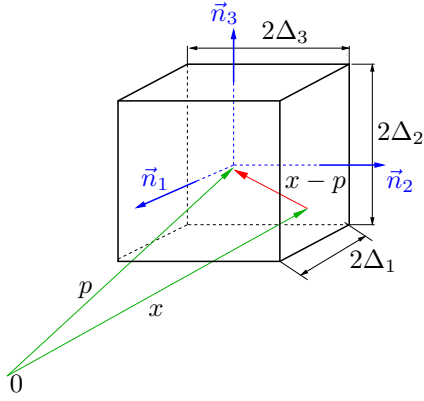} 
    \caption{The regulator $\chi_\Delta(p,x)$}
\end{figure}

If we take the limit $\Delta \rightarrow 0$ (by decreasing $\Delta_\rho\rightarrow 0$ for $\rho=1,2,3$)  we realize that
\be\label{limiting properties of chi delta}
    \lim\limits_{\Delta \rightarrow 0}~ \frac{1}{\vol(\Delta)}~ \chi_\Delta(p,x)= \delta^{(3)}(p,x)   
\ee
where $\delta^{(3)}$ indicates the delta distribution in three dimensions.
Now upon introducing the smeared quantity
\ba\label{regulated volume expression}
   E(p,\Delta,\Delta',\Delta'')
     &:=&\frac{1}{\vol(\Delta)~\vol(\Delta')~\vol(\Delta'')}
        \int\limits_R d^3x \int\limits_R d^3y \int\limits_R d^3z~ 
	     \chi_{\Delta}(p,x)~\chi_{\Delta'}(p,\frac{x+y}{2})~\chi_{\Delta''}(p,\frac{x+y+z}{3})
     \NN
     &&\hspace{6cm}\times
                \frac{1}{3!} \epsilon^{ijk}\epsilon_{abc}E^a_i(x)E^b_j(y)E^c_k(z)
\ea
one notices that
\begin{itemize}
   \item[(i)]{Taking the limits $\Delta_\rho,\Delta'_{\rho'},\Delta''_{\rho''}\rightarrow 0$ in any combination and at any rate with       respect to each other we come back to 
         $\frac{1}{3!} \epsilon^{ijk}\epsilon_{abc}E^a_i(p)E^b_j(p)E^c_k(p)$ due to 
         (\ref{limiting properties of chi delta}).}
   \item[(ii)]{(i) holds for any choice of linearly independent normal vectors 
               $\vec{n}_{\rho},\vec{n}_{\rho'},\vec{n}_{\rho''}$.}
   \item[(iii)]{    \be\label{regularized classical V}
                     V(R)=\lim\limits_{\Delta\rightarrow 0}
                      \lim\limits_{\Delta'\rightarrow 0}
                      \lim\limits_{\Delta''\rightarrow 0}
                      \int\limits_R d^3p \sqrt{\big| E(p,\Delta,\Delta',\Delta'')\big|} \ee}
                 
\end{itemize}

The classical Poisson bracket (\ref{class P A}) contains functional derivatives with respect to the electric field $E^a_i(x)$ and the connection $A_b^j(y)$.
Upon quantization the electric field $E^a_i(x)$ acts as a functional derivative
\be\label{E as functional derivative}
  E^a_i(x)\longrightarrow  \hat{E}^a_i(x):=-\mb{i}\hbar\kappa \frac{\delta}{\delta A^i_a(x)}
\ee
and we have (symbolically), for the action of this functional derivative on a holonomy $h_e(A)$ of an edge $e$  with $x=e(t_x)$ being a point of $e$, (here $\tau_i=-\mb{i}\sigma_i$ denotes a basis for the Lie algebra $su(2)$, $\sigma_i$ are the Pauli matrices) : 
\ba\label{action of functional derivative}
   \frac{\delta h_e(A)}{\delta A^i_a(x)}
   &=&
   \frac{1}{2}
   \Big[\lim\limits_{\tilde{t}\downarrow t_x} +\lim\limits_{\tilde{t}\uparrow t_x}\Big] \int\limits_0^1 dt
               \Big[\delta^3\big(e(\tilde{t}),x \big)\dot{e}^a(t) h_{e[0,t]}\frac{\tau_i}{2}h_{e[t,1] } \Big]
   \NN
   &=&\left\{\begin{array}{lcl}
                \frac{1}{2}~ \dot{e}^a(t_x)~\frac{\tau_i}{2} h_{e[0,1]} &&x=e(t)\big|_{t=0} \\[2mm]
                            ~~~\dot{e}^a(t_x)~h_{e[0,t]}\frac{\tau_i}{2} h_{e[t,1]} &&x=e(t)\big|_{0<t<1}\\[2mm]
                \frac{1}{2}~ \dot{e}^a(t_x)~h_{e[0,1]}\frac{\tau_i}{2}  &&x=e(t)\big|_{t=1}       
             \end{array} \right.
\ea
Then we can formulate the action of a smeared version of (\ref{E as functional derivative}) contained in the regulated expression (\ref{regulated volume expression}) 
\be
   \hat{E}^a_i(p,\Delta):=\frac{1}{\vol(\Delta)}\int\limits_{\sigma} d^3x~\chi_{\Delta}(x)\hat{E}^a_i(x)
\ee
 on a function $f_\gamma(h_{e_1},\ldots,h_{e_N})$ cylindrical with respect to the graph $\gamma$, e.g. for $x=e(t)\big|_{0<t<1}$ as 
 \be\label{regulated volume expression 2}
    \hat{E}^a_i(p,\Delta)f_\gamma=-\frac{\mb{i}\ell_P^2}{\vol(\Delta)}\sum_{e\in E(\gamma)} 
                  \int\limits_0^1dt~ \chi_\Delta\big(p,e(t)\big)
                  ~\dot{e}^a(t)~\tr\Big[h_{e[0,t]}\frac{\tau_i}{2} h_{e[t,1]
                   }\frac{\partial}{\partial h_{e[0,1]}}\Big]~~f_\gamma(h_{e_1},\ldots,h_{e_N})
 \ee
where the trace is taken with respect to the defining representation of $SU(2)$ and we have used the chain rule and Leibnitz rule in order to evaluate the functional derivative on $f_\gamma$.

Now obviously the regulated expression (\ref{regulated volume expression}) contains three expressions of the form (\ref{regulated volume expression 2}), which have to be applied successively if one wants to evaluate the action of (\ref{regulated volume expression}) on a cylindrical function. This involves a careful analysis of the action (\ref{action of functional derivative}) of the individual functional derivatives, in particular if they act on one edge simultaneously, which results in expressions where different $\tau$-matrices have to be inserted into the holonomy at different values of the curve parameter $t$.

The action of $E(p,\Delta,\Delta',\Delta'')$ on a cylindrical function then schematically reads
\ba\label{schematical action of regulated expression}
   \lefteqn{E(p,\Delta,\Delta',\Delta'')f_\gamma= \Bigg[~\sum_{e\,e'e''} M \cdot\epsilon_{abc}~\times}
   \NN
   &\times\!& 
        \int\limits_0^1dt\!\int\limits_0^1dt'\!\int\limits_0^1dt''~
         \chi_\Delta(p,e(t))~
         \chi_{\Delta'}(p,{\textstyle\frac{e(t)+e'(t')}{2}})~
         \chi_{\Delta''}(p,{\textstyle \frac{e(t)+e'(t')+e''(t'')}{3}})~
         \dot{e}(t)^a\dot{e}'(t')^b\dot{e}''(t'')^c
         ~\hat{O}_{e\,e'e''}(t,t',t'') \Bigg]f_\gamma
   \NN          
\ea
Here $M=\frac{\mb{i}~\ell_p^6}{3!\vol(\Delta)\vol(\Delta')\vol(\Delta'')}$ and the sum has to be extended to all triples of (not necessarily distinct) edges $(e\,e'e'')$. By successive application of the three functional derivatives and Leibnitz rule, one obtains expressions $\hat{O}_{e\,e'e''}(t,t',t'')$ which are combinations of traces of holonomies containing inserted $\tau$ matrices and partial derivatives with respect to holonomies of the appropriate edges.
\\   
   
In order to evaluate the characteristic functions $\chi_{\Delta}(p,x),\chi_{\Delta'}(p,\frac{x+y}{2}),\chi_{\Delta''}(p,\frac{x+y+z}{3})$ contained in expression (\ref{regulated volume expression}) one considers a coordinate transformation in the integrals: for a given triple $(e,e',e'')$ of (not necessarily) distinct edges contained in the edge set $ E(\gamma)$, the vector valued function
\be
   x_{ee'e''}(t,t',t''):=\frac{e(t)+e'(t')+e''(t'')}{3} 
\ee
whose Jacobian is given by
\be
   \det\left(\frac{\partial x^a_{ee'e''}}{\partial(t,t',t'')} \right)
   =
   \det\left(\begin{array}{ccc}
                \frac{\partial x^1}{\partial t}
               &\frac{\partial x^2}{\partial t}
               &\frac{\partial x^3}{\partial t}
               \\[2mm]
                \frac{\partial x^1}{\partial t'}
               &\frac{\partial x^2}{\partial t'}
               &\frac{\partial x^3}{\partial t'}
               \\[2mm]
                \frac{\partial x^1}{\partial t''}
               &\frac{\partial x^2}{\partial t''}
               &\frac{\partial x^3}{\partial t''}
             \end{array}
        \right)
   =\frac{1}{3^3}~\epsilon_{abc}~\dot{e}(t)^a~\dot{e}'(t')^b~\dot{e}''(t'')^c
\ee
equals, up to a numerical prefactor, the tangent vectors\footnote{As implied by inserting (\ref{E as functional derivative}), (\ref{action of functional derivative}) into the classical expression (\ref{regulated volume expression}).} contracted with the totally antisymmetric $\epsilon_{abc}$ in (\ref{regulated volume expression}), which is the reason for the choice of arguments in the characteristic function $\chi_{\Delta''}$.

The observation is now that taking the limit $\Delta''_{\rho''}\rightarrow
0$, and simultaneously demanding that $\chi_{\Delta''}(p,{\textstyle
\frac{e(t)+e'(t')+e''(t'')}{3}})\stackrel{!}{=}1$, implies that
$e(t_{int})\stackrel{!}{=}e'(t'_{int})\stackrel{!}{=}e''(t''_{int})\stackrel{!}{=}p$,
that is the edges $e,e',e''$ have to intersect at $p$ at the according curve
parameters $t_{int},t'_{int},t''_{int}$.\footnote{
The decoration `!' in `$\stackrel{!}{=}$' simply indicates that the equality
 is required to hold.}
Therefore the point $p$ is a vertex $v$ in the vertex set $V(\gamma)$ of the
graph $\gamma$. Moreover due to the convention that edges can at most
intersect at their beginning and endpoints,\footnote{This can always be
achieved by subdividing and redirecting the edges of a graph $\gamma$, see
\cite{NumVolSpecLetter}.} we can without loss of generality assume that
$(e,e',e'')$ are outgoing from $p=v$, which consequently serves as their
beginning point and hence $t_{int}=t'_{int}=t''_{int}=0$.

We can thus pull the remaining characteristic functions  $\chi_\Delta(p,e(t))=\chi_{\Delta'}(p,{\textstyle\frac{e(t)+e'(t')}{2}})=1$ out of the integral. The remaining part of the integral then is schematically given as
\ba\label{limit of regulated expression}
   \lefteqn{\lim_{\Delta''\rightarrow 0}E(p,\Delta,\Delta',\Delta'')f_\gamma=}~~~~
   \\
   &=&\Bigg[
   \sum_{e\,e'e''}~\frac{\mb{i}~\ell_p^6~3^3~\chi_\Delta(p,v)~\chi_{\Delta'}(p,v)}{3!\vol(\Delta)\vol(\Delta')}~ \hat{O}_{e\,e'e''}(0,0,0)~
   \int\limits_0^1dt\!\int\limits_0^1dt'\!\int\limits_0^1dt''~
         \delta^{(3)}(p,x_{e\,e'e''})~\det\left(\frac{\partial x^a_{ee'e''}}{\partial(t,t',t'')} \right)
         \Bigg]
         f_\gamma
   \nonumber
\ea
In order to get a non vanishing expression from (\ref{limit of regulated expression}) the triple $(e,e',e'')$ must consist of three distinct edges with linearly independent tangents in $p=v$. 

Now perform a change of variables
\ba 
  d^3t \left|\det\left(\frac{\partial x^a_{ee'e''}}{\partial(t,t',t'')}\right)\right|
  &=& d^3t~~ \underbrace{\sgn\left[\det\left(\frac{\partial x^a_{ee'e''}}{\partial(t,t',t'')}\right)\right]}
  \cdot \det\left(\frac{\partial x^a_{ee'e''}}{\partial(t,t',t'')}\right) 
  \NN
  &=:& d^3t~~~~~~~~~~~~~ \epsilon(e\,e'e'') ~~~~~~~~~\cdot \det\left(\frac{\partial x^a_{ee'e''}}{\partial(t,t',t'')}\right) 
  \NN
  &=& d^3x_{ee'e''}
\ea 
in order to evaluate the $\delta^{(3)}-$distribution in (\ref{limit of regulated expression}).
Note that {\it it is here where the sign factor $\epsilon(e\,e'e'')$ enters}. For this purpose we have to insert a $1=\left[\epsilon(e\,e'e'')\right]^2$ and, observing that integrating the $\delta^{(3)}(p,x_{ee'e''})$ over the positive octant only gives an additional prefactor of $\frac{1}{8}$, we arrive at
\ba\label{limit of regulated expression 2}
   \lim_{\Delta''\rightarrow 0}E(p,\Delta,\Delta',\Delta'')f_\gamma
   &=&\Bigg[
   \sum_{e\,e'e''\in E(\gamma)}~\frac{\mb{i}~\ell_p^6~3^3~\chi_\Delta(p,v)~\chi_{\Delta'}(p,v)}{3!~8\cdot8~\vol(\Delta)~\vol(\Delta')}~\epsilon(e\,e'e'')~ \hat{O}_{e\,e'e''}(0,0,0)~\Bigg]
         f_\gamma
   \NN          
\ea 
where  
\be\label{O operator}
   \hat{O}_{e\,e'e''}(0,0,0)
   =\frac{1}{8}\epsilon_{ijk}X^i_eX^j_{e'}X^k_{e''}
\ee
and $X^i_e:=\tr\big(\tau_i h_{e[0,1]}\frac{\partial}{\partial h_{e[0,1]}}\big)$ denotes the right invariant vector fields  of $\SU(2)$ resulting from the action of the functional derivatives (\ref{action of functional derivative}).
Note that we can choose the order of the right invariant vector fields arbitrarily in (\ref{O operator}), because they commute if they act on distinct edges, that is distinct copies of $\SU(2)$.   

The second prefactor $\frac{1}{8}$ in (\ref{limit of regulated expression 2}) stems from the fact that we only evaluate one sided  functional derivatives  according to (\ref{action of functional derivative}): For a vertex $v$ in the vertex set $V(\gamma)$ of the graph $\gamma$, the point $p$ is the beginning point of the outgoing edges $e,e',e''$, that is $p=e(t)|_{t=0}=e'(t')|_{t'=0}=e''(t'')|_{t''=0}$.

Inserting (\ref{limit of regulated expression 2}) back into (\ref{regularized classical V}) we may synchronize the remaining limits $\Delta,\Delta'\rightarrow 0$ by choosing $\Delta=\Delta'$ and take $\chi_{\Delta}(p,v)=\big[\chi_{\Delta}(p,v)\big]^2$ out of the square root. Taking the limit $\Delta \rightarrow 0$ then results in the operator describing the volume of a spatial region $R$, namely the  volume operator
$\hat{V}(R)_{\gamma}$ acting on the cylindrical function $f_\gamma$ over a graph $\gamma$ as:
\ba
  \hat{V}(R)_{\gamma}f_\gamma&=&\int\limits_R d^3p\widehat{\sqrt{det(q)(p)_{\gamma}}}~~f_\gamma
                     =\int\limits_R d^3p~ \hat{V}(p)_{\gamma}~~f_\gamma 
\ea
where
\ba
   \hat{V}(p)_{\gamma}&=&\ell_P^3 \sum_{v\in V(\gamma)}\delta^{(3)}(p,v)~\hat{V}_{v,\gamma}\\
  \label{erste} \hat{V}_{v,\gamma} &=&\sqrt{\Big| \mb{i}\cdot\tilde{Z} 
                   \sum_{{e_I,e_J,e_K \in E(\gamma) \atop {\vspace{2mm}\atop e_I\cap e_J \cap e_K = v}}} 
		   \epsilon (e_I,e_J,e_K)~ \epsilon_{ijk}X^i_IX^j_JX^k_K\Big|}
\ea
Here $\tilde{Z}$ is a constant depending on the regularization procedure. Its numerical value according to \cite{TT:vopelm} and above is found to be $\tilde{Z}=\frac{\beta^3}{3!\cdot 8}\cdot\left(\frac{3}{4}\right)^3$, however we will keep it unspecified in our calculations,  since the Immirzi parameter $\beta$ can be freely chosen. It only contributes an overall constant scaling to the spectrum of the volume operator. Mostly we will set $Z=1$ below and reinsert it when necessary.

The sum has to be taken over all vertices $v\in V(\gamma)$ of the graph $\gamma$ and at each vertex $v$ over all possible triples $(e_I,e_J,e_K)$ of outgoing edges\footnote{As mentioned before, one can without loss of generality always redirect edges such that there are only outgoing edges at each vertex.} of the graph $\gamma$. Here  $\epsilon (e_I,e_J,e_K)$ is the sign of the cross product of the three tangent vectors of the edges $(e_I,e_J,e_K)$ at the vertex $v$.

Again the  $X^i_I$ are the right invariant vector fields on $SU(2)$ fulfilling the commutator relation\linebreak  
\be\label{rivf commutation relation} 
   [X^i_I,X^j_J]=-2~ \delta_{IJ}~\epsilon^{ijk}~X_I^k
\ee 

\subsection{\label{Right Invariant Vector Fields as Angular Momentum Operators}Right Invariant Vector Fields as Angular Momentum Operators}
One can introduce the self-adjoint right invariant vector fields $Y_J^j:=-\frac{\mb{i}}{2} X^j_J$ fulfilling the usual angular momentum commutation relations  $[Y^i_I,Y^j_J]=\mb{i}~\delta_{IJ}~\epsilon^{ijk}~Y_I^k$.
 
Their action on cylindrical functions $f_\gamma$ and hence spin network functions $T_{\gamma\vec{j}\vec{m}\vec{n}}(A)$ can be associated to the action of ordinary angular momentum operators $J_I^i$ acting on a (recoupled) abstract spin system $\Sket{\vec{a}~ J}{M}{;n}$ as analyzed in \cite{Consistency Check I,Consistency Check II}. This correspondence is summarized in the companion paper \cite{NumVolSpecLetter} and recalled in more detail in section \ref{relation SNF spin system} in the appendix of this paper.
 We can therefore equivalently replace:
\ba
        \epsilon_{ijk}X^i_IX^j_JX^k_K~f_\gamma
       &\leftrightarrow&
        \left(-\frac{2}{\mb{i}} \right)^3\epsilon_{ijk}J^i_IJ^j_JJ^k_K ~~~ \Sket{\vec{a}~J}{M}{;n} 
\ea
Using furthermore the antisymmetry of $\epsilon_{ijk}$ and the fact that $[J_I^i,J_J^j]=0$ whenever $I\ne J$ we can restrict the summation in (\ref{erste}) to $I < J<K$ if we simultaneously write a factor $3!$ in front of the sum. The result is:
\ba
  \label{zweite} \hat{V}_{v,\gamma} &~~=~~&\sqrt{\Big| \tilde{Z}\cdot3! \cdot 2^3
                   \sum_{I<J<K} 
		   \epsilon (e_I,e_J,e_K)~ \epsilon_{ijk}~J^i_IJ^j_JJ^k_K \Big|}
\ea
We then make use of the identity
\ba
  \epsilon_{ijk}~J^i_I J^j_J J^k_K & = & \frac{\mb{i}}{4}\big[(J_{IJ})^2,(J_{JK})^2 \big]                                      
\ea
where $(J_{IJ})^2=\sum_{k=1}^3 (J_I^k+J_J^k)^2$. This relation can be derived by writing down every commutator as
$ \big[ (J_{IJ})^2, (J_{JK})^2\big]=\sum\limits_{i,j=1}^3 \big[(J_I^i+J_J^i)^2,(J_J^j+J_K^j)^2 \big]$, using the
identity $\big[a,bc\big]=\big[a,b\big]c +b\big[a,c \big]$ for the commutator, the angular
momentum commutation relations ($\big[J^i,J^j \big]=\mb{i}\epsilon^{ijk}J^k$) and the fact that $\big[J_I^i,J_J^j \big]=0$ whenever $I \ne J$.

We may then summarize:
\be\label{Volume Definition}
   \hat{V}_{v,\gamma} =\sqrt{\big| Z \cdot\sum_{I<J<K} 
		   \epsilon (e_I,e_J,e_K)~\hat{q}_{IJK} \big|}
		   =:\Big||Z|\cdot Q \Big|^\frac{1}{2}
        	  =\Big||Z|^2\cdot Q^\dagger Q \Big|^\frac{1}{4}
\ee
where $\hat{q}_{IJK} :=\big[(J_{IJ})^2,(J_{JK})^2 \big]$ and
$Z=\tilde{Z}\cdot \frac{\mb{i}}{4}\cdot 3! \cdot 2^3=\mb{i}\cdot\beta^3\cdot\frac{27}{256}$, $\beta$ again denotes the Immirzi parameter. Note that the actual numerical value of $Z$ varies in different regularizations \cite{AL: Quantized Geometry: Volume Operators,TT:vopelm}\footnote{The latter regularization \cite{TT:vopelm}, which we have presented here, differs by a numerical factor of $\frac{27}{8}$ from the former \cite{AL: Quantized Geometry: Volume Operators}. The former exactly reproduces the value of $C_{reg}$ obtained in \cite{Consistency Check I}.}. However \cite{Consistency Check I} shows how it can be fixed to
$Z=\beta^3\cdot\frac{3!\mb{i}}{4}\cdot C_{reg}$, with $C_{reg}=\frac{1}{3!8}$, by demanding consistency of the volume quantization to the usual quantization of the electric fluxes. We will set $Z=1$ in our subsequent analysis, as it only gives an overall numerical scaling of the volume spectrum.\\

Here we have introduced  the shorthand $Q^\dagger Q$. $Q$ is by definition a sum of antisymmetric matrices and hence antisymmetric itself. Multiplying it by its transposed conjugate $Q^\dagger$, one obtains a totally symmetric real matrix. Its eigenvalues $\lambda_{Q^\dagger Q}\ge 0$ are real  and come in pairs $\lambda_{Q^\dagger Q}=\big|\lambda_Q\big|^2$.
Equation (\ref{Volume Definition}) can be understood as follows:
The Volume operator $\hat{V}$ has the same eigenstates as $Q$ or $Q^\dagger Q$, but its eigenvalues are defined as $\lambda_{\hat{V}}:=\big|\lambda_Q\big|^\frac{1}{2}
=\big|\lambda_{Q^\dagger Q}\big|^\frac{1}{4}$.

Our task is then to calculate the spectra of totally antisymmetric real matrices of the form:
\be\label{Q definition}
   Q:=\sum\limits_{I<J<K\le N}\epsilon(IJK)~\hat{q}_{IJK}
\ee
where $\epsilon(IJK)=\mbox{sgn}\big(\det{(\dot{e}_I(v),\dot{e}_J(v),\dot{e}_K(v))}\big)$ denotes the sign of the determinant of the tangents of the three edges $e_I,e_J,e_K$ intersecting at the vertex $v$, and $N$ is the valence of $v$.

\subsection{Matrix Elements in Terms of $3nj$-Symbols}
Now we can apply the recoupling theory of $n$ angular momenta as introduced in the companion paper \cite{NumVolSpecLetter}. Further details are provided in appendix section \ref{Angular Momentum Theory}. 
The aim is to represent $\hat{q}_{IJK}$ in a recoupling
scheme basis. 
We will do this with respect to the standard basis\footnote{See appendix section \ref{Properties of Recoupling Schemes} for the definition.} $\vec{a}(12)$, where we can now easily
restrict our calculations to gauge invariant spin network states by demanding the total angular momentum
$J$ and the total magnetic quantum number $M$ to vanish, i.e.\ 
we will take into account only
recoupling schemes which couple the outgoing spins at the vertex $v$ to resulting angular momentum 0.
In terms of the recoupling schemes these states are given by:
\be\label{Recoupling Basis States} 
  \ket{\vec{g}(IJ)~\vec{j}~J=0~M=0} \::=\, \ket{\vec{g}(IJ)}
\ee
where we have introduced an abbreviation, since the quantum numbers $\vec{j}~,~J=0~,~M=0$ are the same for every
gauge invariant spin network state with respect to a fixed $N$-valent vertex $v$ with edge spins $\vec{j}:=(j_1,\ldots,j_{N}) $. 

We will now represent $\hat{q}_{IJK}:=\big[(J_{IJ})^2,(J_{JK})^2 \big]$ in the standard recoupling 
scheme basis  where 
$\ket{\vec{a}}:=\ket{\vec{a}(12)}\nobreak,\\ \ket{\vec{a}'}:=\ket{\vec{a}'(12)}$. 
The point is that by construction a recoupling scheme basis $\ket{\vec{g}(IJ)}$ diagonalizes the operator
$(G_2)^2 = (J_{IJ})^2 = (J_I+J_J)^2$: 
\be
  (G_2)^2\ket{\vec{g} (IJ)}=g_2(IJ)\big(g_2(IJ)+1 \big)\ket{\vec{g}(IJ)}
\ee      
Furthermore every recoupling scheme $\ket{\vec{g}(IJ)}$ can be expanded in terms of the standard basis via its
expansion coefficients, the $3nj$-symbols.
\begin{samepage}
So it is possible to express \cite{TT:vopelm}
\ba \label{vorformel}
\lefteqn{\big<\vec{a}(12)\big|\hat{q}_{IJK}\big|\vec{a}'(12)\big>=}~~~
   \NN\NN
   & = &\big<\vec{a}(12)\big|\big[(J_{IJ})^2,(J_{JK})^2 \big]\big|\vec{a}'(12)\big> \nonumber\\
   & = &\big<\vec{a}(12)\big|(J_{IJ})^2(J_{JK})^2]\big|\vec{a}'(12)\big> - \big<\vec{a}(12)\big|(J_{JK})^2(J_{IJ})^2\big|\vec{a}'(12)\big>\nonumber\\
   & = &  \sum_{\vec{g}(IJ)}g_2(IJ)(g_2(IJ)+1)
   [\big<\vec{a}(12)\big|\vec{g}(IJ)\big>\big<\vec{g}(IJ)\big|(J_{JK})^2\big|\vec{a}'(12)\big>
   -\big<\vec{a}(12)\big|(J_{JK})^2\big|\vec{g}(IJ)\big>\big<\vec{g}(IJ)\big|\vec{a}'(12)\big>]
   \nonumber\\
   &=& 
   \sum_{\vec{g}(IJ),\vec{g}(JK),\vec{g}''(12)}
   g_2(IJ)(g_2(IJ)+1)g_2(JK)(g_2(JK)+1)
   \big<\vec{g}(IJ)\big|\vec{g}''(12)\big>\big<\vec{g}(JK)\big|\vec{g}''(12)\big>\times\nonumber\\
   &&\times
   [\big<\vec{g}(IJ)\big|\vec{a}(12)\big>\big<\vec{g}(JK)\big|\vec{a}'(12)\big>
   -\big<\vec{a}(12)\big|\vec{g}(JK)\big>\big<\vec{g}(IJ)\big|\vec{a}'(12)\big>]\nonumber\\
   & = & 
   \sum_{\vec{g}''(12)} \left[ \sum_{\vec{g}(IJ)} \right. 
   g_2(IJ)(g_2(IJ)+1)\big<\vec{g}(IJ)\big|\vec{g}''(12)\big>\big<\vec{g}(IJ)\big|\vec{a}(12)\big>\times\nonumber\\
   &&\times \left. \sum_{\vec{g}(JK)}g_2(JK)(g_2(JK)+1)\big<\vec{g}(JK)\big|\vec{g}''(12)\big>\big<\vec{g}(JK)\big|\vec{a}'(12)\big>
   \right] 
   \nonumber\\[2mm]
   \NN
   && - \Big[~~~\vec{a}(12)~~ \rightleftharpoons~~ \vec{a}'(12)~~~ \Big]
   \\[3mm]
   \label{General ME}
   &=&
   \sum_{\vec{g}''(12)} \Big[F^{IJ}\big(\vec{a}(12),\vec{g}''(12)\big)
                  \times     F^{JK}\big(\vec{a}'(12),\vec{g}''(12)\big)                                          \Big]
                  - \Big[~~~\vec{a}(12)~~ \rightleftharpoons~~ \vec{a}'(12)~~~ \Big]
\ea
\end{samepage}
\hspace{-3mm}
where we have introduced the shorthand
\be\label{Constraint Regularization}
   F^{IJ}\big(\vec{a}(12),\vec{g}''(12)\big):=\sum\limits_{\vec{g}(IJ)}g_2(IJ)\Big( g_2(IJ)+1\Big)
                \big<~\vec{g}(IJ)~ \big|~\vec{g}''(12)~ \big>
		\big<~\vec{g}(IJ)~ \big|~\vec{a}(12)~ \big>
\ee
We can nicely see from (\ref{vorformel}) that we obtain
a real antisymmetric matrix possessing purely imaginary eigenvalues (we could alternatively
consider the purely complex version by multiplying all matrix elements by the imaginary unit $\mb{i}$).  
We have inserted the suitable recoupling schemes $\ket{\vec{g}(IJ)},\ket{\vec{g}(JK)}$ diagonalizing 
$(J_{IJ})^2$ and $(J_{JK})^2$ and their expansion in terms of the standard basis $\ket{\vec{g}(12)}$
by using the completeness of the recoupling schemes $\ket{\vec{g}(IJ)}$ for arbitrary $I \ne J $ \footnote{The
summation has to be extended over all possible intermediate recoupling steps $g_2,\ldots,g_{n-1}$,
that is $|j_r-j_q|\le g_k(j_q,j_r) \le j_q+j_r$, allowed by the Clebsch-Gordan Theorem.}
\be
  \mathbbm{1}=\sum_{\vec{g}(IJ)}\ket{\vec{g}(IJ)}\bra{\vec{g}(IJ)} 
\ee

So we have as a first step expressed the matrix elements of $\hat{q}_{IJK}$ in terms of $3nj$-symbols. 

\subsection{\label{Removal of the Arbitrariness of the Edge Labelling}Removal of the Arbitrariness of the Edge Labeling}

We have described how one can represent the volume operator as a matrix acting on a linear vector space whose basis states are labelled by the multilabel $\vec{a}$ of the standard recoupling schemes.
As can be seen from definition \ref{Def 3nj-symbol} in the appendix there exits a unitary basis transformation between any two different orders of recoupling, e.g. $\vec{g}(IJ),\vec{a}(12)$:
\be
   \big|~\vec{g}(IJ)~J~M~;~\vec{n}~\big> 
   = \sum_{\vec{a}(12)}   \big<~\vec{a}(12)~J~M~;~\vec{n}~
                          \big|~\vec{g}(IJ)~J~M~;~\vec{n}~\big>~\cdot~
			  \big|~\vec{a}(12)~J~M~;~\vec{n}~\big>
\ee
where the matrix elements $\big<~\vec{a}(12)~J~M~;~\vec{n}~\big|~\vec{g}(IJ)~J~M~;~\vec{n}~\big>=:U_{\vec{a}\vec{g}}$ of this basis transformation are the so called $3nj$-symbols. Now this basis transformation is unitary as can easily be seen because the recoupling schemes are orthonormal
\be
          \big<~\vec{a}(12)~J~M~;~\vec{n}~\big|~\vec{a}'(12)~J'~M'~;~\vec{n}'~
                          \big>=U_{\vec{a}\vec{a}'}
			  =\delta_{\vec{a}\vec{a}'}\delta_{JJ'}\delta_{MM'}\delta_{\vec{n}\vec{n}'}
\ee
and provide a complete orthonormal basis
\ba 
   \lefteqn{\big<~\vec{a}(12)~J~M~;~\vec{n}~\big|~\vec{a}'(12)~J~M~;~\vec{n}~\big>
   =}\hspace{2cm}
   \nonumber
   \\ 
   &=&
   \delta^{\vec{a}(12)}_{\vec{a}'(12)}
   \nonumber
   \\ 
   &=&
   \sum_{\vec{g}(IJ)}
   \big<~\vec{a}(12)~J~M~;~\vec{n}~\big|~\vec{g}(IJ)~J~M~;~\vec{n}~\big>
   \big<~\vec{g}(IJ)~J~M~;~\vec{n}~\big|~\vec{a}'(12)~J~M~;~\vec{n}~\big>
   \nonumber
   \\    
   &=&\sum_{\vec{g}(IJ)} U_{\vec{a}\vec{g}} U_{\vec{g}\vec{a}'}
\ea
Moreover the $3nj$-symbols are real (the overline denotes complex conjugation) and symmetric
\ba
   \overline{\big<~\vec{a}(12)~J~M~;~\vec{n}~\big|~\vec{g}(IJ)~J~M~;~\vec{n}~\big>}
   =\overline{U_{\vec{a}\vec{g}}}
   &=&U_{\vec{a}\vec{g}}
   = \big<~\vec{a}(12)~J~M~;~\vec{n}~\big|~\vec{g}(IJ)~J~M~;~\vec{n}~\big>
   \NN
   &=&U_{\vec{g}\vec{a}}
   = \big<~\vec{g}(IJ)~J~M~;~\vec{n}~\big|~\vec{a}(12)~J~M~;~\vec{n}~\big>
\ea
Thus we have 
\be
   UU^\dagger=UU^T =U^2 =\mb{1}
\ee

Because the change of the recoupling order can be implemented as a unitary transformation, we have for the matrix representation $Q$ of the volume operator of (\ref{Q definition}), after a transformation from the standard basis $\vec{a}(12)$ to a new basis $\vec{g}(IJ)$:
\be
   Q_{\vec{g}'\vec{g}}:=\big<~\vec{g}'~\big|~Q~\big|~\vec{g}~ \big>=
   \big<~\sum_{\vec{a}'}U_{\vec{g}'\vec{a'}}~\vec{a}'~\big|~Q~\big|~\sum_{\vec{a}}~U_{\vec{g}\vec{a}}~\vec{a}~ \big>=
   \sum_{\vec{a}'\vec{a}}U^\dagger_{\vec{g}'\vec{a'}}U_{\vec{g}\vec{a}}
   \big<~~\vec{a}'~\big|~Q~\big|~\vec{a}~ \big>
   =\sum_{\vec{a}'\vec{a}}U^\dagger_{\vec{g}'\vec{a'}}U_{\vec{g}\vec{a}}
   Q_{\vec{a}'\vec{a}}
\ee
The matrix $Q$ is being transformed as $Q\mapsto U^{-1}QU$ because $U^\dagger=U^{-1}$. Such a unitary transformation does not change the spectrum of $Q$, hence 
{\it the spectrum of the volume operator is independent of the chosen order of recoupling}.

We use this property to drastically decrease the number of assignments of
spins to edges we must use in the numerical analysis, because we can always choose a particular edge labelling $e_1,\ldots,e_N$ in which the spins are sorted 
\be\label{ordered spins}
   j_1\le j_2\le \ldots \le j_N=j_{max}
\ee
Such a labeling then corresponds to
\be\label{spin_degeneracy}
D(\vec{j})=\frac{N}{N_1!\cdot N_2!\cdot\ldots\cdot N_p!}
\ee
arbitrarily labeled spin assignments, where $N_i$ is number of
elements in one of $p$ sets of mutually identical spins, $\sum_{i=1}^p N_i =
N$.  Throughout the remainder of the paper we refer to such a sorted
assignment of spins to vertex edges as a \emph{spin configuration} (or $\vec j$-configuration).

\section{Explicit Matrix Elements of the Volume Operator}
\subsection{Starting Point for Matrix Element Implementation}

The simplified closed expression for (\ref{vorformel}), which has been derived in \cite{Volume Paper}, serves as a starting point for our computations.  Its derivation will not be reproduced here. However, in order to make it implementable on the computer, we will recall the crucial intermediate results of \cite{Volume Paper} and give the detailed outcome of the main formula for the matrix elements 
for all combinations of its arguments.

Let us first state the complete results on $F^{IJ}\big(\vec{a}(12),\vec{g}''(12)\big)$
as contained in (\ref{vorformel}). We find:

\ba   \begin{array}{lcl}
        \fbox{$I=1~~J=2$} && a_2(a_2+1)\displaystyle\prod_{n=2}^N \delta_{g''_n a_n} 
	\\
	\\
	\fbox{$I=1~~J=3$}&& \Bigg[\frac{1}{2}(-1)^{-j_1-j_2}(-1)^{j_3+1} X(j_1~j_3)^\frac{1}{2}A(g''_2~a_2)
	          \SixJ{j_2}{j_1}{g''_2}{1}{a_2}{j_1}
	(-1)^{a_3}\SixJ{a_3}{j_3}{g''_2}{1}{a_2}{j_3}
	~\text{~~~~~~~~~~~~~~~~~~~~~~~~~~}
	\\
	&&+C(j_1~j_3)~\delta_{g''_2 a_2}\Bigg]\displaystyle\prod_{k=3}^N \delta_{g''_k a_k}
	\nonumber
   \end{array}
\ea
\ba\label{Sum Terms I}
   \begin{array}{lcl}	
	\\
	\\
	\fbox{$I=1~~J>3$}&& \Bigg[\frac{1}{2}(-1)^{-j_1-j_2}(-1)^{j_J+1} X(j_1~j_J)^\frac{1}{2}A(g''_2~a_2)
	          \SixJ{j_2}{j_1}{g''_2}{1}{a_2}{j_1}~\times
		  \\&&\times
		  \displaystyle\prod_{n=3}^{J-1} A(g''_n a_n) (-1)^{-j_n+g''_{n-1}+a_{n-1}+1}
		  \SixJ{j_n}{g''_{n-1}}{g''_n}{1}{a_n}{a_{n-1}} \times
	(-1)^{a_J}\SixJ{a_J}{j_J}{g''_{J-1}}{1}{a_{J-1}}{j_J}\\
	&&+C(j_1~j_J)~\displaystyle\prod_{k=2}^{J-1}\delta_{g''_k a_k}\Bigg]\displaystyle\prod_{k=J}^N \delta_{g''_k a_k}
	\\
	\\
	\fbox{$I=2~~J=3$}&& \Bigg[\frac{1}{2}(-1)^{-j_1-j_2}(-1)^{j_3+1}(-1)^{a_2-g''_2} X(j_2~j_3)^\frac{1}{2}A(g''_2~a_2)
	          \SixJ{j_1}{j_2}{g''_2}{1}{a_2}{j_2}
	(-1)^{a_3}\SixJ{a_3}{j_3}{g''_2}{1}{a_2}{j_3}\\
	&&+C(j_2~j_3)~\delta_{g''_2 a_2}\Bigg]\displaystyle\prod_{k=3}^N \delta_{g''_k a_k}
	\\
	\\
	\fbox{$I=2~~J>3$}&& \Bigg[\frac{1}{2}(-1)^{-j_1-j_2}(-1)^{j_J+1}(-1)^{a_2-g''_2} X(j_2~j_J)^\frac{1}{2}A(g''_2~a_2)
	          \SixJ{j_1}{j_2}{g''_2}{1}{a_2}{j_2}~\times
		  \\&&\times
		  \displaystyle\prod_{n=3}^{J-1} A(g''_n a_n) (-1)^{-j_n+g''_{n-1}+a_{n-1}+1}
		  \SixJ{j_n}{g''_{n-1}}{g''_n}{1}{a_n}{a_{n-1}} \times
	(-1)^{a_J}\SixJ{a_J}{j_J}{g''_{J-1}}{1}{a_{J-1}}{j_J}\\
	&&+C(j_2~j_J)~\displaystyle\prod_{k=2}^{J-1}\delta_{g''_k a_k}\Bigg]\displaystyle\prod_{k=J}^N \delta_{g''_k a_k}
	\\
	\\
	\fbox{$I>2~~J=I+1$}&& \Bigg[\frac{1}{2}(-1)^{-2\sum\limits_{n=1}^{I-1}j_n}(-1)^{j_{I+1}-j_I}
	                           (-1)^{a_{I-1}+1}(-1)^{a_I-g''_I} 
	          X(j_I~j_{I+1})^\frac{1}{2}A(g''_I~a_I)~\times
	          \\&&\times~\SixJ{a_{I-1}}{j_I}{g''_I}{1}{a_I}{j_{I}}~\times
		 
	(-1)^{a_{I+1}}\SixJ{a_{I+1}}{j_{I+1}}{g''_{I}}{1}{a_{I}}{j_{I+1}}\\
	&&+C(j_I~j_{I+1})~\displaystyle\prod_{k=2}^{I}\delta_{g''_k a_k}\Bigg]\displaystyle\prod_{l=2}^{I-1} \delta_{g''_l a_l}\prod_{k=I+1}^N \delta_{g''_k a_k}
	\\
	\\
	\fbox{$I>2~~J>I+1$}&& \Bigg[\frac{1}{2}(-1)^{-2\sum\limits_{n=1}^{I-1}j_n}
	                      (-1)^{-\sum\limits_{n=I+1}^{J-1}j_n}
	                      (-1)^{j_J-j_I}(-1)^{a_{I-1}+1}(-1)^{a_I-g''_I} X(j_I~j_J)^\frac{1}{2}A(g''_I~a_I)
	          \SixJ{a_{I-1}}{j_I}{g''_I}{1}{a_I}{j_I}~\times
		  \\&&\times
		  \displaystyle\prod_{n=I+1}^{J-1} A(g''_n a_n) (-1)^{g''_{n-1}+a_{n-1}+1}
		  \SixJ{j_n}{g''_{n-1}}{g''_n}{1}{a_n}{a_{n-1}} \times
	(-1)^{a_J}\SixJ{a_J}{j_J}{g''_{J-1}}{1}{a_{J-1}}{j_J}\\
	&&+C(j_I~j_J)~\displaystyle\prod_{k=2}^{J-1}\delta_{g''_k a_k}\Bigg]\displaystyle\prod_{l=2}^{I-1} \delta_{g''_l a_l}\prod_{k=J}^N \delta_{g''_k a_k}
   \end{array}
\ea
\hrule
~\\~\\
Where we have used the shorthands 

\be\begin{array}{ll}

      C(a,b)=a(a+1)+b(b+1)&~~~~ X(a,b)=2a(2a+1)(2a+2)2b(2b+1)(2b+2) 
      \\\\ 
      A(a,b)=\sqrt{(2a+1)(2b+1)}
\end{array}\ee
\\[3mm]

The explicit expressions $F^{IJ}\big(\vec{a}(12),\vec{g}''(12)\big)$ of (\ref{Sum Terms I}) serve as the building blocks in order to calculate the matrix elements $\big<~\vec{a}(12)~\big|~\hat{q}_{IJK}~\big|~\vec{a}'(12)~\big>$ given in (\ref{General ME}).
One can evaluate (\ref{General ME}) as written down above  by using the appropriate expressions from (\ref{Sum Terms I}).
 
However there is still some simplification possible \cite{Volume Paper}.
When looking at (\ref{Sum Terms I}), obviously each expression comes as a sum of two terms. So multiplying two expressions $F^{IJ}(\cdot)\times F^{JK}(\cdot)$ as in the first term on the right hand side of (\ref{General ME}) would result in four terms. We will now show that by symmetry 
all product terms containing a symmetric $C(\cdot)$ factor  
vanish, as we antisymmetrize with respect to the interchange  $\left[\vec{a}(12) \rightleftharpoons \vec{a}'(12) \right]$.
For every term containing $C(\cdot)$ we effectively get a prefactor of 
          $\prod_{k=2}^N\delta_{g''_k a_k}$ or $\prod_{k=2}^N\delta_{g''_k a'_k}$ if we appropriately concatenate the products of $\delta$-factors in (\ref{Sum Terms I}).
Therefore the product terms $C(j_I,j_J)~\times~C(j_J,j_K)$ are obviously symmetric under $\left[\vec{a}(12) \rightleftharpoons \vec{a}'(12) \right]$ and vanish under antisymmetrization.  \\        

Let us discuss the $\delta$-factors in (\ref{Sum Terms I}) in order to exclude mixed terms containing one $C(\cdot)$. We will present here only the discussion for products containing $C(j_J, j_K)$ in $F^{JK}\big(\vec{a}'(12),\vec{g}''(12) \big)$. The discussion of $C(j_I, j_J)$ in $ F^{IJ}\big(\vec{a}(12),\vec{g}''(12) \big)$ is completely analogous.

For every term containing $C(j_J, j_K)$ we effectively get a prefactor of 
          $\prod_{k=2}^N\delta_{g''_k a'_k}$ if we concatenate the products of $\delta$-factors.
Therefore when considering a product of the form (\ref{General ME}), with a non-$C(j_I,j_J)$-term contributed by $F^{IJ}(\cdot)$ and a $C(j_J,j_K)$-term coming from $F^{JK}(\cdot)$, every $g''_k$ in the $F^{IJ}(\cdot)$-term can be replaced by $a'_k$. 
That is \fbox{$\vec{g}''(12)=\vec{a}'(12)$}.
But then the mixed terms containing one $C(j_J,j_K)$ become symmetric wrt.\ $\left[\vec{a}(12) \rightleftharpoons \vec{a}'(12) \right]$ for the following reasons:
\begin{itemize}
	  \item{$(-1)^{a_2-a'_2}=(-1)^{a'_2-a_2}$ because $(a'_2-a_2)\in \mb{Z}$}
	  \item{$A(a'_k, a_k)$  is symmetric by construction}
	  \item{all (but 2) $6j$-symbols occurring in definition (\ref{Sum Terms I}) for $F^{IJ}(\cdot)$ are of the form \newline  $~~~~~~~~~~~~~~~~~~\SixJ{j_l}{g''_{n-1}}{g''_n}{1}{a_n}{a_{n-1}}
	                                      =\SixJ{j_l}{a_n}{a_{n-1}}{1}{g''_{n-1}}{g''_n}$\\
by the symmetry properties\footnote{$6j$-symbols are invariant under the interchange of 2 columns and the simultaneous flip of two columns, see appendix \ref{Symmetry Properties}.}  
of the $6j$-symbols. But $\vec{g}''(12)=\vec{a}'(12)$.}
	  \item{
due to the $\displaystyle\prod_{k=2}^{I-1}\delta_{g''_k a_k}\prod_{n=J}^{N}\delta_{g''_n a_n}$ in the non-$C(j_I,j_J)$ terms of $F^{IJ}(\cdot)$:
	  \newline	  
$~~~~~~~~~~$ the remaining $6j$-symbols $\SixJ{a_{I-1}}{a_I}{j_I}{1}{j_I}{g''_I}$ and 
	  $(-1)^{a_J}\SixJ{a_J}{j_J}{g''_{J-1}}{1}{a_{J-1}}{j_J}$ 
	  are then symmetric\newline $~~~~~~~~~~$ wrt.\ 
	  $\vec{a}(12) \rightleftharpoons \vec{a}'(12)$, because $\vec{g}''(12)=\vec{a}'(12)$ 
	  \newline $~~~~~~~~~~$and hence,
 $g''_{J-1}=a'_{J-1}$~,~~ $a_{I-1}=g''_{I-1}=a'_{I-1}$~,~~ $g''_I=a'_I,~~a_J=g''_J=a'_J$ . 
	  }
\end{itemize}
Thus when calculating (\ref{General ME}) using (\ref{Sum Terms I}) we are always left with the single product not containing a symmetric $C(j_I,j_J)$ or $C(j_J,j_K)$-part. Note, however, that for cases containing \fbox{$I=1 \; J=2$} in (\ref{Sum Terms I}) we get a contribution.
\\ 

Let us finally take a closer look at the $\delta$-products in the remaining non-symmetric product terms of (\ref{General ME}). In these products we have for $I<J<K$:
\be
  \left[\prod_{k=2}^{I-1}\delta_{g''_k a_k}\prod_{l=J}^N\delta_{g''_l a_l}\right]
  \left[\prod_{m=2}^{J-1}\delta_{g''_m a'_m}\prod_{n=K}^N\delta_{g''_n a'n}\right]
  =
  \prod_{k=2}^{I-1}\delta_{a_k a'_k}\prod_{m=I}^{J-1}\delta_{g''_m a'_m}
  \prod_{l=J}^{K-1}\delta_{g''_l a_l}\prod_{n=K}^N\delta_{a_n a'n}
\ee
We have thus to look at the symmetries between $I\ldots K-1$ only. Here we find:\vspace{-4mm}
\begin{itemize}
   \item[]{\be\label{delta identities}\begin{array}{llllll}
           \mbox{For the first term in (\ref{General ME})}&~~~~&\vec{a}(12)&\vec{a}'(12)&~~~~& 
	   \displaystyle\prod_{m=I}^{J-1}\delta_{g''_m a'_m}
  \prod_{l=J}^{K-1}\delta_{g''_l a_l}
  \\
  \mbox{and the second term in (\ref{General ME})} &~~~~&\vec{a}'(12)&\vec{a}(12)&~~~~& 
	   \displaystyle\prod_{m=I}^{J-1}\delta_{g''_m a_m}
  \prod_{l=J}^{K-1}\delta_{g''_l a'_l}
  \\[-3mm]
  \mbox{ ( $\vec{a}(12)\rightarrow \vec{a}'(12),
                            ~~\vec{a}'(12)\rightarrow \vec{a}(12)$ )}
     \end{array}\ee} 
\end{itemize}
Using again symmetry properties of the $6j$-symbols, and integer-arguments for the $(-1)$-exponents, together with (\ref{delta identities}), we can show, as in the case of symmetric terms, that almost all terms are symmetric with respect to the interchange $\vec{a}(12)\leftrightharpoons \vec{a}'(12)$.
			    
So we are left with the discussion of the following terms: 
One easily sees that for a product $F^{IJ}(\cdot)\times F^{JK}(\cdot)$ (of two terms from (\ref{Sum Terms I}) using the according arguments given in (\ref{General ME}) ) one has:

\be
   (-1)^{a_{I-1}}(-1)^{a_I-a_I'}(-1)^{a_J}\SixJ{a_J}{j_J}{a'_{J-1}}{1}{a_{J-1}}{j_J}
   ~\times~
   (-1)^{a'_{J-1}}(-1)^{a'_J-a_J}\SixJ{a'_{J-1}}{a_J}{j_J}{1}{j_J}{a'_J}
\ee
Here $a_{I-1}=a'_{I-1}$ by (\ref{delta identities}),  $(a_I-a'_I)\in\mb{Z}$, and the $a_J$ exponents cancel. Thus the remaining  part of the product terms in (\ref{General ME}) not symmetric wrt.\ the interchange $\vec{a}(12)\rightarrow \vec{a}'(12),~~\vec{a}'(12)\rightarrow \vec{a}(12)$
is given by
\be
   (-1)^{a'_J}(-1)^{a'_{J-1}}
   \SixJ{a_J}{j_J}{a'_{J-1}}{1}{a_{J-1}}{j_J}
   ~
   \SixJ{a'_{J-1}}{a_J}{j_J}{1}{j_J}{a'_J}
\ee
We will give the explicit matrix element expressions (\ref{General ME}) in the following section.
\subsection{\label{Explicit Matrix Element Expressions}Explicit Matrix Element Expressions}
The discussion above leads us to consider the following 14 special cases for the matrix element $\big<~\vec{a}(12)~\big|~\hat{q}_{IJK}~\big|~\vec{a}'(12)~\big>$:
\be\begin{array}{lll}
	I=1\hspace{5mm}&J=2\hspace{10mm}&K=3
	\\[1mm]
	   &   &K>3
	\\[3mm]
	I=1&J=3&K=J+1
	\\[1mm]
	   &   &K>J+1
	\\[3mm]
	I=1&J>3&K=J+1
	\\[1mm]
	   &   &K>J+1
	\\[5mm]      
	I=2&J=3&K=4
	\\[1mm]
	   &   &K>4
	\\[3mm]
	I=2&J>3&K=J+1
	\\[1mm]
	   &   &K>J+1
	\\[3mm]
	I>2&J=I+1&K=J+1
	\\[1mm]
	   &   &K>J+1
	\\[3mm]
	I>2&J>I+1&K=J+1
	\\[1mm]
	   &   &K>J+1
\end{array}\ee
Using combinations of the according cases (\ref{Sum Terms I}) as written down in (\ref{General ME}) one obtains,\footnote{For completeness all Kronecker-$\delta$'s at the end of the following equations range from $K\ldots N$, however if we consider gauge invariant states, besides  $a'_N=a_N=J\stackrel{!}{=}0$, we automatically have $a'_{N-1}=a_{N-1}\stackrel{!}{=}j_N$ in every case.} using the shorthand \linebreak $\mb{q}_{IJK}:=\big<~\vec{a}(12)~\big|~\hat{q}_{IJK}~\big|~\vec{a}'(12)~\big>$:
\subsubsection[$I=1~~~~J=2~~~~K=3$]{\fbox{$I=1~~~~J=2~~~~K=3$}}
\be\label{q123}\begin{array}{lcll}\displaystyle
    \mb{q}_{123}\hspace{1cm}&=&\displaystyle\Big[a_2(a_2+1)-a_2'(a_2'+1) \Big]\frac{1}{2}(-1)^{-j_1-j_2}(-1)^{j_3+1}(-1)^{a_2'-a_2}(-1)^{a_3'}X(j_2~j_3)^\frac{1}{2}A(a_2'~a_2)~
    \times~\\
    &&\displaystyle\times \SixJ{j_1}{j_2}{a_2}{1}{a_2'}{j_2}\SixJ{a_3'}{j_3}{a_2}{1}{a_2'}{j_3} \times\prod_{n=3}^N \delta_{a'_n~a_n} 
\end{array}\ee

\subsubsection[$I=1~~~~J=2~~~~K>3$]{\fbox{$I=1~~~~J=2~~~~K>3$}}
\be\begin{array}{lcll}
    \mb{q}_{12K}\hspace{1cm}&=&
    \Big[a_2(a_2+1)-a'_2(a'_2+1)\Big] \displaystyle~\frac{1}{2}~(-1)^{-j_1-j_2}(-1)^{j_K+1}(-1)^{a'_2-a_2}~
    X(j_2~j_K)^\frac{1}{2}~A(a_2~a'_2)\times
    \SixJ{j_1}{j_2}{a_2}{1}{a'_2}{j_2}
    \\ [2mm]
    &&\hspace{1cm}\times\displaystyle\prod_{n=3}^{K-1} A(a_n~a'_n)~(-1)^{-j_n+a_{n-1}+a'_{n-1}+1}
       \SixJ{j_n}{a_{n-1}}{a_n}{1}{a'_n}{a'_{n-1}}~\times~
       (-1)^{a'_K}~\SixJ{a'_K}{j_K}{a_{K-1}}{1}{a'_{K-1}}{j_K}
    \\ [2mm]
    &&\hspace{1cm}\times\displaystyle\prod_{n=K}^{N}\delta_{a'_n~a_n}       
\end{array}\ee

\subsubsection[$I=1~~~~J=3~~~~K=J+1$]{\fbox{$I=1~~~~J=3~~~~K=J+1$}}
\be\begin{array}{lcll}
    \mb{q}_{134}\hspace{1cm}&=&
    \Big[\displaystyle\frac{1}{4}~(-1)^{j_1+j_2}(-1)^{j_4}(-1)^{a'_4}~X(j_1~j_3)^\frac{1}{2} X(j_3~j_4)^\frac{1}{2}~
        A(a'_2~a_2)~A(a'_3~a_3) 
	\times
	\SixJ{j_2}{j_1}{a'_2}{1}{a_2}{j_1}
	\SixJ{a'_4}{j_4}{a'_3}{1}{a_3}{j_4}\Big]
    \\[4mm]
    \hspace{1cm}&&
    \times\Bigg[(-1)^{a'_2}(-1)^{a'_3}
               \SixJ{a_3}{j_3}{a'_2}{1}{a_2}{j_3}\SixJ{a'_2}{j_3}{a_3}{1}{a'_3}{j_3}
	      ~ -~
	       (-1)^{a_2}(-1)^{a_3}
               \SixJ{a'_3}{j_3}{a_2}{1}{a'_2}{j_3}\SixJ{a_2}{j_3}{a'_3}{1}{a_3}{j_3}
	       \Bigg]
    \\[4mm]
    \hspace{1cm}&&
    \displaystyle\times\prod_{n=4}^{N}\delta_{a'_n~a_n}
\end{array}\ee

\subsubsection[$I=1~~~~J=3~~~~K>J+1$]{\fbox{$I=1~~~~J=3~~~~K>J+1$}}
\be\begin{array}{lcll}
    \mb{q}_{13K}\hspace{1cm}&=&
    \Big[\displaystyle\frac{1}{4}~(-1)^{j_1+j_2}(-1)^{-\sum\limits_{n=4}^{K-1}j_n}~(-1)^{j_K}~(-1)^{a'_K}~X(j_1~j_3)^\frac{1}{2} X(j_3~j_K)^\frac{1}{2}~
        A(a'_2~a_2)~A(a'_3~a_3) \Big]
    \\[4mm]
    \hspace{1cm}&&
    \times\displaystyle\SixJ{j_2}{j_1}{a'_2}{1}{a_2}{j_1}\times
    \prod_{n=4}^{K-1}A(a_n~a'_n)(-1)^{a_{n-1}+a'_{n-1}+1}
           \SixJ{j_n}{a_{n-1}}{a_n}{1}{a'_n}{a'_{n-1}}
	   ~\times~
	   \SixJ{a'_K}{j_K}{a_{K-1}}{1}{a'_{K-1}}{j_K}
    \\[4mm]
    \hspace{1cm}&&
    \times\Bigg[(-1)^{a'_2}(-1)^{a'_3}
               \SixJ{a_3}{j_3}{a'_2}{1}{a_2}{j_3}\SixJ{a'_2}{j_3}{a_3}{1}{a'_3}{j_3}
	      ~ -~
	       (-1)^{a_2}(-1)^{a_3}
               \SixJ{a'_3}{j_3}{a_2}{1}{a'_2}{j_3}\SixJ{a_2}{j_3}{a'_3}{1}{a_3}{j_3}
	       \Bigg]
    \\[4mm]
    \hspace{1cm}&&
    \displaystyle\times\prod_{n=K}^{N}\delta_{a'_n~a_n}
\end{array}\ee

\subsubsection[$I=1~~~~J>3~~~~K=J+1$]{\fbox{$I=1~~~~J>3~~~~K=J+1$}}
\be\begin{array}{lcll}
    \lefteqn{\mb{q}_{1J\,J+1}\hspace{0cm}=
    \Bigg[\displaystyle\frac{1}{4}~(-1)^{\sum\limits_{n=1}^{J-1}j_n}~
    (-1)^{j_{J+1}}(-1)^{a'_{J+1}}~X(j_1~j_J)^\frac{1}{2} X(j_J~j_{J+1})^{\frac{1}{2}}
    A(a'_2~a_2)A(a_J~a'_J)}
    \\[4mm]
    &&\displaystyle\times
    \SixJ{j_2}{j_1}{a'_2}{1}{a_2}{j_1} \times\prod_{n=3}^{J-1}A(a'_n~a_n)(-1)^{a'_{n-1}+a_{n-1}+1}
    \SixJ{j_n}{a'_{n-1}}{a'_n}{1}{a_n}{a_{n-1}}~\times~\SixJ{a'_{J+1}}{j_{J+1}}{a_J}{1}{a'_J}{j_{J+1}}
    \Bigg]
    \\[4mm]
    &&\displaystyle\times\Bigg[
    (-1)^{a'_{J-1}}(-1)^{a'_J}\SixJ{a_J}{j_J}{a'_{J-1}}{1}{a_{J-1}}{j_J}
                              \SixJ{a'_{J-1}}{j_J}{a_J}{1}{a'_J}{j_J}
    ~-~
    (-1)^{a_{J-1}}(-1)^{a_J}\SixJ{a'_J}{j_J}{a_{J-1}}{1}{a'_{J-1}}{j_J}
                            \SixJ{a_{J-1}}{j_J}{a'_J}{1}{a_J}{j_J}
    \Bigg]			    
    \\[4mm]
    &&\displaystyle\times\prod_{n=J+1}^N \delta_{a'_n~a_n}
\end{array}\ee

\subsubsection[$I=1~~~~J>3~~~~K>J+1$]{\fbox{$I=1~~~~J>3~~~~K>J+1$}}
\be\begin{array}{lcll}
     \lefteqn{\mb{q}_{1JK}\hspace{0cm}=
    \Bigg[\displaystyle\frac{1}{4}~
          (-1)^{\sum\limits_{n=1}^{J-1}j_n}
          (-1)^{-\sum\limits_{n=J+1}^{K-1}j_n}~
          (-1)^{j_K}
	  (-1)^{a'_K}~
	  X(j_1~j_J)^\frac{1}{2} X(j_J~j_K)^{\frac{1}{2}}
          A(a'_2~a_2)A(a_J~a'_J)}
    \\[4mm]
    &&\displaystyle\times
    \SixJ{j_2}{j_1}{a'_2}{1}{a_2}{j_1} \times\prod_{n=3}^{J-1}A(a'_n~a_n)(-1)^{a'_{n-1}+a_{n-1}+1}
    \SixJ{j_n}{a'_{n-1}}{a'_n}{1}{a_n}{a_{n-1}}~
    \\[4mm]
    
    &&\hspace{2.4cm}\displaystyle\times\prod_{n=J+1}^{K-1}A(a'_n~a_n)(-1)^{a'_{n-1}+a_{n-1}+1}
    \SixJ{j_n}{a'_{n-1}}{a'_n}{1}{a_n}{a_{n-1}}~\times~\SixJ{a'_K}{j_K}{a_{K-1}}{1}{a'_{K-1}}{j_K}
    \Bigg]
    \\[4mm]
    &&\displaystyle\times\Bigg[
    (-1)^{a'_{J-1}}(-1)^{a'_J}\SixJ{a_J}{j_J}{a'_{J-1}}{1}{a_{J-1}}{j_J}
                              \SixJ{a'_{J-1}}{j_J}{a_J}{1}{a'_J}{j_J}
    ~-~
    (-1)^{a_{J-1}}(-1)^{a_J}\SixJ{a'_J}{j_J}{a_{J-1}}{1}{a'_{J-1}}{j_J}
                            \SixJ{a_{J-1}}{j_J}{a'_J}{1}{a_J}{j_J}
    \Bigg]			    
    \\[4mm]
    &&\displaystyle\times\prod_{n=K}^N \delta_{a'_n~a_n}
\end{array}\ee

\subsubsection[$I=2~~~~J=3~~~~K=4$]{\fbox{$I=2~~~~J=3~~~~K=4$}}
\be\begin{array}{lcll}
     \mb{q}_{234}\hspace{0cm}&=&\displaystyle
    \Bigg[\frac{1}{4} (-1)^{j_1+j_2+j_4}(-1)^{a'_4}
                      X(j_2~j_3)^\frac{1}{2} X(j_3~j_4)^\frac{1}{2}
                      A(a'_2~a_2) A(a_3~a'_3)
		      \SixJ{j_1}{j_2}{a'_2}{1}{a_2}{j_2}
		      \SixJ{a'_4}{j_4}{a_3}{1}{a'_3}{j_4} ~\Bigg]
    \\[4mm]
    &&\displaystyle\times\Bigg[
      (-1)^{a_2}(-1)^{a'_3}\SixJ{a_3}{j_3}{a'_2}{1}{a_2}{j_3}\SixJ{a'_2}{j_3}{a_3}{1}{a'_3}{j_3}
      -
      (-1)^{a'_2}(-1)^{a_3}\SixJ{a'_3}{j_3}{a'_2}{1}{a_2}{j_3}\SixJ{a_2}{j_3}{a_3}{1}{a'_3}{j_3}
      \Bigg]
    \\[4mm]
    &&\displaystyle\times\prod_{n=4}^N \delta_{a'_n~a_n}

\end{array}\ee

\subsubsection[$I=2~~~~J=3~~~~K>4$]{\fbox{$I=2~~~~J=3~~~~K>4$}}
\be\begin{array}{lcll}
    \mb{q}_{23K}\hspace{0cm}&=&\displaystyle
    \Bigg[\frac{1}{4} (-1)^{j_1+j_2}(-1)^{-\sum\limits_{n=4}^{K-1}j_n}
                      (-1)^{j_K}(-1)^{a'_K}
                      X(j_2~j_3)^\frac{1}{2} X(j_3~j_K)^\frac{1}{2}
                      A(a'_2~a_2) A(a_3~a'_3)
    \\[4mm]
    &&\displaystyle\times
		      \SixJ{j_1}{j_2}{a'_2}{1}{a_2}{j_2}
		   \times
		   \prod_{n=4}^{K-1}A(a_n~a'_n)(-1)^{a_{n-1}+a'_{n-1}+1}
		   \SixJ{j_n}{a_{n-1}}{a_n}{1}{a'_n}{a'_{n-1}}
		   \times   
		   \SixJ{a'_K}{j_K}{a_{K-1}}{1}{a'_{K-1}}{j_K} ~\Bigg]
    \\[4mm]
    &&\displaystyle\times\Bigg[
      (-1)^{a_2}(-1)^{a'_3}\SixJ{a_3}{j_3}{a'_2}{1}{a_2}{j_3}\SixJ{a'_2}{j_3}{a_3}{1}{a'_3}{j_3}
      -
      (-1)^{a'_2}(-1)^{a_3}\SixJ{a'_3}{j_3}{a'_2}{1}{a_2}{j_3}\SixJ{a_2}{j_3}{a_3}{1}{a'_3}{j_3}
      \Bigg]
    \\[4mm]
    &&\displaystyle\times\prod_{n=K}^N \delta_{a'_n~a_n}

\end{array}\ee

\subsubsection[$I=2~~~~J>3~~~~K=J+1$]{\fbox{$I=2~~~~J>3~~~~K=J+1$}}
\be\begin{array}{lcll}
    \lefteqn{\mb{q}_{\,2\,J\,J+1}\hspace{0cm}=\displaystyle
    \Bigg[\frac{1}{4} (-1)^{\sum\limits_{n=1}^{J-1} j_n}
                      (-1)^{a_2-a'_2}
                      (-1)^{j_{J+1}}(-1)^{a'_{J+1}}
                      X(j_2~j_J)^\frac{1}{2} X(j_J~j_{J+1})^\frac{1}{2}
                      A(a'_2~a_2) A(a_J~a'_J)}
    \\[4mm]
    &&\displaystyle\times
		      \SixJ{j_1}{j_2}{a'_2}{1}{a_2}{j_2}
		   \times
		   \prod_{n=3}^{J-1}A(a_n~a'_n)(-1)^{a_{n-1}+a'_{n-1}+1}
		   \SixJ{j_n}{a_{n-1}}{a_n}{1}{a'_n}{a'_{n-1}}
		   \times   
		   \SixJ{a'_{J+1}}{j_{J+1}}{a_J}{1}{a'_J}{j_{J+1}} ~\Bigg]
    \\[4mm]
    &&\displaystyle\times\Bigg[
      (-1)^{a'_J}(-1)^{a'_{J-1}}
      \SixJ{a_J}{j_J}{a'_{J-1}}{1}{a_{J-1}}{j_J}
      \SixJ{a'_{J-1}}{j_J}{a_J}{1}{a'_J}{j_J}
      -
      (-1)^{a_J}(-1)^{a_{J-1}}
      \SixJ{a'_J}{j_J}{a_{J-1}}{1}{a'_{J-1}}{j_J}
      \SixJ{a_{J-1}}{j_J}{a'_J}{1}{a_J}{j_J}
      \Bigg]
    \\[4mm]
    &&\displaystyle\times\prod_{n=J+1}^N \delta_{a'_n~a_n}

\end{array}\ee

\subsubsection[$I=2~~~~J>3~~~~K>J+1$]{\fbox{$I=2~~~~J>3~~~~K>J+1$}}
\be\begin{array}{lcll}
    \lefteqn{\mb{q}_{\,2\,J\,K}\hspace{0cm}=\displaystyle
    \Bigg[\frac{1}{4} (-1)^{\sum\limits_{n=1}^{J-1} j_n}
                      (-1)^{-\sum\limits_{n=J+1}^{K-1}j_n}
                      (-1)^{a_2-a'_2}
                      (-1)^{j_K}(-1)^{a'_K}
                      X(j_2~j_J)^\frac{1}{2} X(j_J~j_K)^\frac{1}{2}
                      A(a'_2~a_2) A(a_J~a'_J)}
    \\[4mm]
    &&\displaystyle\times
		      \SixJ{j_1}{j_2}{a'_2}{1}{a_2}{j_2}
		   \times
		   \prod_{n=3}^{J-1}A(a_n~a'_n)(-1)^{a_{n-1}+a'_{n-1}+1}
		   \SixJ{j_n}{a_{n-1}}{a_n}{1}{a'_n}{a'_{n-1}}
    \\[4mm]
    &&\displaystyle\hspace{2.4cm}\times
		   \prod_{n=J+1}^{K-1}A(a_n~a'_n)(-1)^{a_{n-1}+a'_{n-1}+1}
		   \SixJ{j_n}{a_{n-1}}{a_n}{1}{a'_n}{a'_{n-1}}
		   \times   
		   \SixJ{a'_K}{j_K}{a_{K-1}}{1}{a'_{K-1}}{j_K} ~\Bigg]
    \\[4mm]
    &&\displaystyle\times\Bigg[
      (-1)^{a'_J}(-1)^{a'_{J-1}}
      \SixJ{a_J}{j_J}{a'_{J-1}}{1}{a_{J-1}}{j_J}
      \SixJ{a'_{J-1}}{j_J}{a_J}{1}{a'_J}{j_J}
      -
      (-1)^{a_J}(-1)^{a_{J-1}}
      \SixJ{a'_J}{j_J}{a_{J-1}}{1}{a'_{J-1}}{j_J}
      \SixJ{a_{J-1}}{j_J}{a'_J}{1}{a_J}{j_J}
      \Bigg]
    \\[4mm]
    &&\displaystyle\times\prod_{n=K}^N \delta_{a'_n~a_n}

\end{array}\ee

\subsubsection[$I>2~~~~J=I+1~~~~K=J+1$]{\fbox{$I>2~~~~J=I+1~~~~K=J+1$}}
\be\begin{array}{lcll}
    \lefteqn{\mb{q}_{\,I\,I+1\,I+2}\hspace{0cm}=\displaystyle
    \Bigg[\frac{1}{4} (-1)^{j_I+j_{I+2}}
                      (-1)^{a_{I-1}}
                      (-1)^{a'_{I+2}}
                      X(j_I~j_{I+1})^\frac{1}{2} X(j_{I+1}~j_{I+2})^\frac{1}{2}
                      A(a'_I~a_I) A(a_{I+1}~a'_{I+1})}
    \\[4mm]
    &&\displaystyle
    \times\SixJ{a_{I-1}}{j_I}{a'_I}{1}{a_I}{j_I}
    \times\SixJ{a'_{I+2}}{j_{I+2}}{a_{I+1}}{1}{a'_{I+1}}{j_{I+2}} \Bigg]
    \\[4mm]
    &&\displaystyle\times\Bigg[
      (-1)^{a_I}(-1)^{a'_{I+1}}
      \SixJ{a_{I+1}}{j_{I+1}}{a'_I}{1}{a_I}{j_{I+1}}
      \SixJ{a'_I}{j_{I+1}}{a_{I+1}}{1}{a'_{I+1}}{j_{I+1}}
      -
      (-1)^{a'_I}(-1)^{a_{I+1}}
      \SixJ{a'_{I+1}}{j_{I+1}}{a_I}{1}{a'_I}{j_{I+1}}
      \SixJ{a_I}{j_{I+1}}{a_{I+1}}{1}{a'_{I+1}}{j_{I+1}}
      \Bigg]
    \\[4mm]
    &&\displaystyle\times\prod_{n=2}^{I-1} \delta_{a'_n~a_n} \prod_{n=I+2}^N \delta_{a'_n~a_n}

\end{array}\ee

\subsubsection[$I>2~~~~J=I+1~~~~K>J+1$]{\fbox{$I>2~~~~J=I+1~~~~K>J+1$}}
\be\begin{array}{lcll}
     \lefteqn{\mb{q}_{\,I\,I+1\,K}\hspace{0cm}=\displaystyle
    \Bigg[\frac{1}{4} (-1)^{j_I}(-1)^{-\sum\limits_{n=I+2}^{K-1}j_n}
                      (-1)^{a_{I-1}}
		      (-1)^{j_K}
                      (-1)^{a'_K}
                      X(j_I~j_{I+1})^\frac{1}{2} X(j_{I+1}~j_K)^\frac{1}{2}
                      A(a'_I~a_I) A(a_{I+1}~a'_{I+1})}
    \\[4mm]
    &&\displaystyle
    \times\SixJ{a_{I-1}}{j_I}{a'_I}{1}{a_I}{j_I}
    \times
    \prod_{n=I+2}^{K-1}A(a'_n~a_n)(-1)^{a_{n-1}+a'_{n-1}+1}\SixJ{j_n}{a_{n-1}}{a_n}{1}{a'_n}{a'_{n-1}}
    \times
    \SixJ{a'_K}{j_K}{a_{K-1}}{1}{a'_{K-1}}{j_K}  \Bigg]
    \\[4mm]
    &&\displaystyle\times\Bigg[
      (-1)^{a_I}(-1)^{a'_{I+1}}
      \SixJ{a_{I+1}}{j_{I+1}}{a'_I}{1}{a_I}{j_{I+1}}
      \SixJ{a'_I}{j_{I+1}}{a_{I+1}}{1}{a'_{I+1}}{j_{I+1}}
      -
      (-1)^{a'_I}(-1)^{a_{I+1}}
      \SixJ{a'_{I+1}}{j_{I+1}}{a_I}{1}{a'_I}{j_{I+1}}
      \SixJ{a_I}{j_{I+1}}{a_{I+1}}{1}{a'_{I+1}}{j_{I+1}}
      \Bigg]
    \\[4mm]
    &&\displaystyle\times\prod_{n=2}^{I-1} \delta_{a'_n~a_n} \prod_{n=K}^N \delta_{a'_n~a_n}

\end{array}\ee

\subsubsection[$I>2~~~~J>I+1~~~~K=J+1$]{\fbox{$I>2~~~~J>I+1~~~~K=J+1$}}
\be\begin{array}{lcll}
    \lefteqn{\mb{q}_{\,I\,J\,J+1}\hspace{0cm}=\displaystyle
    \Bigg[\frac{1}{4} (-1)^{\sum\limits_{n=I}^{J-1}j_n}
                      (-1)^{a_{I-1}}
                      (-1)^{a_I-a'_I}
		      (-1)^{j_{J+1}}
		      (-1)^{a'_{J+1}}
                      X(j_I~j_J)^\frac{1}{2} X(j_J~j_{J+1})^\frac{1}{2}
                      A(a'_I~a_I) A(a_J~a'_J)}
    \\[4mm]
    &&\displaystyle
    \times\SixJ{a_{I-1}}{j_I}{a'_I}{1}{a_I}{j_I}
    \times
    \prod_{n=I+1}^{J-1}A(a'_n~a_n)(-1)^{a_{n-1}+a'_{n-1}+1}\SixJ{j_n}{a_{n-1}}{a_n}{1}{a'_n}{a'_{n-1}}
    \times
    \SixJ{a'_{J+1}}{j_{J+1}}{a_J}{1}{a'_J}{j_{J+1}}  \Bigg]
    \\[4mm]
    &&\displaystyle\times\Bigg[
      (-1)^{a'_{J-1}}(-1)^{a'_J}
      \SixJ{a_J}{j_J}{a'_{J-1}}{1}{a_{J-1}}{j_J}
      \SixJ{a'_{J-1}}{j_J}{a_J}{1}{a'_J}{j_J}
      -
      (-1)^{a_{J-1}}(-1)^{a_J}
      \SixJ{a'_J}{j_J}{a_{J-1}}{1}{a'_{J-1}}{j_J}
      \SixJ{a_{J-1}}{j_J}{a'_J}{1}{a_J}{j_J}
      \Bigg]
    \\[4mm]
    &&\displaystyle\times\prod_{n=2}^{I-1} \delta_{a'_n~a_n} \prod_{n=J+1}^N \delta_{a'_n~a_n}

\end{array}\ee

\subsubsection[$I>2~~~~J>I+1~~~~K>J+1$]{\fbox{$I>2~~~~J>I+1~~~~K>J+1$}}
\be\label{end_subQs}\begin{array}{lcll}
    \lefteqn{\mb{q}_{\,I\,J\,K}\hspace{0cm}=\displaystyle
    \Bigg[\frac{1}{4} (-1)^{\sum\limits_{n=I}^{J-1}j_n}
                      (-1)^{-\sum\limits_{n=J+1}^{K-1}j_n}
                      (-1)^{a_{I-1}}
                      (-1)^{a_I-a'_I}
		      (-1)^{j_K}
		      (-1)^{a'_K}
                      X(j_I~j_J)^\frac{1}{2} X(j_J~j_K)^\frac{1}{2}
                      A(a'_I~a_I) A(a_J~a'_J)}
    \\[4mm]
    &&\displaystyle
    \times\SixJ{a_{I-1}}{j_I}{a'_I}{1}{a_I}{j_I}
    \times
    \prod_{n=I+1}^{J-1}A(a'_n~a_n)(-1)^{a_{n-1}+a'_{n-1}+1}\SixJ{j_n}{a_{n-1}}{a_n}{1}{a'_n}{a'_{n-1}}
    \\[4mm]
    &&\displaystyle
    \hspace{3cm}\times
    \prod_{n=J+1}^{K-1}A(a'_n~a_n)(-1)^{a_{n-1}+a'_{n-1}+1}\SixJ{j_n}{a_{n-1}}{a_n}{1}{a'_n}{a'_{n-1}}
    \times
    \SixJ{a'_K}{j_K}{a_{K-1}}{1}{a'_{K-1}}{j_K}  \Bigg]
    \\[4mm]
    &&\displaystyle\times\Bigg[
      (-1)^{a'_{J-1}}(-1)^{a'_J}
      \SixJ{a_J}{j_J}{a'_{J-1}}{1}{a_{J-1}}{j_J}
      \SixJ{a'_{J-1}}{j_J}{a_J}{1}{a'_J}{j_J}
      -
      (-1)^{a_{J-1}}(-1)^{a_J}
      \SixJ{a'_J}{j_J}{a_{J-1}}{1}{a'_{J-1}}{j_J}
      \SixJ{a_{J-1}}{j_J}{a'_J}{1}{a_J}{j_J}
      \Bigg]
    \\[4mm]
    &&\displaystyle\times\prod_{n=2}^{I-1} \delta_{a'_n~a_n} \prod_{n=K}^N \delta_{a'_n~a_n}

\end{array}\ee

\subsection{The Arguments: Special $6j$-Symbols}
In all cases discussed, very special $6j$-symbols occur: All of them contain a 1 as an argument. $6j$-symbols of this kind possess short closed expressions and they restrict the arguments allowed (see \cite{Edmonds}).
\subsubsection{\label{6j symbols in V}Different Cases and their Classification}
The following four  types of $6j$-symbols occur in (\ref{q123}),$\ldots$,(\ref{end_subQs}):
\[\begin{array}{lllllll}
   &~~~&
   I[b,c]&:=&\SixJ{a}{b}{c}{1}{c-1}{b-1}
   &=&\displaystyle
   (-1)^{a+b+c}\Bigg[\frac{(a+b+c)\,(a+b+c+1)\,(-a+b+c-1)\,(-a+b+c)}
                             {(2b-1)\,2b\,(2b+1)\,(2c-1)\,2c\,(2c+1)}\Bigg]^\frac{1}{2}
   \\\nonumber\\
   &~~~&
   II[b,c]&:=&
   \SixJ{a}{b}{c}{1}{c-1}{b}
   &=&\displaystyle
   (-1)^{a+b+c}\Bigg[\frac{2\,(a+b+c+1)\,(-a+b+c)\,(a-b+c)\,(a+b-c+1)}
                             {2b\,(2b+1)\,(2b+2)\,(2c-1)\,2c\,(2c+1)}\Bigg]^\frac{1}{2}
   \\\nonumber\\
    &~~~&
   III[b,c]&:=&
   \SixJ{a}{b}{c}{1}{c-1}{b+1}
   &=&\displaystyle(-1)^{a+b+c}\Bigg[\frac{(a-b+c-1)\,(a-b+c)\,(a+b-c+1)\,(a+b-c+2)}
                             {(2b+1)\,(2b+2)\,(2b+3)\,(2c-1)\,2c\,(2c+1)}\Bigg]^\frac{1}{2}
   \\\nonumber\\
    &~~~&
   IV[b,c]&:=&
   \SixJ{a}{b}{c}{1}{c}{b}
   &=&\displaystyle(-1)^{a+b+c+1}\frac{2\,\Big[b\,(b+1)+c\,(c+1)-a\,(a+1) \Big]}
                             {\Big[2b\,(2b+1)\,(2b+2)\,2c\,(2c+1)\,(2c+2) \Big]^\frac{1}{2}}
\end{array}\]\be\label{4 types of 6j-symbols}\ee
So we always have $6j$-symbols of the general form 
\be
   \SixJ{a}{p}{q}{1}{r}{s}
\ee

This leads us to consider 
nine possible cases of argument combinations:
\be\begin{array}{ccc|c|c|cl}
     p-s&~~~~~~~&&q=r+1 & ~~q=r~~ & q=r-1 
     \\ \cline{1-1} \cline{3-6}
     +1 &&p=s+1 & (1) & (2) & (3) \\\cline{3-6}
    ~~0  &&p=s & (4) & (5) & (6)   \\\cline{3-6}
    -1 &&p=s-1 & (7) & (8) & (9) 
\end{array}\ee
 Using the symmetry properties (\ref{symmetry1}), (\ref{symmetry2}) of the $6j$-symbols, these 9 combinations are covered by the  4 types of $6j$-symbols contained in 
 (\ref{4 types of 6j-symbols})(with the appropriate arguments in brackets): 
 
\be\begin{array}{lllll}
 
   (1) &~~~~& \SixJ{a}{p}{q}{1}{q-1}{p-1} &\longrightarrow& I[b=p,c=q]  
   \\[5mm]
   (2) && \SixJ{a}{p}{q}{1}{q}{p-1}=\SixJ{a}{q}{p}{1}{p-1}{q}&\longrightarrow& II[b=q,c=p]
   \\[5mm]
   (3) && \SixJ{a}{p}{q}{1}{q+1}{p-1}=\SixJ{a}{q}{p}{1}{p-1}{q+1}&\longrightarrow&III[b=q,c=p]
   \\[5mm]
   (4) && \SixJ{a}{p}{q}{1}{q-1}{p} &\longrightarrow &II[b=p,c=q]
   \\[5mm]
   (5) && \SixJ{a}{p}{q}{1}{q}{p} &\longrightarrow &IV[b=p,c=q]
   \\[5mm]
   (6) && \SixJ{a}{p}{q}{1}{q+1}{p}=\SixJ{a}{q+1}{p}{1}{p}{q}=\SixJ{a}{p}{q+1}{1}{q}{p} &\longrightarrow &II[b=p,c=q+1]
   \\[5mm]
   (7) && \SixJ{a}{p}{q}{1}{q-1}{p+1} &\longrightarrow &III[b=p,c=q]
   \\[5mm]
   (8) && \SixJ{a}{p}{q}{1}{q}{p+1}=\SixJ{a}{q}{p+1}{1}{p}{q} &\longrightarrow &II[b=q,c=p+1]
   \\[5mm]
   (9) && \SixJ{a}{p}{q}{1}{q+1}{p+1}=\SixJ{a}{q+1}{p+1}{1}{p}{q} &\longrightarrow &I[b=q+1,c=p+1]   
\end{array}\ee

\subsubsection{Inequalities to be Fulfilled for the Arguments of  $I, II, III, IV$}
Following the general definition (\ref{definition 6j symbol}) of the $6j$-symbols certain inequalities for their arguments must hold. Due to the special form of the arguments these 12 inequalities reduce to a set of 5 inequalities for each case:

\be\begin{array}{llllllllrrll}
      &~~~&
   I[b,c]&=&\SixJ{a}{b}{c}{1}{c-1}{b-1}
   &~~~~~~~&\Delta(a,b,c):&&a+b-c&\ge&0\\[-2mm]
   &&&&&&&&a-b+c&\ge& 0\\
   &&&&&&\Delta(a,c-1,b-1):& &-a+c+b&\ge& 2\\
   &&&&&&\Delta(1,b,b-1):&&b&\ge&1\\
   &&&&&&\Delta(1,c-1,c):&&c&\ge&1
   \\\nonumber\\
    &~~~&
   II[b,c]&=&
   \SixJ{a}{b}{c}{1}{c-1}{b}
   &~~~~~~~&\Delta(a,b,c):&~~~~~&a+b-c& \ge&0\\[-2mm]
   &&&&&   &\Delta(a,c-1,b):& &a-b+c&\ge& 1\\
   &&&&&&&                      &-a+b+c& \ge& 1\\
   &&&&&   &\Delta(1,b,b):&&        b&\ge&\frac{1}{2}\\
   &&&&&   &\Delta(1,c-1,c):&&        c&\ge&1
   \\\nonumber\\
    &~~~&
   III[b,c]&=&
   \SixJ{a}{b}{c}{1}{c-1}{b+1}
   &~~~~~~~&\Delta(a,b,c):&~~~~~&a+b-c& \ge&0\\[-2mm]
   &&&&&&&                      &-a+b+c& \ge& 0\\
   &&&&&   &\Delta(a,c-1,b+1):& &a-b+c&\ge& 2\\
   &&&&&   &\Delta(1,b,b+1):&&        b&\ge&0\\
   &&&&&   &\Delta(1,c-1,c):&&        c&\ge&1
   \\\nonumber\\
    &~~~&
   IV[b,c]&=&
   \SixJ{a}{b}{c}{1}{c}{b}
   &~~~~~~~&\Delta(a,b,c):&~~~~~&a+b-c& \ge&0\\[-2mm]
   &&&&&&&                      &a-b+c& \ge& 0\\
   &&&&&&&                      &-a+b+c& \ge& 0\\
   &&&&&   &\Delta(a,c,b):& & --\\
   &&&&&   &\Delta(1,b,b):&&        b&\ge&\frac{1}{2}\\
   &&&&&   &\Delta(1,c,c):&&        c&\ge&\frac{1}{2}
\end{array}\ee
Moreover we have an integer condition, namely 
in each case \fbox{$a+b+c \in \mathbbm{N}$}
must be fulfilled.
All other $6j$-symbols containing a 1 but different argument ranges as given above or not fulfilling the integer condition vanish identically.

\section{Mathematical Preparations for Implementation on a Computer} 
Here we address the remaining mathematical issues necessary for numerical computation of the volume spectrum.
Recall from (\ref{Volume Definition}) that we have to calculate the spectra of totally antisymmetric real matrices of the form:
\be
   Q:=\sum\limits_{I<J<K\le N}\epsilon(IJK)~\hat{q}_{IJK}
\ee
where again $\epsilon(IJK)=\mbox{sgn}\big(\det{(\dot{e}_I(v),\dot{e}_J(v),\dot{e}_K(v))}\big)$ denotes the sign of the determinant of the tangents of the three edges $e_I,e_J,e_K$ intersecting at the vertex $v$. The actual computation is twofold: firstly we have to compute the recoupling scheme basis in order to apply the matrix element expressions of section 
\ref{Explicit Matrix Element Expressions} for the computation of the constituent  matrices $\hat{q}_{IJK}$. Secondly the possible sign factors $\epsilon(IJK)$ have to be computed.
\subsection{Recoupling Scheme Basis}
\label{Computational Realization of the Recoupling Scheme Basis}
Let us first describe the computation of the recoupling basis.
As it turns out,  the number of gauge invariant recoupling schemes
 contained in all possible recouplings of a set of spins can be surprisingly small.
 To our knowledge an exact result for this number is known only  
for the special case where all spins equal $j=\frac{1}{2}$ \cite{OD-private comm}.
\\

By definition \ref{Def Standard basis} in the appendix  a standard recoupling scheme is the successive recoupling of all $N$ angular momenta $j_1,\ldots,j_N$ at the vertex $v$: Fix a labelling of the edges. Then recouple $j_1,j_2$ to a resulting angular momentum $a_2(j_1~j_2)$. We have due to the Clebsch-Gordan-Theorem $|j_1-j_2|\le a_2(j_1~j_2) \le j_1+j_2$.
Now couple $a_2,j_3$ to a resulting angular momentum $a_3(a_2~j_3)$ where again $|a_2-j_3|\le a_3(a_2~j_3)\le a_2+j_3$.  This continues until we arrive at the last recoupling step where $J\stackrel{!}{=} a_N(a_{N-1}~j_N)$ and thus $|a_{N-1}-j_N|\le J \le a_{N-1}+j_N$. If we work with gauge invariant recoupling schemes we have $J=0$ and therefore must have $j_N\stackrel{!}{=}a_{N-1}$. 

\begin{figure}[hbt!]
\center
    \psfrag{J}{$J$}
    \psfrag{0}{$0$}
    \psfrag{pp}{$\ldots\ldots$}
    
    \psfrag{a2}{$a_2$}
    \psfrag{a2min}{$a_2^{(min)}$}
    \psfrag{a2max}{$a_2^{(max)}$}
    
    \psfrag{a2}{$a_2$}
    \psfrag{a2min}{$a_2^{(min)}$}
    \psfrag{a2max}{$a_2^{(max)}$}
    
    \psfrag{a3}{$a_3$}
    \psfrag{a3min}{$a_3^{(min)}$}
    \psfrag{a3max}{$a_3^{(max)}$}
    
    \psfrag{a4}{$a_4$}
    \psfrag{a4min}{$a_4^{(min)}$}
    \psfrag{a4max}{$a_4^{(max)}$}
    
    \psfrag{a5}{$a_5$}
    \psfrag{a5min}{$a_5^{(min)}$}
    \psfrag{a5man}{$a_5^{(max)}$}
    
    \psfrag{a6}{$a_6$}
    \psfrag{a6min}{$a_6^{(min)}$}
    \psfrag{a6max}{$a_6^{(max)}$}
    
    \psfrag{aN3}{$a_{N-3}$}
    \psfrag{an3m}{$a_{N-3}^{(min)}$}
    \psfrag{an3ma}{$a_{N-3}^{(max)}$}
    
    \psfrag{aN2}{$a_{N-2}$}
    \psfrag{an2m}{$a_{N-2}^{(min)}$}
    \psfrag{an2ma}{$a_{N-2}^{(max)}$}
    
    \psfrag{aN1}{$a_{N-1}$}
    \psfrag{aN1j}{$a_{N-1}=j_N$}
    
    \psfrag{aN}{$a_{N}=J_{\textrm{total}}=0$}
    \includegraphics[height=9cm,width=14cm]{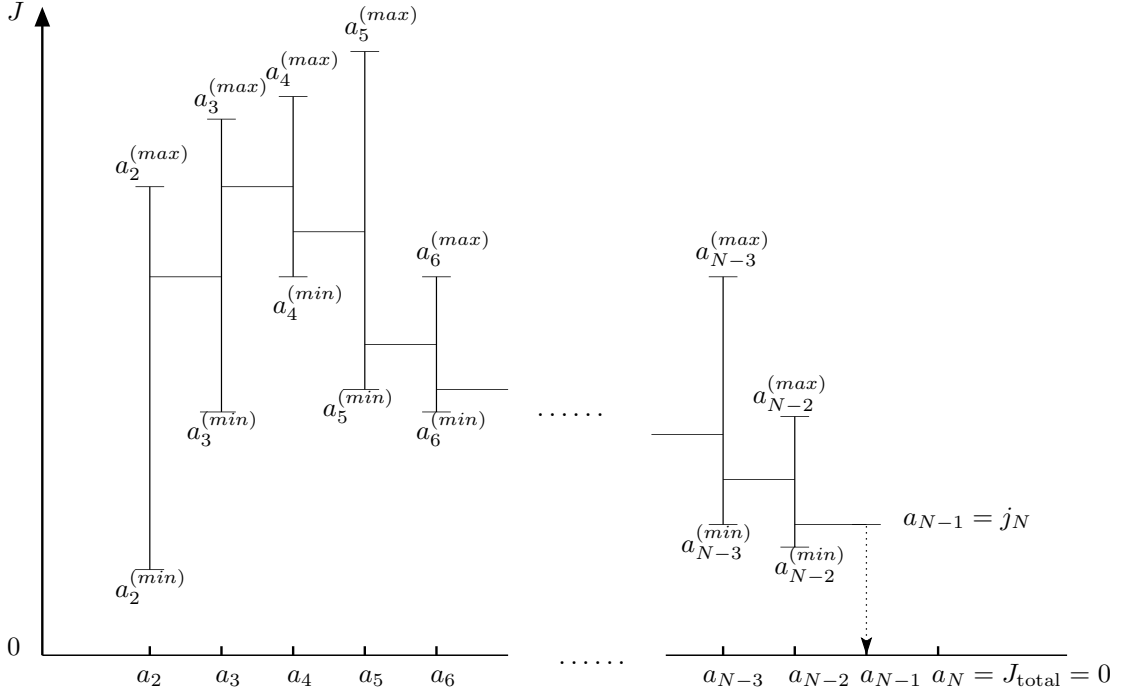}
    \caption{Schematic depiction of the recoupling computation described in
    section \ref{Computational Realization of the Recoupling Scheme Basis}.
    The figure is meant to be read from left to right.  Each vertical line
    represents the range of possible values for the recoupled angular
    momentum $a_i$.  The horizontal lines show a particular choice for each
    recoupling, which subsequently affects the range of possible values for
    the next recoupling $a_{i+1}$.  At the rightmost edge of the diagram we
    see that, to enforce gauge invariance, we restrict
    $a_N\stackrel{!}{=}J=0$.  This in turn requires that
    $a_{N-1}\stackrel{!}{=}j_N$.}
    \label{Schematic Recoupling Scheme}
\end{figure}

The challenge is now to find, for a given spin configuration $j_1,\ldots,j_N$, all possible gauge invariant states, that is all recoupling sequences as illustrated below for which
\be\label{recoupling inequalities}
  a_k^{(min)}:=|a_{k-1}-j_k|\le a_k(a_{k-1}~j_k)\le (a_{k-1}+j_k)=:a_k^{(max)}
\;\;(2 \leq k \leq N-1)
\ee
 (with $a_1=j_1$) and additionally $a_{N-1}\stackrel{!}{=}j_N$. 
Let us moreover define the dimension for each intermediate recoupling step as
 \be
   \dim (a_k) := a_k^{(max)} - a_k^{(min)} +1 
 \ee
 This expression has to be understood iteratively, since for the $k^{\mbox{th}}$ recoupling step the $(k-1)^{\mbox{th}}$ step provides one initial spin as illustrated in figure \ref{Schematic Recoupling Scheme}.
 
\subsubsection{\label{The number of eigenvalues}Estimate for the Number of Eigenvalues}

Throughout our analysis, when referring to numbers of eigenvalues, we of
course take into account their degeneracies.  Thus counting eigenvalues is
equivalent to adding the numbers of recoupling states for each of the vertex
states.

As described in the previous section the construction of a recoupling state 
$\ket{a_2(j_1~j_2)~\ldots~a_{N-1}(a_{N-2}~ j_{N-1})~J(a_{N-1}~ j_N)~M}$
is a successive process in which the possibilities at the $k^{\mbox{th}}$ step are given by the inequality  
(\ref{recoupling inequalities}). Moreover due to (\ref{ordered spins}) in section \ref{Removal of the Arbitrariness of the Edge Labelling} we may assume $j_1\le j_2\le \ldots \le j_N\le j_{max}$.
We can iteratively estimate  $\dim (a_k)$ as
\ba
  \dim(a_2)\le 2j_N+1~~~~\leadsto~~~~\dim(a_3)\le 3j_N+1 
  ~~~~\leadsto~~\ldots~~\leadsto~~~~\dim(a_{N-2})\le (N-2)j_N+1
\ea
Note that due to gauge invariance $\dim(a_{N-1})=\dim(a_{N}=J)=1$. So the dimension of the gauge invariant subspace contained in the $\big[\prod_{k=1}^N (2j_k+1)\big]$-dimensional tensor space is bounded from above by
\ba
  \dim\Big(\big|a_2(j_1~j_2)~\ldots~J(a_{N-1}~ j_N)\!=\!0~M \big>\Big)
  &\le& \prod_{l=2}^{N-2} \big[l\cdot j_N +1 \big]
  \le\prod_{l=2}^{N-2} \big[l\cdot j_N +l \big]
  = \big[j_N+1 \big]^{N-3}\cdot(N-2)!
\ea
which is at the same time a bound for the maximum number of eigenvalues contributed by a fixed spin configuration.
The number of spin configurations $j_1\le j_2\le \ldots \le j_N$ can be calculated as follows:
Assume an $N$-dimensional orthonormal lattice with cubes of edge length 1 as elementary cells. 
Choose a lattice origin, and associate each $j_k$ ($k=1\ldots N$) with a direction of the cube at the origin. If we exclude $j_k=0$ then we can get the number of spin configurations $\numspinconfigs$ for a particular $j_N$ by counting the number of lattice points as:
\be
\numspinconfigs=\sum\limits_{b_{N\!-\!1}=1}^{b_{N}}
             \sum\limits_{b_{N\!-\!2}=1}^{b_{N\!-\!1}} 
             \ldots
             \sum\limits_{b_2=1}^{b_3}
             b_2
            =\frac{1}{(N-1)!} b_N(b_N+1)\cdot \ldots \cdot (b_N+N-2)
            =\left(b_N +N-2 \atop N-1 \right) 
\ee
where we have defined $b_k:=2j_k$ and used that $\sum_{k=1}^L k =\frac{1}{2}L(L+1)$. Thus we can estimate the number $\numevals\big|_{j_N}$ of eigenvalues contributed by spin configurations $j_1\le j_2\le \ldots \le j_N $ with fixed $j_N$:
\ba\label{num evals poly estimate}
  \lefteqn{ \numevals\big|_{j_N}\le \big[j_N+1 \big]^{N-3}\cdot(N-2)!\cdot\left(b_N +N-2 \atop N-1 \right)=}
  \NN\NN
  & =& \frac{1}{N\!-\!1}\left\{ \big[ 2j_N \big]^{N-1}+\big[ 2j_N \big]^{N-2}(1+\ldots+N\!-\!2)
                +\MC{O}\left[\big[ j_N \big]^{N-3} \right]\right\}
       \left\{\big[ j_N \big]^{N-3} + (N\!-\!3)\big[ j_N \big]^{N-4} 
                +\MC{O}\left[\big[ j_N \big]^{N-5} \right]\right\}  
  \NN
  &=&\frac{1}{N-1}\left\{2^{N-1} \big[j_N \big]^{2N-4} 
                       + \Big(2^{N-1}(N-3) + 2^{N-3}(N-1)(N-2)\Big)\big[j_N \big]^{2N-5}
                       + \MC{O}\left[\big[ j_N \big]^{2N-6} \right]   \right\}        
\ea
Here we have computed the coefficients of the two highest orders in $j_N$. We
thus expect that the number of eigenvalues resulting from spin configurations
$j_1\le j_2\le \ldots \le j_N$, $j_N$ fixed 
may be given 
by a polynomial in $j_N$ of degree $2N-4$. This result will be used in order
to give a polynomial fit to the number of eigenvalues obtained in the
eigenvalue computations of section \ref{Numerical Results on Volume
Spectrum}.
\subsection{Sign Factor Combinatorics}
\label{sign_sigma_config_section}
Beside the recoupling part in the definition (\ref{Volume Definition}),  
the sign factors $\epsilon(IJK)$ provide sensitivity of the volume operator to the diffeomorphism invariant structure of the edge embeddings at a particular vertex. The following discussion summarizes this up to now not fully understood problem and sketches the starting point for a detailed analysis. 
We find that there is a fascinating interplay between gauge invariance and the spatial embedding of a given graph, showing a possible link between edge spins and the (spatial) diffeomorphism invariant information encoded in the edge structure of a graph. 

\subsubsection{\label{Gauge Invariance at the Operator Level}Gauge Invariance at the Operator Level}
At the operator level gauge invariance  
\be
  (J_{(tot)})^2:=\sum_{k=1}^3\Big(\sum_{K=1}^N J_K^k\Big)^2=0
\ee
implies that 
\be\label{gauge invariance condition 2}
   J^k_N\stackrel{!}{=}-\sum_{L=1}^{N-1}J^k_L~~~\forall k=1,2,3
\ee
This can be used in (\ref{Volume Definition}) in order to eliminate all $\hat{q}_{IJN}$ containing $N$ as an edge labels as follows:  
Because $\hat{q}_{IJK}=\mathrm{const}\cdot~ \epsilon_{ijk} J^i_IJ_J^jJ_K^k$ we have
\ba
   \label{Volume definition gauge invariant}
   Q
   &=&
   \mathrm{const}\cdot \sum\limits_{I<J<K\le N}\epsilon(IJK)~\epsilon_{ijk} J^i_IJ_J^jJ_K^k 
   \NN
   &=&
   \mathrm{const}\cdot\Big[ 
       \sum\limits_{I<J<K< N}\epsilon(IJK)~\epsilon_{ijk} J^i_IJ_J^jJ_K^k
      +\sum\limits_{I<J<(K\!=\!N)}\epsilon(IJN)~\epsilon_{ijk} J^i_IJ_J^jJ_N^k \Big]
\ea 
Using (\ref{gauge invariance condition 2}) we can rewrite the second sum term on the right hand side of 
(\ref{Volume definition gauge invariant}) as
\ba\label{Volume definition gauge invariant 2}
   \lefteqn{\sum\limits_{I<J<(K\!=\!N)}\epsilon(IJN)~\epsilon_{ijk} J^i_IJ_J^jJ_N^k
   =-\sum_{I<J<N}\sum_{L=1}^{N-1} \epsilon(IJN)~\epsilon_{ijk} J^i_IJ_J^jJ_L^k =}
   \NN
   &=&
   -\sum_{I<J<N}\Big[\sum_{L=1}^{I}\epsilon(IJN)~\epsilon_{ijk} J^i_IJ_J^jJ_L^k
                    +\sum_{L=I+1}^{J}\epsilon(IJN)~\epsilon_{ijk} J^i_IJ_J^jJ_L^k
                    +\sum_{L=J+1}^{N-1}\epsilon(IJN)~\epsilon_{ijk} J^i_IJ_J^jJ_L^k \Big]
   \NN
   &&\fcmt{12}{Configurations with 2 identical edge labels $L=I$ or $L=J$ will not contribute to the sum due to the antisymmetry of $\epsilon_{ijk}$}
   \NN
   &=&
   -\Big[\sum_{L<I<J<N}\epsilon(IJN)~\epsilon_{ijk} J^i_IJ_J^jJ_L^k
        +\sum_{I<L<J<N}\epsilon(IJN)~\epsilon_{ijk} J^i_IJ_J^jJ_L^k
        +\sum_{I<J<L<N}\epsilon(IJN)~\epsilon_{ijk} J^i_IJ_J^jJ_L^k \Big]
   \NN
   &&\fcmt{5}{$\big[J_I^i,J_L^k\big]=0$ for $I\ne L$.}
   \NN
   &=&
   -\Big[\sum_{L<I<J<N}\epsilon(IJN)~\epsilon_{ijk} J_L^kJ^i_IJ_J^j
        +\sum_{I<L<J<N}\epsilon(IJN)~\epsilon_{ijk} J^i_IJ_L^kJ_J^j
        +\sum_{I<J<L<N}\epsilon(IJN)~\epsilon_{ijk} J^i_IJ_J^jJ_L^k \Big]
   \NN
   &=&
   -\Big[\sum_{L<I<J<N}\epsilon(IJN)~\epsilon_{kij} J_L^kJ^i_IJ_J^j
        -\sum_{I<L<J<N}\epsilon(IJN)~\epsilon_{ikj} J^i_IJ_L^kJ_J^j
        +\sum_{I<J<L<N}\epsilon(IJN)~\epsilon_{ijk} J^i_IJ_J^jJ_L^k \Big]
\ea
Upon introducing a suitable relabelling of the sum variables in (\ref{Volume definition gauge invariant 2}), taking each sum term into the form of the first line in (\ref{Volume definition gauge invariant}), we conclude that  the volume operator, if acting on a gauge invariant spin network function, can be rewritten as
\ba\label{Volume definition gauge invariant 3}
   \hat{V}&:=&\sqrt{\Big|Z\cdot 
          \sum\limits_{I<J<K<N}\big[\epsilon(IJK)-\epsilon(JKN)+\epsilon(IKN)-\epsilon(IJN)\big]~\hat{q}_{IJK} \Big|}          
   \NN
   \hat{V}&=:& \sqrt{\Big|Z\cdot 
          \sum\limits_{I<J<K<N}\sigma(IJK)~\hat{q}_{IJK} \Big|} 
          ~~~~\mbox{with}~~~\sigma(IJK):= \epsilon(IJK)-\epsilon(IJN)+\epsilon(IKN)-\epsilon(JKN)
          ~~~~
\ea
which is the final expression for the volume operator acting on gauge invariant recoupling states.
We would like to mention here that (\ref{Volume definition gauge invariant 3}) can be interpreted as a kind of `self-regulating' property of the volume operator when applied to certain `pathological' edge configurations, as illustrated in the following paragraphs.

Consider the following explicit construction of a configuration of $N$ edges outgoing from a vertex $v$, where each ordered edge triple $e_I,e_J,e_K$, $I<J<K$ contributes with a negative sign factor $\epsilon(I,J,K)$. 
Since we are only interested in the sign factor we can make simplifying assumptions, in particular we may arbitrarily choose certain numerical values.
Consider the vertex $v$ as the origin of a 3 dimensional coordinate system with axes $x,y,z$. 
Now consider a circle with radius $r=1$ centered at $y=1$, parallel to the $x$-$z$ plane. Let every edge tangent vector $\dot{e}_K$ end at a point on the circle with coordinates 
$\big(\cos\phi_K,1,\sin\phi_K\big)$ and $\phi_K=2\pi\frac{K}{N}$. Now one may check that for each ordered triple $e_I,e_J,e_K$ with $I<J<K\le N$ we have:
\be
   \det\big({\dot{e}_I,\dot{e}_J,\dot{e}_K}\big)
   = -4\cdot\sin\Big[\pi\frac{K-I}{N}\Big]
            \sin\Big[\pi\frac{K-J}{N}\Big]
	    \sin\Big[\pi\frac{J-I}{N}\Big]
\ee
Since for all arguments $x$ of the sine functions we have $0<x\le\pi$, all of these functions  are $\ge 0$ and therefore we get $\epsilon(I,J,K)=\sgn\Big(\det\big({\dot{e}_I\dot{e}_J,\dot{e}_K}\big)\Big)=-1$ for all ordered edge triples $e_I,e_J,e_K$, $I<J<K$ at $v$.

\begin{figure}[hbt]
    \center
    \psfrag{x}{${x}$}
    \psfrag{y}{${y}$}
    \psfrag{z}{${z}$}
    \psfrag{phik}{$\phi_K$}
    \psfrag{r}{$r$}
    \psfrag{v}{$v$}
    \psfrag{e1}{$\dot{e}_1$}
    \psfrag{e2}{$\dot{e}_2$}
    \psfrag{ek}{$\dot{e}_K$}
    \psfrag{eM}{$\dot{e}_N$}
    \includegraphics[height=5cm]{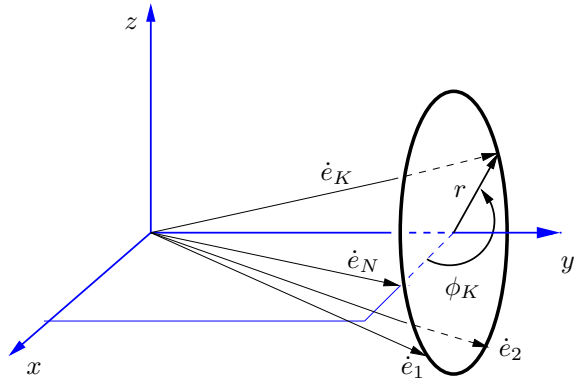} 
    \caption{\label{PathEdge} `Pathological' edge configuration}
\end{figure} 
 
As one can see, this edge configuration is quite special --- it is as if all edges lie in one octant only (if we rotate the coordinate system) or on one 2 dimensional surface as the hull of a cone. 
Such an edge configuration would appear 1 or 2, rather than 3 dimensional.
However, upon imposing gauge invariance, we conclude from (\ref{Volume definition gauge invariant 3}) that such an edge configuration would effectively contribute zero volume, {\em coming from only the orientation of the edges and independent of the spin configuration\footnote{The spin configuration must allow for gauge invariance at $v$ of course.}}.

In \cite{triad I},\cite{triad II} an upper bound for the coherent state expectation value of the operator corresponding to the inverse scale factor in a Friedmann-Robertson-Walker universe was derived. This upper bound provided boundedness of the expectation value, independently of whether 
the coherent states are peaked on classically singular data. However, the authors were unable to exclude the growing of the expectation value as the valence of the vertex on which it is evaluated is increased.   
They supposed this to be an effect of a too rough estimate, assuming all $\epsilon(IJK)$ factors to be identical 1 or -1. However by the explicit construction of the edge configuration shown in figure \ref{PathEdge}, one can see that there indeed exist graphs leading to an unbounded coherent state expectation value of the inverse scale factor operator. 
This problem is now cured upon imposing gauge invariance via (\ref{Volume definition gauge invariant 3}), which is why we refer to this issue as a `self-regulating' property of the volume operator.  

A similar problem arises with respect to a state as presented in figure \ref{PathEdge}: in principle we can produce volume eigenvalues which increase with the number of edge triples at a vertex.  See the discussion in 
section 6 of \cite{NumVolSpecLetter}.

\subsubsection{\label{Sprinkling Sign Factors}Numerical Computation of Sign-Factor Configurations $\vec\epsilon$ Using a Monte Carlo Process  }
Consider an $N$-valent vertex $v$.  For each of the ${N\choose 3}$ triples of edges $1\leq I<J<K\leq
N$ we have a sign factor $\epsilon(IJK)$.  Hereafter we refer to an
assignment of the ${N\choose 3}$ such sign factors at a vertex as a \emph{sign configuration} \linebreak $\vec{\epsilon}:=\{\epsilon(IJK)\}|_{I<J<K\le N}=\big(\epsilon(123),\epsilon(124),\ldots,\epsilon(12N),\epsilon(134),\ldots,\epsilon(13N),\ldots,\epsilon(N\!-\!2,N\!-\!1,N)\big)$.
The remaining question in the definition  (\ref{Volume Definition}) is what
sign configurations  
can be realized in the tangent space at the $N$-vertex\footnote{Recall that $\Sigma$ denotes the 3-dimensional spatial foliation hypersurfaces used in the (3+1)-formulation of General Relativity; see the companion paper \cite{NumVolSpecLetter} for details.} $v\in\Sigma$
at all. This turns out to be  a highly nontrivial problem.

Our first attempt to solve this problem is via a Monte Carlo simulation.
We note that the set of directions of
tangent vectors to the embedded edges at a vertex is equivalent to the set of
points of a unit 2-sphere.
Thus the embedding of an $N$-valent vertex is equivalent to an $N$-tuple of
points on the sphere.
We can explore the collection of such realizable embeddings by randomly
selecting sets of $N$ points on the sphere, via a Poisson process (by which
we mean simply that the points are `sprinkled' at unit density with respect
to the area element on the sphere).  For each such `sprinkling' of the $N$
points, we can compute the sign factors $\epsilon(IJK)$, for each of the ${N
\choose 3}$ triples $1\leq I<J<K\leq N$.  
The particular algorithm we employ is to sprinkle $N$ points, compute the
resulting sign configuration, and then store in a database (hash table) the
number of times $n_{\vec\epsilon}$ this sign configuration has arisen so far.  We continue
this process until the smallest such $n_{\vec\epsilon}$ exceeds a fixed
threshold.  (In practice this threshold can be set at 1, because the number
of sign configurations is enormous.)

One should note that the numbers $n_{\vec\epsilon}$ computed by the above
algorithm have no intrinsic meaning in and of themselves,
since we might do a coordinate transformation on the unit sphere, which changes the probability measure used during the sprinkling process. Moreover this method is unable to give sign configurations  $\vec{\epsilon}$ containing coplanar triples $(IJK)$ with $\epsilon(IJK)=0$, because they form a set of measure zero with respect to area on the 2-sphere.
This is the technical reason why the present work will not consider these coplanar sign configurations, and is restricted to $\epsilon(IJK)=\pm 1$.

Nevertheless, using this method we can compute all the sign configurations $\vec{\epsilon}$ containing only linearly independent triples, as demonstrated in section \ref{Analytical Computation of the Sign-Factors} below.
Using (\ref{Volume definition gauge invariant 3}) it is then straightforward to compute the effective signs $\sigma(IJK)=0,\pm2,\pm4$ resulting from gauge invariance and the restriction  $\epsilon(IJK)=\pm 1$:   

The following table details our results:
\begin{table}[h!]
\center
\begin{tabular}{|c|c|c|c|c|c|c|}
            \hline
            \cmt{1.6}{~\\valence $N$ \\ of  $v$} 
            & \cmt{1.5}{~\\number of triples: $\left(N \atop 3 \right)$\\ } 
            & \cmt{2.2}{maximal number of theoretically possible $\vec{\epsilon}$-sign configs\\ $T_{max}=2^{\left(N \atop 3 \right)}$\\}&   
            \cmt{2.1}{~\\number of realized $\vec{\epsilon}$-sign configs\\\\$~~~~~~T_R$} &
            \cmt{2}{~\\\center{fraction}\\[2mm] $\displaystyle\frac{T_R}{T_{max}}$\\}& 
            \cmt{2}{~\\total number of realized\\ $\vec{\sigma}$-configs}& 
            \cmt{1.7}{~\\multiplicity $\chi_{\vec{\sigma}=\vec{0}}$ of\\ $\vec{\sigma}=\vec{0}$\\-config}
         \\\hline\hline
         4 & 4& 16 & 16 &1 &5&6\\\hline
         5 & 10 & 1024 & 384 & 0.375 &171&24\\\hline
         6 & 20 & $2^{20}$ & 23,808 & 0.023 & 16,413&120\\\hline
         7 & 35 & $2^{35}$ & 3,486,720 &$1.015\cdot 10^{-5}$&3,079,875& 720 \\\hline

\end{tabular}
\caption{Possible sign factor combinations a vertices of valence 4--7.}
\label{Sign factor numbers from sprinkling}
\end{table}

In the next to last column we have listed the number of realizable \emph{$\vec{\sigma}$-configurations}, which is the number of assignments of $\sigma(IJK) = \pm 4, \pm 2$, or 0 to edge triples $e_I,e_J,e_K$ with edge labels $1\leq I<J<K\leq N$ according to (\ref{Volume definition gauge invariant 3}).
They are defined as $\vec{\sigma}:=\{\sigma(IJK)\}|_{I<J<K<N}=\big(\sigma(123),\sigma(124),\ldots,\sigma(12N\!-\!1),\sigma(134),\ldots,\sigma(13N\!-\!1),\ldots,\sigma(N\!-\!3,N\!-\!2,N\!-\!1)\big)$.

Note that there is a unique map $ \MC{S}: \vec{\epsilon}\mapsto \vec{\sigma}$ where 
$\vec{\sigma}:=\{\sigma(IJK) \}\big|_{e_I,e_J,e_K \in E(v) \atop I<J<K}$.
However $\MC{S}$ is not invertible, in particular it will happen that for different $\vec{\epsilon},\vec{\epsilon}~'$-configurations  we have $\MC{S}(\vec{\epsilon})=\MC{S}(\vec{\epsilon}~')=\vec{\sigma}$. We denote the number of $\vec{\epsilon}$-configurations giving the same $\vec{\sigma}$-configuration by $\chi_{\vec{\sigma}}$ and call $\chi_{\vec{\sigma}}$ the redundancy of $\vec{\sigma}$.

In the last column we have written down the number $\chi_{\vec{\sigma}=\vec{0}}$ of distinct sign configurations $\vec{\epsilon}$ which lead to a $\vec{\sigma}$-configuration for which $\sigma(IJK)=0~~\forall~I<J<K<N$. Note that this number seems to be given by $(N-1)!$.

\subsubsection{Analytical Computation of the Sign Configurations $\vec\epsilon$}
\label{Analytical Computation of the Sign-Factors}
The results from the previous section are somewhat unsatisfactory, in that one may be concerned that the Monte Carlo process does not reveal every realizable sign configuration.  For example how do we know that we do not terminate the sprinklings too early?  Additionally, the Monte Carlo technique will never find linearly dependent triples of tangent vectors.
Therefore it is highly desirable to have an analytic proposal at hand in order to compute the number of sign configurations directly, without using Monte Carlo methods.    
Here we present such a proposal, which 
will be worked out in more detail in a forthcoming paper \cite{VertexCombinatorics}.  
 
In what follows we are going to derive a set of inequalities that must be fulfilled by a sign configuration in order for it to correspond to a realizable configuration of tangent vectors,  meaning it must be embeddable in 3-dimensional Riemannian space.
By its definition we have
$\epsilon[IJK]=\mbox{sgn}\big(\det{(\dot{e}_I(v),\dot{e}_J(v),\dot{e}_K(v))}\big)$
and for the tangent vectors $\dot{e}_I(v)$ we introduce the shorthands
$\dot{e}_I(v)=:\vec{v}_I=\left( \begin{array}{c} v_I^1\\v_I^2\\v_I^3 \end{array} \right)$, which are vectors in the tangent space to $v$. Then the determinants  $\det{(\dot{e}_I(v),\dot{e}_J(v),\dot{e}_K(v))}$ can be expressed as
\be\label{definition sign factor}
   \det{(\dot{e}_I(v),\dot{e}_J(v),\dot{e}_K(v))}
   =:\epsilon[IJK]\cdot Q[IJK]=\sum_{i~j~k} \epsilon_{ijk}~v_I^iv_J^jv_K^k
\ee
where $Q[IJK]:=\big|\det{(\dot{e}_I(v),\dot{e}_J(v),\dot{e}_K(v))} \big| \ge 0$ by construction. For an $N$-valent vertex $v$ we have ${N \choose 3}=$\linebreak$\frac{1}{6}N(N-1)(N-2)$ ordered triples with according sign factors. 
\\

Let us for simplicity assume linear independence of all tangent vectors at the vertex $v$, that is $Q[IJK]\ne 0$ $\forall~I<J<K\le N$ in what follows, which implies that $\epsilon[IJK]\ne 0$ and $\epsilon[IJK]=\pm 1$ for all ordered triples $IJK$.
This ignores of course a certain class of configurations, however the calculation presented here can be extended to the linearly dependent case in an obvious way, c.f.\ section \ref{linearly dependent edges}. 
\\

Without loss of generality we may then introduce a basis in the tangent space of $v$ given by the first three tangent vectors $\vec{v}_1,\vec{v}_2,\vec{v}_3$ as
\be\label{basis choice in tangentspace 1}
   \vec{v}_1=\left( \begin{array}{c} 1\\0\\0 \end{array} \right)
   ~~~~~
   \vec{v}_2=\left( \begin{array}{c} 0\\1\\0 \end{array} \right)
   ~~~~~
   \vec{v}_3=\left( \begin{array}{c} 0\\0\\1 \end{array} \right)
\ee 
Note that (\ref{basis choice in tangentspace 1}) implies that $\epsilon[123]=1$.  This is a restriction since it fixes a positive orientation of this triple. However we will express all the other vectors with respect to that choice. Configurations in which $\vec{v}_1,\vec{v}_2,\vec{v}_3$ have negative orientation $\epsilon[123]=-1$ will give a global minus sign to all $\epsilon[IJK]$. We will work with the choice $\epsilon[123]=1$ in what follows but we keep in mind that for every sign configuration we are going to discuss with respect to $\epsilon[123]=1$, there exists a corresponding second one with inverted signs and $\epsilon[123]=-1$. It will be sufficient to discuss only the  $\epsilon[123]=1$ case, apply the found selection rules for realizable sign configurations, and multiply the obtained number of realizable configurations by 2. The second restriction $Q[123]=1$ implied by the choice (\ref{basis choice in tangentspace 1}) is just the normalization of the coordinate volume to 1. This only gives a positive global numerical factor which does not affect the signs $\epsilon[IJK]$.  
\\

Employing all these conventions we can cast (\ref{definition sign factor}) in the following form:
\barr{rrclclclclclclc}\label{basis choice in tangentspace 2}
   &\epsilon[123]~Q[123]&=&~~1  &&\mbox{$\leadsto$ global factor 2}
   \\[0.5cm] 
   (I)&\epsilon[12I]~Q[12I]&=&~~v_I^3&~&\mbox{\fbox{$3<I\le N$}}
   \\
   (II)&\epsilon[13I]~Q[13I]&=&-v_I^2                 &&\mbox{\fbox{$3<I\le N$}}
   \\
   (III)&\epsilon[23I]~Q[23I]&=&~~v_I^1                  &&\mbox{\fbox{$3<I\le N$}}
   \\[0.5cm]
   (1)&\epsilon[1IJ]~Q[1IJ]&=&-v_I^3v_J^2+v_I^2v_J^3 &&\mbox{\fbox{$3<I<J\le N$}}
   \\
   (2)&\epsilon[2IJ]~Q[2IJ]&=&~~v_I^3v_J^1-v_I^1v_J^3 &&\mbox{\fbox{$3<I<J\le N$}}
   \\
   (3)&\epsilon[3IJ]~Q[3IJ]&=&-v_I^2v_J^1+v_I^1v_J^2 &&\mbox{\fbox{$3<I<J\le N$}}
   \\[0.5cm]
   (4)&\epsilon[IJK]~Q[IJK]&=&\multicolumn{3}{l}{
   -v_I^3v_J^2v_K^1 + v_I^3v_J^1v_K^2 - v_I^2v_J^1v_K^3 + v_I^2v_J^3v_K^1 - v_I^1v_J^3v_K^2 + v_I^1v_J^2v_K^3~~~\mbox{\fbox{$3\!<\!I\!<\!J\!<\!K\!\le\! N$}}~~~~~~~\hspace{-1.5cm}~}
   \\
\earr
The relations  (\ref{basis choice in tangentspace 2}) may be rewritten by eliminating the vector components $v_K^k$ in $(1)\ldots (4)$ by plugging into \linebreak $(I),(II),(III)$. Then
\begin{footnotesize}
\barr{rcrclcl}\label{basis choice in tangentspace 3}
   (1)&~~&\epsilon[1IJ]~Q[1IJ]
                       &=&\epsilon[12I]\,\epsilon[13J]~Q[12I]\,Q[13J] - \epsilon[13I]\,\epsilon[12J]~Q[13I]\,Q[12J]
   \\[2mm]
   (2)&&  \epsilon[2IJ]~Q[2IJ]
                       &=&\epsilon[12I]\,\epsilon[23J]~Q[12I]\,Q[23J] - \epsilon[23I]\,\epsilon[12J]~Q[23I]\,Q[12J]
   \\[2mm]
   (3)&&  \epsilon[3IJ]~Q[3IJ]
                       &=&\epsilon[13I]\,\epsilon[23J]~Q[13I]\,Q[23J] - \epsilon[23I]\,\epsilon[13J]~Q[23I]\,Q[13J]
   \\[7mm]
    (4)&&  \epsilon[IJK]~Q[IJK]&=&
     \multicolumn{3}{l}{~~  \epsilon[12I]\,\epsilon[13J]\,\epsilon[23K]~Q[12I]\,Q[13J]\,Q[23K] 
                                         -\epsilon[12I]\,\epsilon[23J]\,\epsilon[13K]~Q[12I]\,Q[23J]\,Q[13K]}
     \\[2mm]
     &&&&\multicolumn{3}{l}{+  \epsilon[13I]\,\epsilon[23J]\,\epsilon[12K]~Q[13I]\,Q[23J]\,Q[12K]
                                         -\epsilon[13I]\,\epsilon[12J]\,\epsilon[23K]~Q[13I]\,Q[12J]\,Q[23K]}
     \\[2mm]
     &&&&\multicolumn{3}{l}{+  \epsilon[23I]\,\epsilon[12J]\,\epsilon[13K]~Q[23I]\,Q[12J]\,Q[13K]
                                         -\epsilon[23I]\,\epsilon[13J]\,\epsilon[12K]~Q[23I]\,Q[13J]\,Q[12K]}
     \\
\earr
\end{footnotesize}
A further simplification can be brought to (\ref{basis choice in tangentspace 3}) if we realize that our assumption 
$\epsilon[IJK]\ne 0$ $~\forall ~I<J<K\le N$ implies that $v_K^k \ne 0$ 
$~\forall ~K=4\ldots N,~k=1,2,3$  due to equations $(I),(II),(III)$ in  (\ref{basis choice in tangentspace 2}).
We are therefore free to set 
$|v_I^3 | = Q[12I]\stackrel{!}{=}1$ which only scales the individual vector\footnote{This can be equivalently thought of as scaling the vector $\vec{v}_I$ by dividing all its components $v_I^k$ by $|v_I^3 |$, giving a new vector $\vec{v}'_I$ pointing in the same direction as $\vec{v}_I$ but with components  ${v'}_I^k=\frac{v_I^k}{|v_I^3 |}$ .} 
$\vec{v}_I$ and thus the moduli $Q[IJK]$ but does not change the direction $\vec{v}_I$ points to. 
That is setting  $|v_I^3 | =1$ does not affect the signs $\epsilon[IJK]$.
Additionally we define the manifestly positive quantities $x_I:=|v_I^1|=Q[23I]$ and $y_I:=|v_I^2|=Q[13I]$.

Using the fact that $\big(~\epsilon[IJK]~\big)^2=1$ we can bring all sign factors to the right hand side of (\ref{basis choice in tangentspace 3}). Now since by construction $Q[IJK]$ is a manifestly positive quantity we can express (\ref{basis choice in tangentspace 3}) as a system of inequalities:
\barr{rcrcllcl}\label{basis choice in tangentspace 5}
   (1)&~~&0&<&\epsilon[1IJ]&\big[ \epsilon[12I]\,\epsilon[13J]~y_J &-& \epsilon[13I]\,\epsilon[12J]~y_I \big]
   \\[2mm]
   (2)&&  0&<&\epsilon[2IJ]&\big[\epsilon[12I]\,\epsilon[23J]~x_J &-& \epsilon[23I]\,\epsilon[12J]~x_I \big]
   \\[2mm]
   (3)&&  0&<&\epsilon[3IJ]&\big[\epsilon[13I]\,\epsilon[23J]~y_I\,x_J  &-& \epsilon[23I]\,\epsilon[13J]~x_I\,y_J \big]
   \\[7mm]
   (4)&&  0&<&\epsilon[IJK]& \multicolumn{3}{l}{\big[~~
                           \epsilon[12I]\,\epsilon[13J]\,\epsilon[23K]~y_J\,x_K 
                          -\epsilon[12I]\,\epsilon[23J]\,\epsilon[13K]~x_J\,y_K }
     \\[2mm]
     &&&&&                 \multicolumn{3}{l}{
                          +~\epsilon[13I]\,\epsilon[23J]\,\epsilon[12K]~y_I\,x_J
                           \,-\epsilon[13I]\,\epsilon[12J]\,\epsilon[23K]~y_I\,x_K}
     \\[2mm]
     &&&&&                 \multicolumn{3}{l}{
                          +~\epsilon[23I]\,\epsilon[12J]\,\epsilon[13K]~x_I\,y_K
                           \,\,-\epsilon[23I]\,\epsilon[13J]\,\epsilon[12K]~x_I\,y_J~~\big]}
     \\
\earr
Now we divide\footnote{Note again that due to our conventions $y_K> 0$ and
$x_K> 0$ ~ $\forall K=4\ldots N$.} $(1)$ by $y_J>0$, $(2)$ by $x_J>0$, $(3)$ by
$(x_J\cdot y_J)>0$, $(4)$ by $(x_K\cdot y_K)>0$
and introduce the quotients
\be
   Y^I_J:=\frac{y_I}{y_J}=\left( Y^J_I\right)^{-1}
   ~~~~~~~~
   X^I_J:=\frac{x_I}{x_J}=\left( X^J_I\right)^{-1}
\ee
with the obvious property that for \fbox{$I<J<K$} we have transitivity, that is
\be
   ~Y^I_K=Y^I_J\cdot Y^J_K~~\mbox{and}~~X^I_K=X^I_J\cdot X^J_K
\ee
We can then bring (\ref{basis choice in tangentspace 5}) into its final form
\barr{rcrcllcl}\label{basis choice in tangentspace 7}
   (1)&~~&0&<&\epsilon[1IJ]&\big[ \epsilon[12I]\,\epsilon[13J] &-& \epsilon[13I]\,\epsilon[12J]~\displaystyle Y^I_J \big]
   \\[4mm]
   (2)&&  0&<&\epsilon[2IJ]&\big[\epsilon[12I]\,\epsilon[23J] &-& \epsilon[23I]\,\epsilon[12J]~ \displaystyle X^I_J \big]
   \\[4mm]
   (3)&&  0&<&\epsilon[3IJ]&\big[\epsilon[13I]\,\epsilon[23J]~\displaystyle Y^I_J  
                    &-& \epsilon[23I]\,\epsilon[13J]~\displaystyle X^I_J \big]
   \\[7mm]
   (4)&&  0&<&\epsilon[IJK]& \multicolumn{3}{l}{\big[\displaystyle~~
                           \epsilon[12I]\,\epsilon[13J]\,\epsilon[23K]~Y^J_K
                          -\epsilon[12I]\,\epsilon[23J]\,\epsilon[13K]~X^J_K}
     \\[4mm]
     &&&&&                 \multicolumn{3}{l}{\displaystyle
                          +~\epsilon[13I]\,\epsilon[23J]\,\epsilon[12K]~Y^I_JY^J_K X^J_K
                           \,-\epsilon[13I]\,\epsilon[12J]\,\epsilon[23K]~Y^I_J Y^J_K}
     \\[4mm]
     &&&&&                 \multicolumn{3}{l}{\displaystyle
                          +~\epsilon[23I]\,\epsilon[12J]\,\epsilon[13K]~X^I_J X^J_K
                           \,\,-\epsilon[23I]\,\epsilon[13J]\,\epsilon[12K]~X^I_J X^J_K Y^J_K~~\big]}
     \\
     \\
\earr

We will subsequently discuss the properties of first the set $(1),(2),(3)$ and secondly $(4)$ as contained in 
(\ref{basis choice in tangentspace 7}) in two separate subsections.
\paragraph{The system $(1),(2),(3)$\\}
If we introduce the short hand signs 
\barr{rclcrclcrcl}\label{Definiton rho's}
         \rho_1^{[IJ]} &=&~~\epsilon[1IJ]\,\epsilon[12I]\,\epsilon[13J]
   &~~~& \rho_3^{[IJ]} &=&~~\epsilon[2IJ]\,\epsilon[12I]\,\epsilon[23J]
   &~~~& \rho_5^{[IJ]} &=&~~\epsilon[3IJ]\,\epsilon[13I]\,\epsilon[23J]
   \\[2mm]
         \rho_2^{[IJ]} &=&-\epsilon[1IJ]\,\epsilon[13I]\,\epsilon[12J]
   &~~~& \rho_4^{[IJ]} &=&-\epsilon[2IJ]\,\epsilon[23I]\,\epsilon[12J]
   &~~~& \rho_6^{[IJ]} &=&-\epsilon[3IJ]\,\epsilon[23I]\,\epsilon[13J]
   \\[2mm]
\earr
we can rewrite the equations $(1),(2),(3)$ of  (\ref{basis choice in tangentspace 7}) as the simple system
 \barr{rcrcllcl}\label{basis choice in tangentspace 8}
   (1)&~~&0&<&\rho_1^{[IJ]} &+& \rho_2^{[IJ]} Y^I_J
   \\[2mm]
   (2)&&  0&<&\rho_3^{[IJ]} &+& \rho_4^{[IJ]} X^I_J
   \\[2mm]
   (3)&&  0&<&\rho_5^{[IJ]} Y^I_J &+& \rho_6^{[IJ]} X^I_J
\earr
Each of the 3 inequalities in (\ref{basis choice in tangentspace 8}) has 4 different sign-combinations:
{\renewcommand{\arraystretch}{1.8}
\arraycolsep=0.12em
\barr{ccc}\label{solutions to (1)-(3)}
   \begin{array}{|r||c|c||l|} 
      \multicolumn{4}{l}{\mbox{Solutions to inequality (1)}} \\
      \cline{1-4}
      \mbox{case}& \rho_1^{(IJ)} & \rho_2^{(IJ)} &\mbox{condition}
      \\\hline\hline
       1.1&+ &+ & 0<Y^I_J
      \\\hline
       1.2&+ &- & 0<Y^I_J<1
      \\\hline
       1.3&- &+ & 1<Y^I_J
      \\\hline
       1.4&- &- & \mbox{no solution}
      \\\hline
   \end{array}
   &~~~~
   \begin{array}{|r||c|c||l|} 
   \multicolumn{4}{l}{\mbox{Solutions to inequality (2)}} \\
   \cline{1-4}
      \mbox{case}& \rho_3^{(IJ)} & \rho_4^{(IJ)} &\mbox{condition}
      \\\hline\hline
       2.1&+ &+ & 0<X^I_J
      \\\hline
       2.2&+ &- & 0<X^I_J<1
      \\\hline
       2.3&- &+ & 1<X^I_J
      \\\hline
       2.4&- &- & \mbox{no solution}
      \\\hline
   \end{array}
   &~~~~
   \begin{array}{|r||c|c||l|} 
   \multicolumn{4}{l}{\mbox{Solutions to inequality (3)}} \\
   \cline{1-4}
      \mbox{case}& \rho_5^{(IJ)} & \rho_6^{(IJ)} &\mbox{condition}
      \\\hline\hline
       3.1&+ &+ & 0\!<\!X^I_J ~~0\!<\!Y^I_J 
      \\\hline
       3.2&+ &- & X^I_J<Y^I_J
      \\\hline
       3.3&- &+ & Y^I_J<X^I_J
      \\\hline
       3.4&- &- & \mbox{no solution}
      \\\hline
   \end{array}~~~~~~~~~
\earr
}

It remains to discuss the complete set of solutions\footnote{That is, consistent combinations of the signs.} for the coupled inequality system $(1),(2),(3)$ in (\ref {basis choice in tangentspace 8}), where we will exclude those sign configurations which contain the cases $1.4$ or $2.4$ or $3.4$ of (\ref{solutions to (1)-(3)}) since they do not contribute a solution.
One finds the following table of solutions. It has to be read as follows: 
Each row gives a combination of 3 solutions contained in (\ref{solutions to (1)-(3)}) (for example the first row indicates a combination of solution 1.1 for inequality (1), 2.1 for inequality (2), 3.1 for inequality (3)   as given in (\ref{solutions to (1)-(3)})). 
Then the resulting sign combinations are written as $'+'\equiv 1$ and $'-'\equiv-1$.
Finally the conditions implied by the given set of solutions are written down.

{\renewcommand{\arraystretch}{1.6}
\arraycolsep=0.1em
\be\label{general solution (1)-(3)}
\begin{array}{cc}
\cmt{15}{
\begin{footnotesize}
   \[\begin{array}{|c|c|c||c|c|c|c|c|c||l|l|} 
      \hline
      (1)& (2) & (3) &\,\rho_1^{[IJ]}\,&\,\rho_2^{[IJ]}\,&\,\rho_3^{[IJ]}\,& \,\rho_4^{[IJ]}\,&\, \rho_5^{[IJ]}\,& \,\rho_6^{[IJ]}\,&\,\mbox{conditions given by (1),(2),(3)}\,&\,\mbox{final condition}\,
      \\\hline\hline
      1&1&1&+&+&+&+&+&+&~0\!<\!Y^I_J\big|0\!<\!X^I_J~ & ~0\!<\!Y^I_J\big|0\!<\!X^I_J~\\\hline
      1&1&2&+&+&+&+&+&-&~0\!<\!Y^I_J\big|0\!<\!X^I_J\big|X^I_J\!<\!Y^I_J & ~0\!<X^I_J\!<Y^I_J\\\hline
      1&1&3&+&+&+&+&-&+&~0\!<\!Y^I_J\big|0\!<\!X^I_J\big|Y^I_J\!<\!X^I_J  &~0\!<Y^I_J\!<X^I_J\\\hline
      1&2&1&+&+&+&-&+&+&~0\!<\!Y^I_J\big|0\!<X^I_J\!<\!1&~0\!<\!Y^I_J\big|0\!<X^I_J\!<\!1~\\\hline
      1&2&2&+&+&+&-&+&-&~0\!<\!Y^I_J\big|0\!<\!X^I_J\!<\!1\big|X^I_J\!<\!Y^I_J
      &~0\!<\!X^I_J\!<\!1\big|X^I_J\!<\!Y^I_J~\\\hline
      1&2&3&+&+&+&-&-&+&~0\!<\!Y^I_J\big|0\!<\!X^I_J\!<\!1\big|Y^I_J\!<\!X^I_J~
      &~0\!<\!Y^I_J\!<\!X^I_J\!<\!1\\\hline
      1&3&1&+&+&-&+&+&+&~0\!<\!Y^I_J\big|1\!<\!X^I_J~&~0\!<\!Y^I_J\big|1\!<\!X^I_J~\\\hline
      1&3&2&+&+&-&+&+&-&~0\!<\!Y^I_J\big|1\!<\!X^I_J\big|X^I_J\!<\!Y^I_J~
      &~1\!<\!X^I_J\!<\!Y^I_J\\\hline
      1&3&3&+&+&-&+&-&+&~0\!<\!Y^I_J\big|1\!<\!X^I_J\big|Y^I_J\!<\!X^I_J~
      &~0\!<\!Y^I_J\!<\!X^I_J\big|1\!<\!X^I_J\\\hline
      2&1&1&+&-&+&+&+&+&~0\!<\!Y^I_J\!<\!1\big|~0\!<\!X^I_J~
      &~0\!<\!Y^I_J\!<\!1\big|~~0\!<\!X^I_J~\\ \hline
      2&1&2&+&-&+&+&+&-&~0\!<\!Y^I_J\!<\!1\big|~0\!<\!X^I_J\big|X^I_J\!<\!Y^I_J~
      &~0\!<\!X^I_J\!<\!Y^I_J\!<\!1~\\\hline
      2&1&3&+&-&+&+&-&+&~0\!<\!Y^I_J\!<\!1\big|~0\!<\!X^I_J\big|Y^I_J\!<\!X^I_J~
      &~0\!<\!Y^I_J\!<\!1\big|~Y^I_J\!<\!X^I_J~\\\hline
      2&2&1&+&-&+&-&+&+&~0\!<\!Y^I_J\!<\!1\big|~0\!<\!X^I_J\!<\!1~\
      &~0\!<\!Y^I_J\!<\!1\big|~0\!<\!X^I_J\!<\!1~\\\hline
      2&2&2&+&-&+&-&+&-&~0\!<\!Y^I_J\!<\!1\big|~0\!<\!X^I_J\!<\!1\big|X^I_J\!<\!Y^I_J
      &~0\!<\!X^I_J\!<\!Y^I_J\!<\!1\\\hline
      2&2&3&+&-&+&-&-&+&~0\!<\!Y^I_J\!<\!1\big|~0\!<\!X^I_J\!<\!1\big|Y^I_J\!<\!X^I_J
      &~0\!<\!Y^I_J\!<\!X^I_J\!<\!1\\\hline
      2&3&1&+&-&-&+&+&+&~0\!<\!Y^I_J\!<\!1\big|~1\!<\!X^I_J&~0\!<\!Y^I_J\!<\!1\!<\!X^I_J\\\hline
      2&3&2&+&-&-&+&+&-&~0\!<\!Y^I_J\!<\!1\big|~1\!<\!X^I_J\big|~X^I_J\!<\!Y^I_J
      &\mbox{~\bf no solution}\\\hline
      2&3&3&+&-&-&+&-&+&~0\!<\!Y^I_J\!<\!1\big|~1\!<\!X^I_J\big|~Y^I_J\!<\!X^I_J&
      ~0\!<\!Y^I_J\!<\!1\!<\!X^I_J\\\hline
      3&1&1&-&+&+&+&+&+&~1<Y^I_J\big|~0\!<\!X^I_J&~1\!<\!Y^I_J\big|~0\!<\!X^I_J\\\hline
      3&1&2&-&+&+&+&+&-&~1<Y^I_J\big|~0\!<\!X^I_J\big|~X^I_J\!<\!Y^I_J
      &~1\!<\!Y^I_J~\big|~0\!<\!X^I_J\!<\!Y^I_J\\\hline
      3&1&3&-&+&+&+&-&+&~1\!<\!Y^I_J\big|~0\!<\!X^I_J\big|~Y^I_J\!<\!X^I_J
      &~1\!<\!Y^I_J\!<\!X^I_J\\\hline
      3&2&1&-&+&+&-&+&+&~1\!<\!Y^I_J\big|~0\!<\!X^I_J\!<\!1&~0\!<\!X^I_J\!<\!1\!<\!Y^I_J\\\hline
      3&2&2&-&+&+&-&+&-&~1\!<\!Y^I_J\big|~0\!<\!X^I_J\!<\!1\big|~X^I_J\!<\!Y^I_J
      &~0\!<\!X^I_J\!<\!1\!<\!Y^I_J\\\hline
      3&2&3&-&+&+&-&-&+&~1\!<\!Y^I_J\big|~0\!<\!X^I_J\!<\!1\big|~Y^I_J\!<\!X^I_J
      &\mbox{~\bf no solution}\\\hline
      3&3&1&-&+&-&+&+&+&~1\!<\!Y^I_J\big|~1\!<\!X^I_J&~1\!<\!Y^I_J\big|~1\!<\!X^I_J\\\hline
      3&3&2&-&+&-&+&+&-&~1\!<\!Y^I_J\big|~1\!<\!X^I_J\big|~X^I_J\!<\!Y^I_J
      &~1\!<\!X^I_J\!<\!Y^I_J\\\hline
      3&3&3&-&+&-&+&-&+&~1\!<\!Y^I_J\big|~1\!<\!X^I_J\big|~Y^I_J\!<\!X^I_J
      &~1\!<\!Y^I_J\!<\!X^I_J\\
      \hline
   \end{array}\]
   \end{footnotesize}}&
   \end{array}
\ee
}

By inspection of the definition of the $\rho$-signs in (\ref{Definiton 
rho's}), due to the arguments of sign factors we notice that
\be\label{rho identity 1}
  \rho_1^{[IJ]}\cdot\rho_2^{[IJ]}\cdot\ldots\cdot\rho_6^{[IJ]}=-1
\ee
and (ignoring the cases $2-3-2$ and $3-2-3$ which give no solution) we will only have 12 cases contained in (\ref{general solution (1)-(3)}) with an odd number of minus signs among the $\rho$'s.  The remaining solutions are given in the following table:

\begin{footnotesize}
{\renewcommand{\arraystretch}{1.6}
\arraycolsep=0.1em
\be\label{special solution (1)-(3)}
   \begin{array}{|l||c|c|c||c|c|c|c|c|c||l|l|} 
      \hline
      \mbox{~case~}&(1)& (2) & (3) &\,\rho_1^{[IJ]}\,&\,\rho_2^{[IJ]}\,&\,\rho_3^{[IJ]}\,& \,\rho_4^{[IJ]}\,&\, \rho_5^{[IJ]}\,& \,\rho_6^{[IJ]}\,&\,\mbox{final condition}\,
      \\\hline\hline
      1&1&1&2&+&+&+&+&+&-& ~0\!<X^I_J\!<Y^I_J\\\hline
      2&1&1&3&+&+&+&+&-&+&~0\!<Y^I_J\!<X^I_J\\\hline
      3&1&2&1&+&+&+&-&+&+&~0\!<\!Y^I_J\big|0\!<X^I_J\!<\!1~\\\hline
      4&1&3&1&+&+&-&+&+&+&~0\!<\!Y^I_J\big|1\!<\!X^I_J~\\\hline
      5&2&1&1&+&-&+&+&+&+&~0\!<\!Y^I_J\!<\!1\big|~~0\!<\!X^I_J~\\ \hline
      6&2&2&2&+&-&+&-&+&-&~0\!<\!X^I_J\!<\!Y^I_J\!<\!1\\\hline
      7&2&2&3&+&-&+&-&-&+&~0\!<\!Y^I_J\!<\!X^I_J\!<\!1\\\hline
      8&2&3&3&+&-&-&+&-&+&~0\!<\!Y^I_J\!<\!1\!<\!X^I_J\\\hline
      9&3&1&1&-&+&+&+&+&+&~1\!<\!Y^I_J\big|~0\!<\!X^I_J\\\hline
      10&3&2&2&-&+&+&-&+&-&~0\!<\!X^I_J\!<\!1\!<\!Y^I_J\\\hline
      11&3&3&2&-&+&-&+&+&-&~1\!<\!X^I_J\!<\!Y^I_J\\\hline
      12&3&3&3&-&+&-&+&-&+&~1\!<\!Y^I_J\!<\!X^I_J\\
      \hline
   \end{array}
\ee
}
\end{footnotesize}
\\[3mm]
Note that for a 5-valent vertex ($N=5$) we only need to consider the
solutions $1-12$ as provided by (\ref{special solution (1)-(3)}) for the pair
$(I,J)=(4,5)$, however for $N>5$ we have for every triple $I<J<K<N$ three
relevant copies of the system (\ref{basis choice in tangentspace 8}) which
read as
 \barr{rcrcllccrcrcllc}\label{triple system}
   (1)^{[IJ]}&~~&0&<&\rho_1^{[IJ]} &+& \rho_2^{[IJ]} Y^I_J
   &~~~~~~~&
   (1)^{[JK]}&~~&0&<&\rho_1^{[JK]} &+& \rho_2^{[JJ]} Y^J_K
   \\[2mm]
   (2)^{[IJ]}&&  0&<&\rho_3^{[IJ]} &+& \rho_4^{[IJ]} X^I_J
   &&
   (2)^{[JK]}&&  0&<&\rho_3^{[JK]} &+& \rho_4^{[JK]} X^J_K
   \\[2mm]
   (3)^{[IJ]}&&  0&<&\rho_5^{[IJ]} Y^I_J &+& \rho_6^{[IJ]} X^I_J
   &&
   (3)^{[JK]}&&  0&<&\rho_5^{[JK]} Y^J_K &+& \rho_6^{[JK]} X^J_K
   \\[5mm] 
     
   (1)^{[IK]}&~~&0&<&\rho_1^{[IK]} &+& \rho_2^{[IK]} Y^I_JY^J_K
   \\[2mm]
   (2)^{[IK]}&&  0&<&\rho_3^{[IK]} &+& \rho_4^{[IK]} X^I_JX^J_K
   \\[2mm]
   (3)^{[IK]}&&  0&<&\rho_5^{[IK]} Y^I_JY^J_K &+& \rho_6^{[IK]} X^I_JX^J_K  
\earr
wherein for the pair $(IK)$ we have again used that $Y^I_K=Y^I_JY^J_K$ and $X^I_K=X^I_JX^J_K$ by definition. Moreover we have used definition (\ref{Definiton rho's}) in order to introduce the $\rho$-signs 
which can be seen as a coordinate transformation from the sign factors $\epsilon[IJK]$ to the $\rho$'s. However, the definitions of the $\rho$-signs  are redundant as the following three identities are obvious:
\ba\label{rho identity 2}
   \rho_1^{[IJ]}\rho_2^{[IJ]}\rho_1^{[JK]}\rho_2^{[JK]}
   &=\epsilon[12I]~\epsilon[13I]~\epsilon[13K]~\epsilon[12K]
   &=-\rho_1^{[IK]}\rho_2^{[IK]}
   \NN
   \rho_3^{[IJ]}\rho_4^{[IJ]}\rho_3^{[JK]}\rho_4^{[JK]}
   &=\epsilon[12I]~\epsilon[23I]~\epsilon[23K]~\epsilon[12K]
   &=-\rho_3^{[IK]}\rho_4^{[IK]}
   \NN
   \rho_5^{[IJ]}\rho_6^{[IJ]}\rho_5^{[JK]}\rho_6^{[JK]}
   &=\epsilon[13I]~\epsilon[23I]~\epsilon[23K]~\epsilon[13K]
   &=-\rho_5^{[IK]}\rho_6^{[IK]}
\ea

A short MATHEMATICA calculation reveals that the coupled inequality  system (\ref{triple system}) has 312 solutions if we impose the restrictions (\ref{rho identity 2}).
Each  is given by a consistent combination of solutions for each of the three subsystems $[IJ],[JK],[IK]$: Each such solution of the subsystems can be labelled according to the cases of the first column in  (\ref{special solution (1)-(3)}).

\paragraph{The inequality $(4)$\\}
Let us now additionally consider inequality number (4) contained in (\ref{basis choice in tangentspace 7}).
This can be rewritten, upon using the obvious identities induced by (\ref{Definiton rho's})
and collecting terms with equal prefactors as follows:
\barr{rcrcllcl}\label{basis choice in tangentspace 10}
     &&  0&<&\epsilon[IJK]& \multicolumn{3}{l}{\Big[\displaystyle~~
                           \epsilon[12I]\,\epsilon[3JK]\big(\,\rho_5^{[JK]}~Y^J_K+\,\rho_6^{[JK]}~X^J_K\big)}
     \\[4mm]
     &&&&&                 \multicolumn{3}{l}{\displaystyle
                          -~\epsilon[13I]\,\epsilon[2JK]\,Y^I_JY^J_K\big(\,\rho_4^{[JK]}~X^J_K
                         \,+\rho_3^{[JK]}~\big)}
     \\[4mm]
     &&&&&                 \multicolumn{3}{l}{\displaystyle
                          +~\epsilon[23I]\,\epsilon[1JK]\,X^I_J X^J_K~\big(\rho_2^{[JK]}Y^J_K+ \rho_1^{[JK]}
			  ~~\big)\Big]}
     \\
\earr
Here we notice that the terms in the round brackets precisely resemble the terms on the right hand side of the inequalities $(1)^{[JK]},(2)^{[JK]},(3)^{[JK]}$ contained in the inequality system (\ref{triple system}), and are thus positive by construction if we want to obtain a solution.
Let us abbreviate the sign-prefactors in (\ref{basis choice in tangentspace 10}) by
\be\label{definition kappa}
\begin{array}{lcrcr}
  \kappa_3^{[IJK]}&:=& \epsilon[12I]~\epsilon[3JK]
  &{=}& \epsilon[12I]~\epsilon[13J]~\epsilon[23K]~\rho_5^{[JK]}
  \\[2mm]
  \kappa_2^{[IJK]}&:=&-\epsilon[13I]~\epsilon[2JK]
  &{=}&-\epsilon[13I]~\epsilon[12J]~\epsilon[23K]~\rho_3^{[JK]}
  \\[2mm]
  \kappa_1^{[IJK]}&:=&\epsilon[23I]~\epsilon[1JK]
  &{=}&\epsilon[23I]~\epsilon[12J]~\epsilon[13K]~\rho_1^{[JK]}
\end{array}
\ee 
where we have again used (\ref{Definiton rho's}) on the right hand side. Let us take a closer look at products of sign factors on the right hand sides of (\ref{definition kappa}). 
According to (\ref{Definiton rho's}) the following identities hold:
\barr{lclcl}
   \rho_5^{[IK]}~\rho_6^{[IK]}&=&-\epsilon[13I]~\epsilon[23K]~\epsilon[23I]~\epsilon[13K]
   &~~~~~\leadsto~~&
   \epsilon[23I]~\epsilon[13K]=-\epsilon[13I]~\epsilon[23K]~\rho_5^{[IK]}~\rho_6^{[IK]}~~~~~
   \\[2mm]
   \rho_1^{[IJ]}~\rho_2^{[IJ]}&=&-\epsilon[12I]~\epsilon[13J]~\epsilon[13I]~\epsilon[12J]
   &~~~~~\leadsto~~&
   \epsilon[13I]~\epsilon[12J]=-\epsilon[12I]~\epsilon[13J]~\rho_1^{[IJ]}~\rho_2^{[IJ]}~~~~~
\earr 
Let us define the new sign-factor
\ba\label{new sign-factor}
   \lambda^{[IJK]}:=
   \epsilon[23I]~\epsilon[12J]~\epsilon[13K]&=&\epsilon[12J]~\epsilon[23I]~\epsilon[13K]
   \NN
   &=&-\epsilon[13I]~\epsilon[12J]~\epsilon[23K]~\rho_5^{[IK]}~\rho_6^{[IK]}
   \NN
   &=&-\epsilon[23K]~\rho_5^{[IK]}~\rho_6^{[IK]}~\epsilon[13I]~\epsilon[12J]
   \NN
   &=&\epsilon[12I]~\epsilon[13J]~\epsilon[23K]~\rho_1^{[IJ]}~\rho_2^{[IJ]}~\rho_5^{[IK]}~\rho_6^{[IK]}
\ea
Using (\ref{new sign-factor}) we can now express the $\kappa$'s of (\ref{definition kappa}) completely in terms of $\rho$-signfactors and the overall sign $\lambda^{[IJK]}$:
\barr{lclcr}\label{definition kappa 2}
\kappa_1^{[IJK]}&=&\lambda^{[IJK]}~\rho_1^{[JK]}
  \\[2mm]
  \kappa_2^{[IJK]}&=&\lambda^{[IJK]}~\rho_5^{[IK]}~\rho_6^{[IK]}~\rho_3^{[JK]}
  \\[2mm]
  \kappa_3^{[IJK]}
  &=&\lambda^{[IJK]}~\rho_1^{[IJ]}~\rho_2^{[IJ]}~\rho_5^{[IK]}~\rho_6^{[IK]}~\rho_5^{[JK]}
\earr
In the last step we use (\ref{definition kappa}), (\ref{definition kappa 2}) to finally express (\ref{basis choice in tangentspace 10}) in terms of $\rho$-signfactors and the overall sign $\lambda^{[IJK]}$:
   
\barr{rcll}\label{basis choice in tangentspace 11}
 
       0&<&\epsilon[IJK]~\lambda^{[IJK]}\Big[&\displaystyle
                          \rho_1^{[IJ]}~\rho_2^{[IJ]}~\rho_5^{[IK]}~\rho_6^{[IK]}~\rho_5^{[JK]}
			  ~~\big(\,\rho_5^{[JK]}~Y^J_K+\,\rho_6^{[JK]}~X^J_K\big)
     \\[4mm]
      &&\multicolumn{2}{l}{ \displaystyle
                          +~\rho_5^{[IK]}~\rho_6^{[IK]}~\rho_3^{[JK]}~Y^I_J~Y^J_K~\big(\,\rho_4^{[JK]}~X^J_K
                         \,+\rho_3^{[JK]}~\big)
                         +~\rho_1^{[JK]}\,X^I_J X^J_K~\big(\rho_2^{[JK]}Y^J_K+ \rho_1^{[JK]}
			  ~~\big)\Big]}
 
\earr

The advantage of reformulating (\ref{basis choice in tangentspace 10}) as (\ref{basis choice in tangentspace 11}) is that we can give now a general tree of solutions to the total system (\ref{basis choice in tangentspace 7}) for an arbitrary edge triple $I<J<K\le N$: Using the $\rho$-signs we can label each such solution again by a solution to the coupled triple inequality system  (\ref{triple system}) and additionally by an {\it overall} sign $S:=\epsilon[LJK]\lambda^{[IJK]}$.
For every solution of the triple system  (\ref{triple system}) we seek compatible signs  $S:=\epsilon[IJK]\lambda^{[IJK]}=\pm 1$. If a solution exists the sign $S:=\epsilon[IJK]\lambda^{[IJK]}$ is written down in the subsequent table in the fourth column. We find that the total inequality system (\ref{basis choice in tangentspace 7}) obeys 372 solutions given in the table below.

\be\label{General Triple Solutions}\hspace{-5mm}
\begin{array}{cc}
\cmt{14.6}{
\begin{footnotesize}
\[\begin{array}{cccccc}
\begin{array}{|r|r|r|r|} \hline
\mbox{IJ}&\mbox{JK}&\mbox{IK}&S\\\hline\hline
1&1&6&-1\\
1&1&10&-1\\
1&1&11&-1\\
1&2&6&-1\\
1&2&6&1\\
1&2&7&1\\
1&2&8&1\\
1&2&10&-1\\
1&2&11&-1\\
1&2&12&-1\\
1&2&12&1\\
1&3&5&1\\
1&3&9&1\\
1&4&5&1\\
1&4&9&-1\\
1&4&9&1\\
1&5&3&1\\
1&5&4&1\\
1&6&1&-1\\
1&7&1&-1\\
1&7&1&1\\
1&7&2&1\\
1&8&1&1\\
1&8&2&1\\
1&9&3&1\\
1&9&4&1\\
1&10&1&-1\\
1&11&1&-1\\
1&11&1&1\\
1&12&1&1\\
1&12&2&1\\\hline
2&1&6&-1\\
2&1&7&-1\\
2&1&7&1\\
2&1&8&1\\
2&1&10&-1\\
2&1&11&-1\\
2&1&11&1\\
2&1&12&1\\
2&2&7&1\\
2&2&8&1\\
2&2&12&1\\
2&3&5&1\\
2&3&9&1\\
2&4&5&1\\
2&4&9&1\\
2&5&3&1\\
2&5&4&1\\
2&6&1&-1\\
2&6&2&-1\\
2&6&2&1\\
2&7&2&1\\
2&8&2&1\\
2&9&3&1\\
2&9&4&-1\\
2&9&4&1\\
2&10&1&-1\\
2&10&2&-1\\
2&11&1&-1\\
2&11&2&-1\\
2&12&2&-1\\
2&12&2&1\\
\hline
\end{array}
&~
\begin{array}{|r|r|r|r|} \hline
\mbox{IJ}&\mbox{JK}&\mbox{IK}&S\\\hline\hline
3&1&5&1\\
3&1&9&1\\
3&2&5&-1\\
3&2&5&1\\
3&2&9&1\\
3&3&6&-1\\
3&3&7&-1\\
3&3&10&-1\\
3&4&6&-1\\
3&4&7&-1\\
3&4&8&-1\\
3&4&8&1\\
3&4&10&-1\\
3&4&10&1\\
3&4&11&1\\
3&4&12&1\\
3&5&1&-1\\
3&5&2&-1\\
3&6&3&1\\
3&7&3&-1\\
3&7&3&1\\
3&8&3&-1\\
3&8&4&-1\\
3&9&1&-1\\
3&9&2&-1\\
3&10&3&1\\
3&11&3&-1\\
3&11&3&1\\
3&11&4&-1\\
3&12&3&-1\\
3&12&4&-1\\\hline
4&1&5&1\\
4&1&9&1\\
4&2&5&1\\
4&2&9&1\\
4&3&6&-1\\
4&3&7&-1\\
4&3&7&1\\
4&3&8&1\\
4&3&10&-1\\
4&3&11&-1\\
4&3&11&1\\
4&3&12&1\\
4&4&8&1\\
4&4&11&1\\
4&4&12&1\\
4&5&1&-1\\
4&5&2&-1\\
4&5&2&1\\
4&6&3&1\\
4&6&4&1\\
4&7&3&1\\
4&7&4&1\\
4&8&4&-1\\
4&8&4&1\\
4&9&1&-1\\
4&9&2&-1\\
4&10&3&1\\
4&10&4&-1\\
4&10&4&1\\
4&11&4&-1\\
4&12&4&-1\\
\hline
\end{array}
&~
\begin{array}{|r|r|r|r|} \hline
\mbox{IJ}&\mbox{JK}&\mbox{IK}&S\\\hline\hline

5&1&3&-1\\
5&1&3&1\\
5&1&4&1\\
5&2&3&1\\
5&2&4&1\\
5&3&1&1\\
5&3&2&1\\
5&4&1&1\\
5&4&2&1\\
5&5&6&1\\
5&5&7&1\\
5&5&8&1\\
5&6&5&-1\\
5&6&5&1\\
5&7&5&1\\
5&8&5&1\\
5&9&6&1\\
5&9&7&1\\
5&9&8&-1\\
5&9&8&1\\
5&9&10&-1\\
5&9&10&1\\
5&9&11&-1\\
5&9&12&-1\\
5&10&5&-1\\
5&10&9&-1\\
5&11&5&-1\\
5&11&9&-1\\
5&12&5&-1\\
5&12&5&1\\
5&12&9&-1\\\hline
6&1&1&-1\\
6&1&1&1\\
6&2&1&-1\\
6&2&2&-1\\
6&3&3&-1\\
6&4&3&-1\\
6&4&4&-1\\
6&5&5&-1\\
6&6&6&-1\\
6&6&6&1\\
6&7&6&-1\\
6&7&7&-1\\
6&8&6&-1\\
6&8&7&-1\\
6&8&8&-1\\
6&9&5&-1\\
6&9&9&-1\\
6&9&9&1\\
6&10&6&1\\
6&10&10&1\\
6&11&6&1\\
6&11&10&1\\
6&11&11&1\\
6&12&6&-1\\
6&12&6&1\\
6&12&7&-1\\
6&12&8&-1\\
6&12&10&1\\
6&12&11&1\\
6&12&12&-1\\
6&12&12&1\\
\hline
\end{array}
&~
\begin{array}{|r|r|r|r|} \hline
\mbox{IJ}&\mbox{JK}&\mbox{IK}&S\\\hline\hline

7&1&1&1\\
7&1&2&1\\
7&2&2&-1\\
7&2&2&1\\
7&3&3&-1\\
7&4&3&-1\\
7&4&4&-1\\
7&4&4&1\\
7&5&5&-1\\
7&6&6&1\\
7&6&7&1\\
7&7&7&-1\\
7&7&7&1\\
7&8&7&-1\\
7&8&8&-1\\
7&9&5&-1\\
7&9&9&-1\\
7&10&6&1\\
7&10&7&1\\
7&10&10&1\\
7&11&6&1\\
7&11&7&-1\\
7&11&7&1\\
7&11&8&-1\\
7&11&10&1\\
7&11&11&-1\\
7&11&11&1\\
7&11&12&-1\\
7&12&7&-1\\
7&12&8&-1\\
7&12&12&-1\\\hline
8&1&1&1\\
8&1&2&1\\
8&2&2&1\\
8&3&3&-1\\
8&3&3&1\\
8&3&4&1\\
8&4&4&1\\
8&5&5&-1\\
8&5&5&1\\
8&6&6&1\\
8&6&7&1\\
8&6&8&1\\
8&7&7&1\\
8&7&8&1\\
8&8&8&-1\\
8&8&8&1\\
8&9&5&-1\\
8&9&9&-1\\
8&10&6&1\\
8&10&7&1\\
8&10&8&-1\\
8&10&8&1\\
8&10&10&-1\\
8&10&10&1\\
8&10&11&-1\\
8&10&12&-1\\
8&11&8&-1\\
8&11&11&-1\\
8&11&12&-1\\
8&12&8&-1\\
8&12&12&-1\\
\hline
\end{array}
&~
\begin{array}{|r|r|r|r|} \hline
\mbox{IJ}&\mbox{JK}&\mbox{IK}&S\\\hline\hline

9&1&3&1\\
9&1&4&1\\
9&2&3&1\\
9&2&4&1\\
9&3&1&-1\\
9&3&1&1\\
9&3&2&1\\
9&4&1&1\\
9&4&2&1\\
9&5&6&-1\\
9&5&6&1\\
9&5&7&1\\
9&5&8&1\\
9&5&10&-1\\
9&5&11&-1\\
9&5&12&-1\\
9&5&12&1\\
9&6&5&1\\
9&6&9&1\\
9&7&5&1\\
9&7&9&1\\
9&8&5&1\\
9&8&9&-1\\
9&8&9&1\\
9&9&10&-1\\
9&9&11&-1\\
9&9&12&-1\\
9&10&9&-1\\
9&10&9&1\\
9&11&9&-1\\
9&12&9&-1\\\hline
10&1&1&-1\\
10&2&1&-1\\
10&2&2&-1\\
10&3&3&-1\\
10&3&3&1\\
10&4&3&-1\\
10&4&4&-1\\
10&5&5&-1\\
10&5&5&1\\
10&5&9&1\\
10&6&6&-1\\
10&6&10&-1\\
10&7&6&-1\\
10&7&7&-1\\
10&7&10&-1\\
10&8&6&-1\\
10&8&7&-1\\
10&8&8&-1\\
10&8&8&1\\
10&8&10&-1\\
10&8&10&1\\
10&8&11&1\\
10&8&12&1\\
10&9&9&1\\
10&10&10&-1\\
10&10&10&1\\
10&11&10&1\\
10&11&11&1\\
10&12&10&1\\
10&12&11&1\\
10&12&12&1\\
\hline
\end{array}
&~
\begin{array}{|r|r|r|r|} \hline
\mbox{IJ}&\mbox{JK}&\mbox{IK}&S\\\hline\hline

11&1&1&-1\\
11&2&1&-1\\
11&2&2&-1\\
11&2&2&1\\
11&3&3&1\\
11&3&4&1\\
11&4&4&-1\\
11&4&4&1\\
11&5&5&1\\
11&5&9&1\\
11&6&6&-1\\
11&6&10&-1\\
11&6&11&-1\\
11&7&6&-1\\
11&7&7&-1\\
11&7&7&1\\
11&7&8&1\\
11&7&10&-1\\
11&7&11&-1\\
11&7&11&1\\
11&7&12&1\\
11&8&8&1\\
11&8&11&1\\
11&8&12&1\\
11&9&9&1\\
11&10&10&-1\\
11&10&11&-1\\
11&11&11&-1\\
11&11&11&1\\
11&12&11&1\\
11&12&12&1\\\hline
12&1&1&-1\\
12&1&1&1\\
12&1&2&1\\
12&2&2&1\\
12&3&3&1\\
12&3&4&1\\
12&4&4&1\\
12&5&5&1\\
12&5&9&1\\
12&6&6&-1\\
12&6&6&1\\
12&6&7&1\\
12&6&8&1\\
12&6&10&-1\\
12&6&11&-1\\
12&6&12&-1\\
12&6&12&1\\
12&7&7&1\\
12&7&8&1\\
12&7&12&1\\
12&8&8&1\\
12&8&12&1\\
12&9&9&-1\\
12&9&9&1\\
12&10&10&-1\\
12&10&11&-1\\
12&10&12&-1\\
12&11&11&-1\\
12&11&12&-1\\
12&12&12&-1\\
12&12&12&1\\\hline
\end{array}
\end{array}\]
\end{footnotesize}
}&
\end{array}
\ee

\subsubsection{Results}
Table (\ref{General Triple Solutions}) can be taken as a starting point to compute all possible sign configurations $\vec{\epsilon}$ for a given $N$-valent vertex $v$. Here it will be of particular interest whether one can get conclusions for high valent vertices from the consistency of all the triple solutions, such that as more and more $\epsilon$ signs must be equal 
most of the $\sigma[IJK]$ prefactors must vanish. 
We expect that it will be possible to develop
an effective algorithm starting from (\ref{General Triple Solutions}) \cite{VertexCombinatorics}.
A preliminary MATHEMATICA calculation reveals, for the numbers of linearly independent triples at valences $N=4,5,6$, the results shown in table \ref{Analytic check of the sprinkling results}.

\begin{table}[h!]
\center
\begin{tabular}{|c|c|c|c|c|}
\hline
   n &           \cmt{3}{\center Number of triples $\left(N\atop 3\right)$}  &
                 \cmt{3.5}{\center Number of potential sign configurations 
                 $T_{\mathrm{max}} = 2^{\left(N\atop 3\right)}$\\[-3mm]~} & 
                 \cmt{3}{\center Number of realizable sign configurations $T_R$}&
                 \cmt{1.5}{\center~\\[-1mm]~ $\displaystyle\frac{T_{R}}{T_{\mathrm{max}}}$} 
\\\hline\hline
4 & 4& 16 & 16 &1 \\\hline
5 & 10 & 1024 & 384 & 0.375 \\\hline
6 & 20 & $2^{20}$ & 23,808 & 0.023 \\\hline
\end{tabular}
\caption{\label{Analytic check of the sprinkling results}Analytic check of the sprinkling results of table \ref{Sign factor numbers from sprinkling} using the inequality system method of this section.}
\end{table}

Comparing this table with the results of the Monte Carlo sprinkling process reveals that, in the case of 4, 5, and 6-valent vertices, we have found all possible 
(non-coplanar) sign configurations from the sprinkling.

\subsubsection{Remark on Linearly Dependent Tangents}
\label{linearly dependent edges}
In this section we sketch how edges having tangents which are linearly dependent 
can be treated. 
Given the set  
of tangent vectors of the edges 
at a vertex
$v$, consider first only the maximal set of edges having linearly independent tangents using the procedure described above\footnote{Such a set will always exist, since in order to have non-vanishing volume we must always have at least one linearly independent triple.}. Calculate the possible sign configurations for this set of edges.
Then consider the remaining edges having linearly dependent tangents for each of the obtained sign configurations as follows:
      
For three linearly dependent tangent vectors $\vec{v}_{K},\vec{v}_{L},\vec{v}_{M}$ we find constants $\alpha_{MK},\alpha_{ML}\in\mb{R}$ such that
$\vec{v}_M=\alpha_{MK}\cdot\vec{v}_K+\alpha_{ML}\cdot\vec{v}_L$. In particular we have for the components 
$v_M^i$, $i=1,2,3$ 
\ba
  v_M^i &=& \alpha_{MK} \cdot v_K^i+\alpha_{ML} \cdot v_L^i
  \NN
  &\downarrow&
  \NN
  \sgn[v_M^i]\cdot \big|v_M^i\big|
  &=&\sgn[\alpha_{MK}]\sgn[v_K^i]\cdot \big|\alpha_{MK} v_K^i\big| + 
     \sgn[\alpha_{ML}]\sgn[v_L^i]\cdot \big|\alpha_{ML} v_L^i\big|
\ea
These conditions then have to be imposed in the procedure above, giving further restrictions to possible sign \mbox{configurations} in (\ref{General Triple Solutions}) on the complete set of triples. A detailed numerical computation based on the analysis presented here will be given in \cite{VertexCombinatorics}.

\section{Computational Details}
In this section we discuss the details of our implementation of the numerical
computation of the volume spectrum.

Our code takes the valence $N$ of a vertex and the maximum spin $\jmax$ as input parameters.
It  
computes eigenvalues for the $N$-vertex for
all spin configurations $j_1 \leq j_2 \leq
\ldots \leq j_N \leq \jmax$, subject to the constraint that $\sum_{i=1}^N
j_i$ is an integer, since otherwise there is no gauge invariant recoupling
scheme.
Since the edge spins take on a
countable set of values, we store doubled spins as (1 byte) integers on the
computer.

\subsection{Recoupling Scheme Implementation}
\label{recoupling_implementation}
As discussed in section \ref{Computational Realization of the Recoupling 
Scheme Basis}, we must compute the set of all possible recouplings of the
edge spins to total angular momentum $J=0$.  We do this with a recursive
procedure, as follows.  Each basis vector $\vec a$ begins with $j_1$ as its
first component.  The second component $a_2$ can potentially take any value
$|j_1-j_2| \leq a_2 \leq j_1+j_2$ which differs from those bounds by an
integer.  For each of these potential values, the third component $a_3$ can
take any value $|a_2-j_3| \leq a_2 \leq a_2+j_3$ (which also differs by an
integer from the bounds), and so on.  Thus the recursive procedure adds a
value for the next component of a potential basis vector $\vec a$.  When it
reaches $a_{N-2}$ it considers only values for which it is possible to arrive at
$a_{N-1}=j_{N}$, which is required to get $a_N=J=0$.  Any instances of the
recursion for which this is not possible are abandoned at this stage.

The particular algorithm employed stores each of the components of $\vec a$
obtained so far for each instance of the recursion.  Since the majority of
these instances will be abandoned at `stage $a_{N-2}$', this is somewhat
inefficient, and can cause problems when dealing with high valences (or
particularly large spins).  A more efficient use of memory would be to employ
some tree-like algorithm which only stores, for each instance of the
recursion, the potential values of the `newest' component of $\vec a$.

Another issue with this algorithm is that it traverses every potential
sequence of values for the spin recouplings, and only very close to the end
does it abandon sequences which cannot possibly recouple to total angular
momentum zero.  It may be more efficient to make this determination as soon
as possible, so the computer does not waste cycles traversing `not possibly
gauge invariant' recoupling schemes.

It turns out that in practice neither of these issues make any difference,
because the computable values of valence and spin are much more tightly
bounded by consideration of the vast number of sign configurations $\vec
\epsilon$.  The CPU time required to compute eigenvalues, and disk space
required to store them, vastly outweigh any concerns about the efficiency of
the recoupling algorithm.

\subsection{Realizable Sign Factors via Poisson Sprinkling Process}
The Monte Carlo computation of the set of realizable sign configurations is
described in section \ref{Sprinkling Sign Factors}.  Here we comment on the
Poisson sprinkling into $S^2$.  

We describe points on $S^2$ via the usual $(\theta, \phi)$ coordinates, and
wish to select $N$ points at random such that, for any region of the
sphere of area $A$, the number of points sprinkled into that region is
Poisson distributed with mean $\frac{NA}{4\pi}$.  Given random variables $x, y$
which are uniformly distributed in the unit interval, we select $N$ points
from
\barr{rcl}
\theta &=& \arccos(2x-1)\\
\phi &=& 2\pi y
\label{sprinkle_2-sphere}
\earr
This gives $\theta \in [0,\pi]$ and $\phi \in [0, 2\pi]$.  
If we form an area element by taking the product of their 
differentials,  
we get $(-)\sin\theta d\theta
d\phi = 4\pi dx \; dy$, indicating that this gives the correct area element on the sphere.  (The sign clearly has no significance).\footnote{
To say this another way, we can regard (\ref{sprinkle_2-sphere}) as
  a coordinate transformation on the 2-sphere.  The area element in the
  ($\theta, \phi$) coordinates is related to that in the Cartesian
  coordinates by
\be
\tilde{d}x \wedge \tilde{d}y = J \tilde d\theta \wedge \tilde d\phi
\ee
where $J$ is the Jacobian of the coordinate transformation.  
Inverting (\ref{sprinkle_2-sphere}) and expanding leads to the same equality as above.
}

\subsection{\label{Eval Computation}Eigenvalue Computation and Numerical Errors}
As explained in section \ref{Right Invariant Vector Fields as Angular Momentum Operators}, we
have a real antisymmetric matrix $Q$, or a real symmetric matrix $Q^\dagger
Q$, whose eigenvalues we wish to compute.
LAPACK \cite{lapack} is an extensive, efficient linear algebra library
written in Fortran 77, available from Netlib \cite{netlib}.
One of the computations it performs is a singular value decomposition of a real matrix $A$ into $U S V^T$, where $U$ and $V$ are orthogonal, and $S$ is diagonal.  The diagonal elements of $S$ are the \emph{singular values}.
The singular decomposition of $Q$ is
\be
Q = U S V^T
\ee
Multiplying on the right with $Q^T$ gives
\begin{eqnarray*}
Q Q^T &=& U S V^T V S^T U^T= U SS^T U^T= U S^2 U^{-1}
\end{eqnarray*}
which is the diagonalization of $QQ^T=QQ^\dagger$.  Thus the eigenvalues of
$QQ^\dagger$ are simply the square of the singular values of $Q$.
As discussed in section
\ref{Right Invariant Vector Fields as Angular Momentum Operators}, these eigenvalues are simply
the square of those of $Q$ (which are pure imaginary and come in complex
conjugate pairs).  Thus the singular values of $Q$ are non-positive real
numbers and come in pairs, and their modulus is the square of that of
the eigenvalues of $\hat V$.  (Note that due to the definition of $\hat V$ we
are only concerned with the modulus of its eigenvalues.)

LAPACK employs an efficient recursive algorithm to compute the singular
values, which it returns with fixed relative accuracy.
(e.g.\ as compared to the `symmetric eigenvalue problem' (c.f.\ section
\ref{eigenproblem}), for which the errors scale with the 1-norm of the input matrix.)
The absolute error of the singular values is bounded by
\be
\delta s_i \leq c \epsilon s_{\mathrm{max}}
\label{errors}
\ee
where $\epsilon$ is machine epsilon (around $10^{-16}$), $s_{\mathrm{max}}$
is the largest singular value for that matrix, and $c$ is a constant which governs the
convergence criterion of the algorithm.  For our calculations $c\approx 90$.

The effect of these numerical errors on our dataset is extremely small, and
is described in detail in section \ref{numerical_errors}.

\subsection{The Cactus Framework} 
The Cactus framework \cite{cactus} is a modular, open source, problem solving
environment for high performance computing.  It was originally developed to
deal with the complex and computationally intensive problem of numerical
relativity, where one seeks to extract the detailed form of gravitational
waves which are emitted from astrophysical events such as binary star or
black hole coalescence.  The extensive needs of that computational task have
led to what is now a very generic and extensible framework, which enables a
wide class of applications to be easily placed on supercomputers.  For this
reason it seemed a natural framework to use for the development of code to
explore Loop Quantum Gravity, and in particular to perform this volume
spectrum computation.  

The name `Cactus' is meant to describe the framework
metaphorically as a collection of modules, called `thorns', which are bound
together by the `flesh'.  The thorns declare their characteristics, e.g.\
public variables called `grid functions', to the
flesh, which then generates code to allow the thorns to communicate and
interoperate cleanly.  To maintain portability Cactus defines its own
datatypes, so it can guarantee their size and presence on all architectures it
supports.

Perhaps the greatest benefit which Cactus affords is automatic
parallelization.  Cactus contains a software layer called a \emph{driver},
which hides the details of implementing parallel code from the application
programmer.  For our computation it allowed us to easily run on a distributed
memory supercomputer, without much more difficulty than that required to
develop the code to run on a single processor.  

The volume spectrum computation described above is composed of three stages.
First, given a spin configuration, 
one must compute the recoupling basis, as described in sections
\ref{Computational Realization of the Recoupling Scheme Basis} and
\ref{recoupling_implementation}.  Next, with the recoupling basis in hand,
one computes the ${N-1 \choose 3}$ sub-matrices $\hat q_{IJK}$ of $Q$ for
each triple $1 \leq I < J < K < N$, shown in (\ref{Q definition}).  These
submatrices are shown in full detail in equations (\ref{q123}) through
(\ref{end_subQs}).  The final task is to combine the resulting submatrices,
according to (\ref{Q definition}), for each $\vec{\sigma}$-configuration (c.f.\
section \ref{Sprinkling Sign Factors}).  This last task is easily
parallelized, by splitting the set of $\vec{\sigma}$-configurations among the
processors.  This is implemented within Cactus by storing a
$\vec{\sigma}$-configuration within the 64 bit integer type
`\code{CCTK\_INT8}'\footnote{It turns out 64 bits is enough to store the
$\vec{\sigma}$-configuration for a 7-vertex, but not for an 8-vertex.  The
code could easily be generalized to handle higher valences.}.  The thorn
which implements the volume computation then declares a one dimensional
\code{CCTK\_INT8} grid function, which is automatically distributed among the
processors by the standard Cactus driver \code{PUGH}.

A second benefit of using Cactus is that
it provides a degree of language independence.  
The LAPACK library is written in Fortran 77, a language which is now quite
old and inconvenient to work with.  
With Cactus one can write thorn code in
C, C++, or any variant of Fortran, and Cactus takes care of all issues
regarding inter-language communication.  All of the code developed for this Volume computation is written in
C, save one routine in Fortran 90 which performs all the calls to the LAPACK
library.  

Additionally Cactus is `familiar with' external packages such as LAPACK and
BLAS (upon which LAPACK depends).  It provides thorns which automatically
detect and link to these libraries, which makes it extremely easy to port the
code to a new machine.

Last, but not least, is the modularity and extensibility of Cactus.  The
underlying philosophy behind its design is to facilitate code sharing and
building a `community code base' upon which all workers in a field can draw.
A user is not required to release his or her thorns, of course, but the
modularity of the thorn-flesh design makes it easy to incorporate code
developed by others in one's own code, so that one can build on the
developments of others rather than each user having to start from scratch.
We intend to release the code which generated the results of this
paper.\footnote{Instructions for downloading the code will appear shortly.}
It is hoped that this software will be a first step in constructing a
community code base for performing high performance computations within Loop
Quantum Gravity.

\section{\label{Numerical Results on Volume Spectrum}Numerical Results on Volume Spectrum}
We will now show the results of our computations. It is in general remarkable that the $\vec{\sigma}$-configuration will determine the shape of the spectrum as well as the behavior of the smallest non-zero eigenvalues (in particular the presence of a volume gap), and the largest eigenvalues as we increase the maximum spin $j_{max}$.

\subsection{General Remarks}
\label{NumericalResults-Remarks}
Let us begin by briefly reviewing the particular computation we have
performed.
Recall that given a graph $\gamma$ we have a set of  
edges $E(\gamma)$ and a set of  
vertices $V(\gamma)$.
Each edge $e\in E(\gamma)$ carries a representation of $SU(2)$ with weight
$j_e$, that is we have a set $\vec{j}:=\{j_e\}_{e \in E(\gamma)}$ of
spins. In the gauge invariant setup we furthermore have at every vertex $v\in
V(\gamma)$ an intertwiner $I_v$ projecting on to one trivial representation
contained in the reduction of the tensor product of the
$SU(2)$-representations carried by the edges outgoing\footnote{Recall that we
  can always redirect edges such that at every $v\in V(\gamma)$ with valence
  greater or equal 3 we only have outgoing edges. } from $v$.
Thus we have a set $\vec{I}:=\{I_v\}\big|_{v \in V(\gamma)}$. 
We denote the set of $N$ edges outgoing from $v$ by $E(v)$,  
and the set of spins incident at the vertex $v$ by $\vec{j}_v:=\{j_I\}|_{e_I \in E(v)}$.
By definition (\ref{Volume Definition}), (\ref{Q definition}), the action of the volume operator $\hat{V}$ can be separated to the single vertices $v\in V(\gamma)$. We can thus restrict our analysis to $\hat{V}$ acting on a single but arbitrary vertex $v$, since we can determine the action of $\hat{V}$ on  arbitrary graphs from that.

$\hat{V}$ is sensitive to the sign factors $\epsilon(IJK):=\sgn[\det(\dot{e}_I,\dot{e}_J,\dot{e}_K)] $ which are the sign of the determinant of the tangents $\dot{e}_I,\dot{e}_J,\dot{e}_K$ of each ordered triple $(e_I,e_J,e_K),~~(I<J<K)$ of edges outgoing from $v$. 
A particular sign configuration $\vec{\epsilon}:=\{\epsilon(IJK) \}\big|_{e_I,e_J,e_K \in E(v) \atop I<J<K}$ is fixed by the chosen graph $\gamma$ (as a representative of its orbit under the diffeomorphism group). Note again that, upon imposing gauge invariance due to (\ref{Volume definition gauge invariant 3}),
there is a unique map $ \MC{S}: \vec{\epsilon}\mapsto \vec{\sigma}$ where 
$\vec{\sigma}:=\{\sigma(IJK) \}\big|_{e_I,e_J,e_K \in E(v) \atop I<J<K}$.
Note that $\MC{S}$ is not invertible, in particular it will happen that for different $\vec{\epsilon},\vec{\epsilon}~'$-configurations  we have $\MC{S}(\vec{\epsilon})=\MC{S}(\vec{\epsilon}~')=\vec{\sigma}$. Let us denote the number of $\vec{\epsilon}$-configurations giving the same $\vec{\sigma}$-configuration by $\chi_{\vec{\sigma}}$. We call $\chi_{\vec{\sigma}}$ the redundancy of $\vec{\sigma}$.
In the end we will compute the volume spectrum for diffeomorphism
equivalence classes of vertex embeddings, each of which has a corresponding sign
configuration $\vec\epsilon$.  Since only the $\vec\sigma$-configurations
enter into the eigenvalue computation, we must simply keep in mind that 
each $\vec\sigma$-configuration comes with degeneracy $\chi_{\vec{\sigma}}$.

Given a fixed $N$-valent vertex $v$ with fixed spin configuration 
$\vec{j}_v=\{j_1\le j_2\le\ldots\le j_{N}\}$ and fixed sigma configuration $\vec{\sigma}$,
we compute the
resulting matrix $Q$ given in (\ref{Q definition}). As mentioned earlier
we will employ Planck units, with $\ell_P^3$, and the $Z$ appearing in (\ref{Volume
  Definition}) (which is a function of the Immirzi parameter $\beta$ and the regularization procedure), set to 1.
We then compute the eigenvalues $\lambda_Q$. 
The square root of the modulus of the $\lambda_Q$ gives the eigenvalues of $\lambda_{\hat{V}}$ of $\hat{V}$. Note that the eigenvalues  $\lambda_{\hat{V}}=\sqrt{|\pm \lambda_Q|}$ come in pairs or are identical to 0.
We store one number from each of the eigenvalues pairs, along with the corresponding spin and sigma configurations, in binary data files.\footnote{These are stored in a native binary format.  The results described in this paper come from 188GB of data.}

Ideally we would like to compute the full spectrum of the volume operator.
Since its action on spin networks can be reduced to that on a single vertex,
we consider only the latter.
While in principle one must compute the eigenvalues above for all
valences $N$ and 
spin
configurations $\vec{j}$, it will be necessary, due to finite computational resources, to introduce a cutoff $j_{max}$
on the maximum allowed spin in the spin configurations, and restrict ourselves to valences $N \leq 7$.
We thus compute, for each valence $N \leq 7$, for every
spin configuration $\vec{j}_v=\{j_1\le j_2\le\ldots\le j_{N}\}$,
for every sigma configuration $\vec{\sigma}$ described in section \ref{sign_sigma_config_section}, 
  the set of eigenvalues $\lambda_{Q}$ of the resulting matrix $Q$ as given in (\ref{Q definition}). From that one arrives at the set of individual eigenvalues $\lambda_{\hat{V}}$ for $\hat{V}$.
Each eigenvalue thus obtained is regarded as coming with a redundancy of
$D(\vec{j})\cdot\chi_{\vec{\sigma}}$.  The first factor arises from the
degeneracy of the spin configuration (\ref{spin_degeneracy}) as described in section \ref{Removal of the Arbitrariness of the Edge Labelling}, and the second from the degeneracy of the sigma configuration mentioned above.

\begin{figure}[htbp]
  \center
  \psfrag{frequency}{$\numevals$}
  \psfrag{eigenvalue}{$\eval$}
  \includegraphics{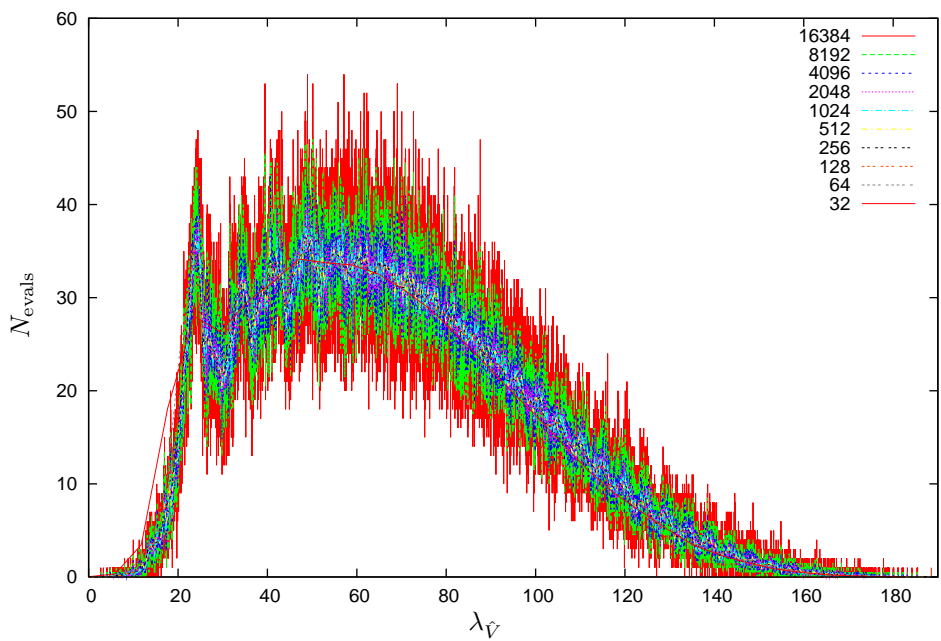}
  \caption{Illustration of effect of bin width on histograms.}
  \label{coarse_graining}
\end{figure}

In the following we display the collection of eigenvalues we have computed
in a number of graphs.  Most of these take the form of histograms: the real
line between 0 and the maximum occurring eigenvalue $\maxev$ is divided into
$\nbins$ equal width bins.  The number of eigenvalues falling into each bin, taking into account
the degeneracies mentioned above, is plotted.  We ignore zero eigenvalues in
all histograms.  (In general there are \emph{many} more zero eigenvalues than
the number of eigenvalues in any histogram bin that we display.)
We have attempted to choose the number of bins for the histograms such that
we capture as much fine structure of this `spectral density function' as
possible, without using 
a too fine  bin width such that the
histogram appears `noisy'.
Figure \ref{coarse_graining} illustrates the effect of bin width on the
histogram.  The histogram is like those described in \S
\ref{5v_fixed_sig_conf_hists}, showing the density of eigenvalues for a
5-vertex with \sigconf = (0 -2 -4 -4) and all spins up to $\jmax=\frac{23}{2}$.
We see that decreasing the number of bins reduces the apparent noise in the
curve.  For a very small number of bins some fine detail of the density curve
is washed out.  Here, when increasing the bin width by a factor of 2, we also
multiply the number of eigenvalues in that bin by $\frac{1}{2}$, so that the
curves line up with the same vertical scale.  
Throughout the rest of the paper, whenever two histograms of varying bin
width appear in the same plot, we employ this same normalization technique,
so that the histograms can be displayed with the same vertical scale.

Additionally we will present various fits, which measure characteristic parameters of the eigenvalue set we have computed. For the number of eigenvalues as a function of $\jmax$ we employ a polynomial fit according to the estimate in section \ref{The number of eigenvalues}. Unless otherwise stated we have used for all other fits linear regression for logarithmic (in case of exponential dependencies) or double logarithmic (in case of possible polynomial dependencies) plots. This results from the fact that there is currently no applicable analytic model for the spectrum at hand, which could give us at least the functional dependencies of different characteristic spectral parameters. Therefore the fits we present here have to be seen rather as a suggestion to a possible behavior than as a proof. Also the error ranges we give should be considered within this context.

\subsection{Gauge Invariant 4-Vertex}
The 4-vertex has been studied extensively in \cite{Volume Paper}.  New analytic results concerning the degeneracy of the spectrum and on the smallest non-zero eigenvalue are presented in section \ref{sec Analytical Results}. Here we present only a histogram of the eigenvalues in figure \ref{4 vertex 1}, for
completeness, as well as a fit of the total number of eigenvalues.

Note that the 4-vertex has five realizable $\vec\sigma$-configurations, with
$\sigma(123)$ able to take any of its possible values -4, -2, 0, 2, or 4.
The sign has no effect, so the only role of the $\vec\sigma$-configuration,
in this case, is to scale the overall spectrum by 0, 2, or 4.  For
consistency with the higher valences we include all eigenvalues contributed from these five
$\vec\sigma$-configurations in our analysis, whereas in \cite{Volume Paper} only the bare matrix $\hat{q}_{123}$ was analyzed. Our numerical analysis extends the criterion for the smallest non-zero eigenvalue $\mineval$ given in \cite{Volume Paper} to a higher maximum spin of $\jmax=\frac{126}{2}=63$. This sequence is contributed by spin configurations $j_1=j_2=\frac{1}{2}$ and $j_3=j_4=j_{max}$. As shown in \cite{Volume Paper} these configurations give rise to an eigenvalue sequence $\mineval(j_{max})\propto \sqrt{2\sqrt{j_{max}(j_{max}+1)}}$.
This eigenvalue sequence as well as its spin configuration are in good agreement to the analytic lower bound (\ref{gauge inv 4-vertex: lower bound non zero eval}) for the eigenvalues\footnote{Recall that  $\eval=\sqrt{|\pm \lambda_Q|}$.} $\lambda_Q$, as $\lambda_Q\ge j_{max}$.

\subsubsection{Histograms}

\begin{figure}[hbt!]
\center
    \center
    \psfrag{frequency}{$\numevals$}
    \psfrag{eigenvalue}{$\eval$}
    \includegraphics[height=12cm]{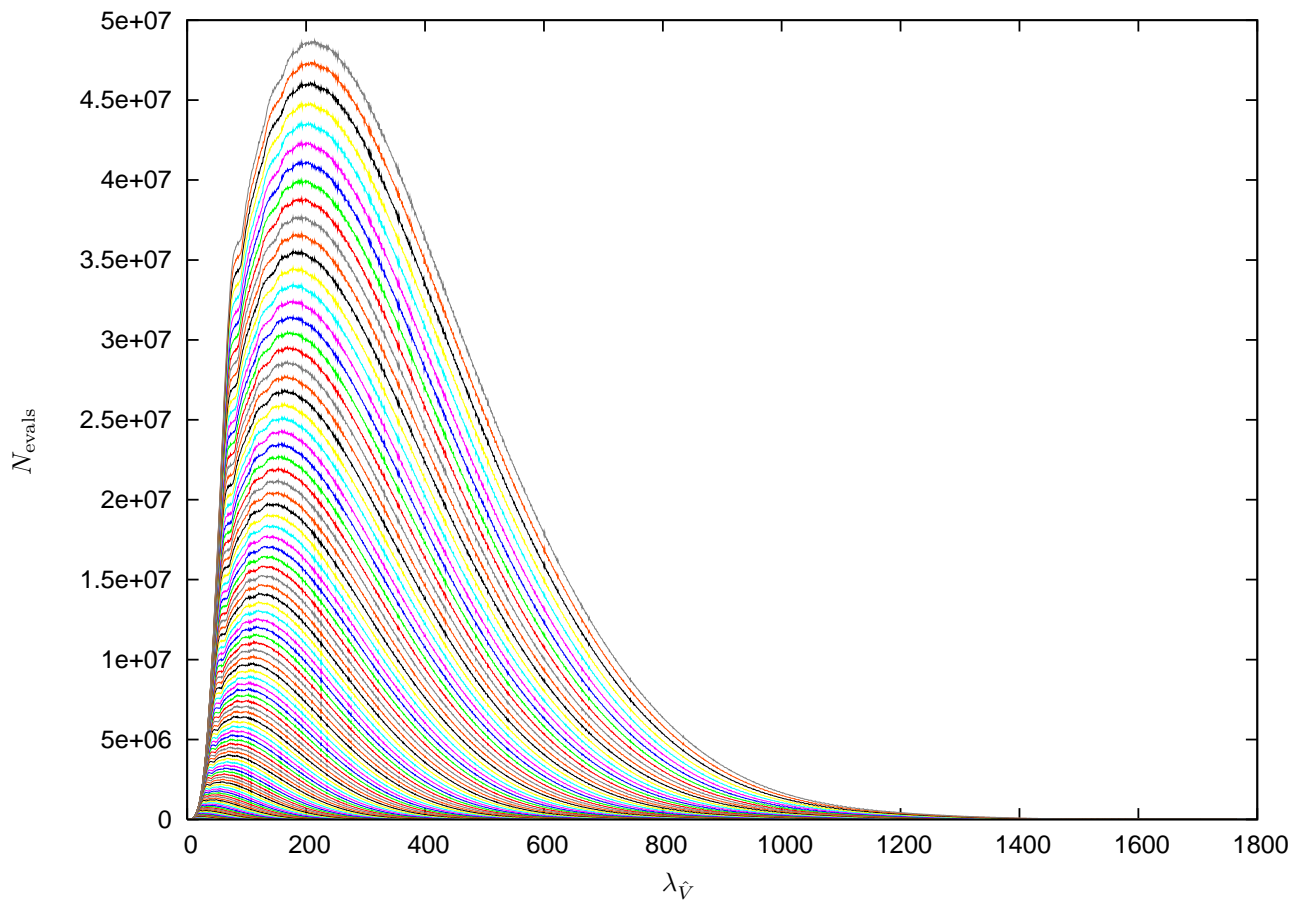}
    \caption{\label{4 vertex 1}Overall 2048 bin histograms at the gauge invariant
    4-vertex up to $j_{max}=\frac{126}{2}$.  There are 44,893,393,776
    eigenvalues in all, of which 17,364,833,136 are zero.}
\end{figure}
Our results for the 4-vertex are depicted in figure \ref{4 vertex 1}, which
was generated with $\nbins=2048$.  Each line represents the histogram of the
eigenvalues for all \sigconfs{} and all values of edge spins up to a maximum spin
$\jmax$ (and all five \sigconfs{}).   
Thus the histogram for a given $\jmax$
includes all eigenvalues which were used to generate the histogram for
$\jmax-\frac{1}{2}$.  
We will refer to this diagram as an `overall' histogram, for the given valence.
Here the topmost line (gray) is the histogram for
$\jmax=\frac{126}{2}$, the red line beneath it is the histogram for
$\jmax=\frac{125}{2}$, and so forth.
There are 17,364,833,136 zero eigenvalues not included in the histogram.

\subsubsection{Number of Eigenvalues}
As a consistency check with the results in \cite{Volume Paper}, we present, according to section \ref{The number of eigenvalues}, the total number number of eigenvalues (including zero eigenvalues)  
for all vertex configurations $j_1\le j_2\le j_3 \le j_4=\jmax$.  Fitting to a polynomial of degree four in $\jmax$ we obtain
\be
   \nevals^{(fit)}(\jmax)=\sum_{k=0}^4 r_k ~\cdot~(\jmax)^k
\ee
with coefficients (and their $95\%$ confidence interval)
\[\left(
\begin{array}{lll}
 r_0 &= \text{54.4937} & \pm\text{5.66522} \\
 r_1 &= -\text{117.403} & \pm\text{1.22582} \\
 r_2 &= -\text{10.6641} & \pm\text{0.07806} \\
 r_3 &= \text{213.333} & \pm\text{0.00184} \\
 r_4 &= \text{106.667} & \pm\text{0.000014}
\end{array}
\right)\]
For this fit the quantity $\chi^2:=\sum_{\jmax=\frac{1}{2}}^{\frac{126}{2}} \big(\nevals(\jmax) -\nevals^{(fit)}(\jmax) \big)^2  $ takes the numerical value $\chi^2=4533.14$ for the set of 126 data points.

\begin{figure}[htbp]
 \begin{minipage}[t]{8.6cm}
    \psfrag{NumEval}{$\numevals(\jmax)$}
    \psfrag{jmax}{$\jmax$}
    \psfrag{FourvNumEvals}{}
    \includegraphics[width=8.7cm]{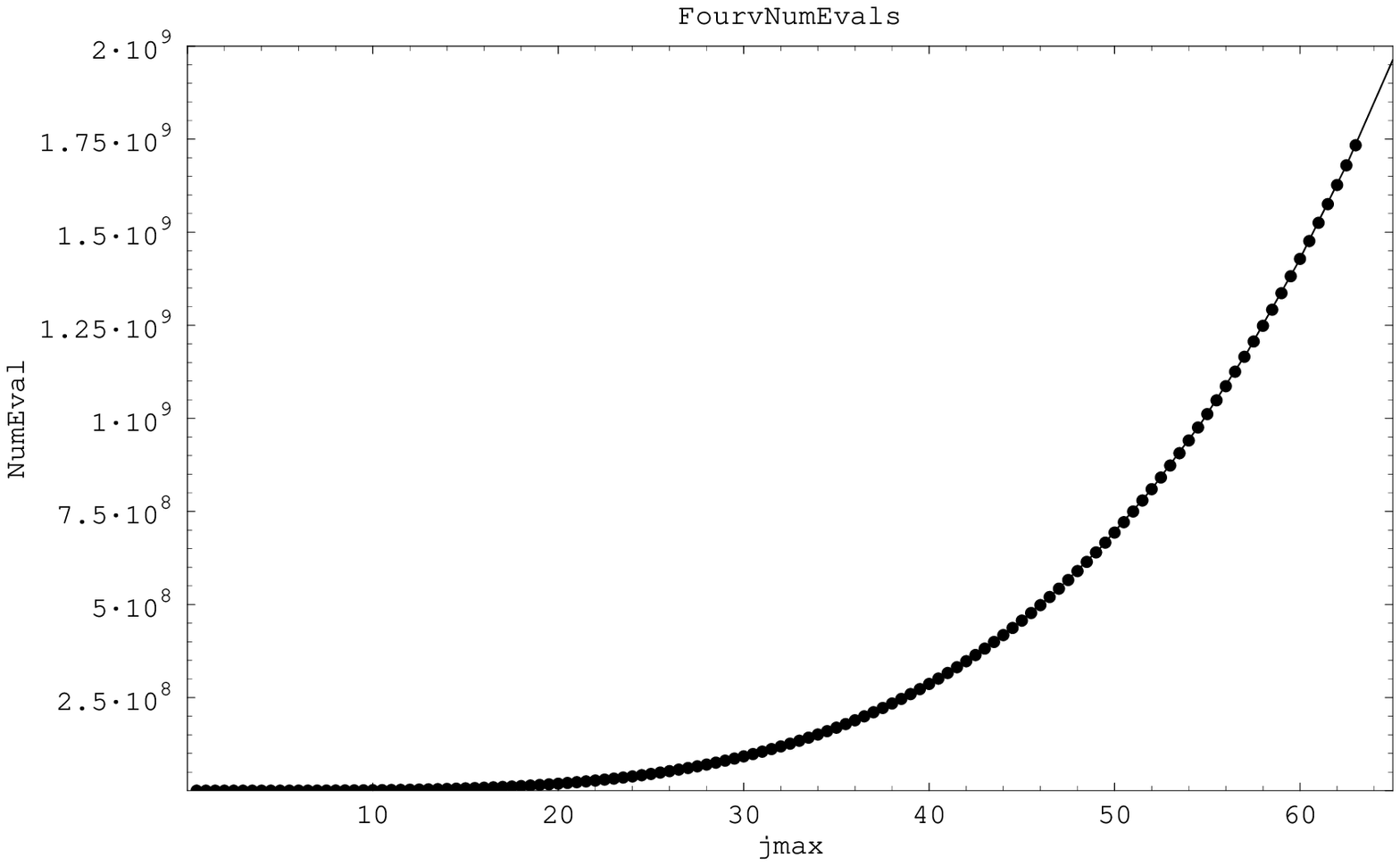}
    \caption{Number of eigenvalues at the 4-vertex.  The solid line is a fit to a fourth order polynomial as given in the text.}
\end{minipage}
~~~~
 \begin{minipage}[t]{8.6cm}
    \psfrag{NumEval}{$\numevals(\jmax)$}
    \psfrag{jmax}{$\jmax$}
    \psfrag{FourvNumEvals}{}
    \includegraphics[width=8.7cm]{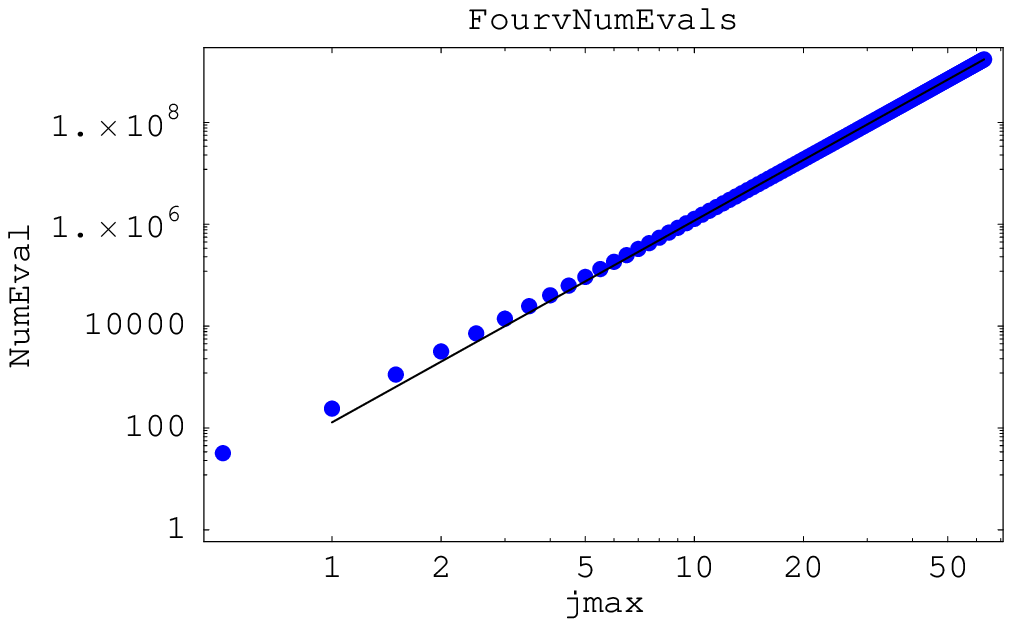}
    \caption{Number of eigenvalues at the 4-vertex in a double logarithmic plot using a two-parameter model. Again the computed data is represented by dots, the fit is a solid line.}
 \end{minipage}
\end{figure}
As the number of eigenvalues is governed by the construction rules for the recoupling scheme basis, one can also give a model having a minimal number of parameters in order to get the behavior of $\numevals$ in the leading order in $\jmax$ as $\nevals^{(fit)}(\jmax)= r \cdot (\jmax)^{s}$. In this case we obtain (using the non linear fit routine of Mathematica) $r=128.981\pm 0.428$ and $s=3.96158\pm 0.00082$, $\chi^2=1.18\cdot 10^{13}$ which approximates the data quite well.  

Note that in \cite{Volume Paper} the cumulative number of eigenvalues was analyzed, that is all eigenvalues, including zero eigenvalues, contributed from spin configurations $0\le j_1\le j_2\le j_3 \le j_4\le\jmax$.
In the present paper we have excluded all spin configurations with $0=j_1$, because setting the spin of an edge to 0 corresponds to deleting that edge from the graph it belongs to. Moreover the present analysis considers the redundancy of the     
$\sigma_{123}=0 (6),\pm 2 (4), \pm 4 (1)$ prefactor: In brackets we have given the number $\chi_{\vec{\sigma}}$ of $\vec{\epsilon}=(\epsilon(123),\epsilon(124),\epsilon(134),\epsilon(234))$ configurations of linearly independent edge tangents which give rise to the according $\sigma_{123}$ prefactor.
As the sum over all $\chi_{\vec{\sigma}}$ is 16, we include this additional factor in our number of eigenvalue counting, as opposed to \cite{Volume Paper} which does not consider this factor. By taking into account this difference we exactly reproduce equation (95) in  \cite{Volume Paper}.

\subsection{Gauge Invariant 5-Vertex}

\subsubsection{Histograms}
\label{5v_histograms}

According to section \ref{Sprinkling Sign Factors} we have 171 $\vec{\sigma}$-configurations. Due to the modulus taken in the definition (\ref{Volume definition gauge invariant 3}), configurations which differ by an overall sign can be identified.  Therefore (excluding the $\sigma(IJK)=0~~\forall~I<J<K$ configuration) we are effectively left with 85 nontrivial $\vec{\sigma}$-configurations.
(There is clearly a symmetry in the \sigconfs{} whereby if one \sigconf{} is
realizable then its pair with all opposite signs is also realizable.)

\begin{figure}[htbp]
    \center
    \psfrag{frequency}{$\numevals$}
    \psfrag{eigenvalue}{$\eval$}
    \includegraphics[height=11.3cm]{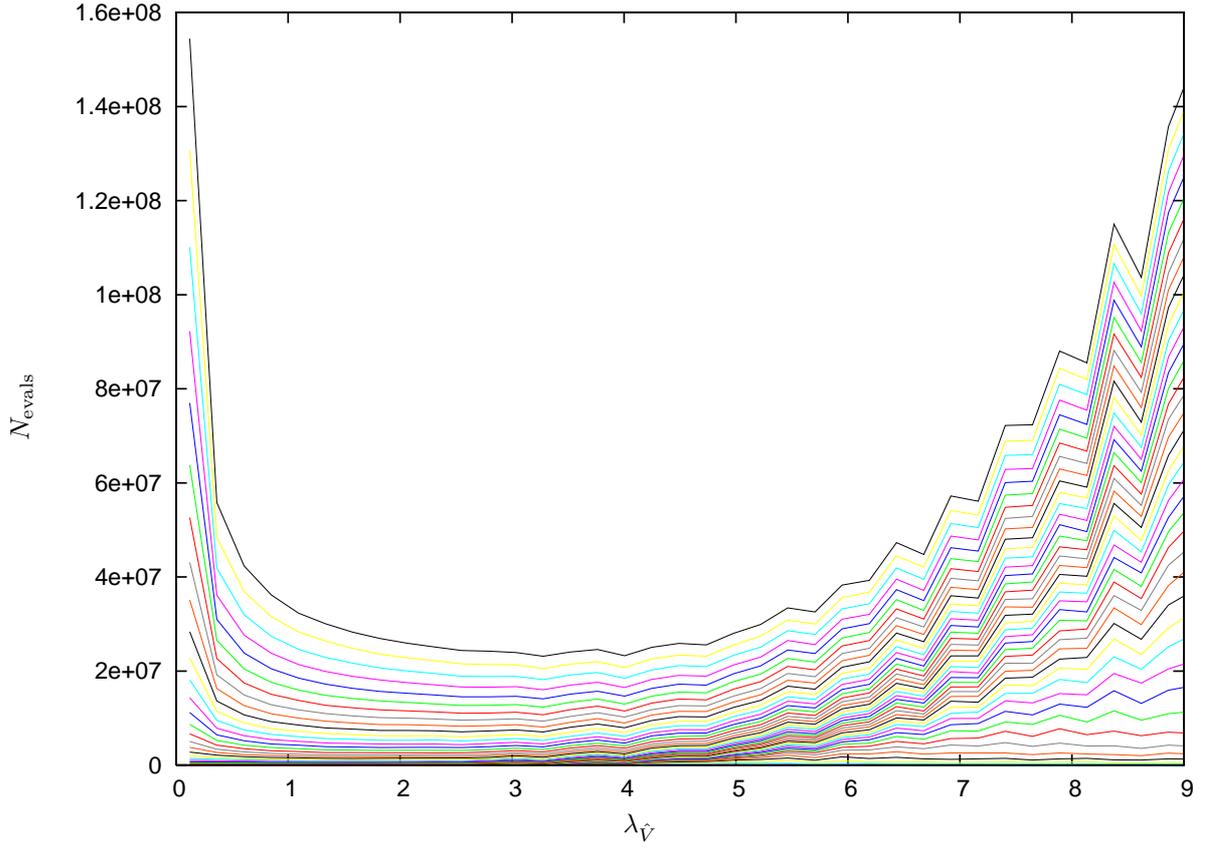}
    \caption{\label{5 vertex 2}\label{lip_figure}Zoom of eigenvalues close to 0  at the gauge invariant 5-vertex up to $j_{max}=\frac{44}{2}$.}
\end{figure}

\begin{figure}[htbp]
    \center
    \psfrag{frequency}{$\frac{N_{\mathrm{bin}}}{\numevals(\jmax})$}
    \psfrag{eigenvalue}{$\eval$}
    \psfrag{5-vertex:VertNormalizedFullSpectrumUpToVariousJmax}{}
    \includegraphics[height=11.2cm]{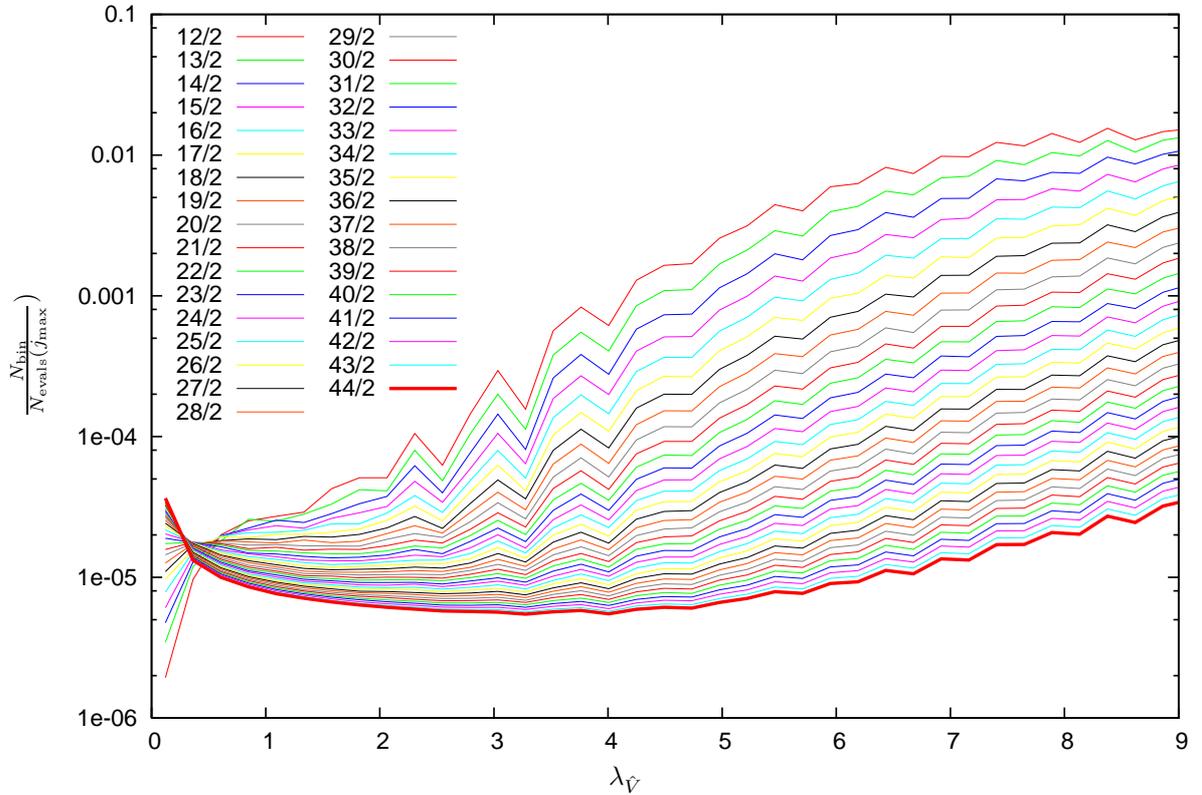}
    \caption{\label{5 vertex lip logplot}Zoom of log-plot of eigenvalues close to 0 at the gauge invariant 5-vertex up to $j_{max}=\frac{44}{2}$. We display the normalized histogram here: The number of eigenvalues per bin $N_{\mathrm{bin}}$
is divided by the total number of eigenvalues $\numevals(\jmax)$ (as given in section \ref{Number of Eigenvalues 5v}), such that the sum of all histogram bin occupation numbers gives 1. For larger $\jmax$ we clearly see a drastically increasing eigenvalue density close to zero.}
\end{figure}

We present here the histograms for the 5-vertex, for each maximum spin
$j_{max}\leq \frac{44}{2}$.  These include all 85
$\vec{\sigma}$-configurations.
It turns out that, as opposed to the gauge invariant 4-vertex, the smallest eigenvalue 
does not always increase with $j_{max}$. Rather this property is dictated by the particular $\vec{\sigma}$-configuration, as can be seen in the discussion of the smallest eigenvalues below. Here this manifests itself in the growth of spectral density with $\jmax$ at zero, as can been seen 
in figure  \ref{5 vertex 2}. Figure \ref{5 vertex lip logplot} demonstrates that, 
upon dividing the number of eigenvalues $\numevals$ in each histogram bin by the total number of eigenvalues $\numevals(\jmax)$ (as given in section \ref{Number of Eigenvalues 5v}), one finds that this growth of the spectral density becomes more distinct at larger $j_{max}$.

Note that this cannot be traced back to
numerical errors (see section \ref{numerical_errors} for a detailed discussion), but is a property of the volume operator itself: Eigenvalues are accumulated close to zero, resulting in a non-vanishing
and in the limit $j_{max}\rightarrow\infty$ possibly infinite spectral density there.
As we go from $\eval=0$ to larger $\eval$ the eigenvalue density approaches a minimum and then goes over into a rising edge. This edge then finally decays into a long tail of the eigenvalue distribution, which can be traced back to the finite cutoff in $\jmax$.   
\begin{figure}[htbp]
 \center
    \psfrag{frequency}{$\numevals$}
    \psfrag{eigenvalue}{$\eval$}
    \includegraphics[height=12.0cm]{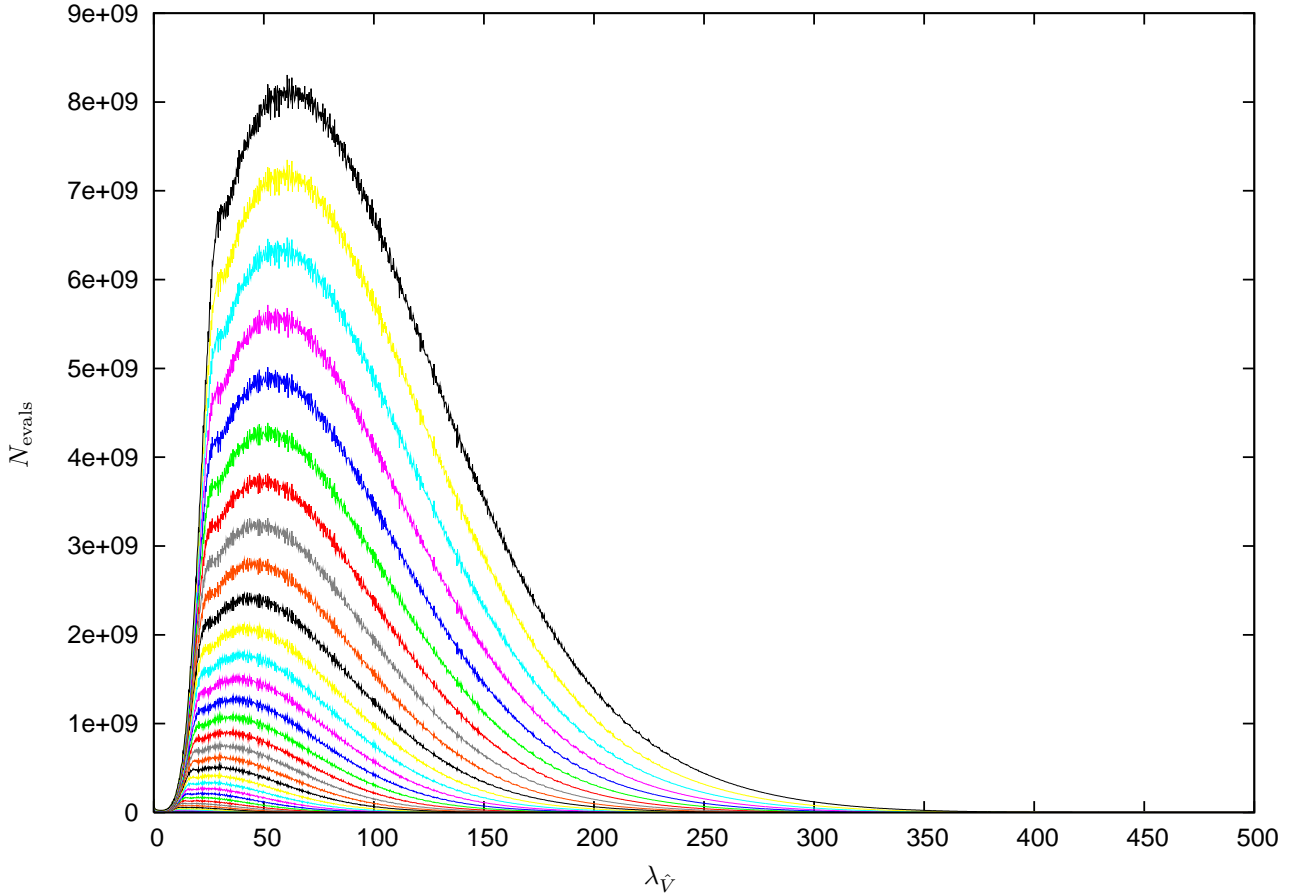}
    \caption{\label{5 vertex 1}Overall 2048 bin histograms at the gauge
    invariant 5-vertex up to $j_{max}=\frac{44}{2}$.  There are 4,680,292,624,128
    eigenvalues in all, of which 450,641,502,288 are regarded as zero.}
\end{figure}

\begin{figure}[htbp]

\begin{minipage}{8.5cm}    
    \center
    \psfrag{NumEvals}{$\frac{N_{\mathrm{bin}}}{N_{\mathrm{bin}}^{\mathrm{max}}(\jmax)}$}
    \psfrag{lV}{$\frac{\eval}{\eval^{max}(\jmax)}$}
    \includegraphics[height=6.5cm]{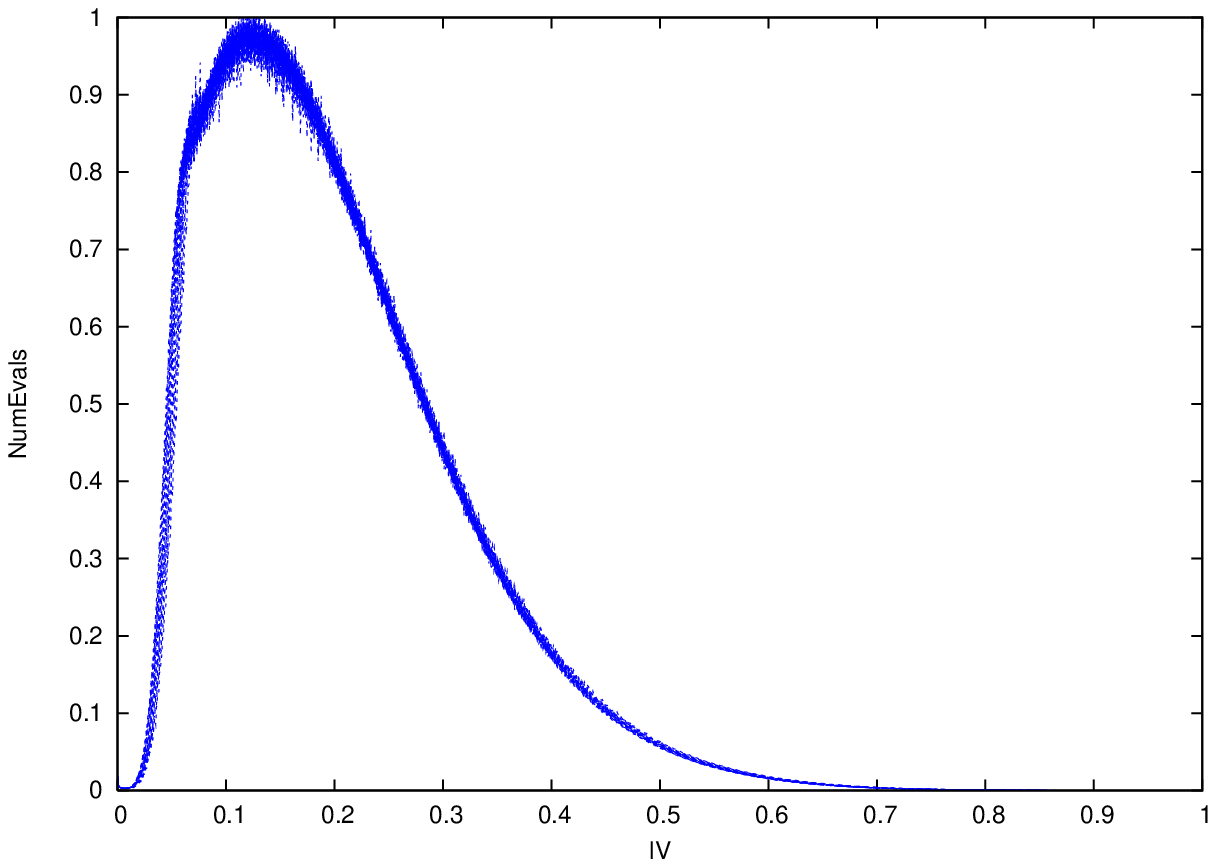}
\end{minipage}
\begin{minipage}{5cm}   
   \center
    \psfrag{NumEval}{$\frac{N_{\mathrm{bin}}}{N_{\mathrm{bin}}^{\mathrm{max}}(\jmax)}$}
    \psfrag{lV}{$\frac{\eval}{\eval^{max}(\jmax)}$}
    \psfrag{2jmax}{$2\jmax$}
    \includegraphics[height=6.5cm]{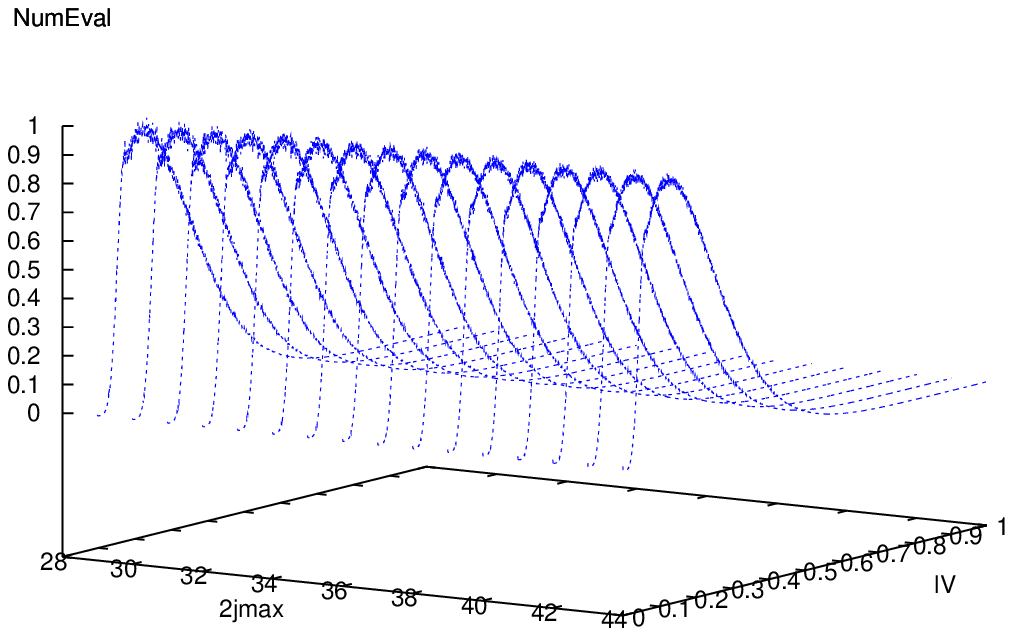}    
\end{minipage}
    \caption{\label{5v_overall_normalized}Normalized fixed-$\jmax$ histograms
    for the 5-vertex for $\jmax=\frac{29}{2},\ldots,\frac{44}{2}$ (right figure). All histograms superposed are shown in the left figure.}
\end{figure}

Figure \ref{5v_overall_normalized}  portrays 16 histograms, one for each value of $\jmax=\frac{29}{2},\ldots,\frac{44}{2}$.  Here
each bin occupation number $N_{\mathrm{bin}}$ is divided by the maximum bin occupation  number $N_{\mathrm{bin}}^{\mathrm{max}}(\jmax)$ of eigenvalues for
that $\jmax$, and in addition each eigenvalue is divided by the maximum
eigenvalue at that $\jmax$, $\eval^{max}(\jmax)$.  Thus each of the histograms is normalized, such
that we may see if they possess a universal shape which is independent of
$\jmax$.  Here one can see that the 16 normalized histograms each have a quite
similar shape.

\subsubsection{Effect of Different $\vec{\sigma}$-configurations}
\label{5v_fixed_sig_conf_hists}
\begin{figure}[hbtp]
\center
\begin{minipage}{8.2cm}
    \center
    \psfrag{frequency}{$\numevals$}
    \psfrag{eigenvalue}{$\eval$}
    \psfrag{5-vertex; sigmas =  0  0  2  0}{\large $\vec{\sigma}=(0~~0~~2~~0$)}
    \includegraphics[width=8.5cm]{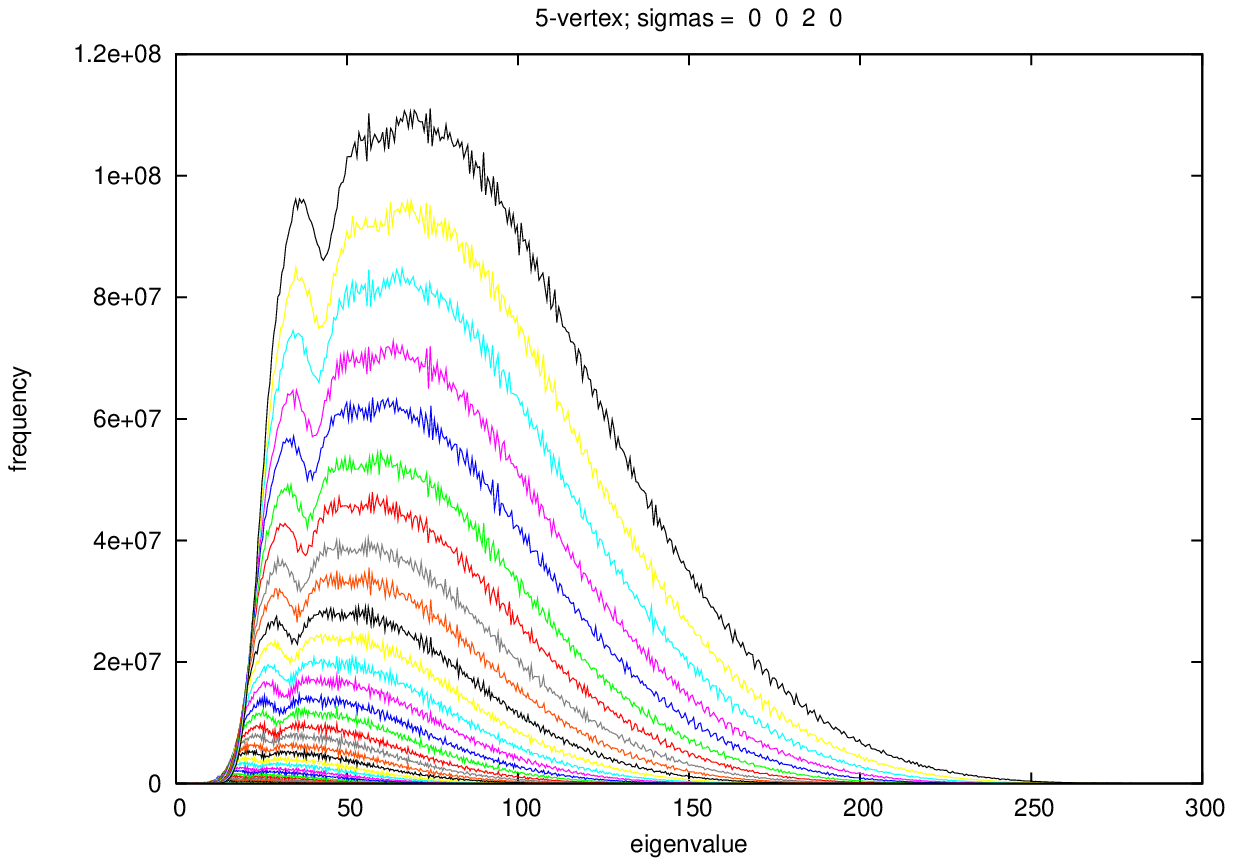}
    \end{minipage}
~
\begin{minipage}{8.2cm}
    \center
    \psfrag{frequency}{$\numevals$}
    \psfrag{eigenvalue}{$\eval$}
    \psfrag{5-vertex; sigmas 60 =  2  0  0  0}{\large $\vec{\sigma}=(2~~0~~0~~0$)}
    \includegraphics[width=8.5cm]{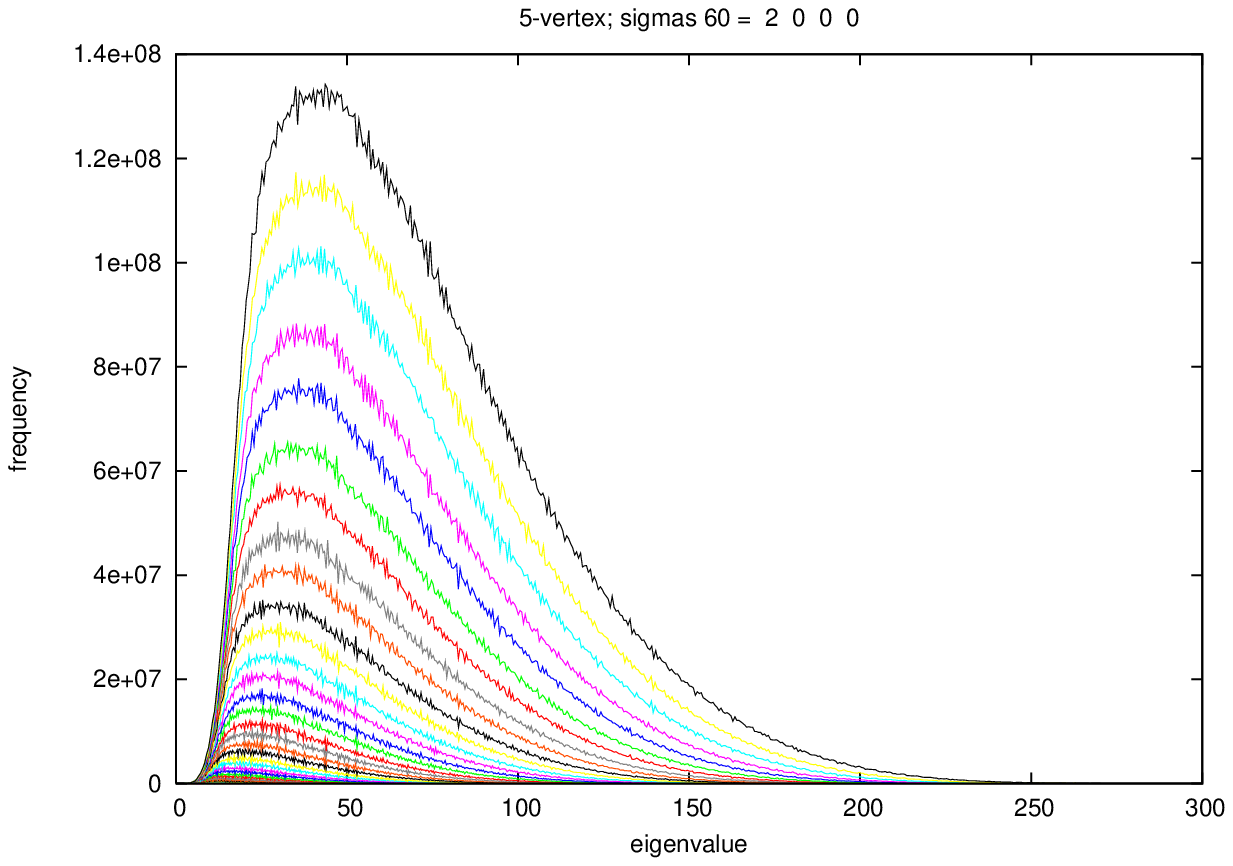}
\end{minipage}
\caption{Histograms for individual \sigconfs{} of the 5-vertex.}
\label{5v_fixed_sigma_conf_hists}
\end{figure}

To give a feel for the effect of the \sigconf, we show in figure
\ref{5v_fixed_sigma_conf_hists} a sampling of
histograms as in figure \ref{5 vertex 1}, but for various fixed \sigconfs.
The first three \sigconfs{} shown (which have at least one $\sigma=\pm 4$)
give a smallest non-zero eigenvalue which decreases with $\jmax$ (c.f.\
section \ref{5v_extremal_eigenvalues}).  The next two \sigconfs{} give a
smallest non-zero eigenvalue which increases with $\jmax$.  The last
histogram is for the \sigconf{} which gives a constant smallest non-zero
eigenvalue.  Some details regarding these histograms are given in table
\ref{5v_fixed_sigconf_hist_details}.  The $\chi_{\vec{\sigma}}$ is defined in
\S \ref{NumericalResults-Remarks}.  $N_{\mathrm{zeros}}$ is the total number
of zero eigenvalues which occur for each \sigconf{}.
The last three histograms look sharper with fewer bins because there are more
\signconfs{} corresponding to these \sigconfs, which causes more jitter in
the spectrum.
\begin{table}[htbp]
\center
\begin{tabular}{|rrrr|rrrc|}
\hline
\multicolumn{4}{|c|}{\sigconf} & $\chi_{\vec{\sigma}}$ & $N_{\mathrm{zeros}}$ & $\nbins$ & behavior \\
\hline
4 & -2 &  0 & 4 &  2 & 2,631,232,052 & 2048 & decreasing $\mineval$\\ 
2 & -4 & -2 & 0 &  2 & 1,155,745,232 & 2048 & decreasing $\mineval$\\ 
2 &  0 &  4 & 2 &  2 &   540,623,092 & 2048 & decreasing $\mineval$\\ 
0 &  0 &  2 & 2 & 12 & 4,298,843,832 &  512 & increasing $\mineval$\\ 
0 &  0 &  2 & 0 & 12 & 7,439,949,072 &  512 & increasing $\mineval$\\ 
2 &  0 &  0 & 0 & 12 & 9,473,449,392 &  512 &   constant $\mineval$\\ 
\hline
\end{tabular}
\caption{Details of some fixed \sigconf{} histograms for the 5-vertex.}
\label{5v_fixed_sigconf_hist_details}
\end{table}

Note how the \sigconf{} affects many aspects of the spectrum, including the
accumulation at zero, a dip in the spectral density at small eigenvalues
around 10, 
the presence of a secondary peak at 30 or so, the
presence of jagged edges for larger eigenvalues, and the magnitude of the largest eigenvalues which arise.

\begin{figure}[hbtp]
\center
\begin{minipage}{8.2cm}
  \center
  \psfrag{frequency}{$\numevals$}
  \psfrag{eigenvalue}{$\eval$}
  \psfrag{5-vertex; sigmas =  4 -2  0  4}{\large $\vec{\sigma}=(4~-\!2~~0~~4$)}
  \includegraphics[width=8.5cm]{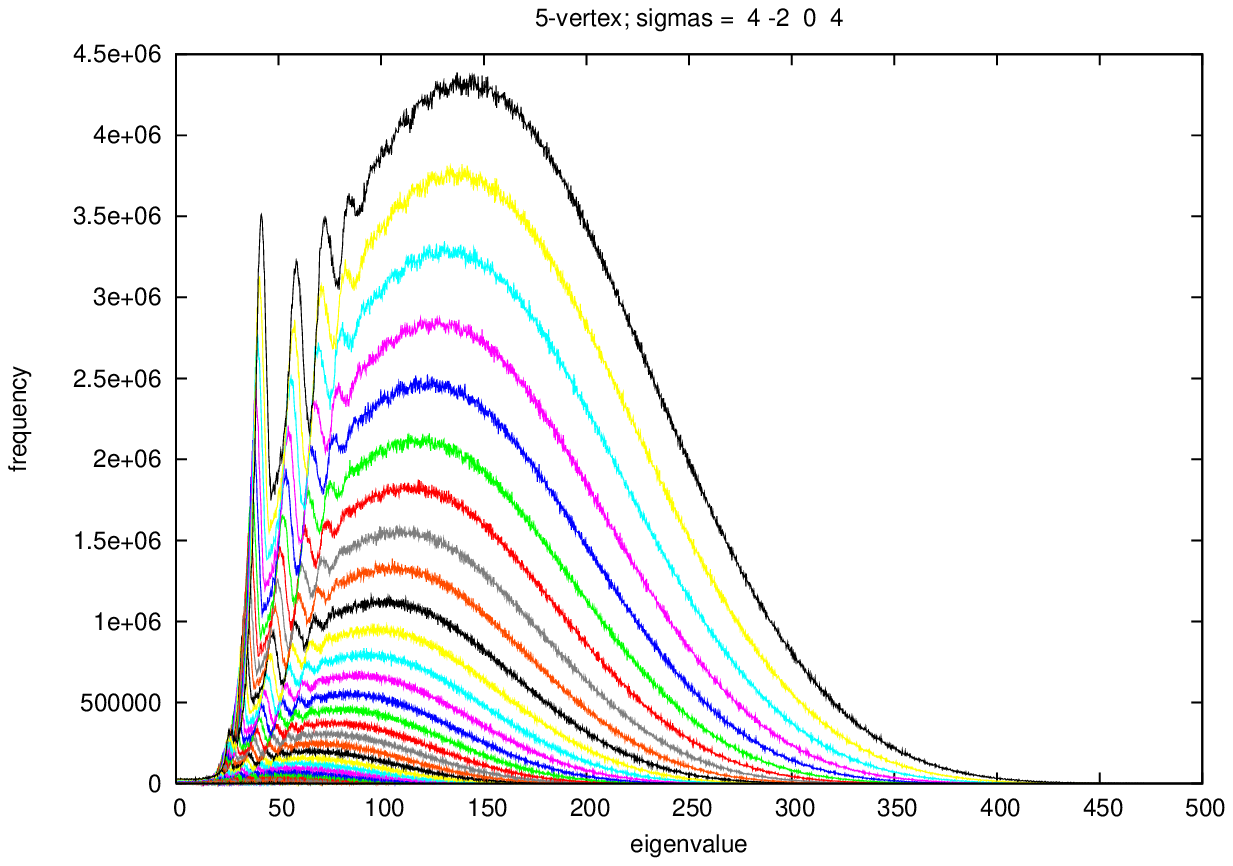}
\end{minipage}
~
\begin{minipage}{8.2cm}
    \center
    \psfrag{frequency}{$\numevals$}
    \psfrag{eigenvalue}{$\eval$}
    \psfrag{5-vertex; sigmas =  2 -4 -2  0}{\large $\vec{\sigma}=(2~-\!4~-\!2~~0$)}
    \includegraphics[width=8.5cm]{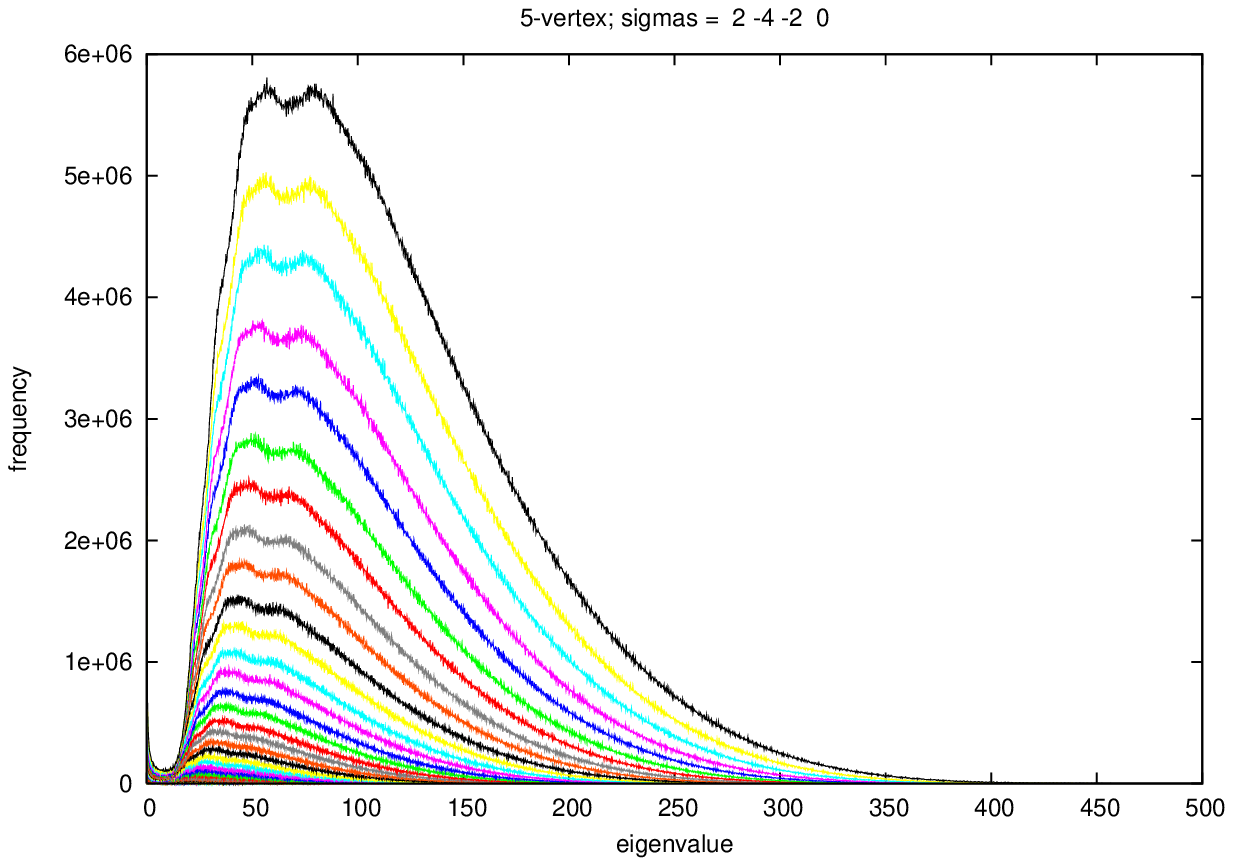}
\end{minipage}
\end{figure}
\begin{figure}[hbtp]
\center
\begin{minipage}{8.2cm}
    \center
    \psfrag{frequency}{$\numevals$}
    \psfrag{eigenvalue}{$\eval$}
    \psfrag{5-vertex; sigmas 37 =  2  0  4  2}{\large $\vec{\sigma}=(2~~0~~4~~2$)}
    \includegraphics[width=8.5cm]{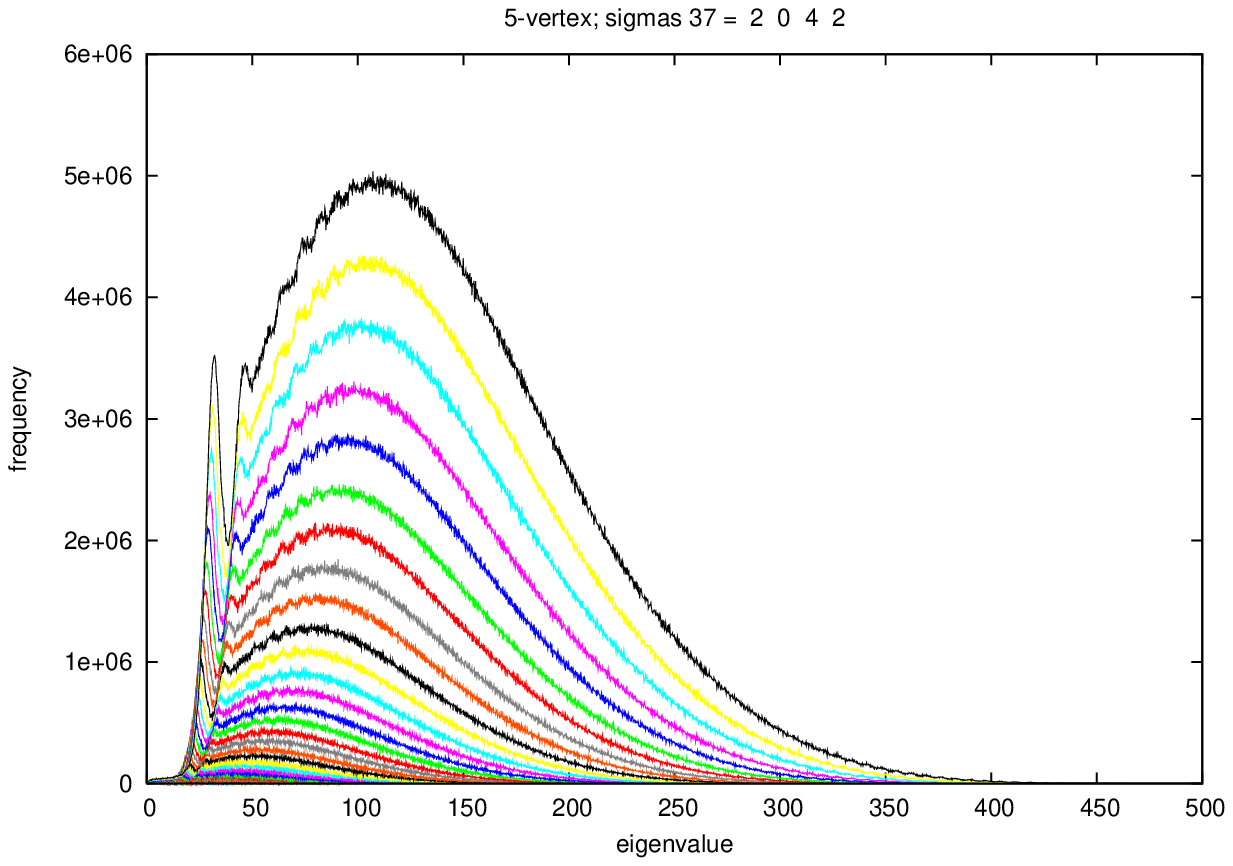}
\end{minipage}
~
\begin{minipage}{8.2cm}
  \center
    \psfrag{frequency}{$\numevals$}
    \psfrag{eigenvalue}{$\eval$}
    \psfrag{5-vertex; sigmas =  0  0  2  2}{\large $\vec{\sigma}=(0~~0~~2~~2$)}
    \includegraphics[width=8.5cm]{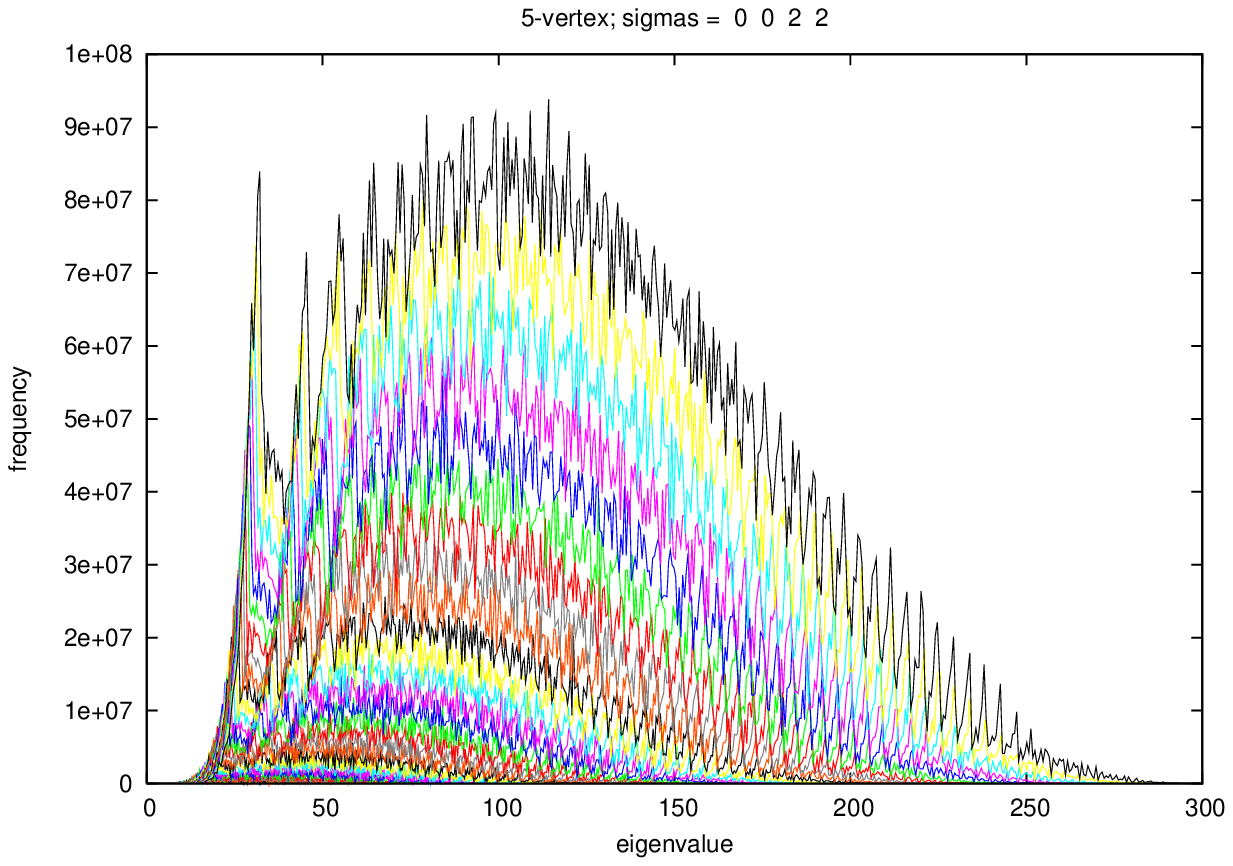}
\end{minipage}
\end{figure}

\begin{figure}[htb]
\begin{center}
  \psfrag{frequency}{$\numevals$}
  \psfrag{eigenvalue}{$\eval$}
  \includegraphics[width=17.7cm]{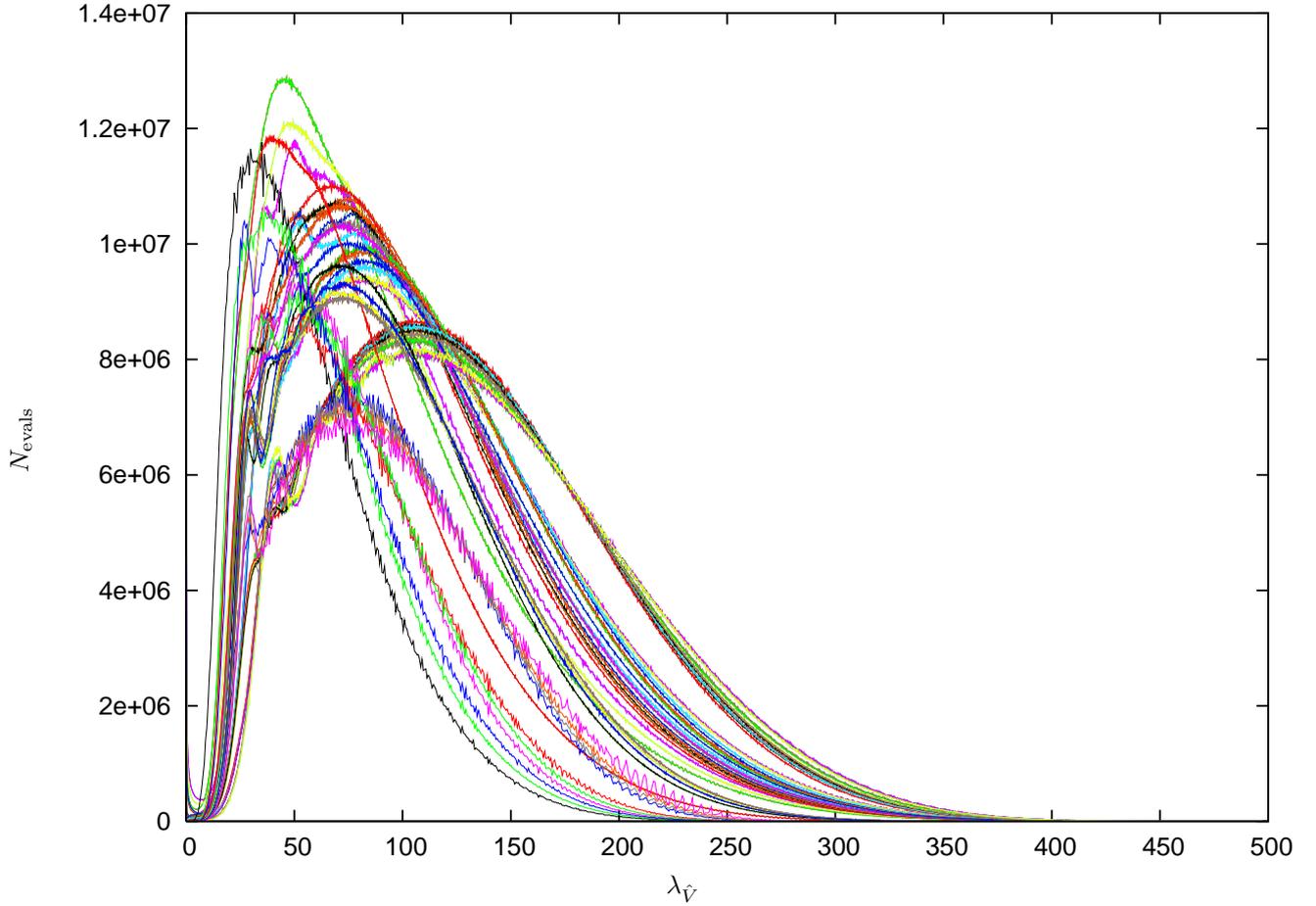}
\end{center}
\caption{Histogram for each of the eighty-five \sigconfs{} of the 5-vertex.}
\label{5v_all_sigconfs}
\end{figure}

In figure \ref{5v_all_sigconfs} we show the histograms for each \sigconf{},
without imposing any constraints on $\jmax$ (save that it is
$\leq\frac{44}{2}$).  Here we have also ignored the \sigconf{} redundancy
($\chi_{\vec{\sigma}}$ of \S \ref{NumericalResults-Remarks}), and weighted
each individual \sigconf{} equally.  Each histogram has 2048 bins, save those
of 
the \sigconfs{} corresponding to the $\vec\sigma$-indices 30, 32--34, 36, and
40 from table \ref{min_evals_sig_conf_equivalence_classes} of \S
\ref{5v_extremal_eigenvalues}, which have only 512.  These
later ten histograms have their $\numevals$ divided by 4, so they appear on the
same scale as the other 75 histograms of figure \ref{5v_all_sigconfs}.
This `coarse graining' improves the clarity of the figure, which is otherwise
obscured by the rapidly fluctuating $\numevals$ for these ten histograms.

\subsubsection{Fitting of Histograms} 
\label{fit_histogram5}

\begin{figure}[hbt!]
\center
    \center
    \psfrag{lambda}{$\eval$}
    \psfrag{l}{$\ln[\numevals]$}
    \psfrag{FiveVFitRisingEdgeLog}{}
    \includegraphics[height=6.2cm]{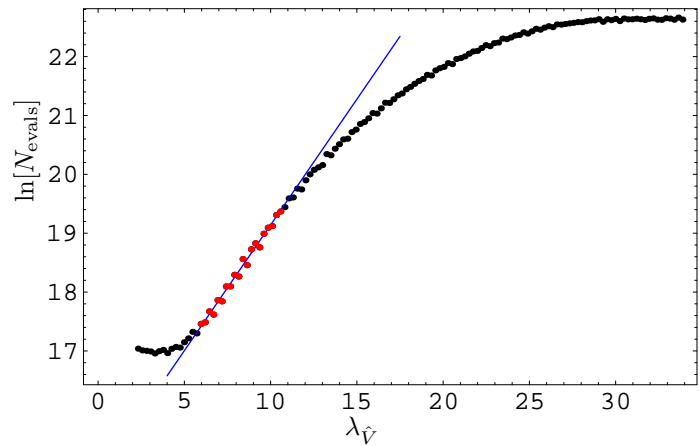}
    \caption{\label{5 vertex Fit 1} Logarithmic plot of the fit to the rising
    edge of the spectral density for the gauge invariant 5-vertex.  The
    points shown in red were used in the fitting process, the blue solid line is the fitted function.}
\end{figure}

The overall histogram for the 5-vertex, shown in figure \ref{5 vertex 1},
seems to possess an exponentially rising edge for eigenvalues around $\eval=$ 10 to 30 or
so.  Note that as we consider larger and larger $\jmax$, this edge does not
change much as compared to the increase in the remainder of the histogram,
save being extended to larger eigenvalues.  Thus this exponentially rising
edge may be a feature of the spectrum in the limit of $\jmax\to\infty$.
Figure \ref{5 vertex Fit 1} shows an exponential fit to this rising edge.  The range of data points to fit, as well as the form of the fitting function, are chosen arbitrarily,
as there is no directly applicable analytic model for the eigenvalue distribution available at the moment (though in \cite{Meissner} an expression for the 4-vertex was derived, see comments in \cite{NumVolSpecLetter}). 
Thus we would like to emphasize that the errors given for fits in the spectrum throughout this paper, although meaningful within the chosen set of data points, have to be taken in the context of the  
choice of fitting model and data point set we have made by hand.
Our point here is to demonstrate that the spectrum in principle allows for an exponentially increasing number of eigenvalues.
The fit gives the logarithm of the number of eigenvalues $\numevals(\eval)$ as a function of the eigenvalue $\eval$ by:
\be
   \ln[\numevals(\eval)]^{(fit)}=( 14.87 \pm 0.20) + (0.43\pm 0.02) \eval
   ~~~~~~~~~~\longrightarrow~~~~~~~~~~~
   \numevals^{(fit)}(\eval) \sim 2.87\cdot 10^6 \cdot \mb{e}^{0.43~\eval} 
\ee
The quantity $\chi^2:=\sum_{k=1}^{20} \big( \numevals^{(fit)}(\eval^{(k)}) - \numevals^{(k)}(\eval)\big)^2 $ has the numerical value $\chi^2=1.06\cdot10^{15}$ on the chosen set of 20 data points.

\subsubsection{Extremal Eigenvalues}
\label{5v_extremal_eigenvalues}

\begin{figure}[hbtp]
\center
\begin{minipage}{8cm}
    \center
    \psfrag{sci}{$\vec{\sigma}$-index}
    \psfrag{mineval}{$\mineval $}
    \psfrag{jmax}{$2 j_{max}$}
    \includegraphics[width=9cm]{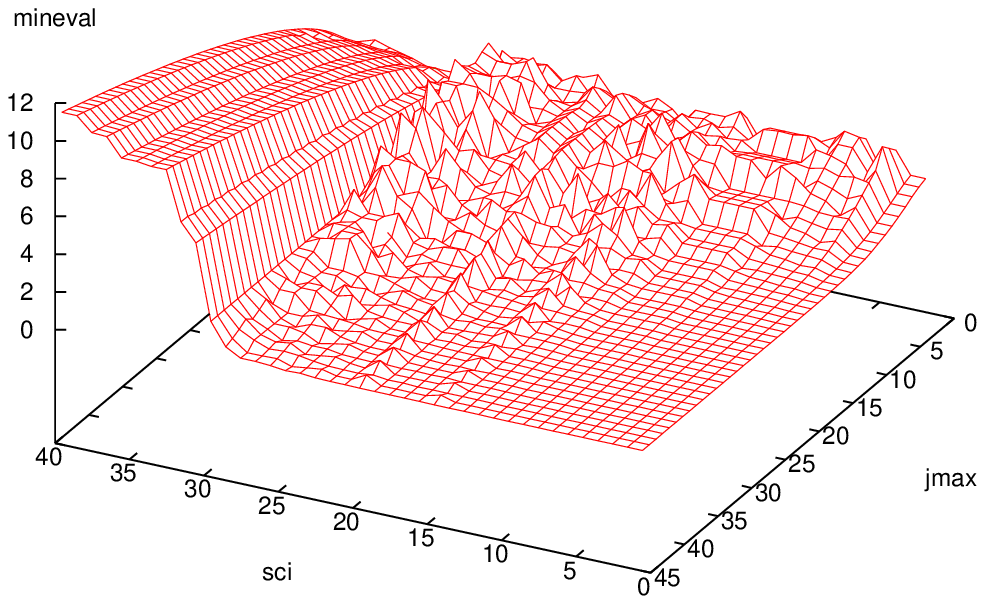}
    \caption{\label{5 vertex min}Smallest non-zero eigenvalues at the gauge
    invariant 5-vertex up to $j_{max}=\frac{44}{2}$. The $\vec{\sigma}$-index
    is defined in table \ref{min_evals_sig_conf_equivalence_classes}.}
\end{minipage}~~~~
\begin{minipage}{8cm}
    \center
    \psfrag{sci}{$\vec{\sigma}$-index}
    \psfrag{mineval}{$\maxeval $}
    \psfrag{jmax}{$2 j_{max}$}
    \includegraphics[width=9cm]{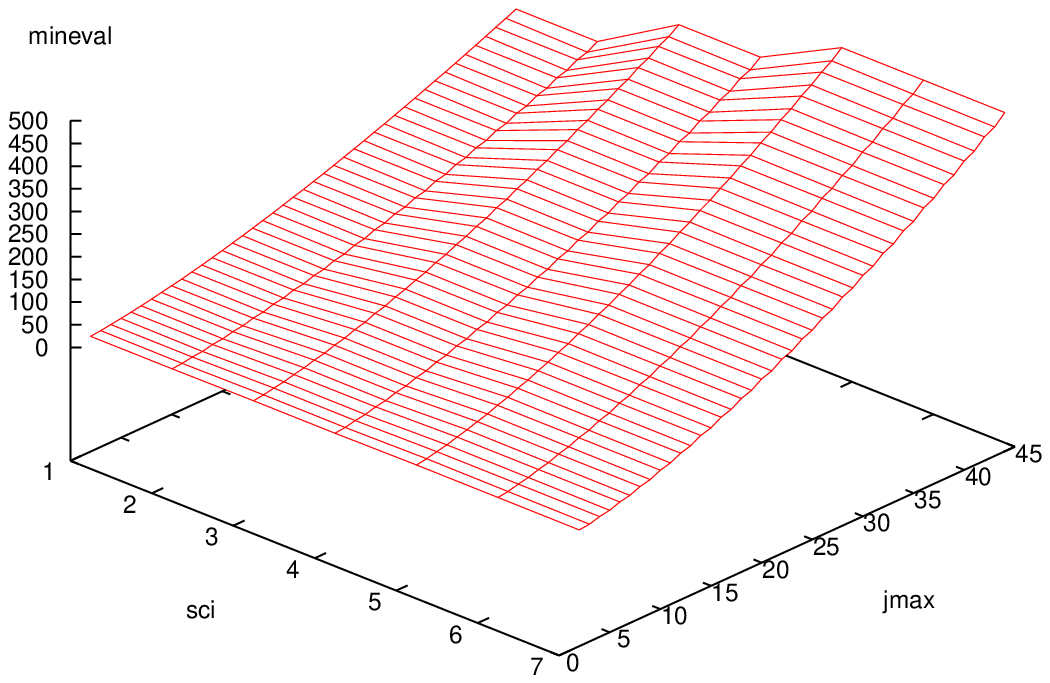}
    \caption{\label{5 vertex max}Largest Eigenvalues at the gauge invariant
    5-vertex up to $j_{max}=\frac{44}{2}$. The $\vec{\sigma}$-index is
    defined in table \ref{max_evals_sig_conf_equivalence_classes}.}
\end{minipage}
\end{figure}

We have seen in section \ref{5v_fixed_sig_conf_hists} that the \sigconf{} has a
considerable effect on the shape of the distribution of eigenvalues.  Here we
examine the largest and smallest eigenvalues which arise in the spectrum, as
a function of $\jmax$ and \sigconf.  Figures \ref{5 vertex min} and \ref{5
 vertex max} show the minimum and maximum non-zero eigenvalues of the
spectrum, for a given \sigconf{} and $\jmax$.

As it turns out from the numerical analysis many of the \sigconfs{} give identical minimum eigenvalues $\mineval$ for each value of
$\jmax$,
and we have separated them into equivalence classes,
such that any two \sigconfs{} which have the same $\mineval(\jmax)$ will lie in the
same equivalence class. 
The decision whether two  \sigconfs{} belong to the same equivalence class is made as follows: Along with the smallest eigenvalue data we have recorded the numerical error bound of each eigenvalue according to section \ref{Eval Computation}.
We take the maximum numerical error which occurs among all the computed smallest eigenvalues as a threshold and look at the absolute differences in an elementwise subtraction of two eigenvalue lists. If the maximum numerical value of all the differences does not exceed the threshold then we consider the two \sigconfs{} as elements of the same equivalence class. 
These equivalence classes, which are detailed in
tables \ref{min_evals_sig_conf_equivalence_classes} and
\ref{max_evals_sig_conf_equivalence_classes}, 
are used in the figures \ref{5 vertex min} and \ref{5 vertex max}.

By inspection of figure \ref{5 vertex min}, one can clearly see that some
\sigconfs{} give rise to a minimum eigenvalue $\minev$ which decreases with $\jmax$,
some give rise to a $\minev$ which increases with $\jmax$, and there is one in the
middle for which $\minev$ does not vary with $\jmax$.  These three behaviors
of the minimum eigenvalues serve to classify the \sigconfs{} into three groups.  In
the following table we count the number of \sigconfs{} in each of these
groups, and also the number of inequivalent sequences $\mineval(\jmax)$
which arise from the \sigconfs{} in each group.  (We will often refer to ordered sequences of eigenvalues, such as $\mineval(\jmax)$, $\maxeval(\jmax)$, as $\jmax\to\infty$, as \emph{eigenvalue sequences}.)
We find:\footnote{
Note that the list of equivalence classes given in tables
\ref{min_evals_sig_conf_equivalence_classes} and
\ref{max_evals_sig_conf_equivalence_classes} may be regarded as preliminary,
because the eigenvalue sequences $\mineval(\jmax)$ for two different
\sigconfs{} may match only up to some finite $\jmax$.  
As observed already in \cite{Volume Paper}, the structure of the volume spectrum becomes richer as we increase the cutoff $\jmax$, in particular new eigenvalue sequences may show up.  Thus the number of equivalence classes could in principle increase at yet larger $\jmax$.}
\begin{center}
\begin{tabular}{|l|l||r|}
  \hline
  increasing $\sigma$-conf. & equiv. classes & 10\\\cline{2-3}
                            & total          & 24\\\hline
  decreasing $\sigma$-conf. & equiv. classes & 29\\\cline{2-3}
                            & total          & 60\\\hline
  constant $\sigma$-conf.   & equiv. classes & 1\\\cline{2-3}
                            & total          & 1\\\hline
\end{tabular}
\end{center}
The analysis of the data for the smallest non-zero eigenvalues reveals that increasing smallest eigenvalues are contributed from $\vec{\sigma}$-configurations containing only $(0,\pm 2)$ but not $\pm4$, whereas decreasing eigenvalues are contributed by $\vec{\sigma}$-configurations containing $(0,\pm 2,\pm 4)$. Table \ref{min_evals_sig_conf_equivalence_classes} lists the equivalence classes of $\vec{\sigma}$-configurations showing a particular behavior for $\mineval(\jmax)$ in detail.  Recall the definition $\vec{\sigma}:=\{\sigma(123),\sigma(124),\sigma(134),\sigma(234)\}$ where the arguments of  $\sigma(IJK)$ are the edge labels for the particular edge triple $(e_I,e_J,e_K)$.  
We have assigned an index `$\vec{\sigma}$-index' to every equivalence class of $\vec{\sigma}$-configurations which relates it to the according curve in figure \ref{5 vertex min}.

\begin{table}[htbp]
\center
\begin{footnotesize}
\begin{tabular}{cccccc}
\cmt{3}{
\[\begin{array}{|l||rrrr|}
       \hline 
       \cmt{0.8}{$\vec{\sigma}-$\\ index } & \cmt{0.5}{\mbox{Decreasing}\\$ \vec{\sigma}$\mbox{-config}.} &&&
       \\\hline
        1 &4 & 2 & 2 & 0 \\
	  &4 & -2 & -2 & 0
       \\\hline
        2 &4 & 4 & 2 & 0 \\
	  &4 & 4 & -2 & 0 \\
	  &2 & -4 & -2 & 0 \\
	  &2 & 4 & -2 & 0
	\\\hline
	3 &2 & 2 & 4 & 0 \\
	  &2 & 2 & -4 & 0
	\\\hline
        4 &4 & -2 & 0 & 2 \\
	  &4 & 2 & 0 & -2
	\\\hline
	5 &4 & 2 & -4 & 0 \\
	  &4 & -2 & -4 & 0
	\\\hline
	6 & 2 & 4 & 4 & 0 \\
	  &2 & -4 & -4 & 0
	\\\hline
	7 & 4 & 0 & 2 & 2 \\
	  & 4 & 0 & -2 & -2
	\\\hline
	8 & 2 & -4 & -2 & 4 \\
          &2 & 4 & -2 & -4
	\\\hline
	9 & 0 & 4 & -2 & -4 \\
	  &0 & 4 & 2 & -4
	\\\hline
	10& 2 & -4 & -4 & 2 \\
	  &2 & 4 & 4 & 2
	\\\hline
	11& 0 & 4 & 4 & -2 \\
	  & 0 & 4 & 4 & 2
	\\\hline
	12& 2 & -4 & 0 & 2 \\
          & 2 & 4 & 0 & 2
	\\\hline
	13& 2 & 2 & 0 & 4 \\
	  & 2 & 2 & 0 & -4
	\\\hline
	14& 4 & 0 & -4 & -2 \\
	  & 4 & 0 & -4 & 2
	\\\hline
\end{array}\] }
&
\cmt{3}{
\[\begin{array}{|l||rrrr|}
       \hline 
       \cmt{0.8}{$\vec{\sigma}-$\\ index } & \cmt{0.5}{\mbox{Decreasing}\\$ \vec{\sigma}$\mbox{-config}.} &&&
       \\\hline	
	15& 4 & -2 & -4 & 2 \\
	  &4 & 2 & -4 & -2
	\\\hline
	16& 2 & 0 & -2 & 4 \\
	  &2 & 0 & -2 & -4
	\\\hline
	17& 4 & 4 & 0 & -2 \\
	  &4 & 4 & 0 & 2
	\\\hline
	18& 2 & 0 & -4 & 2 \\
	  & 2 & 0 & 4 & 2
	\\\hline
	19& 0 & 4 & 2 & 2 \\
	  & 0 & 4 & -2 & -2
	\\\hline
	20& 4 & -2 & -2 & 4 \\
	  & 4 & 2 & 2 & 4
	\\\hline
	21& 0 & 2 & 4 & -2 \\
	  & 0 & 2 & -4 & -2
	\\\hline
	22& 0 & 2 & 2 & -4 \\
	  & 0 & 2 & 2 & 4
	\\\hline
	23& 4 & 4 & -2 & -2 \\
	  & 4 & 4 & 2 & 2
	\\\hline
	24& 4 & -2 & 0 & 4 \\
	  & 4 & 2 & 0 & 4
	\\\hline
	25& 4 & 0 & -2 & 4 \\
	  & 4 & 0 & 2 & 4
	\\\hline
	26& 2 & 4 & 0 & -4 \\
	  & 2 & -4 & 0 & 4
	\\\hline
	27& 0 & 2 & -4 & -4 \\
	  & 0 & 2 & 4 & 4
	\\\hline
	28& 2 & 2 & 4 & 4 \\
	  & 2 & 2 & -4 & -4
	\\\hline
	29& 2 & 0 & -4 & -4 \\
	  & 2 & 0 & 4 & 4\\\hline
\end{array}\] }
&~~~~&
\cmt{3}{
\[\begin{array}{|l||rrrr|}
       \hline 
       \cmt{0.8}{$\vec{\sigma}-$\\ index } & \cmt{0.5}{\mbox{Constant}\\$ \vec{\sigma}$\mbox{-config}.} &&&
       \\\hline
       30&2 & 0 & 0 & 0\\\hline 
\end{array}\] }
&~~~~&
\cmt{3}{
\[\begin{array}{|l||rrrr|}
       \hline 
       \cmt{0.8}{$\vec{\sigma}-$\\ index } & \cmt{0.5}{\mbox{Increasing}\\$ \vec{\sigma}$\mbox{-config}.} &&&
       \\\hline
       31& 2 & 2 & -2 & 0 \\
	 &2 & 2 & 2 & 0 \\
	 &2 & -2 & -2 & 0
       \\\hline
       32&0 & 2 & 0 & 0
       \\\hline
       33&0 & 0 & 0 & 2
       \\\hline
       34&0 & 0 & 2 & 0
       \\\hline
       35&2 & -2 & 0 & 2 \\
	 &2 & 2 & 0 & -2 \\
	 &2 & 2 & 0 & 2
       \\\hline
       36&0 & 2 & 0 & -2 \\
	 &2 & 0 & 0 & 2 \\
	 &0 & 2 & 2 & 0 \\
	 &2 & 0 & -2 & 0 \\
	 &2 & 2 & 0 & 0
       \\\hline
       37&0 & 2 & 2 & -2 \\
	 &0 & 2 & -2 & -2 \\
	 &0 & 2 & 2 & 2
       \\\hline
       38&2 & -2 & -2 & 2 \\
	 &2 & 2 & 2 & 2 \\
	 &2 & 2 & -2 & -2
       \\\hline
       39& 2 & 0 & -2 & 2 \\
	 &2 & 0 & -2 & -2 \\
	 &2 & 0 & 2 & 2
       \\\hline
       40&0 & 0 & 2 & 2
       \\\hline
\end{array}\] }
\end{tabular}
\end{footnotesize}
\caption{Equivalence classes of \sigconfs{} with regard to $\minev(\jmax)$.
  The \sigconfs{} are grouped in accordance with the behavior of the
  smallest eigenvalues.}
\label{min_evals_sig_conf_equivalence_classes}
\end{table}

\begin{figure}[hbtp]
 \center
    \psfrag{2}{$2 j_{max}$}
    \psfrag{jmax}{}
    \psfrag{M}{$\mineval$}
    \psfrag{5}{}
    \psfrag{V12firstMinEvalSeriesLogPlot}{}
    \includegraphics[width=16cm]{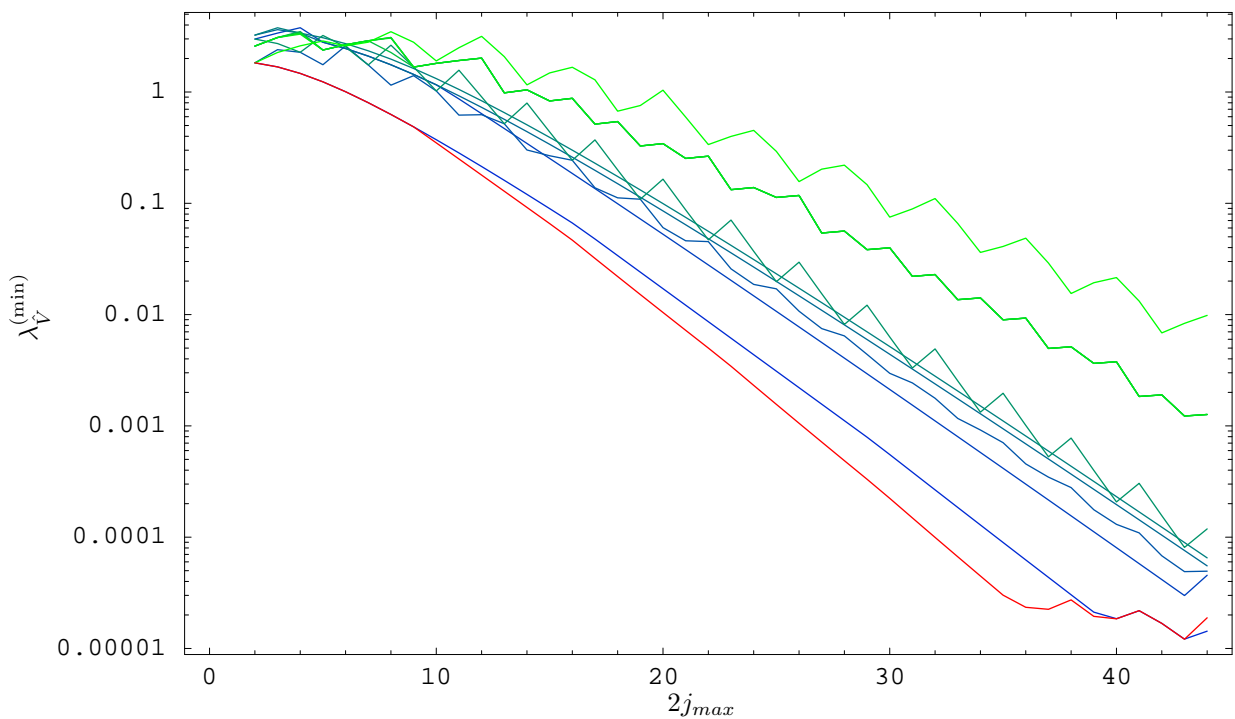}
    \caption{\label{5 vertex smallest 12 eval series}The twelve smallest eigenvalue sequences at the gauge invariant 5-vertex. The red line indicates $\vec{\sigma}_{min}$.}
\end{figure}

The fact that there is a $\vec\sigma$-configuration giving constant smallest eigenvalues, independent of the value of $j_{max}$, seems to be puzzling,  
however this can be easily understood when we realize that this is the configuration 
$\vec{\sigma}:=\{\sigma(123),\sigma(124),\sigma(134),\sigma(234)\} = \{2, 0, 0, 0\}$, as follows.
If we look at equation (\ref{q123}), which gives the matrix elements for
$q_{123}$, then the  corresponding matrix only depends on the spins $j_1, j_2
, j_3$  and the intermediate recoupling $a_2$ and $a_3$. So it is effectively
the matrix of the gauge invariant 4-vertex. Now even if $j_5$ becomes large, there still
exist spin configurations with a large $j_4$. Recall that the intermediate recouplings
in a recoupling scheme at the gauge invariant 5-vertex are defined as:
$a_2(j_1~j_2), a_3(a_2~j_3), a_4(a_3~j_4)\stackrel{!}{=}j_5$
where one has to apply the rules for the recoupling according to section \ref{Computational Realization of the Recoupling Scheme Basis}.
Now the smallest absolute numerical value for eigenvalues of the gauge invariant 
4-vertex was found in \cite{Volume Paper}  to be given by a configuration with
$j_1\sim j_2 \sim 1$ and  $j_3\sim a_3=j_4 \sim j_{max}$.\footnote{This result is proved below in section \ref{Lower Bound}.}
But now we have this case embedded in a 5-vertex.
This means that if $j_3$ is tiny then $a_3$ can be tiny as well 
(since $j_1\le j_2\le j_3\le j_4\le j_5$) whereas 
it might still happen that $j_3+j_4 \sim j_5$.
This case can occur for every $j_5=j_{max}$, since we can still freely choose
$j_4~j_5$ (because we consider all $j_1,\ldots,j_4$-combinations up to $j_5$).
So we always get the same value.  
In the case of the 5-vertex, since the number of spins is odd,
there is a sort of `struggling' going on in the spin configuration which gives rise to the smallest eigenvalues, due to the condition 
that the sum of the 5 spins must be an integer
(because otherwise we cannot recouple them to resulting zero spin).
Thus different spin combinations can contribute the smallest eigenvalue: At the gauge invariant 4-vertex we saw that the smallest eigenvalue is contributed by spin configurations $j_1=j_2=\frac{1}{2}$ and $j_3=j_4=j_{max}$. Now assume a similar statement is true for the gauge invariant 5-vertex, for example\footnote{This is just a fictive example to illustrate our point.} something like $j_1=j_2=\frac{1}{2}$ and $j_3=j_4=j_5=j_{max}$. Then it is easy to see that while in case of the 4-vertex it does not matter if $j_{max}$ is integer or half integer, this does matter at the 5-vertex. The combination here can only be realized if $j_{max}$ is integer, otherwise there is no recoupling to resulting spin 0 possible. Thus for $j_{max}$ half integer, a different spin combination will contribute the smallest non-zero eigenvalues. This is nicely demonstrated by the oscillating curves in figure  \ref{5 vertex smallest 12 eval series}.

\paragraph{Overall smallest eigenvalues}
As can be seen from the analysis of the data, the overall smallest non-zero eigenvalue, for each value of $\jmax$, is
always contributed by
$\vec{\sigma}:=\{\sigma(123),\sigma(124),\sigma(134),\sigma(234)\}=\{4, 2, 2, 0\}\equiv\{4, -2, -2, 0\}$ $=:\vec{\sigma}_{min}$.
It is contributed by the spin configurations listed in table \ref{5v smallest eval spin config} (here $j_5=\jmax$).
\begin{table}[htb]
\center
\begin{footnotesize}
\begin{tabular}{|r|r|r|r|r|}
\hline
$2\cdot j_1$&$2\cdot j_2$&$2\cdot j_3$&$2\cdot j_4$&$2\cdot j_5$\\\hline
1&1&2&2&2\\
1&2&3&3&3\\
1&3&4&4&4\\
1&4&5&5&5\\
1&5&6&6&6\\
1&6&7&7&7\\
1&7&8&8&8\\
1&8&9&9&9\\
3&8&9&10&10\\
3&9&10&11&11\\
3&10&11&12&12\\
3&11&12&13&13\\
3&12&13&14&14\\
3&13&14&15&15\\
5&13&14&16&16\\
5&14&15&17&17\\
5&15&16&18&18\\
5&16&17&19&19\\
5&17&18&20&20\\
5&18&19&21&21\\
5&19&20&22&22\\
7&19&20&23&23\\
7&20&21&24&24\\
7&21&22&25&25\\
7&22&23&26&26\\
7&23&24&27&27\\
7&24&25&28&28\\
7&25&26&29&29\\
9&25&26&30&30\\
9&26&27&31&31\\
9&27&28&32&32\\
9&28&29&33&33\\
9&29&30&34&34\\
9&30&31&35&35\\
7&32&33&36&36\\
5&34&35&37&37\\
3&36&37&38&38\\
3&37&38&39&39\\
3&37&40&40&40\\
5&37&38&41&41\\
1&41&42&42&42\\
1&42&43&43&43\\
3&40&43&44&44\\\hline
\end{tabular}
\end{footnotesize}
\caption{The spin configurations contributing the smallest eigenvalues for $\vec{\sigma}_{min}$.}
\label{5v smallest eval spin config}
\end{table}

\begin{figure}[hbtp]
\center
\begin{minipage}[t]{8.6cm}
    \psfrag{2jmax}{$2 j_{max}$}
    \psfrag{mineval}{$\mineval$}
    \includegraphics[width=8.8cm]{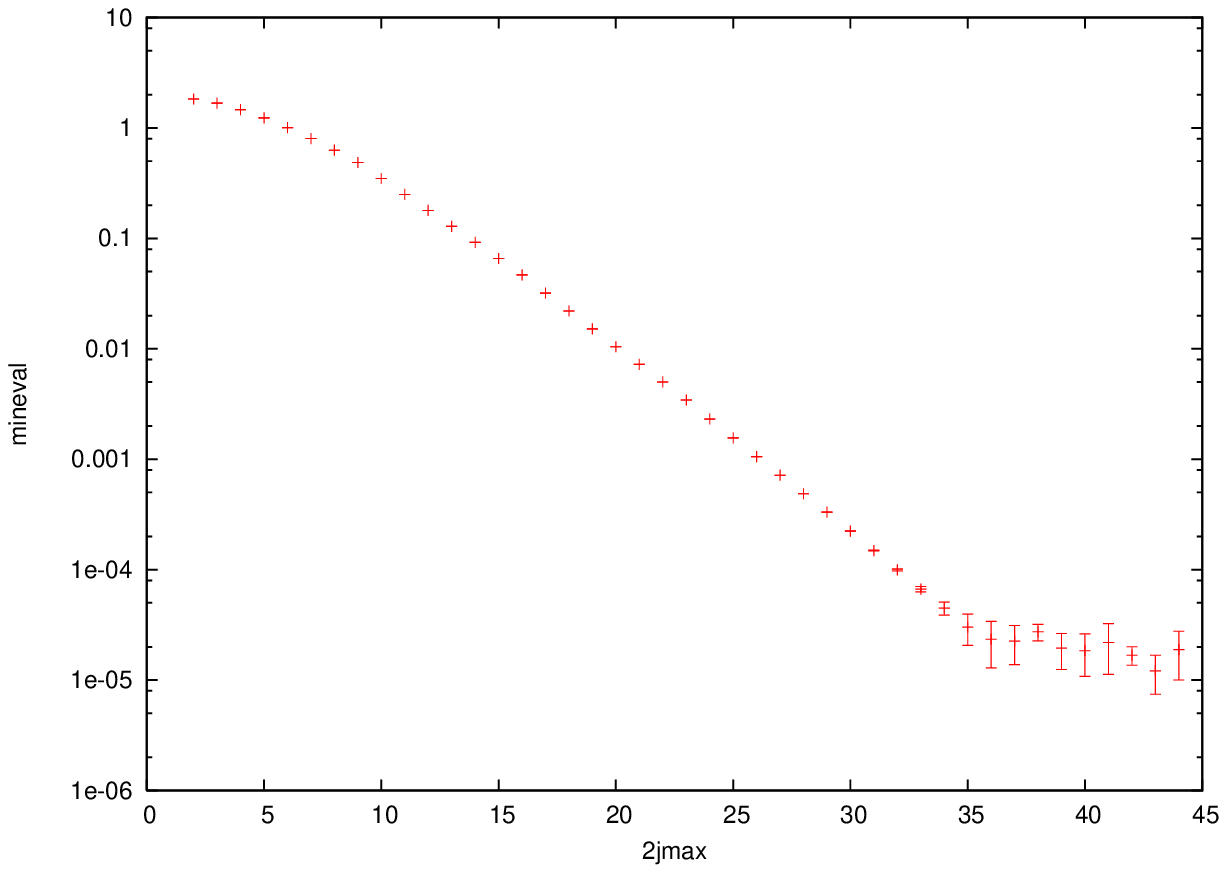}
    \caption{\label{5 vertex smallest evals}Smallest eigenvalue sequence at the gauge invariant 5-vertex, contributed by $\vec{\sigma}_{min}:=\{\sigma(123),\sigma(124),\sigma(134),\sigma(234)\}=\{4, 2, 2, 0\}$.  This is $\vec\sigma$-index number 1 in figure \ref{5 vertex min}.}    
\end{minipage}
~~
\begin{minipage}[t]{8.6cm}
    \psfrag{Twojmax}{$2 j_{max}$}
    \psfrag{M}{$\ln[\mineval]$}
    \psfrag{5}{}
    \psfrag{V12firstMinEvalSeriesLogPlot}{}
    \includegraphics[width=8.8cm]{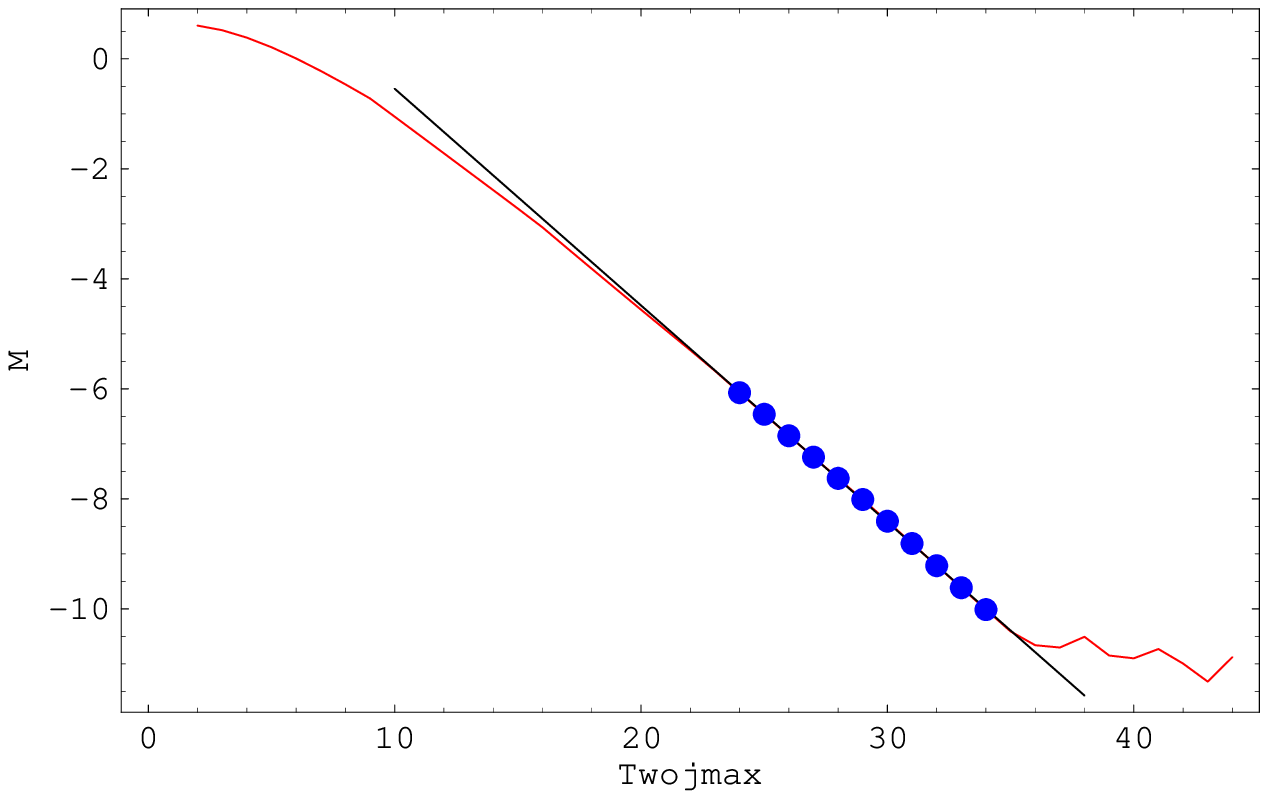}
    \caption{\label{5 vertex smallest eval series fit} Fit of the logarithm $\ln[\mineval]$ of the smallest eigenvalue sequence of figure \ref{5 vertex smallest evals}. We have used $\mineval(\jmax)$ for $\jmax=\frac{24}{2},\ldots \frac{34}{2}$ for the fitting (blue dots). The red line indicates the eigenvalue sequence of $\vec{\sigma}_{min}$, the black line the fit function.} 
\end{minipage}
\end{figure}

These minimum eigenvalues are plotted in figure \ref{5 vertex smallest evals}, with the eigenvalue axis in log scale.  Note that the eigenvalues become comparable to the numerical noise at around $\jmax=\frac{35}{2}$. This can also be seen in the breaking of the spin-pattern in table \ref{5v smallest eval spin config} for $j_5=\jmax>\frac{35}{2}$.
By inspection of figure \ref{5 vertex smallest evals} one can extract information on the behavior of $\mineval$ depending on $\jmax$. A linear fit for a chosen set of data (see caption of figure \ref{5 vertex smallest eval series fit}) reveals that 
$\lambda_{\hat{V}}^{(min),(fit)}(2\cdot\jmax) \approx \mb{e}^{(3.395\pm 0.073)}~ \mb{e}^{- (0.394\pm 0.003 ) 2\cdot\jmax}$. We find that $\chi^2:=\sum_{k=1}^{11}\big(\lambda_{\hat{V}}^{(min),(fit)}(\jmax^{(k)})-\mineval(\jmax^{(k)})\big)^2=7.7\cdot10^{-10}$ for the chosen set of 11 data points.

In order to show the generic behavior of the smallest eigenvalue sequences we have included figure
\ref{5 vertex smallest 12 eval series}.  Note that the eigenvalue sequences are not always clearly separated from each other numerically,
even when the eigenvalues are large compared to the numerical errors. They are also not always monotonically decreasing. This can be traced back to the following two reasons: Firstly certain eigenvalue subsequences show up only above a certain value of $\jmax$, therefore we can expect a `change of the smallest eigenvalue sequence'  
as $\jmax$ increases. Secondly, gauge invariance demands the sum of the five spins be an integer. Hence the spin configuration which gives rise to the matrix contributing the smallest eigenvalue can be forced to change, depending on the parity of $\jmax$:
we may `jump' between different eigenvalue sequences which contribute the smallest eigenvalue. These effects give rise to a sort of `oscillating' curve, whose enveloping curve nevertheless still decreases.  
 As the numerical noise is approached at $\jmax>\frac{35}{2}$ a clean distinction between the smallest two eigenvalue sequences is lost. 
Comparison with the numerical errors shown in figure \ref{5 vertex smallest evals}, and figure \ref{5v_min_eval_error1th-zoom}, indicates that this is caused by numerical noise.

\begin{table}[htbp]
\center
\begin{footnotesize}
\center
\begin{tabular}{ccccccccc}
\cmt{3}{
\[\begin{array}{|l||rrrr|}
       \hline 
       \cmt{0.8}{$\vec{\sigma}-$\\ index } & \cmt{0.5}{\mbox{ }\\$ \vec{\sigma}$\mbox{-config}.} &&&
       \\\hline
       1&0 & 0 & 2 & 2 \\
	&0 & 2 & 0 & -2 \\
	&2 & 0 & 0 & 2 \\
	&0 & 0 & 0 & 2
       \\\hline
       2&0 & 2 & 2 & 0 \\
	&2 & 0 & -2 & 0 \\
	&0 & 0 & 2 & 0 \\
	&2 & 2 & 0 & 0 \\
	&0 & 2 & 0 & 0 \\
	&2 & 0 & 0 & 0
       \\\hline
       3&2 & -2 & -2 & 2 \\
	&2 & 2 & 2 & 2 \\
	&2 & 2 & -2 & -2 \\
	&0 & 2 & 2 & -2 \\
	&0 & 2 & -2 & -2 \\
	&0 & 2 & 2 & 2 \\
	&2 & 0 & -2 & 2 \\
	&2 & 0 & -2 & -2 \\
	&2 & 0 & 2 & 2 \\
	&2 & -2 & 0 & 2 \\
	&2 & 2 & 0 & -2 \\
	&2 & 2 & 0 & 2
       \\\hline
       4&2 & 2 & -2 & 0 \\
	&2 & 2 & 2 & 0 \\
	&2 & -2 & -2 & 0
       \\\hline
       \end{array}\] }
&~~~~&
\cmt{3}{
\[\begin{array}{|l||rrrr|}
       \hline 
       \cmt{0.8}{$\vec{\sigma}-$\\ index } & \cmt{0.5}{\mbox{ }\\$ \vec{\sigma}$\mbox{-config}.} &&&
       \\\hline
       5&2 & 2 & 4 & 4 \\
	&2 & 2 & -4 & -4 \\
	&0 & 2 & -4 & -4 \\
	&0 & 2 & 4 & 4 \\
	&2 & 0 & -4 & -4 \\
	&2 & 0 & 4 & 4 \\
	&2 & -4 & -2 & 4 \\
	&2 & 4 & -2 & -4 \\
	&0 & 4 & -2 & -4 \\
	&0 & 4 & 2 & -4 \\
	&4 & -2 & -2 & 4 \\
	&4 & 2 & 2 & 4 \\
	&2 & 4 & 0 & -4 \\
	&2 & -4 & 0 & 4 \\
	&4 & 0 & -2 & 4 \\
	&4 & 0 & 2 & 4 \\
	&0 & 2 & 2 & -4 \\
	&0 & 2 & 2 & 4 \\
	&4 & -2 & 0 & 4 \\
	&4 & 2 & 0 & 4 \\
	&2 & 0 & -2 & 4 \\
	&2 & 0 & -2 & -4 \\
	&2 & 2 & 0 & 4 \\
	&2 & 2 & 0 & -4
       \\\hline
       \end{array}\] }
&~~~~&
\cmt{3}{
\[\begin{array}{|l||rrrr|}
       \hline 
       \cmt{0.8}{$\vec{\sigma}-$\\ index } & \cmt{0.5}{\mbox{ }\\$ \vec{\sigma}$\mbox{-config}.} &&&
       \\\hline
       6&2 & -4 & -4 & 2 \\
	&2 & 4 & 4 & 2 \\
	&0 & 4 & 4 & -2 \\
	&0 & 4 & 4 & 2 \\
	&4 & -2 & -4 & 2 \\
	&4 & 2 & -4 & -2 \\
	&4 & 0 & -4 & -2 \\
	&4 & 0 & -4 & 2 \\
	&0 & 2 & 4 & -2 \\
	&0 & 2 & -4 & -2 \\
	&4 & 4 & -2 & -2 \\
	&4 & 4 & 2 & 2 \\
	&2 & 0 & -4 & 2 \\
	&2 & 0 & 4 & 2 \\
	&0 & 4 & 2 & 2 \\
	&0 & 4 & -2 & -2 \\
	&4 & 4 & 0 & -2 \\
	&4 & 4 & 0 & 2 \\
	&4 & 0 & 2 & 2 \\
	&4 & 0 & -2 & -2 \\
	&2 & -4 & 0 & 2 \\
	&2 & 4 & 0 & 2 \\
	&4 & -2 & 0 & 2 \\
	&4 & 2 & 0 & -2
       \\\hline
       \end{array}\] }
&~~~~&
\cmt{3}{
\[\begin{array}{|l||rrrr|}
       \hline 
       \cmt{0.8}{$\vec{\sigma}-$\\ index } & \cmt{0.5}{\mbox{ }\\$ \vec{\sigma}$\mbox{-config}.} &&&
       \\\hline
       7&2 & 4 & 4 & 0 \\
	&2 & -4 & -4 & 0 \\
	&4 & 2 & -4 & 0 \\
	&4 & -2 & -4 & 0 \\
	&2 & 2 & 4 & 0 \\
	&2 & 2 & -4 & 0 \\
	&4 & 4 & 2 & 0 \\
	&4 & 4 & -2 & 0 \\
	&2 & -4 & -2 & 0 \\
	&2 & 4 & -2 & 0 \\
	&4 & 2 & 2 & 0 \\
	&4 & -2 & -2 & 0\\\hline
\end{array}\] }
\end{tabular}
\end{footnotesize}
\caption{Equivalence classes of \sigconfs{} with regard to $\maxev(\jmax)$.}
\label{max_evals_sig_conf_equivalence_classes}
\end{table}

\paragraph{Overall largest eigenvalues}

In the same way as for the minimum eigenvalues, we place the \sigconfs{} into
equivalence classes according to the maximum eigenvalue sequence
$\maxev(\jmax)$.  In this case there are only seven equivalence classes, out
of the 85 \sigconfs.  They are listed in table
\ref{max_evals_sig_conf_equivalence_classes}.  The $\vec\sigma$-indices
listed there are used to plot the maximum eigenvalue sequences $\maxev(\jmax)$
in figure \ref{5 vertex max}.

\begin{figure}[hbtp]
\center
\begin{minipage}[t]{8.0cm}
    \psfrag{2jmax}{$2 j_{max}$}
    \psfrag{maxev}{$\maxeval$}
    \psfrag{{4,2,2,0}}{$\vec{\sigma}_{min}$}
    \includegraphics[width=8.5cm]{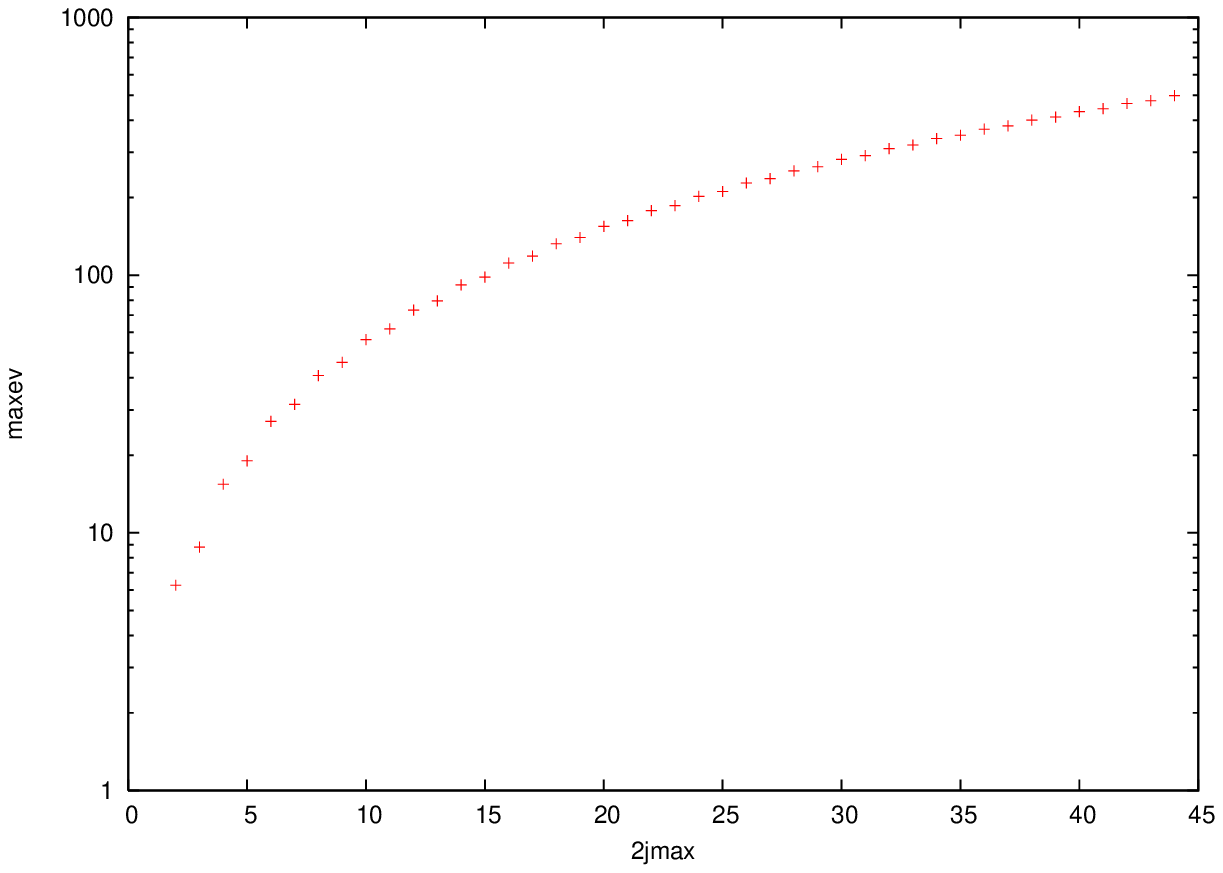}
    \caption{\label{5 vertex largest evals}Largest eigenvalue sequence at the gauge invariant 5 vertex contributed by  $\vec{\sigma}_{min}:=\{\sigma(123),\sigma(124),\sigma(134),\sigma(234)\}$ contained in the equivalence class with index number 7 in figure \ref{5 vertex max}.}
\end{minipage}
~~
\begin{minipage}[t]{8.0cm}
    \psfrag{Twojmax}{\cmt{2}{\vspace{0.5mm}$\ln[2 j_{max}]$}}
    \psfrag{M}{$\ln[\maxeval]$}
    \includegraphics[width=8.5cm]{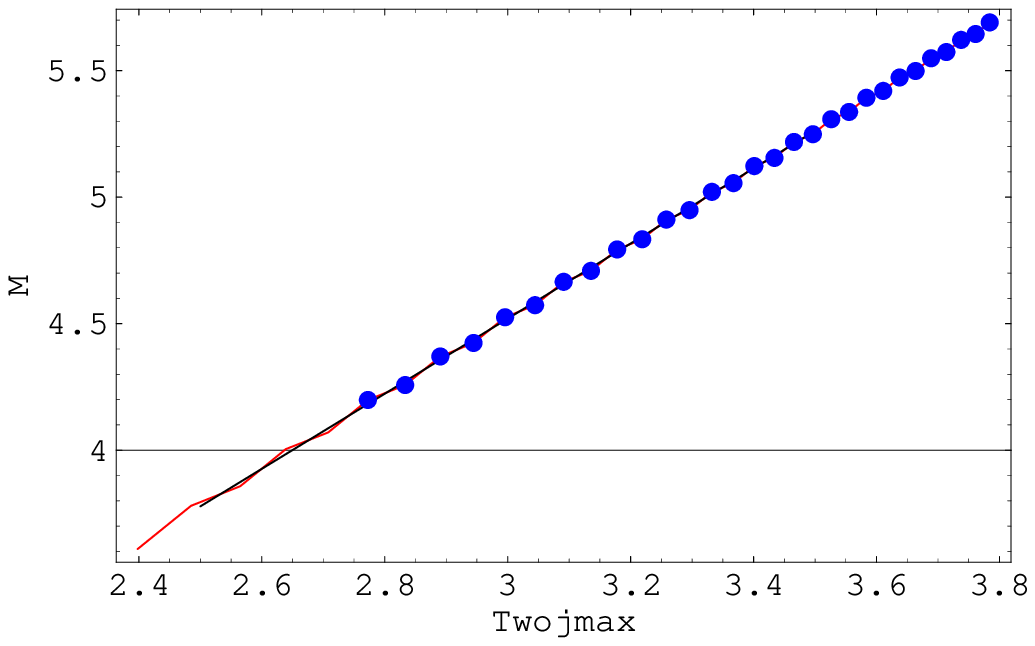}
    \caption{\label{5 vertex largest eval fit}Fit of the logarithm $\ln[\maxeval]$ of the largest eigenvalue sequence of figure \ref{5 vertex largest evals} versus $\ln[2 \jmax]$. We have used $\maxeval(\jmax)$ for $\jmax=\frac{16}{2},\ldots \frac{44}{2}$ for the fitting (blue dots). The red line indicates the eigenvalue sequence, the black line the fit function given in the text.}
\end{minipage}
\end{figure}

The overall largest eigenvalue sequence is contributed by those
$\vec{\sigma}$-configurations which are contained in the equivalence class of
$\vec{\sigma}$-index 7.  These maximum eigenvalues are plotted in figure
\ref{5 vertex largest evals}, with the eigenvalue axis in log scale.  It is
interesting to note
that the \sigconf{} which gives rise to the smallest eigenvalues, $\vec{\sigma}=\{4, 2, 2, 0\}\equiv\{4, -2, -2, 0\}$, also lies in
the equivalence class of \sigconfs{} which gives rise to the largest eigenvalues.
Figure \ref{5 vertex largest eval fit} shows a fit to $\maxev$ as
$\maxevfit(2\cdot \jmax)=\mb{e}^{0.067\pm 0.043} \cdot (2\cdot\jmax)^{1.484 \pm 0.013}$, with $\chi^2 :=\sum_{k=1}^{29}\big(\lambda_{\hat{V}}^{(max),(fit)}(\jmax^{(k)})-\maxev(\jmax^{(k)})\big)^2=57.3 $ for the chosen set of 29 data points.
  
\subsubsection{\label{Number of Eigenvalues 5v}Number of Eigenvalues}
According to (\ref{num evals poly estimate}) the number of eigenvalues $\numevals(\jmax)$ for configurations with $j_1\le j_2\le j_3 \le j_4 \le j_5=\jmax$ will be given by a polynomial of the order of $[j_{max}]^{2N-4}\equiv [j_{max}]^{6}$:   
\be
   \nevals^{(fit)}(\jmax)=\sum_{k=0}^6 r_k ~\cdot~(\jmax)^k
\ee
with coefficients (and their $95\%$ confidence interval) 
\[\left(
\begin{array}{lll}
 r_0 &= \text{106.59} & \pm\text{22515.70} \\
 r_1 &= \text{1523.19} & \pm\text{26095.85} \\
 r_2 &= -\text{164.42} & \pm\text{9685.98} \\
 r_3 &= -\text{8467.44} & \pm\text{1581.95} \\
 r_4 &= \text{9014.52} & \pm\text{126.55} \\
 r_5 &= \text{15678.82} & \pm\text{4.86} \\
 r_6 &= \text{5226.69} & \pm\text{0.07}
\end{array}
\right)\]
For this fit the quantity $\chi^2:=\sum_{\jmax=\frac{1}{2}}^{\frac{44}{2}} \big(\nevals(\jmax) -\nevals^{(fit)}(\jmax) \big)^2  $ takes the numerical value $\chi^2=2.26\cdot 10^9$ for the set of 44 data points.

\begin{figure}[htbp]
 \center
\begin{minipage}{8cm}
    \center
    \psfrag{N}{$\numevals(\jmax)$}
    \psfrag{jmax}{$\jmax$}
    \psfrag{FiveVNumEvals}{}
    \includegraphics[width=8.6cm]{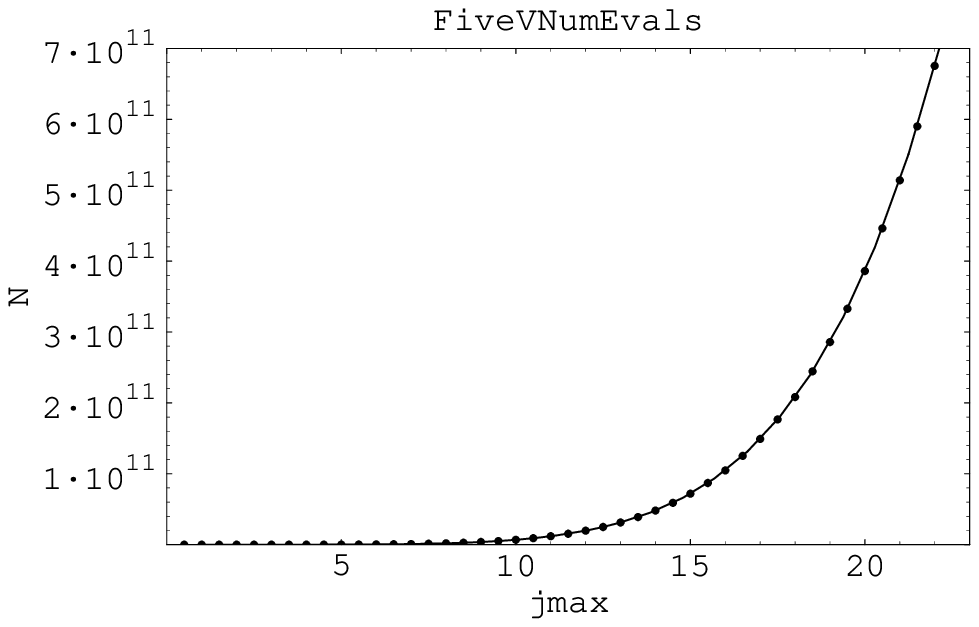}
    \caption{Number of eigenvalues (dots) at the 5-vertex, fitted by a sixth order polynomial (solid curve).}
\end{minipage}
~~~~
\begin{minipage}{8cm}
    \center
     \psfrag{NumEval}{$\numevals(\jmax)$}
    \psfrag{jmax}{$\jmax$}
    \psfrag{FivevNumEvals}{}
    \includegraphics[width=8.6cm]{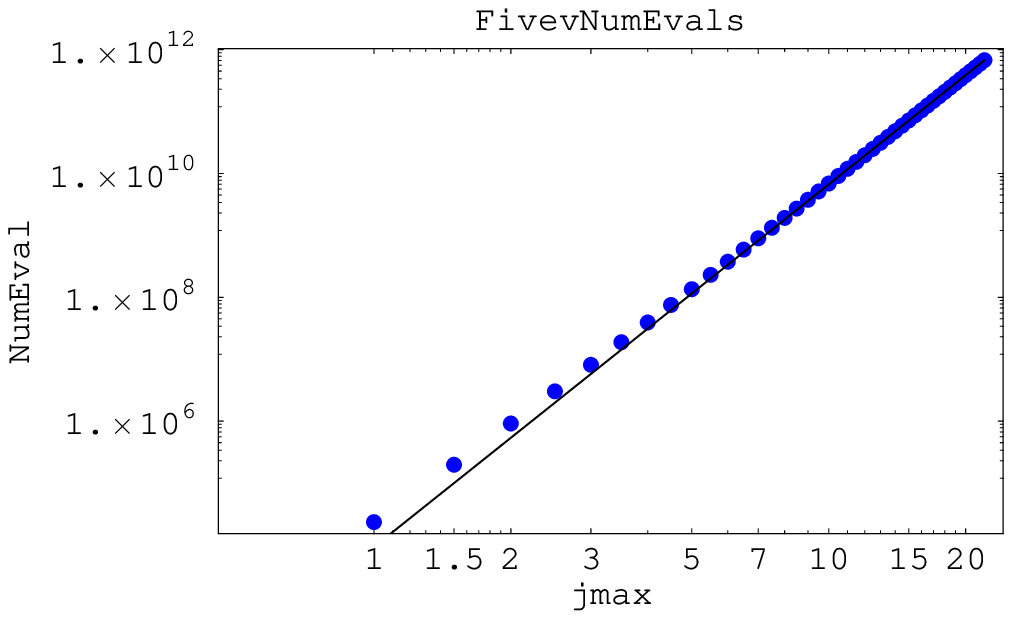}
    \caption{\label{Num Evals 5 V loglog plot}Number of eigenvalues at the 5-vertex in a double logarithmic plot for the two parameter minimum model.}
\end{minipage}
\end{figure}

As in the case for the 4-vertex, one can also try a non linear 
two parameter fit $\nevals^{(fit)}(\jmax)= r \cdot (\jmax)^{s}$, in order to measure the behavior of $\numevals$ in the leading order in $\jmax$. In this case we obtain $r=9296.6\pm 96.02$ and $s=5.85582 \pm 0.00341$, $\chi^2:=\sum_{\jmax=\frac{1}{2}}^{\frac{44}{2}} \big(\nevals(\jmax) -\nevals^{(fit)}(\jmax) \big)^2  =1.34\cdot 10^{18}$ for the set of 44 data points (see figure \ref{Num Evals 5 V loglog plot}).

 \newpage
\subsection{Gauge Invariant 6-Vertex}

There are 16413 $\vec{\sigma}$-configurations for the 6-vertex, according to
section \ref{Sprinkling Sign Factors}.  As mentioned in \S
\ref{5v_extremal_eigenvalues}, 
configurations which differ by an overall sign can be identified. Therefore (excluding $\sigma(IJK)=0~~\forall~I<J<K$) we are left with 8206 non-trivial $\vec{\sigma}$-configurations.

\subsubsection{Histograms}
\begin{figure}[hbtp]
 \center
     \psfrag{frequency}{$\numevals$}
    \psfrag{eigenvalue}{$\eval$}
    \psfrag{6-vertex: full spectrum up to various jmax}{}
    \includegraphics[width=15cm]{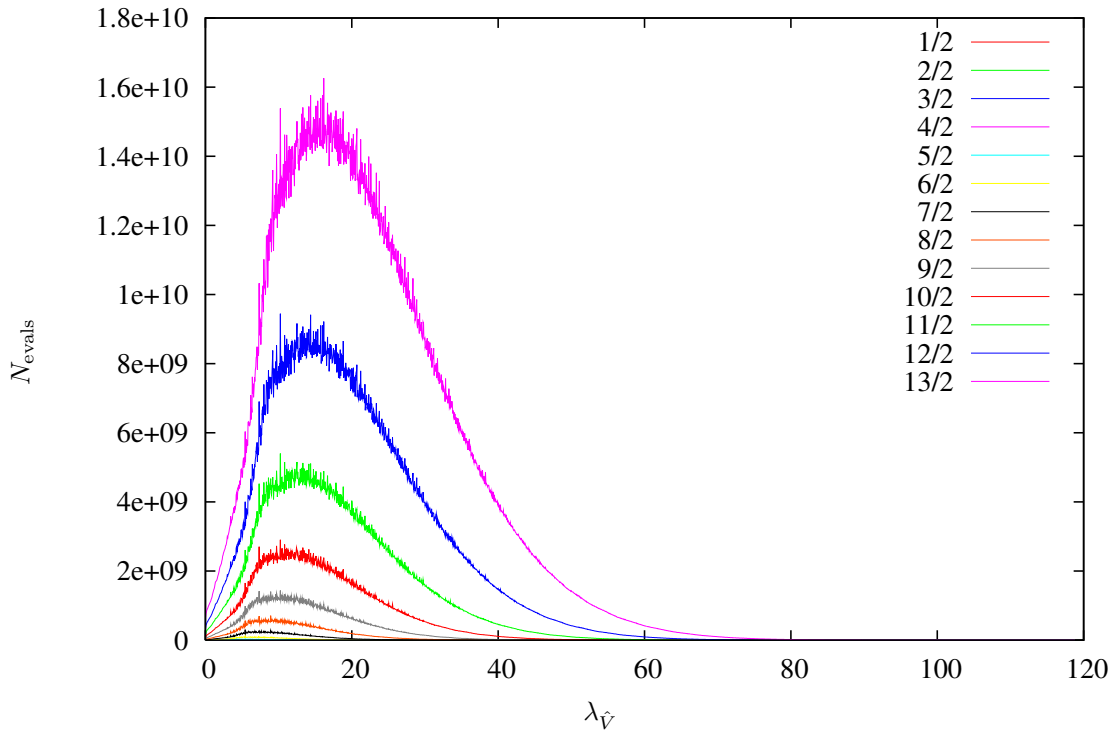}
    \caption{\label{6 vertex 1}Overall histograms at the gauge invariant
    6-vertex up to $j_{max}=\frac{13}{2}$.  There are 7,349,844,794,112
    eigenvalues in all, of which 212,946,944,688 are zero.  $\nbins=2048$.}
\end{figure}
As one can see in figure \ref{6 vertex 1}, the overall
histograms for the gauge invariant 6-vertex show a similar behavior as those of the 4- and 5-vertex. However, there is no `lip' in the spectral density close to zero visible 
(and quadrupling the number of bins does not alter this).
Nevertheless the number of eigenvalues near zero increases with $\jmax$ and,
as can be seen from section \ref{Extremal Evals 6v} below, the behavior of the
smallest eigenvalue of a particular vertex configuration depends on its
$\vec{\sigma}$-configuration.  
Thus it seems that upon taking all eigenvalues from all \sigconfs{} the eigenvalues
at the 6-valent vertex are rather equally spaced from 0 on, whereas the
spectrum of the 5-vertex has an accumulation point at 0. Nevertheless there are also single  \sigconfs{} which show a lip in their spectral density, as demonstrated in figure \ref{6v_fixed_sigconf_zoom}.

\begin{figure}[hbtp]
 \center
    \psfrag{frequency}{$\frac{N_{\mathrm{bin}}}{\numevals(\jmax)}$}
    \psfrag{eigenvalue}{$\eval$}
    \includegraphics[width=15cm]{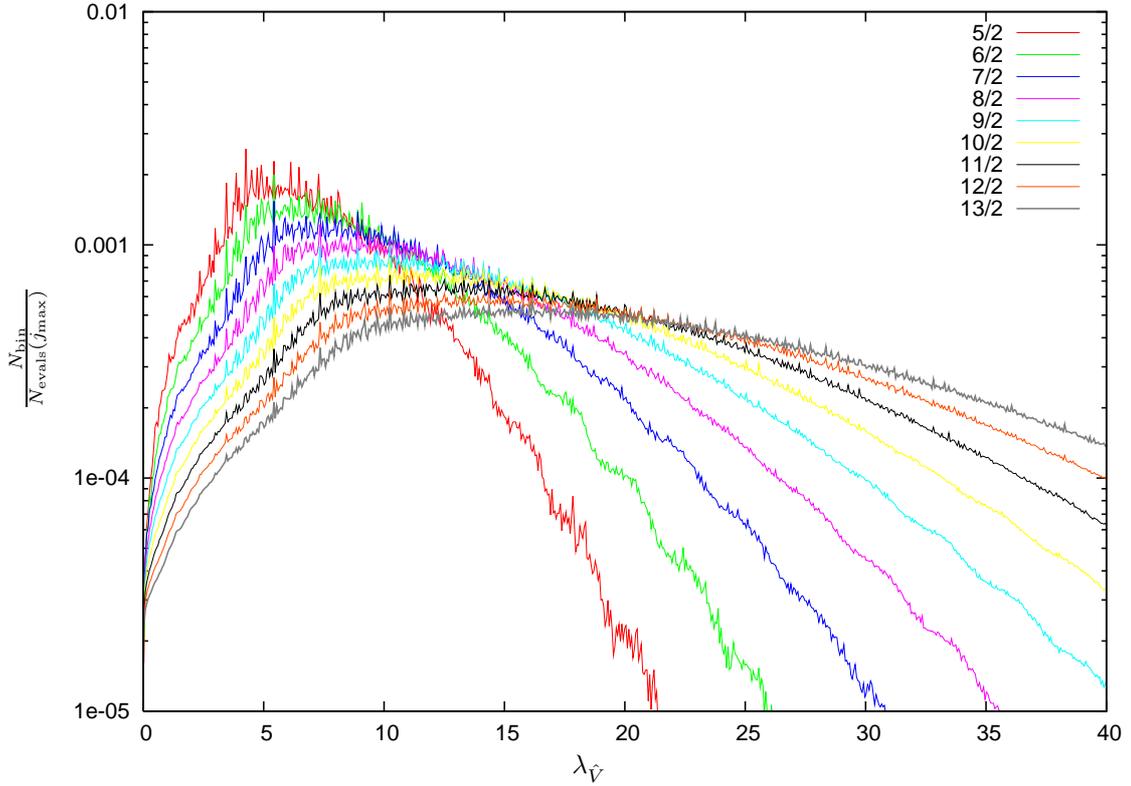}
    \caption{Normalized overall histogram for the 6-vertex.  The vertical
      axis is in logscale.  The histograms for the smallest four values of
      $\jmax$ have been removed to give a clearer picture of the behavior at
      larger $\jmax$.  $\nbins=2048$.}
    \label{6v_overall_normalized}
\end{figure}
In figure \ref{6v_overall_normalized} we show the normalized overall
histogram for the 6-vertex, analogous to figure \ref{5 vertex lip logplot} of
section \ref{5v_histograms}, in which each bin occupation number $N_{\mathrm{bin}}$ has been
divided by the total number of eigenvalues $\numevals(\jmax)$ at that $\jmax$.  There we see
that, as opposed to the situation with the 5-vertex, the normalized spectral density does decrease close to zero.

\begin{figure}[hbtp]
  \center
  \begin{minipage}{8.5cm}
    \center
    \psfrag{freq}{$\numevals$}
    \psfrag{eval}{$\eval$}
    \includegraphics[width=8.5cm]{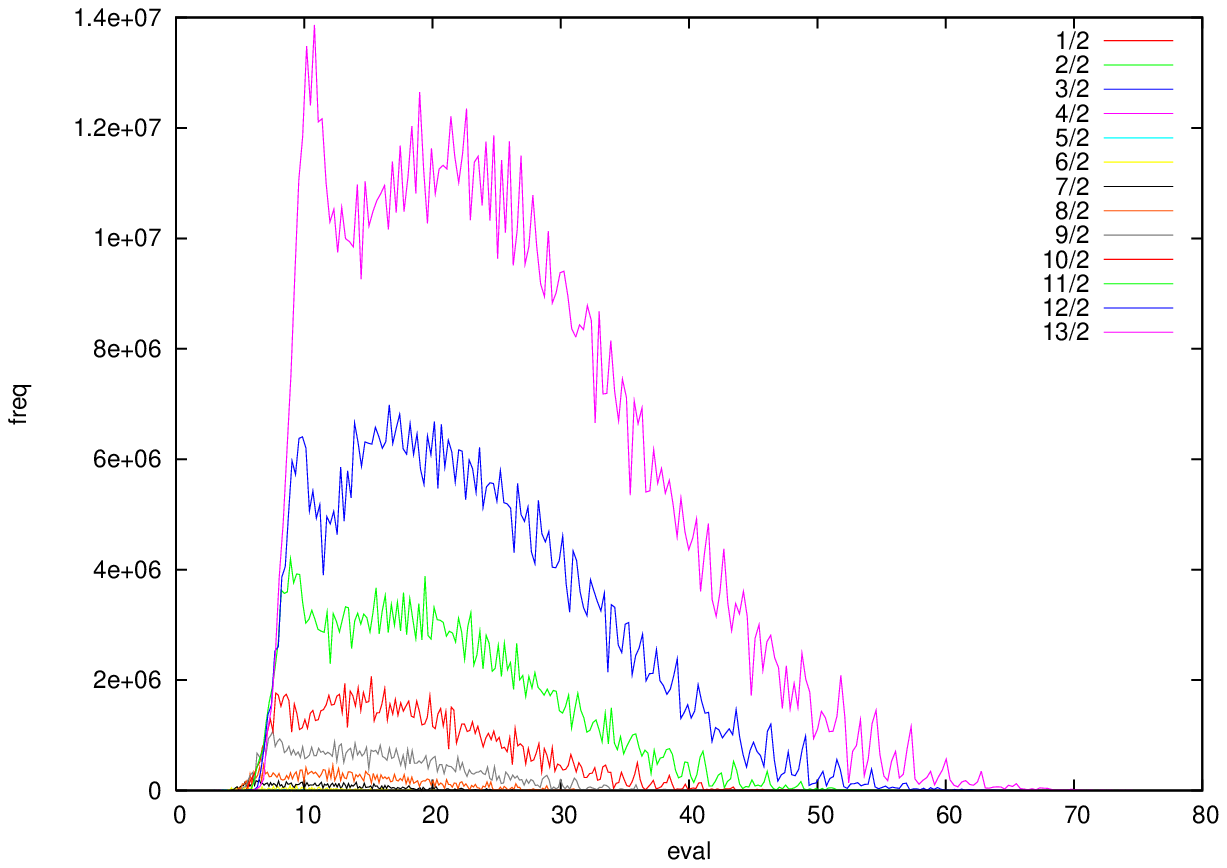}
  \end{minipage}
  \begin{minipage}{8.5cm}
    \center
    \psfrag{freq}{$\numevals$}
    \psfrag{eval}{$\eval$}
    \includegraphics[width=8.5cm]{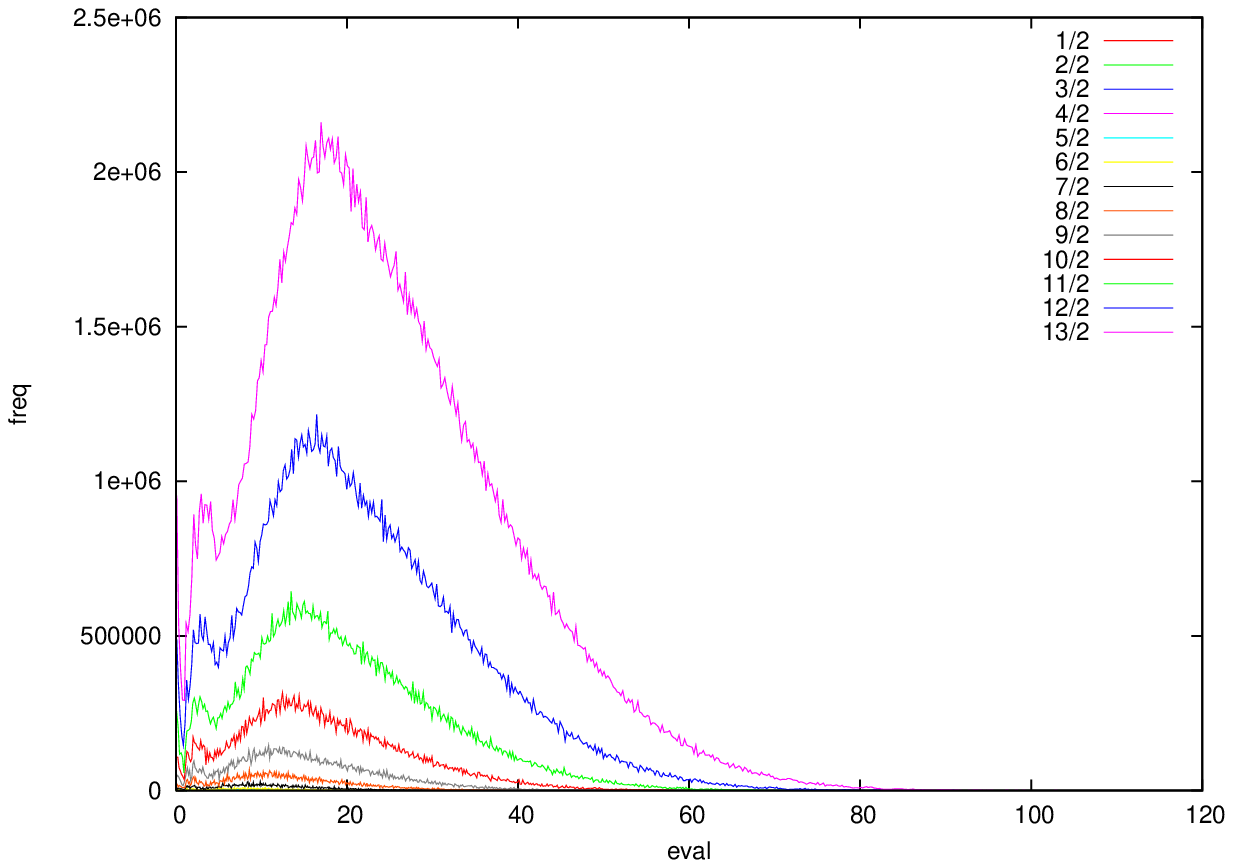}
  \end{minipage}
  \caption{Two sample histograms for fixed \sigconf. On  
the left
  $\vec\sigma$ = (0, 0, 0, 2, 2, 2, 2, 2, 2, 2) and
  on the right 
  $\vec\sigma$ = (2, -4, -4, 0, 0, 0, 4, 2, 2, 0).} 
\label{6v_fixed_sig_conf}
\end{figure}
Figure \ref{6v_fixed_sig_conf} shows two histograms for fixed
\sigconf{}
$\vec{\sigma}=\big(\sigma(123),\sigma(124),\sigma(125),
                                                 \sigma(134),\sigma(135),
                                                             \sigma(145),
                                                 \sigma(234),$ $\sigma(235),
                                                             \sigma(245),
                                                             \sigma(345)\big)$.  The
\sigconf{} on the left yields a minimum eigenvalue which increases with
$\jmax$, while the \sigconf{} on the right yields a decreasing minimum
eigenvalue.  Some details regarding these histograms are listed in table
\ref{6v_fixed_sigconf_hist_details}.
\begin{table}[htbp]
\center
\begin{tabular}{|rrrrrrrrrr|rrrc|}
\hline
\multicolumn{10}{|c|}{\sigconf} & $\chi_{\vec{\sigma}}$ & $N_{\mathrm{zeros}}$ & $\nbins$ & behavior \\
\hline
0 &  0 &  0 & 2 & 2 & 2 & 2 & 2 & 2 & 2 & 8 & 169,083,736 & 256 & increasing $\mineval$\\ 
2 & -4 & -4 & 0 & 0 & 0 & 4 & 2 & 2 & 0 & 2 &   2,998,262 & 512 & decreasing $\mineval$\\ 
\hline
\end{tabular}
\caption{Details of two fixed \sigconf{} histograms for the 6-vertex.}
\label{6v_fixed_sigconf_hist_details}
\end{table}
Figure \ref{6v_fixed_sigconf_zoom} shows the portion of the histograms on the
right side of figure \ref{6v_fixed_sig_conf} for $\eval \leq 10$.  There we
see the lip quite clearly, which indicates that it is the `averaging' over
all \sigconfs{} which causes the lip to be flattened out in figure
\ref{6 vertex 1}.
\begin{figure}[hbtp]
 \center
    \center
    \psfrag{freq}{$\numevals$}
    \psfrag{eval}{$\eval$}
    \includegraphics[width=13cm]{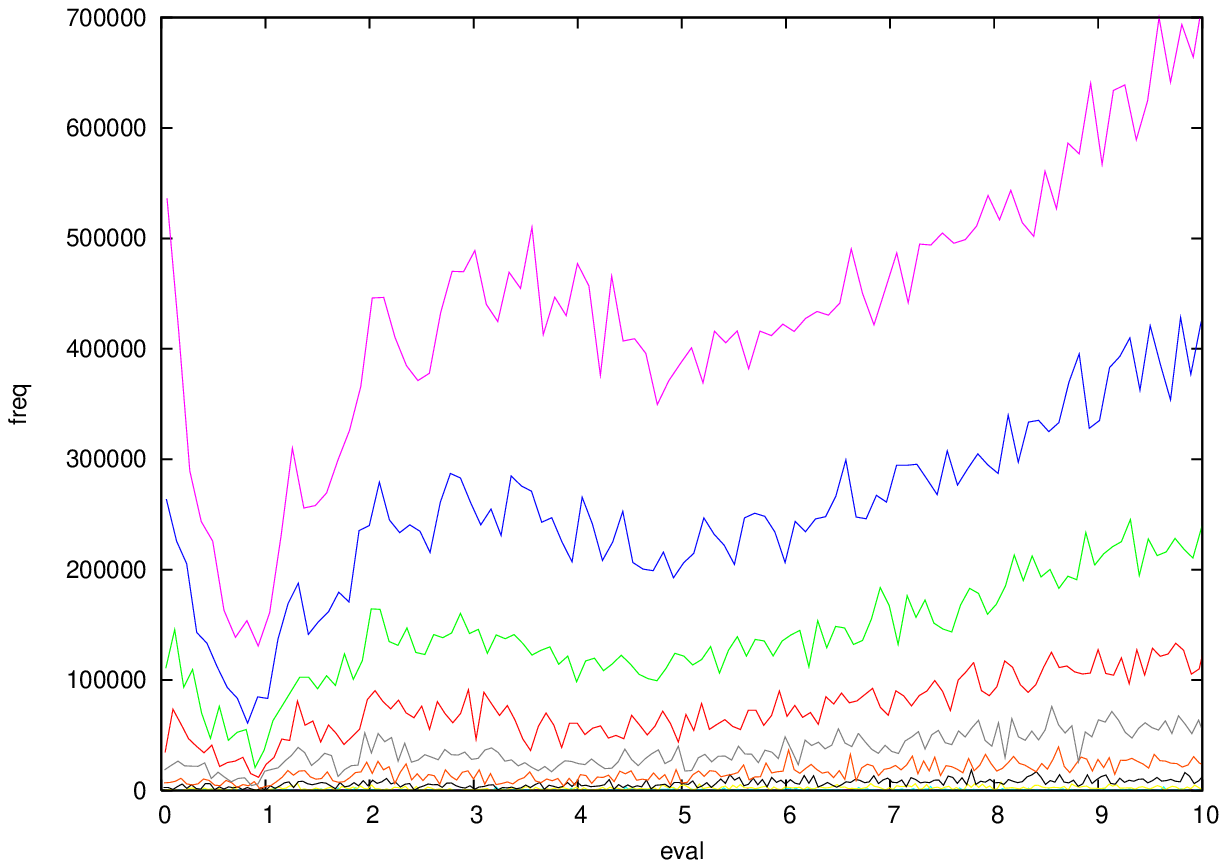}
    \caption{Zoom of the lip of the decreasing $\mineval$ histogram of figure
    \ref{6v_fixed_sig_conf}.  Here we use $\nbins=1024$ to show more detail.}
    \label{6v_fixed_sigconf_zoom}
\end{figure}

\subsubsection{Fitting of Histograms}

\begin{figure}[h!btp]
\center
    \center
    \psfrag{lambda}{$\eval$}
    \psfrag{l}{$\ln[\numevals]$}
    \psfrag{SixVFitRisingEdgeLog}{}
    \includegraphics[height=5.0cm]{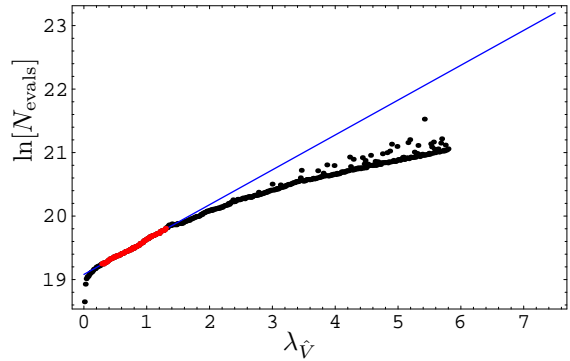}
     \caption{\label{6 vertex Fit 1} Logarithmic plot of the fit to the
     rising edge at the gauge invariant 6-vertex. The points that were used
     in the fitting process are indicated in red, the fitted function is the solid blue line.}
\end{figure}

As in \S \ref{fit_histogram5}, figure \ref{6 vertex 1} seems to possess a
spectral density which rises exponentially with $\jmax$.
A fit to this rising edge is shown in figure \ref{6 vertex Fit 1}, in which the logarithm of the number $\numevals$ of eigenvalues  as a function of the eigenvalue $\eval$  is given by:
\be
   \ln[\numevals(\eval)]^{(fit)}=(19.081\pm 0.007) + (0.549\pm 0.008) \eval
   ~~~~~~~~~~\longrightarrow~~~~~~~~~~~
   \numevals^{(fit)}(\eval) \sim 0.19\cdot 10^9 \cdot \mb{e}^{0.55~\eval} 
\ee
The quantity $\chi^2:=\sum_{k=1}^{71} \big( \numevals^{(fit)}(\eval^{(k)})- \numevals^{(k)}(\eval)\big)^2 $ has the numerical value $\chi^2=7.257\cdot10^{14}$ on the chosen set of 71 data points.

\subsubsection{\label{Extremal Evals 6v}Extremal Eigenvalues}

Table \ref{6v_equiv_classes} shows the number of equivalence classes of
\sigconfs{} for both the decreasing and increasing minimum eigenvalues, as in
\S \ref{5v_extremal_eigenvalues}.

\begin{table}
\center
\begin{tabular}{|l|l||l|l|}
   \cline{3-4} \multicolumn{2}{c|}{} & $\mineval$  &$\maxeval$\\
   \cline{3-4} \hline
   increasing $\sigma$-conf.&equiv. classes &58               & 30     
   \\\cline{2-4}   
                            &total          &323              & 8206
   \\\hline
   decreasing $\sigma$-conf.&equiv. classes &3645             &  -
   \\\cline{2-4}   
                            &total          &7856             & -
   \\\hline
   constant $\sigma$-conf.&equiv. classes   &5                & -
   \\\cline{2-4}   
                            &total          &27               & -
   \\\hline                                                  
\end{tabular}
\caption{Numbers of equivalence classes of \sigconfs{} for increasing and
 decreasing minimum eigenvalue, and maximum eigenvalue, for the 6-vertex. The information whether a configuration is increasing, decreasing or constant is obtained by comparing the (minimum) eigenvalue for $\jmax=2$ with that from $\jmax=\frac{13}{2}$.} 
\label{6v_equiv_classes}
\end{table}

\paragraph{Smallest Eigenvalue\\}

\begin{figure}[hbtp]
\center
\begin{minipage}[t]{8.6cm}
    \psfrag{eval}{$\mineval$}
    \psfrag{jmax}{$\jmax$}
    \psfrag{sigconf}{\cmt{3}{~\\[-1mm]\sigconf}}
    \includegraphics[width=8.6cm]{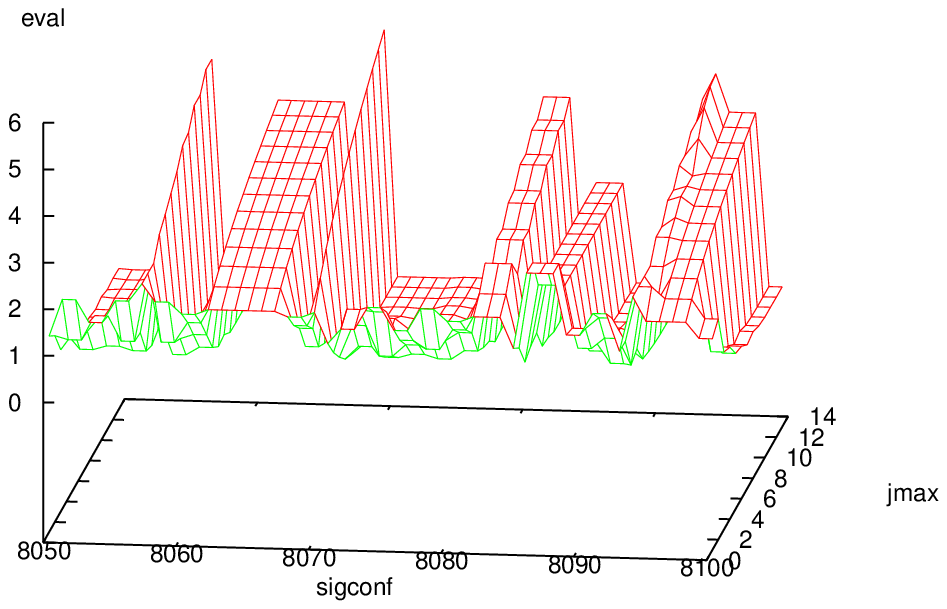}
    \caption{\label{6v_rising_min_eval} Minimum eigenvalue vs.\ $\jmax$ for
    an arbitrary sampling of 21 \sigconfs.  (In all plots for the 6-vertex, the integers labeling the
    \sigconfs{} have no intrinsic meaning.)}
\end{minipage}
~
\begin{minipage}[t]{8.6cm}
    \psfrag{min eval}{$\mineval$}
    \psfrag{jmax}{$\jmax$}
    \psfrag{sigma config}{$\vec\sigma$-config}
    \includegraphics[width=8.6cm]{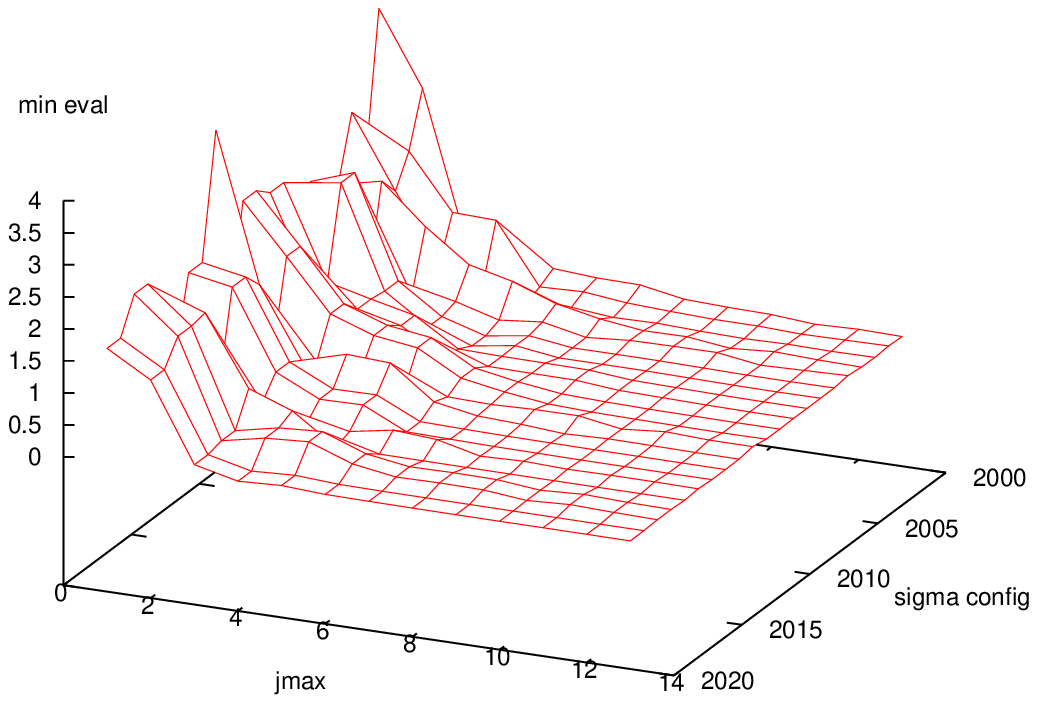}
    \caption{\label{6v_decreasing_min_eval} Minimum eigenvalue vs.\ $\jmax$ for
    a sampling of 21 \sigconfs{} for which all
    $\mineval$ decrease with $\jmax$.}
\end{minipage}~~~
\end{figure}

Figure \ref{6v_rising_min_eval} shows how the smallest eigenvalue varies with
$\jmax$ for a sample of 21 \sigconfs{}.  As is the case for the 5-vertex,
some \sigconfs{} lead to rising minimum eigenvalues, and some to falling.
Figure \ref{6v_decreasing_min_eval} shows the same for a collection of
\sigconfs{} which only yield a decreasing minimum eigenvalues.

\begin{figure}[h!tbp]
 \center
\begin{minipage}[t]{8.7cm}
    \psfrag{Twojmax}{$2\jmax$}
    \psfrag{Ml}{$\mineval$}
    \psfrag{SixV50firstMinEval}{}
    \includegraphics[width=8.7cm]{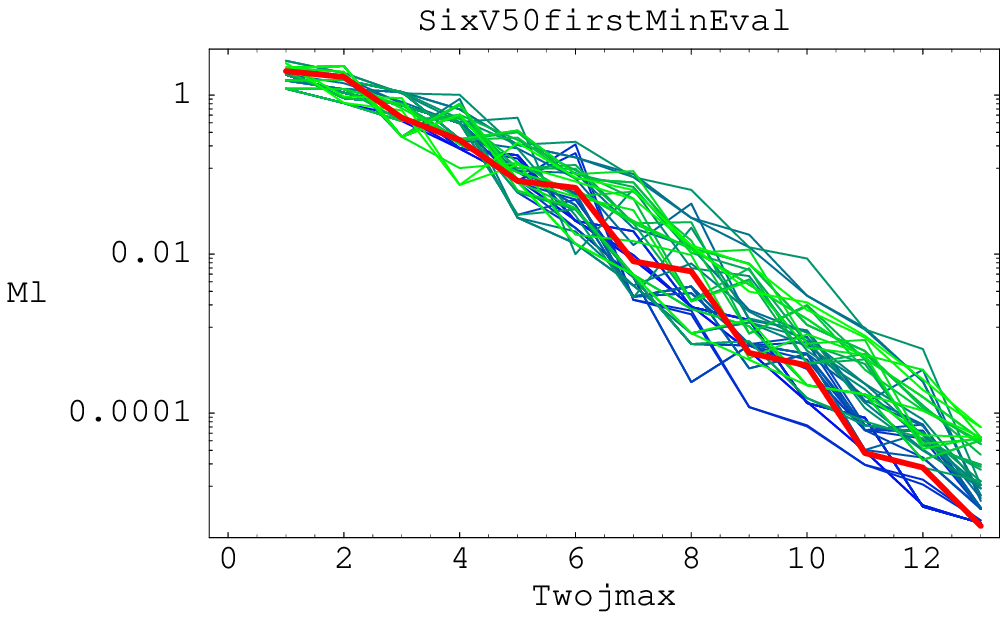}
    \caption{\label{6 vertex smallest eval series}The 50 smallest eigenvalue sequences (sorted with respect to the smallest eigenvalue at $\jmax=\frac{13}{2}$) at the gauge invariant 6-vertex. The red line indicates the smallest eigenvalues from $\vec{\sigma}_{min}$ in this ordering.}
\end{minipage}
~~~
\begin{minipage}[t]{8.7cm}
    \psfrag{Twojmax}{$2 \jmax$}
    \psfrag{M}{$\ln[\mineval]$}
    \psfrag{SixV50firstMinEval}{}
    \includegraphics[width=8.7cm]{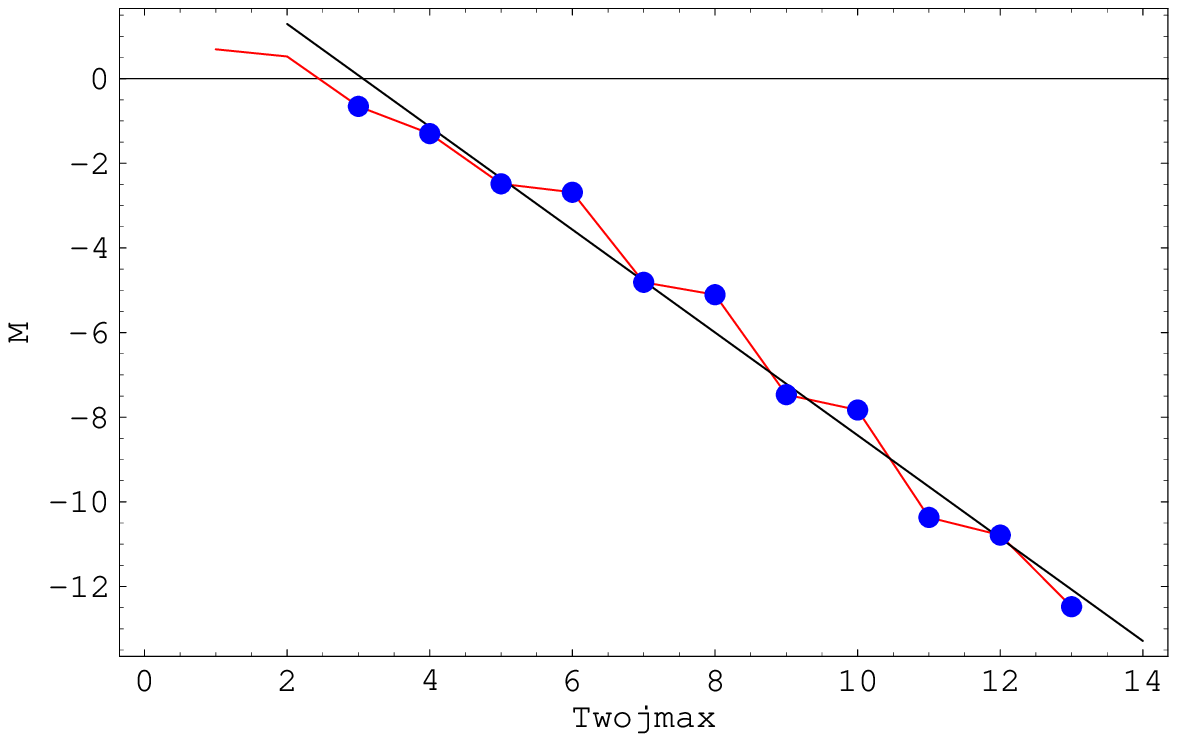}
    \caption{\label{6 vertex fit smallest eval series}A curve for for the smallest 
eigenvalue sequence of the 6-vertex, which occurs for
    $\vec{\sigma}_{min}=(2, 0, 0, 0, 0, 0, 0, 0, -2, 0) ,(2, 0, 0, 0, 0, 0, 0, 0, 2, 0)$.  We have used $\mineval(\jmax)$ for $\jmax=\frac{3}{2},\ldots \frac{13}{2}$ for the fitting (blue dots). The red line indicates the eigenvalue sequence of $\vec{\sigma}_{min}$, the black line the fit function.  The error bars for these data points appear in figure \ref{6v_min_eval_error1th-zoom}.}
\end{minipage}
\end{figure}

As it turns out, it is not possible to identify a genuine smallest eigenvalue sequence for the 6-vertex, at least not in the computed spin parameter range. This is due to the fact that the different eigenvalue sequences intersect each other and can hardly be separated. 
Figure \ref{6 vertex smallest eval series} illustrates this property.   
Nevertheless we can define a smallest eigenvalue sequence by taking into account the ordering of the smallest eigenvalue sequences at a fixed $\jmax$. The chosen sequence is displayed as red curve in figure \ref{6 vertex smallest eval series}, it is obtained by comparing all eigenvalue sequences at $\jmax=\frac{13}{2}$.
It is contributed by the spin configurations shown in table \ref{6v smallest eval spin config} (here $j_6=\jmax$).
\begin{table}[htbp]
\center
\center
\begin{footnotesize}
\begin{tabular}{|r|r|r|r|r|r|}
\hline
$2\cdot j_1$&$2\cdot j_2$&$2\cdot j_3$&$2\cdot j_4$&$2\cdot j_5$ & $2\cdot j_6$\\\hline
1&1&1&1&1&1\\
1&1&1&1&2&2\\
2&2&3&3&3&3\\
1&1&3&3&4&4\\
1&1&3&3&5&5\\
1&1&5&5&6&6\\
1&2&2&3&7&7\\
1&1&7&7&8&8\\
1&1&7&7&9&9\\
1&1&9&9&10&10\\
1&1&9&9&11&11\\
1&1&11&11&12&12\\
1&1&11&11&13&13
\\\hline
\end{tabular}
\end{footnotesize}
\caption{The spin configurations contributing the smallest eigenvalues for 
$\vec{\sigma}_{min}=$ (2, 0, 0, 0, 0, 0, 0, 0, -2, 0), (2, 0, 0, 0, 0, 0, 0, 0, 2, 0).}
\label{6v smallest eval spin config} 
\end{table}

We can then (as in case if the 5-vertex) apply a fitting of $\ln[\mineval]$ versus $2\jmax$.
A linear fit for a chosen set of data (see caption of figure \ref{6 vertex fit smallest eval series}) reveals that 
$\mineval(\jmax)=\mb{e}^{(3.723\pm 1.119)} \mb{e}^{- (1.215\pm 0.130 ) 2\cdot\jmax}$. Here $\chi^2:=\sum_{k=1}^{11}\big(\lambda_{\hat{V}}^{(min),(fit)}(\jmax^{(k)}) - \mineval(\jmax^{(k)}) \big)^2 =0.32$ for the chosen set of 11 spin configurations.

\begin{figure}[htb]
    \center
\begin{minipage}[t]{8.2cm}
   \psfrag{MaxEval}{$\maxeval$}
    \psfrag{2jmax}{$2 \jmax$}
    \psfrag{sigconf}{\sigconf}
    \includegraphics[width=8.5cm]{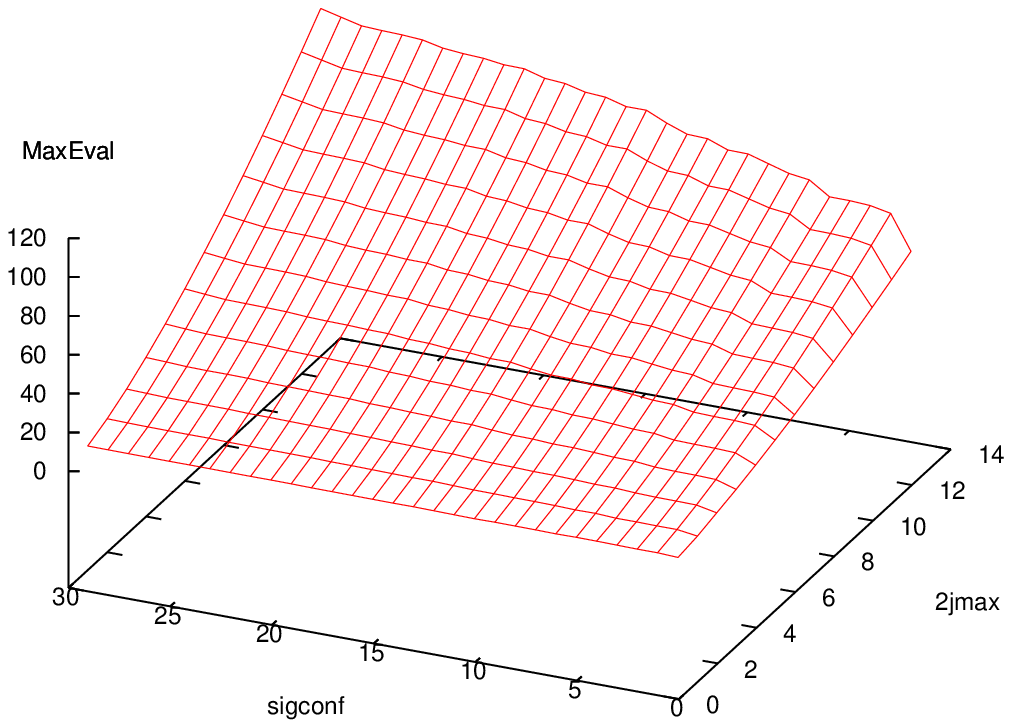}
    \caption{\label{6v_max_eval} Maximum eigenvalue vs.\ $\jmax$ for the 30 \sigconf{} equivalence classes at the 6-vertex.}
\end{minipage}
~~
\begin{minipage}[t]{8.2cm}
    \psfrag{LogM}{$\ln[\maxeval]$}
    \psfrag{Log2jmax}{$\ln[2 \jmax]$}
    \includegraphics[width=8.5cm]{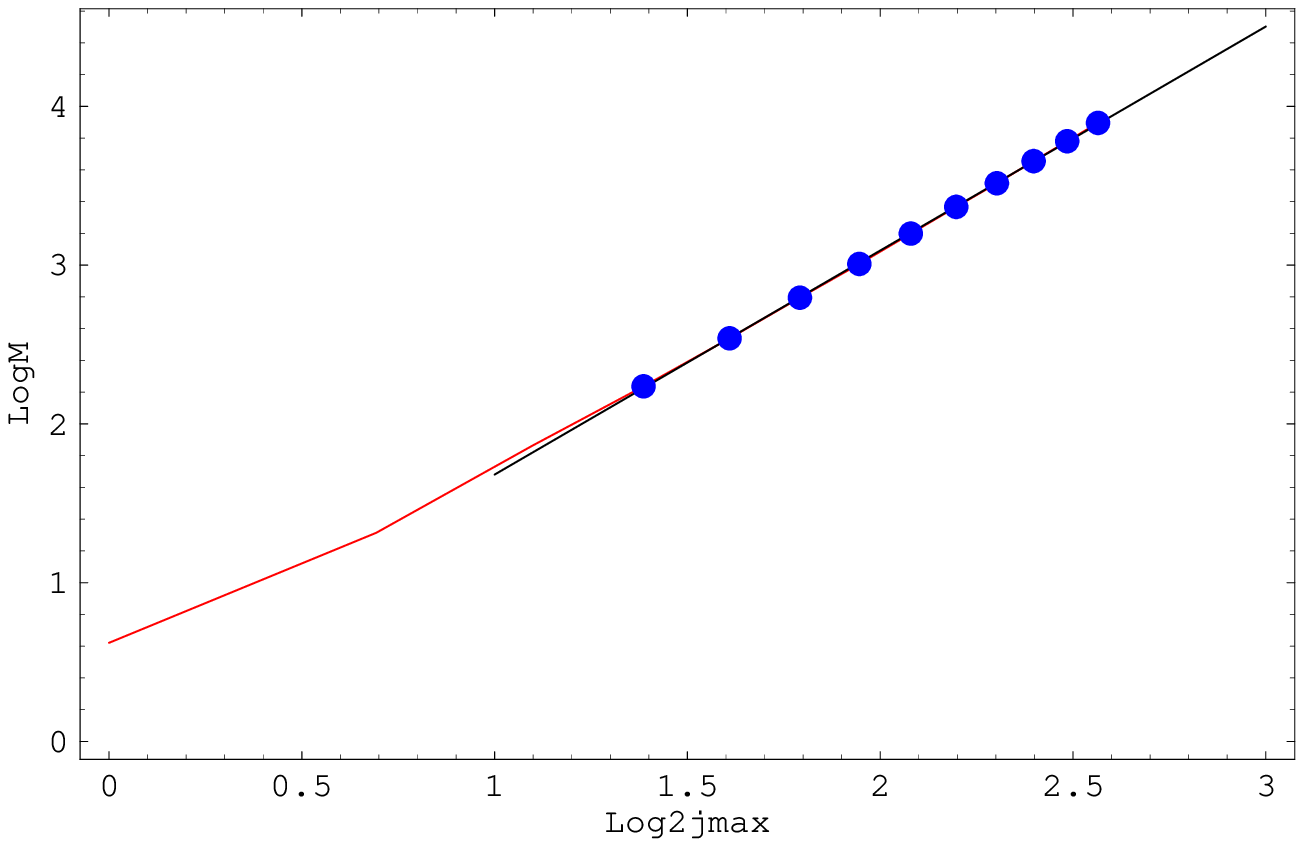}
    \caption{\label{6v_max_eval fit} 
    Fit of the logarithm $\ln[\maxeval]$ of the largest eigenvalue sequence of figure \ref{6v_max_eval} (\sigconf{} index 30) versus $\ln[2 \jmax]$. We have used $\maxeval(\jmax)$ for $\jmax=\frac{4}{2},\ldots \frac{13}{2}$ for the fitting (blue dots). The red line indicates the eigenvalue sequence, the black line the fit function given in the text.}
\end{minipage}
\end{figure}

\paragraph{Largest Eigenvalue\\}
In figure \ref{6v_max_eval} we show the maximum eigenvalues for the 30
equivalence classes (according to the maximum eigenvalues) of \sigconfs.

Figure \ref{6v_max_eval fit} displays a fit to the largest of the maximum
eigenvalue sequences.  
The fitting function is $\maxev(\jmax)=\mb{e}^{0.270\pm 0.026} \cdot
(2\cdot\jmax)^{1.4108 \pm 0.012}$. Moreover 
$\chi^2:=\sum_{k=1}^{10}\big(\lambda_{\hat{V}}^{(max),(fit)}(\jmax^{(k)}) -\maxev(\jmax^{(k)})\big)^2=0.23$ for the chosen set of 10 values of $\jmax$.

\subsubsection{Number of Eigenvalues}
According to (\ref{num evals poly estimate}) the number of eigenvalues $\numevals(j_{max})$ for configurations with $j_1\le j_2\le j_3 \le j_4 \le j_5 \le j_6 =\jmax$ will be given by a polynomial of the order of $[j_{max}]^{2N-4}\equiv [j_{max}]^{8}$:   
\be
   \nevals^{(fit)}(\jmax)=\sum_{k=0}^8 r_k ~\cdot~(\jmax)^k
\ee
with coefficients (and their $95\%$ confidence interval) 
\[\left(
\begin{array}{lll}
 r_0 &= \text{5.39}  \cdot 10^6  &\pm \text{2.20}  \cdot 10^7 \\
 r_1 &= -\text{2.42}  \cdot 10^7 &\pm \text{9.63}  \cdot 10^7 \\
 r_2 &= \text{4.05}  \cdot 10^7  &\pm \text{1.56}  \cdot 10^8 \\
 r_3 &= -\text{3.38}  \cdot 10^7 &\pm \text{1.28}  \cdot 10^8 \\
 r_4 &= \text{1.55}  \cdot 10^7  &\pm \text{5.89}  \cdot 10^7 \\
 r_5 &= -\text{4.74}  \cdot 10^6 &\pm \text{1.60}  \cdot 10^7 \\
 r_6 &= \text{3.78}  \cdot 10^6  &\pm \text{2.55}  \cdot 10^6 \\
 r_7 &= \text{2.50}  \cdot 10^6  &\pm \text{0.22 } \cdot 10^6 \\
 r_8 &= \text{644618.00 }        &\pm \text{7799.21 }
\end{array}
\right)\]
For this fit the quantity $\chi^2:=\sum_{\jmax=\frac{1}{2}}^{\frac{44}{2}} \big(\nevals(\jmax) -\nevals^{(fit)}(\jmax) \big)^2  $ takes the numerical value $\chi^2=3.64\cdot 10^{11}$.

\begin{figure}[htbp]
\begin{minipage}[t]{8.3cm}
    \psfrag{N}{$\numevals(\jmax)$}
    \psfrag{jmax}{$\jmax$}
    \psfrag{SixVNumEvals}{}
    \includegraphics[width=8.7cm]{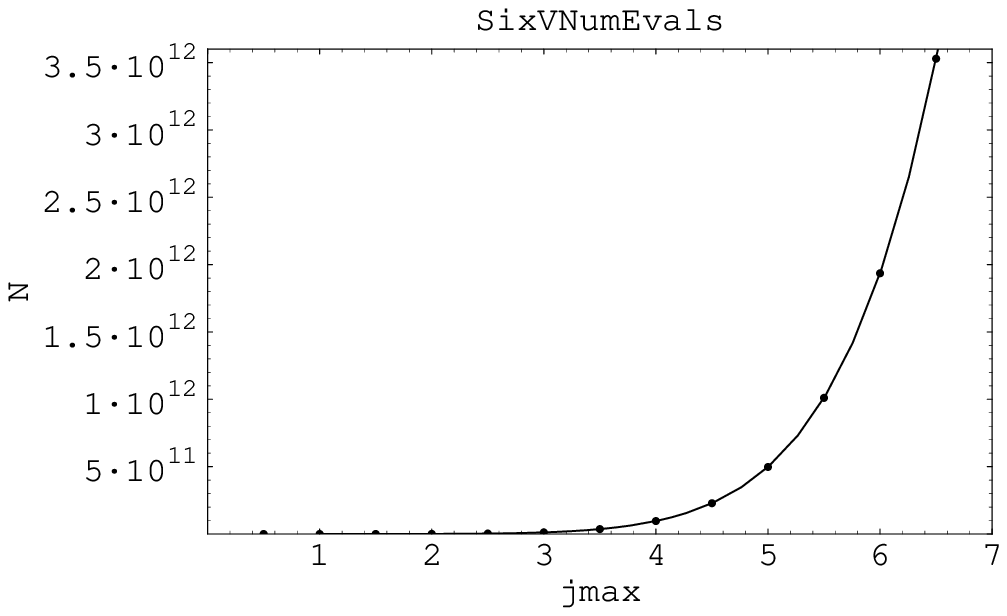}
    \caption{Number of eigenvalues at the 6-vertex fitted by a polynomial of degree eight (solid line).}
\end{minipage}
~~
\begin{minipage}[t]{8.3cm}
    \psfrag{SixvNumEvals}{}
    \psfrag{NumEval}{$\nevals$}
    \psfrag{jmax}{$\jmax$}
    \psfrag{SixVNumEvals}{}
    \includegraphics[width=8.7cm]{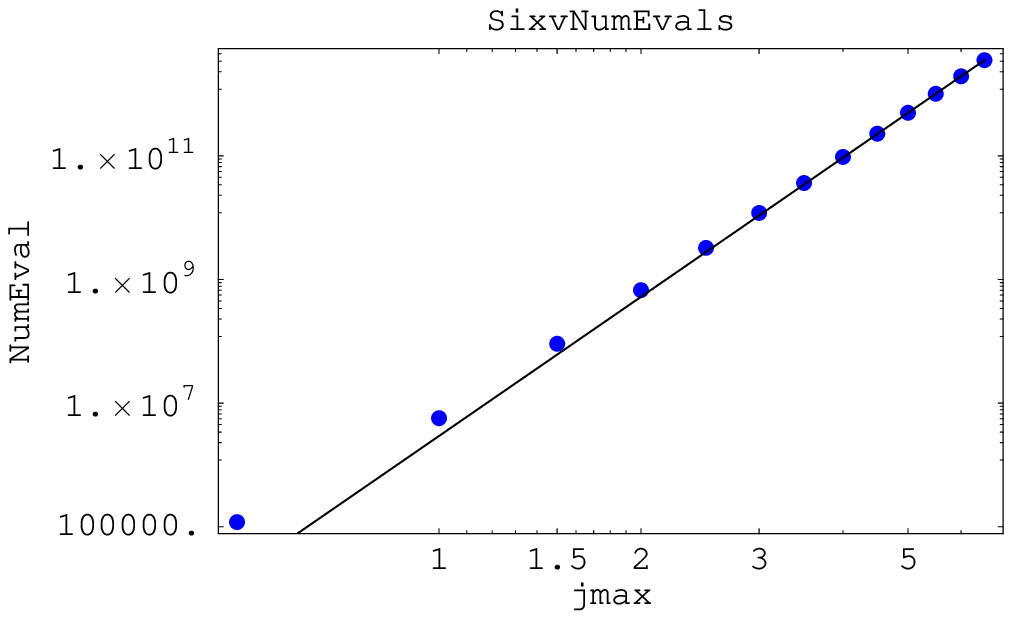}
    \caption{\label{Num Evals 6 V loglog plot}Log-log plot of the number of eigenvalues at the 6-vertex.  The solid line shows a two parameter fit to the data.}
\end{minipage}
\end{figure}

As in the case for the 4 and 5-vertex, one can also do a non linear fit to a minimal model $\nevals^{(fit)}(\jmax)= r \cdot (\jmax)^{s}$, in order to get the behavior of $\numevals$ in the leading order in $\jmax$. In this case we obtain $r=2.94389\cdot 10^6\pm 85970$ and $s=7.47748\pm 0.01586$, $\chi^2:=\sum_{\jmax=\frac{1}{2}}^{\frac{44}{2}} \big(\nevals(\jmax) -\nevals^{(fit)}(\jmax) \big)^2=3.49\cdot 10^{19}$ for the set of 13 data points (see figure \ref{Num Evals 6 V loglog plot}).

\subsubsection{Cubic 6-Vertex}
\label{cube}

Here we include results for  
a 6-vertex which has linearly dependent edge tangent vectors. The cubic 6-vertex arises due to its convenient topology and geometry, 
and is frequently used in order to construct arbitrarily large periodic graphs with definite vertex structure. 
It is used in particular when working with coherent states, as in \cite{Sahlmann:2002qk}, as well in the recently initiated program \cite{ALQG}. As it turns out cubic 6-vertex configurations provide a volume gap: the smallest non-zero eigenvalue grows with the maximum spin. 

\begin{figure}[h!bt]
\center
\begin{minipage}{8.5cm}
    \center
    \psfrag{1}{$e_1$}
    \psfrag{2}{$e_2$}
    \psfrag{3}{$e_3$}
    \psfrag{4}{$e_4$}
    \psfrag{5}{$e_5$}
    \psfrag{6}{$e_6$}
    \includegraphics[width=4.0cm]{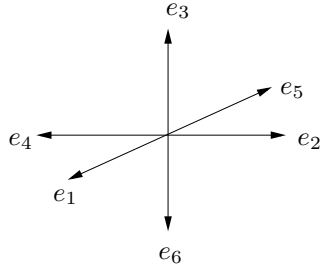}
    \caption{\label{6 vertex cubic setup}Cubic  6-vertex:  setup and sign configuration.}
\end{minipage}
\begin{minipage}{8cm}
    \center
\begin{footnotesize}
  \barr{llll}
   \epsilon(123)=1 &    &\\
   \epsilon(124)=0  &\epsilon(134)=1  &\\
   \epsilon(125)=0   &\epsilon(135)=0  &\epsilon(145)=0\\
   \epsilon(126)=-1&\epsilon(136)=0&\epsilon(146)=1&\epsilon(156)=0\\
   \\
   &\epsilon(234)=0&\\
   &\epsilon(235)=-1&\epsilon(245)=0\\
   &\epsilon(236)=0&\epsilon(246)=0&\epsilon(256)=-1\\
   \\
   && \epsilon(345)=-1\\
   &&\epsilon(346)=0&\epsilon(356)=0\\
   \\
   &&&\epsilon(456)=1\\
   \\[-15mm]
   \sigma(123)=2&\\
   \sigma(124)=2&\\
   \sigma(125)=2&\\
   \\
   \sigma(134)=2&\sigma(234)= 0\\
   \sigma(135)=0&\sigma(235)=  -2\\
   \\
   \sigma(145)=-2&\sigma(245)=-2&\sigma(345)=-2
   \nonumber
\earr
\end{footnotesize}
\end{minipage}
\end{figure}

\subsubsection*{Histogram}
\begin{figure}[h!]
  \center
    \center
    \psfrag{frequency}{$\numevals$}
    \psfrag{eigenvalue}{$\eval$}
    \includegraphics{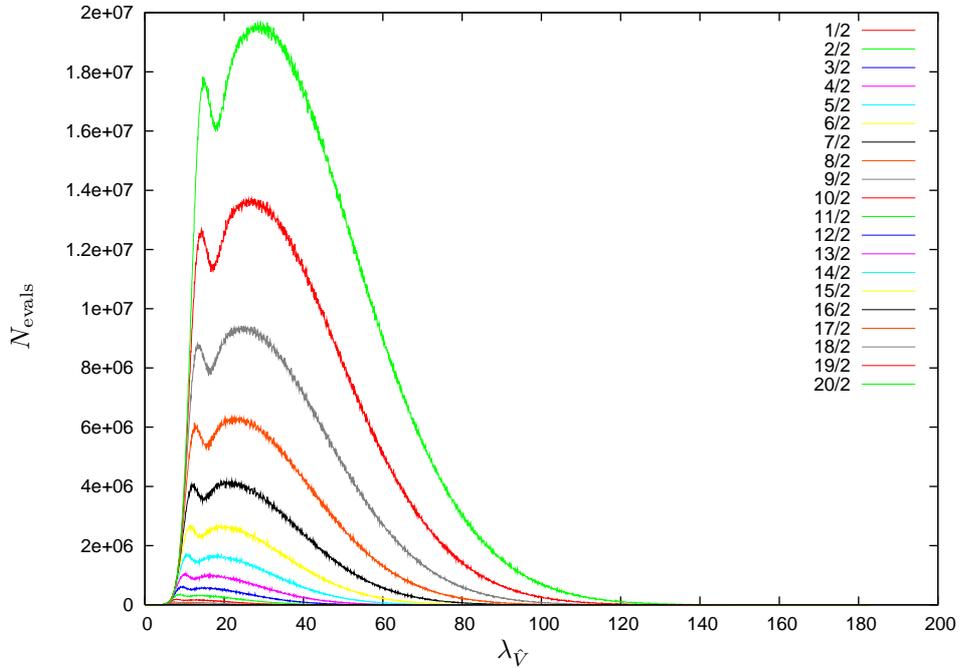}
    \caption{\label{6 vertex cubic 1}Overall histograms at the gauge
    invariant cubic 6-vertex up to $j_{max}=\frac{20}{2}$.  There are
    11,121,868,100 eigenvalues in all, of which 535,933,540 are zero.  $\nbins=2048$.}
\end{figure}

The cumulative histogram in figure \ref{6 vertex cubic 1} nicely illustrates two main parts of the spectrum: a
rising edge for small $\eval\sim 10$, which at larger $\eval$ is dominated by the cutoff imposed by finite $\jmax$.

\subsubsection*{Extremal Eigenvalues}
\begin{figure}[h!]
\center
\begin{minipage}{8cm}
    \center
    \psfrag{mineval}{$\mineval$}
    \psfrag{2jmax}{${\atop \displaystyle 2 \jmax}$}
    \psfrag{minevals}{min evals}
    \includegraphics[width=8cm]{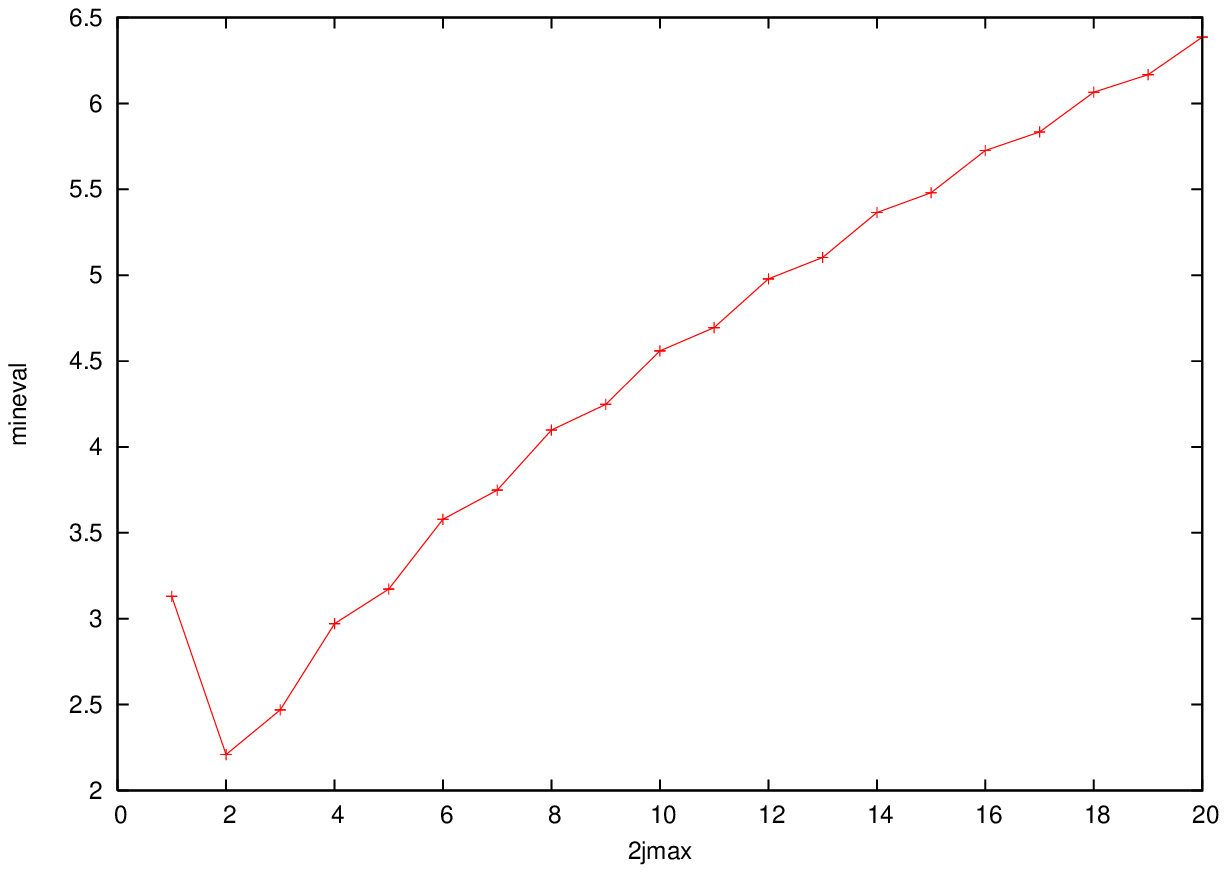}
    \caption{\label{6 vertex cubic min}Smallest eigenvalues at the gauge invariant cubic 6-vertex up to $\jmax=\frac{20}{2}$.}
\end{minipage}~~~~
\begin{minipage}{8cm}
    \center
    \psfrag{maxeval}{$\maxeval$}
    \psfrag{2jmax}{${\atop \displaystyle 2 \jmax}$}
    \psfrag{maxevals}{max evals}
    \includegraphics[width=8cm]{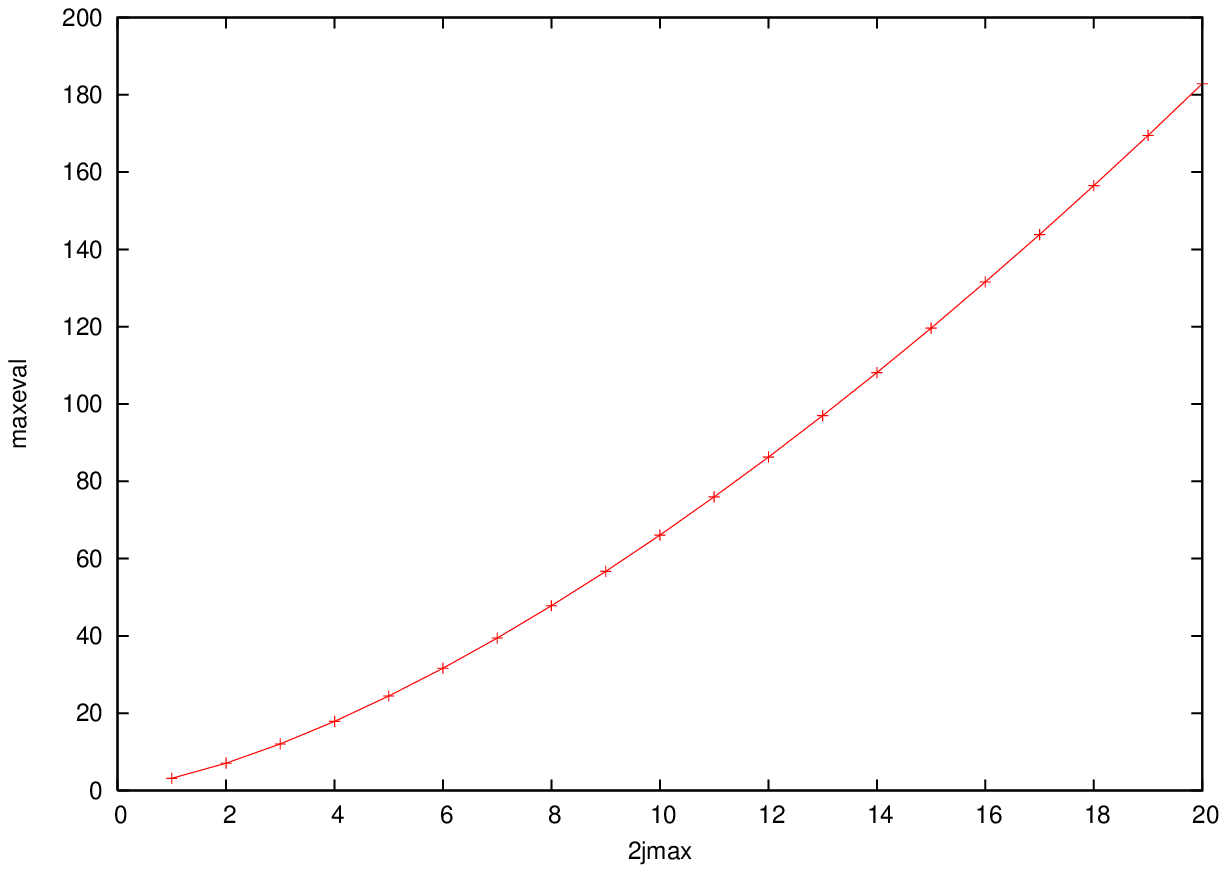}
    \caption{\label{6 vertex cubic max}Largest eigenvalues at the gauge invariant cubic 6-vertex up to $\jmax=\frac{20}{2}$.}
\end{minipage}
\end{figure}

As one can see from the plot in figure \ref{6 vertex cubic min}, the cubic gauge invariant 6-vertex belongs to the class of $\vec{\sigma}$-configurations having an increasing smallest non-zero eigenvalue $\mineval$ as the maximal spin  $\jmax$ is increased.
The oscillatory behavior of the smallest non-zero eigenvalue can be traced back to different spin configurations which contribute the smallest eigenvalue for fixed $j_6=\jmax$. We find (except for $j_1=j_2=j_3=j_4=j_5=j_6=\jmax=\frac{1}{2}$): 
\begin{center}
\begin{tabular}{|l||c|c|c|c|c|c|c|}
\hline
                        &$j_1$&$j_2$&$j_3$&$j_4$&$j_5$&$j_6$
\\\hline\hline
$\jmax$ half integer  &$\jmax-1          $&$\jmax-\frac{1}{2}$
                        &$\jmax-\frac{1}{2}$&$\jmax-\frac{1}{2}$
                        &$\jmax-\frac{1}{2}$&$\jmax$
\\ 
$\jmax$      integer  &$\jmax-\frac{1}{2}$&$\jmax-\frac{1}{2}$
                        &$\jmax-\frac{1}{2}$&$\jmax-\frac{1}{2}$
                        &$\jmax$            &$\jmax$
\\\hline
\end{tabular}
\end{center}
This nicely illustrates the effect of gauge invariance, namely that the sum of all spins must be an integer in order to have the chance to recouple them to resulting 0 angular momentum.

The plot for the largest eigenvalues is given in figure \ref{6 vertex cubic max}.

\begin{figure}[h!]
\center
\begin{minipage}[t]{8cm}
    \psfrag{M}{$\ln[\mineval]$}
    \psfrag{Twojmax}{${\atop \displaystyle \ln[ 2 \jmax]}$}
    \includegraphics[width=8cm]{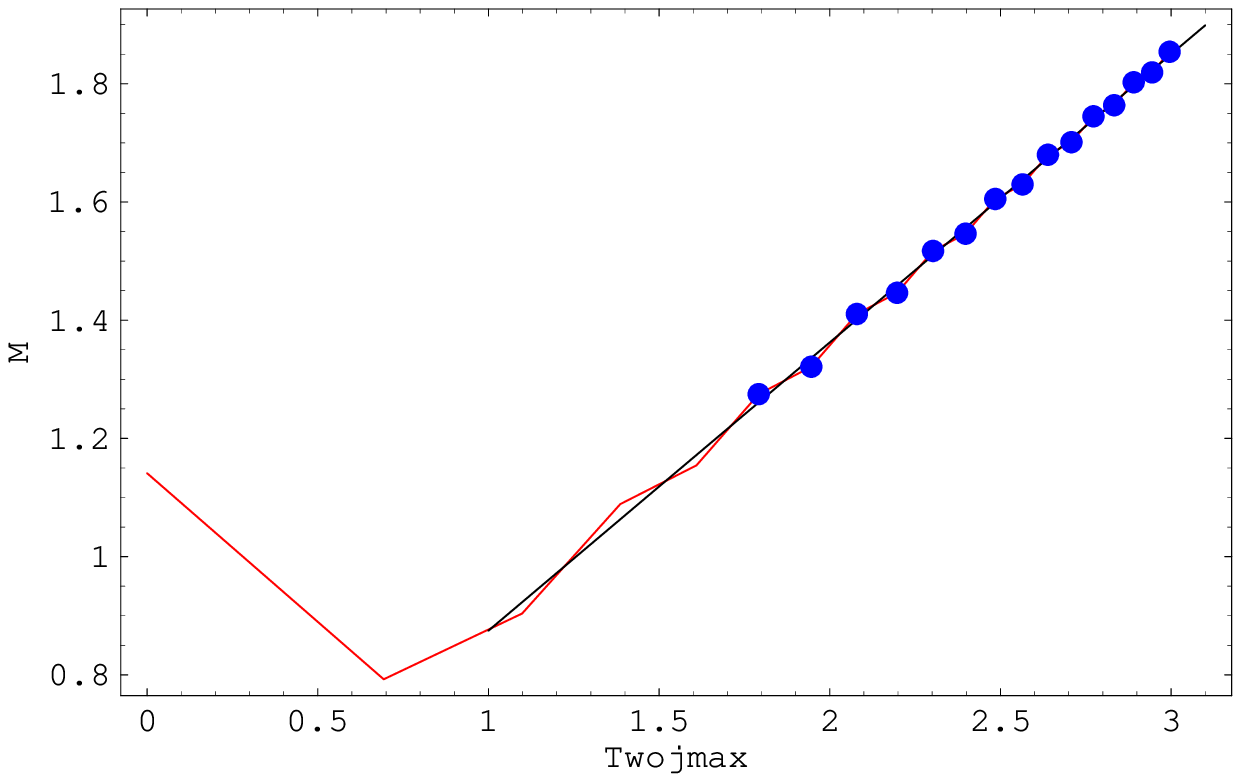}
    \caption{\label{6 vertex fit cubic min}Fit of the smallest eigenvalue sequence at the gauge invariant cubic 6-vertex up to $\jmax=\frac{20}{2}$. We have used $\mineval(\jmax)$ for $\jmax=\frac{6}{2},\ldots \frac{20}{2}$ for the fitting (blue dots). The red line indicates the minimum eigenvalue sequence, the black line the fit function.}
\end{minipage}~~~~
\begin{minipage}[t]{8cm}
    \psfrag{LogM}{$\ln[\maxev]$}
    \psfrag{Log2jmax}{${\atop \displaystyle \ln[ 2 \jmax]}$}
    \includegraphics[width=8cm]{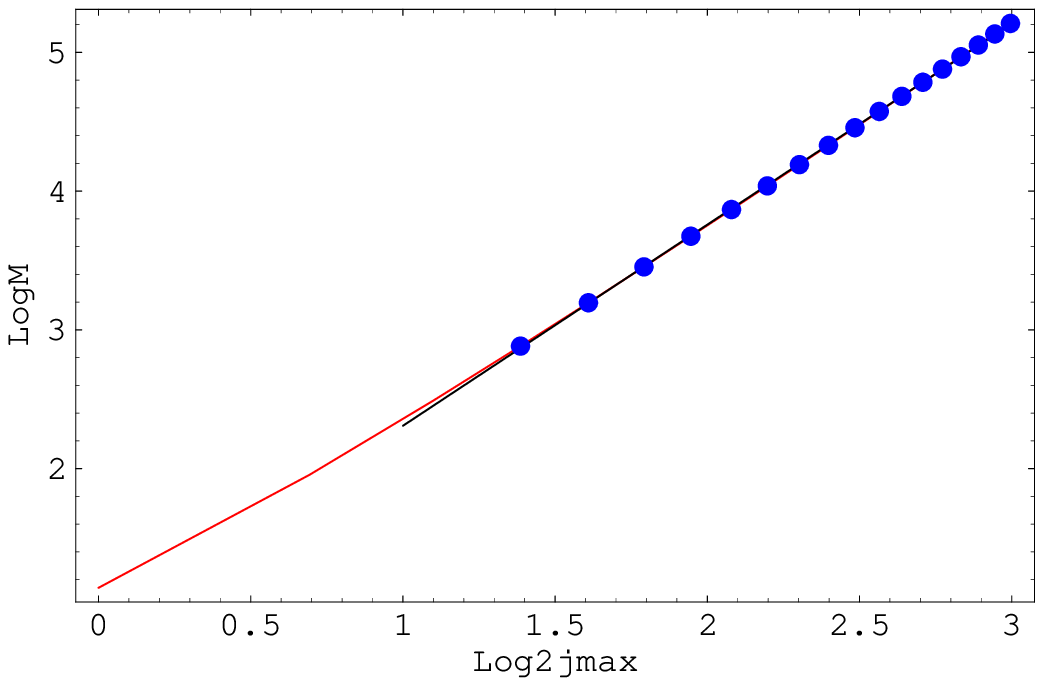}
    \caption{\label{6 vertex fit cubic max}Fit of the largest eigenvalues at the gauge invariant cubic 6-vertex up to $\jmax=\frac{20}{2}$.  We have used $\maxev(\jmax)$ for $\jmax=\frac{4}{2},\ldots \frac{20}{2}$ for the fitting (blue dots). The red line indicates the maximum eigenvalue sequence, the black line the fit function.}
\end{minipage}
\end{figure}
In figures \ref{6 vertex fit cubic min} and \ref{6 vertex fit cubic max} we
show fits of these data.  The fitting functions are:
\ba
\lambda_{\hat{V}}^{(min),(fit)}(\jmax)&=&\mb{e}^{0.387\pm 0.036}  (2\jmax)^{0.4877 \pm 0.0144 }
\ea
Here $\chi^2:=\sum_{k=1}^{15} \Big(\lambda_{\hat{V}}^{(min),(fit)}(\jmax^{(k)}) -\mineval(\jmax^{(k)}\big)\Big)^2=0.02$ for the chosen set of 15 data points. For the largest eigenvalue sequence we obtain
\ba
\lambda_{\hat{V}}^{(max),(fit)}(\jmax)&=&\mb{e}^{0.859\pm 0.0157} (2\jmax)^{1.44954 \pm 0.0064}
\ea
Here $\chi^2:=\sum_{k=1}^{17} \big(\lambda_{\hat{V}}^{(max),(fit)}(\jmax^{(k)}) -\maxev(\jmax^{(k)})\big)^2=4.04$ for the chosen set of 17 data points. 

\newpage
\subsection{Gauge Invariant 7-Vertex}
The 7-vertex has 3,079,875 \sigconfs, of which 1,912,373 are `non-trivial' (not all zero, and not equivalent to another up to an overall sign).  This, along with the septic number of spin configurations for a given $\jmax$, 
makes it extremely expensive computationally.  Thus we have 7-vertex data only for much smaller spins than for the other valences (though simply in terms of number of eigenvalues our 7-vertex data vastly overwhelms that for smaller valences!).

\subsubsection{Histograms}
\begin{figure}[h!]
 \center
    \center
    \psfrag{frequency}{$\numevals$}
    \psfrag{eigenvalue}{$\eval$}
    \includegraphics{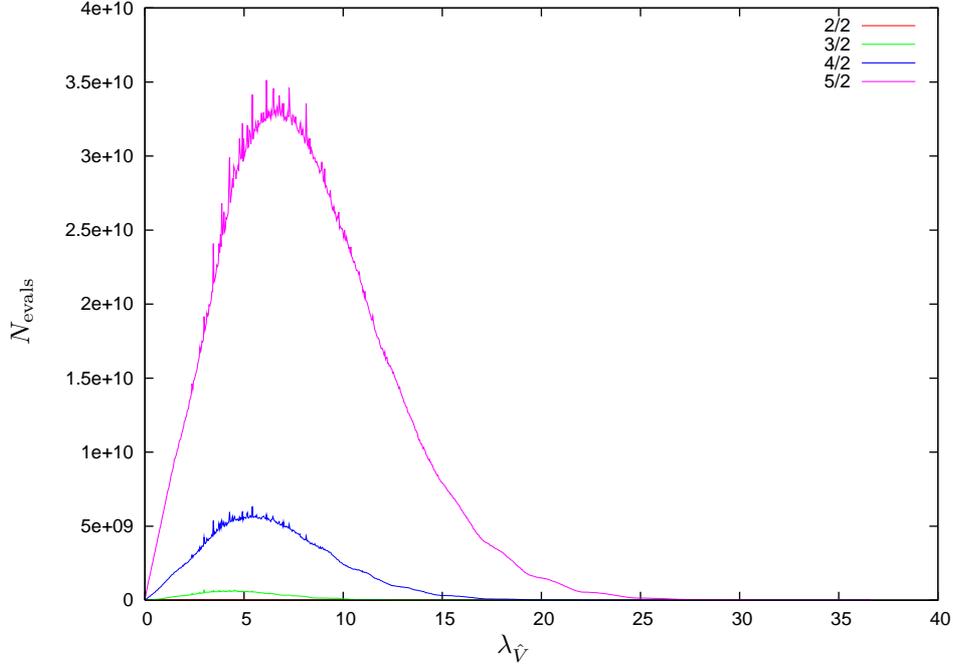}
    \caption{\label{7 vertex 1}Overall histograms for the gauge invariant
    7-vertex up to $\jmax=\frac{5}{2}$.  There are 9,475,991,439,360 eigenvalues
    in all, of which 146,895,794,400 are zero.  $\nbins=2048$.}
 \end{figure}
In figure \ref{7 vertex 1} we again see no lip in the spectral density close to zero, and furthermore notice that the density
appears to be linearly increasing for eigenvalues $\lesssim 3$.
Both of these features are likely due to the fact that we have computed these
spectra only up to the relatively small spin $\jmax=\frac{5}{2}$.  At the
5-vertex, for small
values of $\jmax$, we observe that there is no lip present (e.g.\ from figure
\ref{5 vertex lip logplot}).
Additionally, at the 6-vertex, for extremely small values of $\jmax$, we also fail to observe
an exponentially rising edge in the spectral density, but rather see linear
growth as above.
It seems likely, then, that both the exponentially increasing `edge', and the
lip at zero, will arise with larger $\jmax$, as it does at smaller valences.

\subsubsection{Fitting of Histograms}
\begin{figure}[hbt!]

   \center
    \psfrag{lambda}{$\eval$}
    \psfrag{l}{$\ln[\numevals]$}
    \psfrag{SevenVFitRisingEdgeLog}{}
    \includegraphics[height=6.4cm]{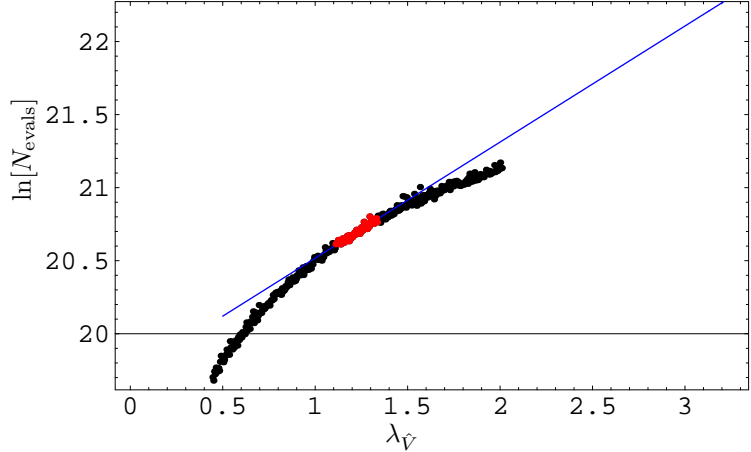}
     \caption{\label{7 vertex Fit 1} Logarithmic plot of the fit to the rising edge at the gauge invariant 7-vertex.  The points that were used in the fitting process are colored red.}
\end{figure}

As mentioned above, figure \ref{7 vertex 1} seems to possess a spectral density which rises linearly rather than exponentially with $\eval$. 
A fit to this rising edge is shown in figure \ref{7 vertex Fit 1}, in which the logarithm of the number $\numevals$ of eigenvalues  as a function of the eigenvalue $\eval$  is given by:
\be
   \ln[\numevals(\eval)]^{(fit)}=(19.72\pm 0.084) + (0.7943\pm 0.067) \eval
   ~~~~~~~~~~\longrightarrow~~~~~~~~~~~
   \numevals^{(fit)}(\eval) \sim 0.37\cdot 10^9 \cdot \mb{e}^{0.79~\eval} 
\ee
The quantity $\chi^2:=\sum_{k=1}^{51} \big( \numevals^{(fit)}(\eval^{(k)})-  \numevals(\eval^{(k)})\big)^2$
has the numerical value $\chi^2=1.25\cdot 10^{16}$ on the chosen set of 51 data points. We find that,  although the overall shape of the rising edge does not resemble a line in the logarithmic plot, the exponential fit is better than a linear fit, at least for the chosen set of data points.

Figure \ref{7 vertex Fit 1}  seems to indicate that we have not
yet entered the regime of large $\jmax$.
As the exponential growth of the spectral density is expected to be a
property of the spectrum at large $\jmax$,
the exponential fitting is not obviously applicable for the computed range of
spins.

\subsubsection{Extremal Eigenvalues}
\label{7v_extremal_evals}

As for the 5 and 6-vertices, by inspection of the eigenvalue data, we find smallest eigenvalue sequences which are decreasing, constant, or increasing as $\jmax$ is increased.

Since we can go only to a maximum spin of $\jmax=\frac{5}{2}$, we only have 4 data points (the case where all spins are equal to $\frac{1}{2}$ is excluded by the odd valence and the requirement that the sum of spins has to be integer in order to obtain gauge invariance), which is too few to identify an overall smallest eigenvalue sequence.
However, we can take the overall smallest eigenvalue in our numerical data, which arises e.g.\ from \sigconf{} \mbox{(2,~2,~2,~-4,~0,~0,~0,~0,~-2,~-2,~2,~2,~4,~-2,~4,~4,~0,~0,~0,~-2)},\footnote{The sigmas are again ordered in the
same pattern as for the smaller valences, i.e.\ 123 124 125 126 134 135 136
145 146 156 234 235 236 245 246 256 345 346 356 456.} 
 and take this as the smallest eigenvalue 
 \sigconf{}. The smallest eigenvalues of this sequence 
are contributed by the following spin configurations (here $j_7=\jmax$):
\begin{table}[htbp]
\center
\cmt{13}{
\center
\begin{footnotesize}
\begin{tabular}{|r|r|r|r|r|r|r|}
\hline
$2\cdot j_1$&$2\cdot j_2$&$2\cdot j_3$&$2\cdot j_4$&$2\cdot j_5$ & $2\cdot j_6$
& $2\cdot j_7$\\\hline
2&2&2&2&2&2&2\\
2&2&2&2&2&3&3\\
2&2&2&2&3&3&4\\
2&2&2&2&2&3&5
\\\hline
\end{tabular}
\end{footnotesize}
\caption{The spin configurations contributing the overall smallest eigenvalue for 
$\vec{\sigma}_{min}=(2, 2, 2, -4, 0, 0, 0, 0, -2, -2, 2, 2, 4, -2, 4, 4, 0, 0, 0, -2)$.}
\label{7v smallest eval spin config}}
\end{table} \\
The overall smallest eigenvalue is contributed from the spin config with $j_7=\jmax=\frac{5}{2}$, its numerical value is found to be $\mineval|_{\jmax=\frac{5}{2}}=0.000597922$. 
By inspection of the data we find that this eigenvalue is in fact obtained from 120 different $\vec{\sigma}$-configurations, each with the spin configuration in the last line of table 
\ref{7v smallest eval spin config}. This already indicates the importance of understanding the orbits of the $\vec{\sigma}$-configurations under permutations of the edges at a vertex \cite{VertexCombinatorics}: We have partially fixed this symmetry by demanding ordered spins (\ref{ordered spins}), however as 5 spins are equal in the spin configuration leading to $\mineval|_{\jmax=\frac{5}{2}}$, we are free to permute 5 out of 7 edges, that is we have $5!=120$ possibilities, each giving a different $\vec{\sigma}$-configuration.

\subsection{Numerical Errors}
\label{numerical_errors}

An interesting feature of our data, which is discussed in section
\ref{5v_histograms}, is a `lip' in the spectral density of the 5-vertex at
eigenvalues near zero, as manifested in figure \ref{lip_figure}.  One may be
concerned whether this is a genuine feature of the theory, or if it arises
from numerical noise.
A look at figure \ref{5 vertex min} causes alarm, in that it shows the
smallest non-zero eigenvalue decreasing monotonically to values arbitrarily
close to zero.  Since the numerical errors are obviously bounded away from
zero by machine epsilon, for some finite $\jmax$ the minimum eigenvalues must
inevitably become comparable with the numerical noise.

\subsubsection{Error bound provided by LAPACK}

In presenting our data we regard as zero any singular value
\be
s_i \leq \thresh \delta s_i
\label{error_threshold}
\ee
where $\delta s_i$ is given by eqn.\ (\ref{errors}) and $\thresh$ is a constant numerical threshold.  As we reduce $\thresh$,
we will regard smaller and smaller eigenvalues as non-zero.  For the data
presented up to this point $\thresh = 1$.
From \S \ref{Eval Computation} we recall that the eigenvalues are related to the singular values by
$\eval = \sqrt{s_i}$, so in terms of the eigenvalues, our error bound is
\be
\delta \eval = \frac{\delta s_i}{2 \sqrt{s_i}}
 = \frac{\delta s_i}{2 \eval}
\label{err_V}
\ee
The $\delta s_i$ are proportional to the largest singular value of
the matrix which gives rise to $s_i$.  This is bounded by the square of the
largest eigenvalues which occur.  A more significant effect is that,
because of the square root,
the errors in the eigenvalues are inversely proportional to the eigenvalues
themselves.  Thus at large $\jmax$, as $\eval$ converges to zero,
our errors grow without bound.  Because of this the computation of volume
eigenvalues for valences $N>4$ is bounded by machine precision, rather than
computation time or system memory.

The error bars in figure \ref{5 vertex min}
reveal that, for the 5-vertex, the smallest non-zero eigenvalues are
comparable with our error bound for $\jmax \gtsim \frac{35}{2}$.
\begin{figure}[htbp]
\center
\center
\psfrag{sci}{\sigconf}
\psfrag{jmax}{$2 \jmax$}
\includegraphics[width=15cm]{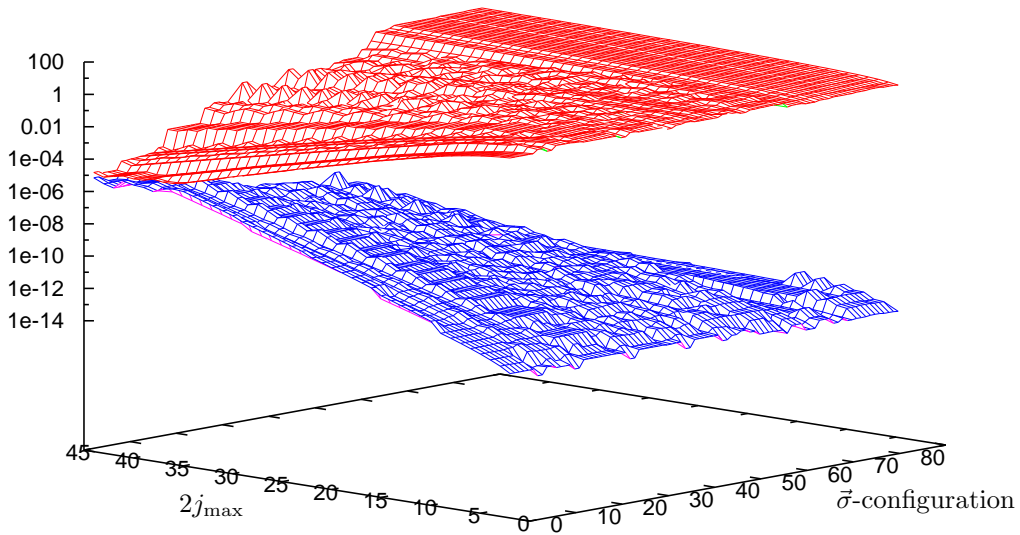}
\caption{Smallest non-zero eigenvalues of 5-vertex (top, red), along with their LAPACK error
  bound (bottom, blue).  The vertical axis is in logscale.}
\label{5v_min_eval_error1th}
\end{figure}
Figure \ref{5v_min_eval_error1th} portrays the closeness of the remainder of our 5-vertex
dataset to the LAPACK error bound.  The top (red) surface is the minimum
eigenvalues for each \sigconf{} and $\jmax$. (This is equivalent to figure
\ref{5 vertex min}.\footnote{The \sigconfs{} of figure \ref{5v_min_eval_error1th} are
  labeled by a sequential index (starting from zero) over the \sigconfs{} shown in table
  \ref{min_evals_sig_conf_equivalence_classes}, rather than the
  $\vec\sigma$-indices of figure \ref{5 vertex min}.})  The bottom (blue) surface is the LAPACK error bound for
these minimum eigenvalues.  The vertical axis is in logscale.  
Note that the errors grow in inverse proportion to the eigenvalues, as
expected by (\ref{err_V}).
They
become comparable only for a tiny `corner' of the dataset, with $\jmax \gtsim
\frac{35}{2}$, and $\vec\sigma$-indices (from table
\ref{min_evals_sig_conf_equivalence_classes}) 1--4.
(The smallest eigenvalues are always greater than the error bound by
construction, c.f.\ (\ref{error_threshold}) with $\thresh=1$.)
\begin{figure}[htb]
\center
\psfrag{sci}{\cmt{3}{\sigconf}}
\psfrag{jmax}{\cmt{3}{$2 \jmax$}}
\includegraphics[width=17cm]{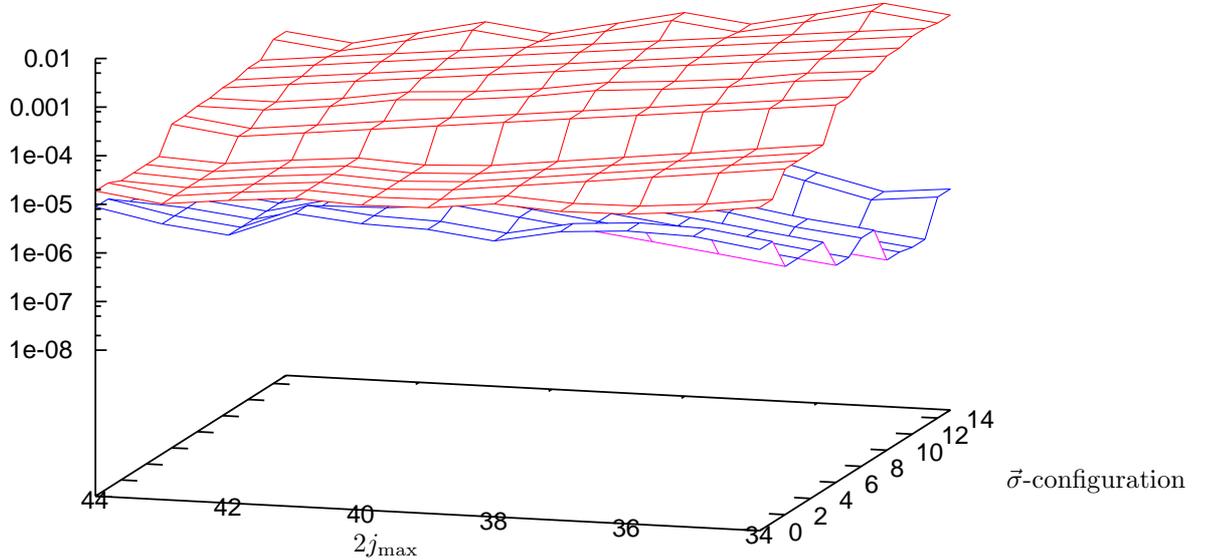}
\caption{Smallest eigenvalues of 5-vertex (top, red), along with their LAPACK error
  bound (bottom, blue), for `large $\jmax$ small \sigconf{} corner' of dataset.  The vertical axis is in logscale.}
\label{5v_min_eval_error1th-zoom}
\end{figure}
Zooming into this corner, as shown in figure \ref{5v_min_eval_error1th-zoom},
illustrates the effect of the error bound.  Consider the \sigconfs{} labeled
6 and 7.  (These are $\vec\sigma$-index 3 in table
\ref{min_evals_sig_conf_equivalence_classes}.)  The smallest non-zero
eigenvalue decreases steadily until $\jmax=\frac{44}{2}$, at which point it
rises suddenly.  For \sigconfs{} 2--4 we see similar behavior, with the
smallest non-zero eigenvalue rising at $\jmax=\frac{41}{2}$, and \sigconfs{}
0 and 1 see a rise at both $\jmax=\frac{41}{2}$ and $\frac{38}{2}$.  This is
likely due to the smallest non-zero eigenvalue dipping below the error
threshold, such that it is regarded as zero from inequality
(\ref{error_threshold}).  Then the next-smallest eigenvalue is regarded as
zero, resulting in an apparent jump in figure
\ref{5v_min_eval_error1th-zoom}.  Thus we can see that some small number of
our non-zero eigenvalues, for $\jmax \gtsim \frac{36}{2}$, are descending
beneath our error bound.  In section \ref{noise_effect} we will estimate the
number of eigenvalues that are thus `lost'.

\subsubsection{Characterization of true numerical noise}

How well does the error bound provided by LAPACK describe our numerical
noise?  To explore this question, we can vary $\thresh$ in
(\ref{error_threshold}), and watch what happens to our data.

\begin{figure}[htbp]
\center
\begin{minipage}{8.5cm}
\center
\psfrag{thresh}{$\thresh$}
\psfrag{nzeros}{$N_{\mathrm{zeros}}$}
\includegraphics[width=8.7cm]{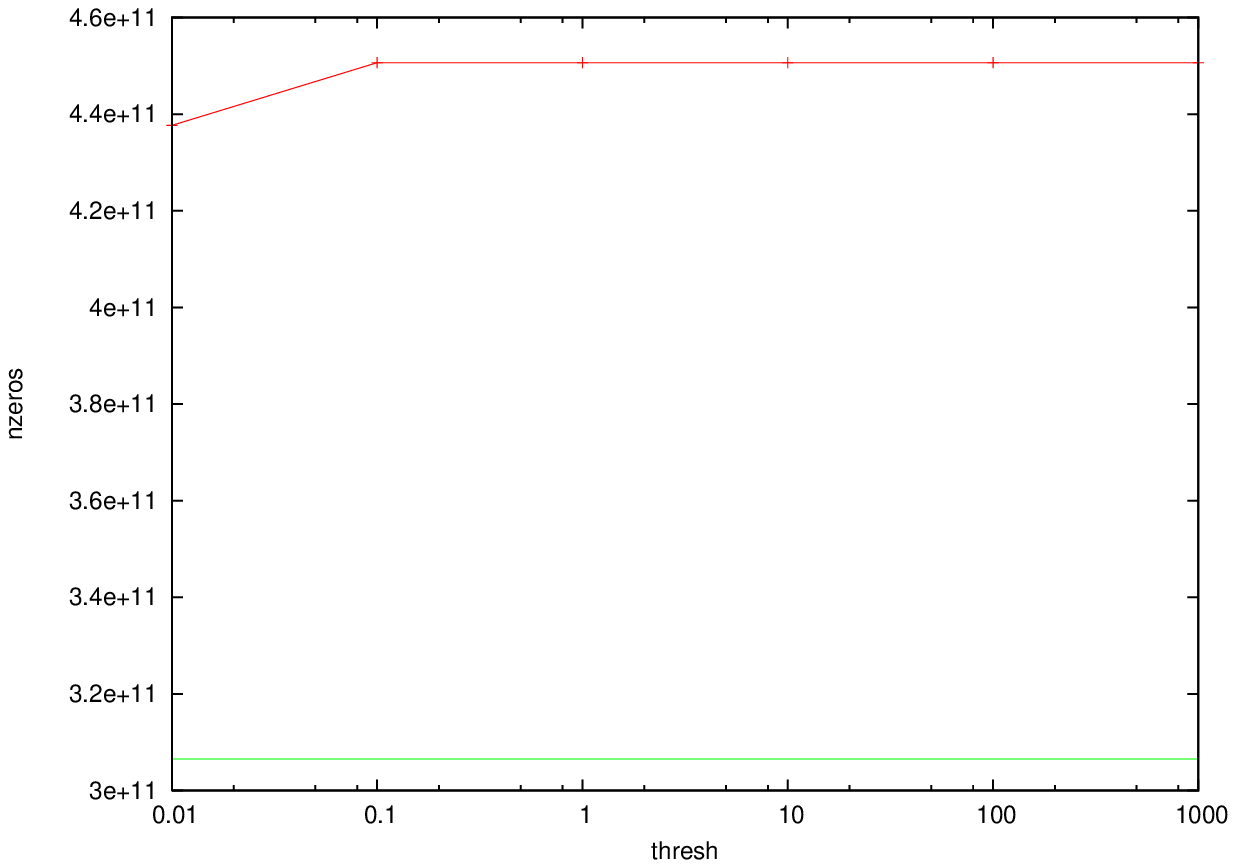}
\end{minipage}
\begin{minipage}{8.5cm}
\center
\psfrag{thresh}{$\thresh$}
\psfrag{nzeros}{$N_{\mathrm{zeros}}$}
\includegraphics[width=8.7cm]{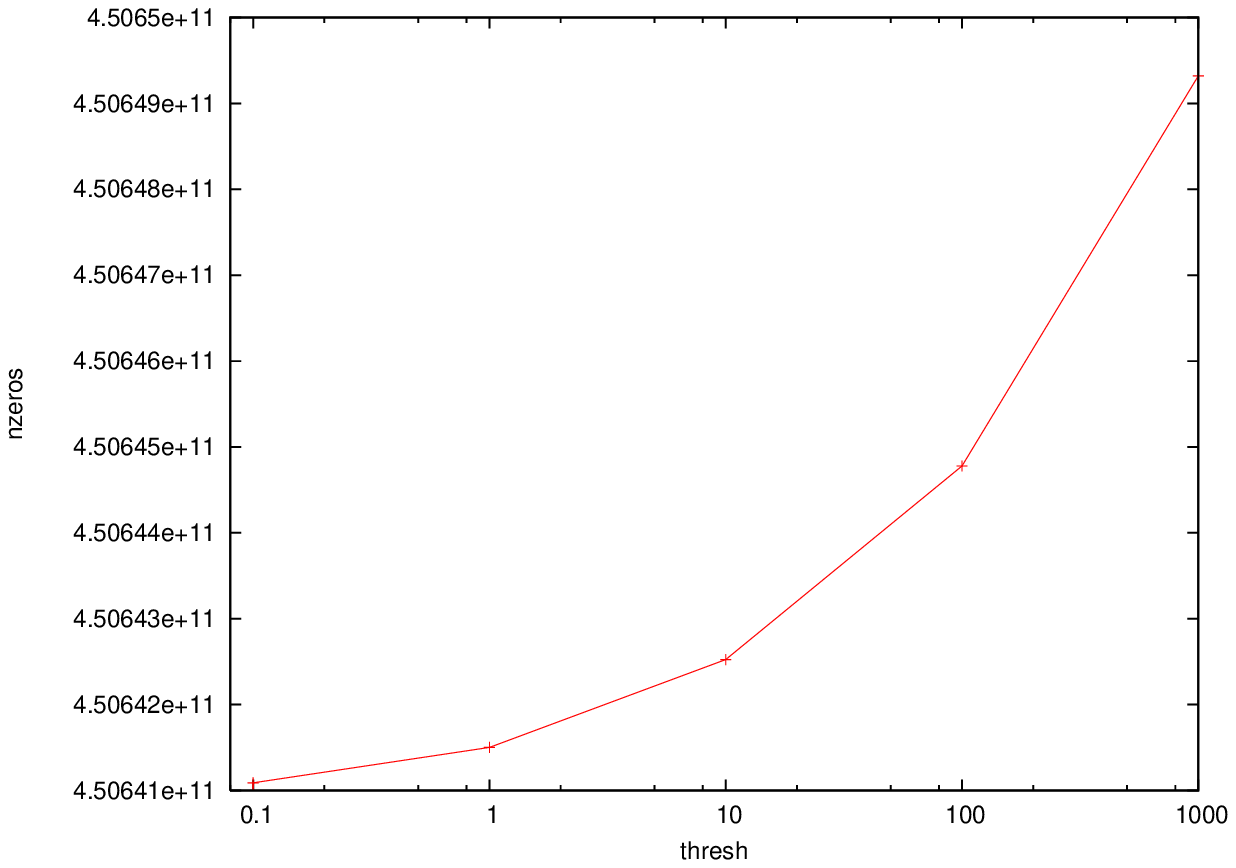}
\end{minipage}
\caption{Number of zero eigenvalues for 5-vertex, as a function of error
  threshold $\thresh$.  The green line at the bottom of the left plot indicates the
  number of `obvious' zeros, with $\thresh=0$.  The plot on the right zooms
  in on the data points for $\thresh>0.08$.}
\label{nzeros_vs_thresh}
\end{figure}

The simplest measure of the effect of varying the error threshold is to count
the number of zero eigenvalues.  Figure \ref{nzeros_vs_thresh} plots the
overall number of zero eigenvalues as a function of $\thresh$.  The left
figure shows that the number of zeros is essentially constant for any
$\thresh \geq 0.1$.  Below that value the number of zeros drops off rapidly,
indicating that a significant number of zero eigenvalues are being regarded
as non-zero due to the threshold being less than the numerical noise.  The
plot on the right shows a relatively slow increase in the number of zeros for
larger thresholds, which is expected.  Here genuine non-zero eigenvalues are
being regarded as zero due to the large error threshold.

\begin{figure}[htbp]
\center
\psfrag{sci}{\cmt{3}{~\\[-1mm]\sigconf}}
\psfrag{jmax}{\cmt{3}{~\\[-1mm]$2 \jmax$}}
\includegraphics[width=19cm]{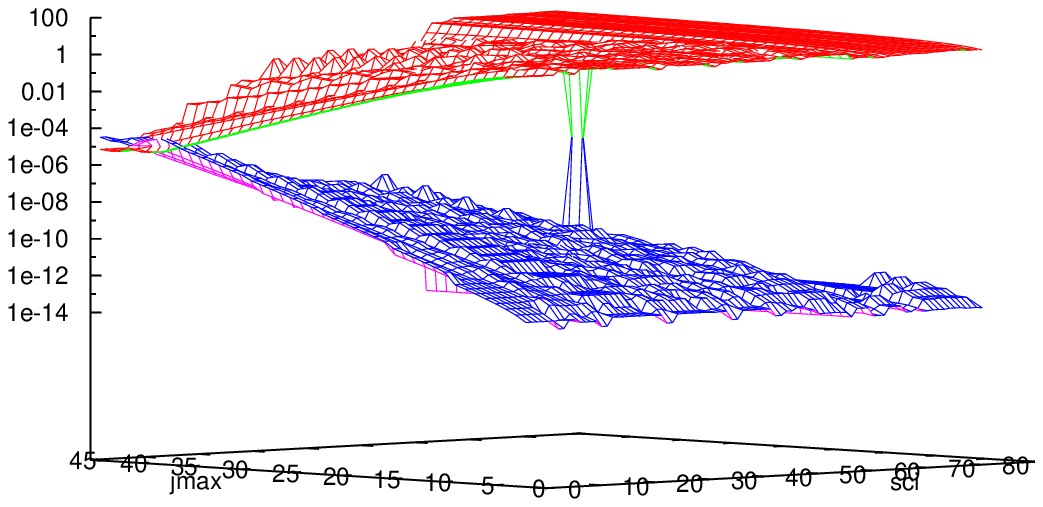}
\caption{Smallest eigenvalues of 5-vertex (top, red), along with their LAPACK error
  bound (bottom, blue), for $\thresh=0.1$.  The vertical axis is in logscale.}
\label{5v_min_eval_error.1th}
\end{figure}

Given the results of the above paragraph, we may be tempted to use
$\thresh=0.1$, so as to catch as many genuine small but non-zero eigenvalues
as possible.  Figure \ref{5v_min_eval_error.1th} shows what happens to figure
\ref{5v_min_eval_error1th} when $\thresh$ is reduced to 0.1.  Two features
are of note.  In our `large $\jmax$ small \sigconf{} corner' we see that with
$\thresh<1$ the error now can exceed the smallest non-zero eigenvalues, which
it does.  We also note a huge spike for \sigconf{} 72 ($\vec\sigma$-index 36), $\jmax= \frac{35}{2}$--$\frac{36}{2}$,
for which the smallest non-zero eigenvalue and its error jump to meet (in
fact cross) at around $10^{-5}$.  This occurs because numeric noise has
exceeded the error threshold, and is now being regarded as a genuine smallest
non-zero eigenvalue.  Note that this essentially random event can occur at
any point in our dataset.  We thus see that our true numerical error can
exceed one tenth of LAPACK's bound, and thus stick with their reported bound
with $\thresh=1$.  The fact that this problem occurs just below the reported
bound testifies to it being an accurate and tight estimate of the true
numerical error.

At $\thresh=0.01$ the analog of figure \ref{5v_min_eval_error.1th} becomes
mostly noise:  The smallest non-zero eigenvalues are a constant $\sim
10^{-5}$ for most \sigconfs{} and values of $\jmax$.
\subsubsection{Effect of noise on results}
\label{noise_effect}

\begin{figure}[htbp]
\center
\psfrag{sci}{\cmt{3}{~\\[-1mm]\sigconf}}
\psfrag{jmax}{\cmt{3}{~\\[-1mm]$2 \jmax$}}
\psfrag{minev}{$\mineval$}
\includegraphics[width=15cm]{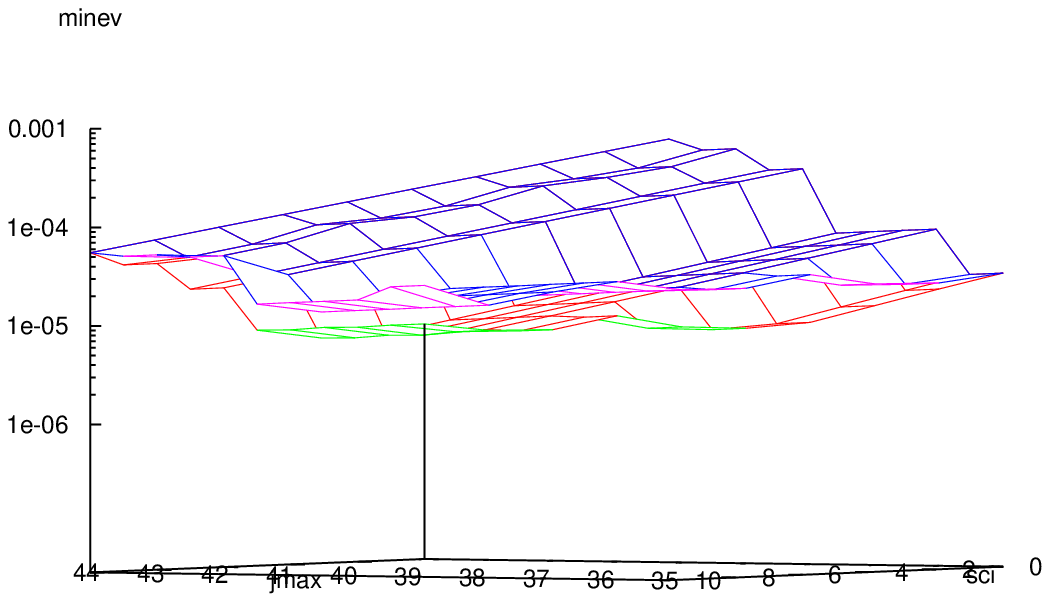}
\caption{Smallest eigenvalues of 5-vertex, for $\thresh=1$ (top surface: upper
  side blue, bottom side magenta), and $\thresh=0.1$ (bottom surface: upper
  side red, bottom side green).  The vertical axis is in logscale.}
\label{5v_min_eval_error1.1th}
\end{figure}

Reducing $\thresh$ allows us to probe the `problematic corner' a bit closer
to the numerical noise.
Figure \ref{5v_min_eval_error1.1th} portrays the effect of smaller $\thresh$
on the smallest non-zero eigenvalues.  Clearly some valid non-zero
eigenvalues are being washed out due to the noise.

The number of eigenvalues `between the two surfaces' of figure
\ref{5v_min_eval_error1.1th} is the difference between the two leftmost data
points on the right hand plot of figure \ref{nzeros_vs_thresh}, which is
413,160.  The largest difference between data points on the right hand plot
is 8,230,920.  It is impossible to determine how many more non-zero eigenvalues
there are below the bottom surface, but we expect it to be approximately of this larger
order of magnitude.
Now the bin width of the histogram in figure \ref{5 vertex 2} is
$\frac{497.223680}{2048}=0.242785$.  Clearly this is much larger than any of
the eigenvalues close to the error threshold that we are considering, so the
only effect of noise on the 5-vertex histogram is in the bin for smallest
$\eval$.  The population of that bin, for $\thresh=1$, is
154,430,880.  The true value is slightly larger than this.
An addition of 8,230,920 to this number will have an
indiscernible effect.  In any event the lip is a genuine feature of the volume
operator, and not a result of numerical error.

\subsubsection{Valences $N \neq 5$}

\begin{figure}[htbp]
\center
\psfrag{sci}{\sigconf}
\psfrag{jmax}{$2 \jmax$}
\includegraphics[width=17cm]{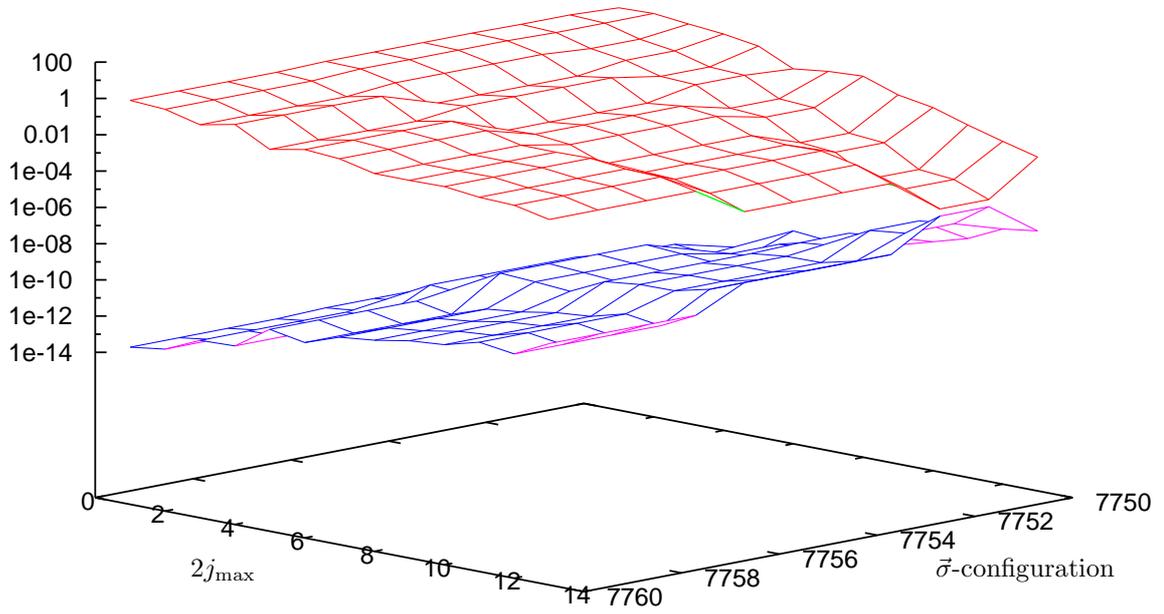}
\caption{Some of the smallest non-zero eigenvalues of the 6-vertex (top,
  red), along with their LAPACK error bound (bottom, blue).  The vertical
  axis is in logscale.}
\label{6v_min_eval_error1th}
\end{figure}

Figure \ref{6v_min_eval_error1th} depicts the distance between the error
bound and smallest non-zero eigenvalues for the 6-vertex.  The minimal
eigenvalue is closest to the error bound for the \sigconfs{}
7751 and 7752 shown, out of 8207 \sigconfs{} in total.
\begin{figure}[htbp]
\center
\psfrag{minev}{$\lambda_{\mathrm{min}}$}
\psfrag{jmax}{$2 \jmax$}
\includegraphics[width=15cm]{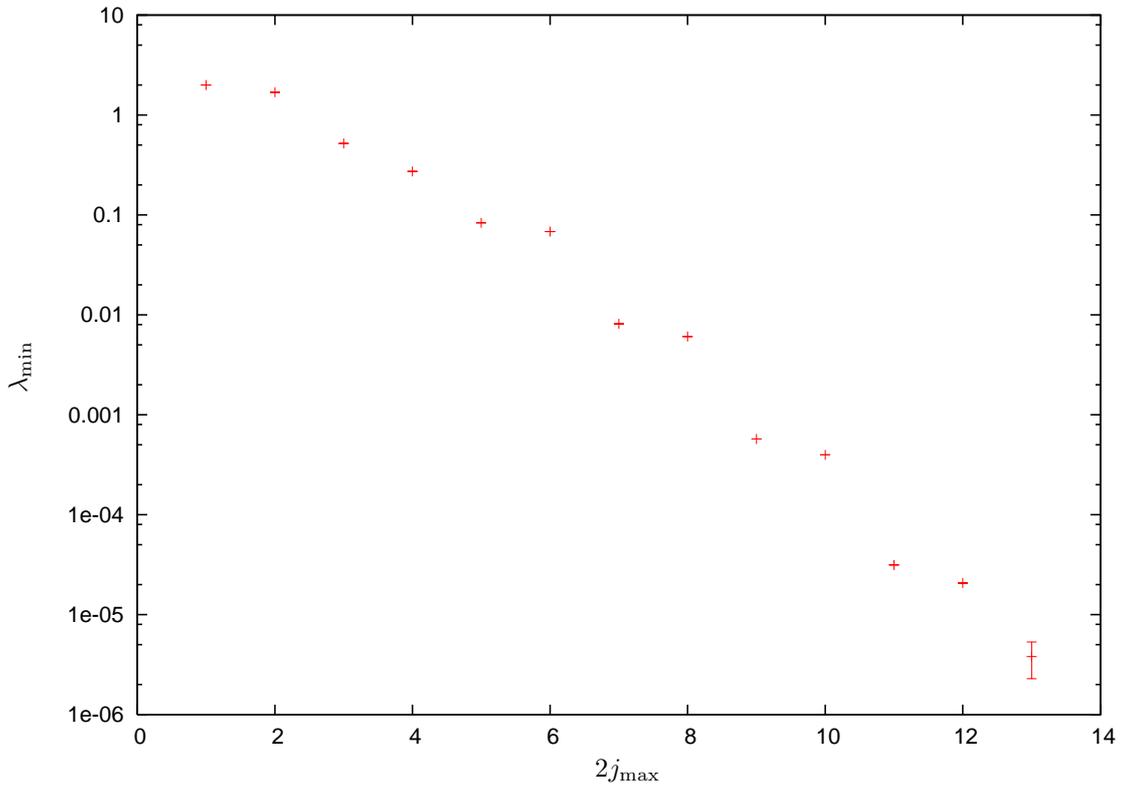}
\caption{Minimal eigenvalues for \sigconf{} 7751 = (0,  2, -2,  0,  0,  0, -2,  0,  2,  0),
which yields the overall smallest non-zero eigenvalue at the 6-vertex up to $\jmax=\frac{13}{2}$.  $\thresh=1$.}
\label{6v_min_eval_error1th-zoom}
\end{figure}
Figure \ref{6v_min_eval_error1th-zoom} shows the same numbers for
\sigconf{} 7751, with error bars.  This curve also appears as one of the fifty
shown in figure \ref{6 vertex smallest eval series}.  It appears that we are
just at the threshold of the numerical noise for the 6-vertex.  Given the
analysis of the 5-vertex errors above, we expect the errors to have
negligible effect on the 6-vertex results.

(One of) the smallest eigenvalue(s) for our 7-vertex data, as described in section \ref{7v_extremal_evals},
arises from \sigconf{} \mbox{(2,~2,~2,~-4,~0,~0,~0,~0,~-2,~-2,~2,~2,~4,~-2,~4,~4,~0,~0,~0,~-2)},
with spins \mbox{(2, 2, 2, 2, 2, 3, 5)}.  Its value is 0.000597922, with error bound $2.54519 \cdot 10^{-9}$.  Clearly numerical error is not a significant issue for the
7-vertex.

Since the smallest non-zero eigenvalues for the 4-vertex rise with $\jmax$,
we do not expect any effect from numerical noise.  The relative error for
every eigenvalue at the 4-vertex will be negligible.
The same situation holds for the cubic 6-vertex of section \ref{cube}.

\subsubsection{Symmetric Eigenproblem}
\label{eigenproblem}
Earlier we computed eigenvalues more directly using subroutines
LAPACK provides for the `symmetric eigenproblem'.  To cast the matrices into
symmetric form we computed eigenvalues for $QQ^\dagger$ rather than $Q$.
In this case numerical error is much more prominent, for a number of
reasons.  

For the symmetric eigenproblem, LAPACK 
employs the `relatively robust representation algorithm', which involves an
$LDL^T$ factorization and a number of `translates', to 
return the matrix eigenvalues $e_i$ with a
fixed absolute error
\be
\delta e_i = \epsilon \|QQ^\dagger \|_1
\ee
where $\|QQ^\dagger\|_1$ is the 1-norm (maximum column sum) of $QQ^\dagger$.
This 1-norm is bounded from below by the largest eigenvalue.
For singular value decomposition (SVD) the relative error is fixed and thus
proportional to the largest eigenvalue.

We see that the relative error resulting from any algorithm is bounded below
by the machine precision.
Thus the error in the smallest eigenvalues can never be less than the maximum
eigenvalue times machine epsilon, so that,
when considering numerical error,
what is important is the ratio between the largest and smallest eigenvalues
which arise from a given matrix.
When we square the matrix, we double this ratio, which in effect halves the
relative precision of the computation.

Note that, as manifested in figures \ref{5v_min_eval_error1th} and
\ref{6v_min_eval_error1th}, the error bound in the smallest eigenvalues appears to be growing exponentially with $\jmax$.
This occurs simply because we are taking the square root of the singular
values to get the eigenvalues, and the minimum eigenvalues decay exponentially toward
zero (as one can see from figure \ref{5 vertex smallest evals}), so,
according to eqn.\ \ref{err_V}, the error in the square root must grow exponentially.
In the case of the symmetric eigenproblem, since we square $Q$ before
computing eigenvalues, we must take a 4th root to
get the volume eigenvalues.  In this case the analog of (\ref{err_V}) has the
volume eigenvalue \emph{cubed} in the denominator.  Thus the 
log of the errors in the volume eigenvalues will grow three times as fast as
for SVD.  
This effect, along with the effect of squaring $Q$ discussed in
the previous paragraph, represents a considerable increase in numerical error.
We are thus able to explore the spectrum to much higher values of $\jmax$ with
the singular value decomposition.

\subsection{Histograms of Entire Data Set}

\begin{figure}[htbp]
  \center
\hspace{-2mm}
\begin{minipage}{8.8cm}
  \psfrag{frequency}{$\nevals$}
  \psfrag{eigenvalue}{$\eval$}
  \psfrag{4-vertex}{\hspace{-10mm}\tiny 4-vertex}
  \psfrag{5-vertex}{\hspace{-10mm}\tiny 4,5-vertices}
  \psfrag{6-vertex}{\hspace{-10mm}\tiny 4,5,6-vertices}
  \psfrag{7-vertex}{\hspace{-10mm}\tiny 4,5,6,7-vertices}
  \includegraphics[width=8.8cm]{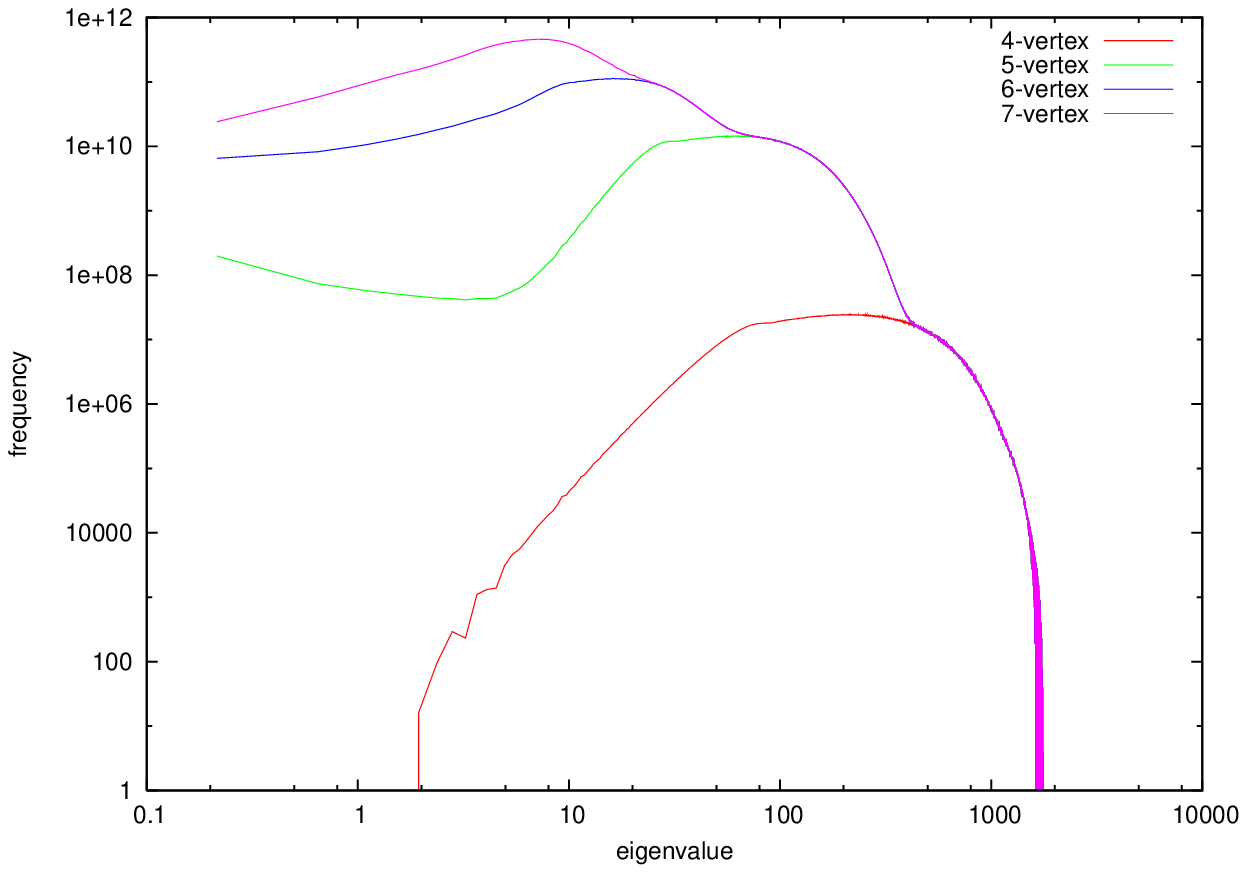}
  \caption{Histograms for entire dataset, in log-log scale.  $\nbins = 4096$.
  Each successive curve adds eigenvalues for the next larger valence.  Here
  the cutoffs for $\jmax$ are: $\frac{126}{2}$ for the 4-vertex,
  $\frac{44}{2}$ for the 5-vertex, $\frac{13}{2}$ for the 6-vertex, and
  $\frac{5}{2}$ for the 7-vertex.}
  \label{overall_hist}
\end{minipage}
~
\begin{minipage}{8.8cm}
  \psfrag{numevals}{$\nevals$}
  \psfrag{eigenvalue}{$\eval$}
  \psfrag{4vertex}{\hspace{-4mm}\tiny 4-vertex}
  \psfrag{5vertex}{\hspace{-4mm}\tiny 5-vertex}
  \psfrag{6vertex}{\hspace{-4mm}\tiny 6-vertex}
  \psfrag{7vertex}{\hspace{-4mm}\tiny 7-vertex}
  \includegraphics[width=8.8cm]{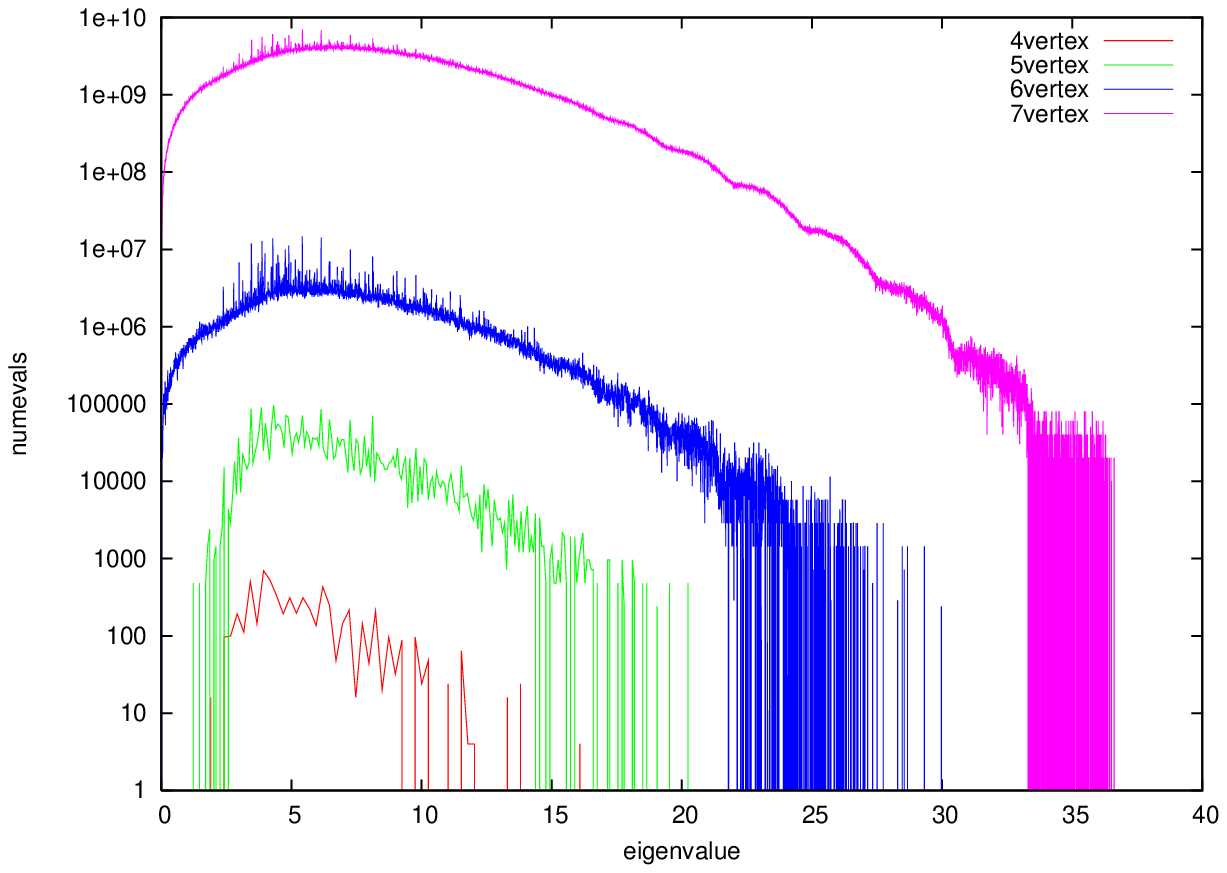}
  \caption{Histograms for entire dataset, in log-linear scale, with $\jmax$
  fixed at $\frac{5}{2}$.  Here the histograms are not cumulative: each
  valence appears separately.\newline $\nbins = 64, 256, 4096, 8192$, for
  valences 4 -- 7 respectively.\newline}
  \label{fixed_jmax_overall_hist}
\end{minipage}
\end{figure}

Figure \ref{overall_hist} presents four histograms of our entire data set.
Each represents all eigenvalues we have computed up to a given valence.
(Thus the blue curve, for example, includes all 4-, 5-, and 6-vertex
eigenvalues we have computed.)
There are 21,551,022,251,376 eigenvalues in all, of which
 827,849,074,512 are
(regarded as, c.f.\ discussion on errors in \S \ref{numerical_errors} above) zero.

Figure \ref{fixed_jmax_overall_hist} shows histograms for the entire dataset
truncated to $\jmax=\frac{5}{2}$.  We see that, as one might
expect, larger valences are able to contribute larger eigenvalues, even for
comparable values of spin.  Note that $\jmax=\frac{5}{2}$ is not large enough
to see the absence of a volume gap for the 5-vertex.

\section{\label{sec Analytical Results}Analytical Results at the Gauge Invariant 4-Vertex}

In this section we present three analytical results on the spectral properties of the volume operator at the gauge invariant 4-valent vertex. 
We first prove the spectral simplicity and second give a general formula for the eigenvectors of the volume operator at the 4-vertex depending on the eigenvalues, which agrees with the earlier publication \cite{Carbone Carfora Marzouli}. Third we analyze how the smallest non-zero eigenvalue scales with the maximal spin $\jmax$ at the vertex.

For the methods we are using here the particular shape of the operator, when represented as a matrix, will be crucial. In fact many theorems developed for Jacobi matrices\footnote{That is symmetric tridiagonal matrices.} can be carried over to the particular case under investigation. Here \cite{Courant Hilbert Approach Book} provides the crucial theorems and proofs.

\subsection{Setup: Explicit Expression for the Volume Operator as $D$-dimensional Matrix $\widehat{Q}_D$}

The general case of the gauge invariant 4-vertex deals with 4 edges $e_1, \ldots,e_4$ outgoing from the vertex $v$.  Each edge $e_K$ ($K=1\ldots 4$) carries a $2j_K+1$ dimensional representation of $SU(2)$.
Due to gauge invariance the standard recoupling basis states (\ref{Definition recoupling scheme}) are given in this case as
\be\label{gauge invariant recoupling basis for 4-vertex}
   \ket{a_2(j_1\,j_2)~a_3(a_2\,j_3)\!\stackrel{!}{=}\!j_4~J(a_3\,j_4)\!=\!0~M\!=\!0}
\ee
where the intermediate recoupling $a_3(a_2\,j_3)$ has to equal $j_4$ due to the Clebsch-Gordan Theorem and as a result the intermediate recoupling $a_2(j_1\,j_2)$ is the only degree of freedom for fixed spins $j_1,\ldots,j_4$.
Therefore expression (\ref{Volume definition gauge invariant 3}) for the matrix representation of the volume operator simplifies dramatically to give 
\be
  \hat{V}_v=\sqrt{\big|Z\cdot \sigma(123)~\hat{q}_{123} \big|}=\sqrt[4]{|Z|^2\cdot[\sigma(123)]^2 \cdot \widehat{Q}^{\dagger} \widehat{Q}}
\ee
where we have introduced the shorthand $\widehat{Q}:=\hat{q}_{123}$. Note that $\sigma(123)=\epsilon(123)-\epsilon(124)+\epsilon(134)-\epsilon(234)=0,\pm 2,\pm 4$ gives a constant numerical prefactor depending on the relative orientations of the 4 edges only. In the following we will assume that $\sigma(123)\ne 0$ but leave its numerical value unspecified, that is we will drop it from our formulae and reinsert it at the end. We set the regularization prefactor $Z=\mb{i}$, c.f.\ section \ref{Right Invariant Vector Fields as Angular Momentum Operators}.\footnote{
Note that due to the definition of the eigenvalue $\eval$ of $\hat{V}$ as $\eval:=\sqrt{|\lambda_Q|}$, this does not change the volume spectrum.  Because $Q$ is an antisymmetric matrix, setting
its matrix elements 
to be real or imaginary
only rotates its spectrum in the complex plane either 
to the complex axis or the real axis.}
Using (\ref{q123}) together with the definitions of section  \ref{6j symbols in V} we obtain for the matrix $\widehat{Q}=\widehat{Q}_D$ ($D$ denotes the dimension)  
the following shape \cite{de Pietri,Volume Paper}:

\setlength{\arraycolsep}{1.2mm}
\be\label{general matrix 4 vertex}
   \widehat{Q}_D=\left( 
     \begin{array}{cccccccccccccccc}
 0     & -q_1 &  0   &\cdots&  0     &  0     &0   \\ 
 q_1   &   0  & -q_2 &  \cdots&  0     &  0     &  0        \\ 
 0     & q_2  &  0   &\cdots&  0     &  0     &  0        \\ 
 \vdots&\vdots&\vdots&    \ddots&\vdots&\vdots&\vdots \\
  0&0&0&\cdots& 0    &-q_{D-2}&  0       \\ 
    0  &   0  &   0   &\cdots&q_{D-2}&  0     &-q_{D-1}      \\ 
    0  &   0  &   0  &\cdots&   0   &q_{D-1} &   0    \\  
\end{array} 
   \right)
\ee
where the matrix elements are given by 
\be\label{general matrix element 4 vertex}\begin{array}{lllll}
   q_k&=&\frac{\mb{i}}{\sqrt{(2a_2+1)(2a_2-1)}}&
   \Big[(j_1+j_2+a_2+1)(-j_1+j_2+a_2)(j_1-j_2+a_2)(j_1+j_2-a_2+1)\times \\
   &&&\!\!\times(j_3+j_4+a_2+1)(-j_3+j_4+a_2)(j_3-j_4+a_2)(j_3+j_4-a_2+1)\Big]^\frac{1}{2}
\end{array}\ee
and the intermediate recoupling step $a_2=a_2(j_1\,j_2)$ is given as $a_2=a_2^{(min)}+k$,
that is
\be \label{range of a2}
\max{\big[|j_2-j_1|,|j_4-j_3|}\big]=a_2^{(min)}\le a_2 \le a_2^{(max)}=\min{\big[j_1+j_2,j_3+j_4 \big]} 
\ee
and 
$D=\dim \widehat{Q}_D =a_2^{(max)}-a_2^{(min)}+1$.
 
\subsubsection{Known Results} 

\begin{enumerate}
  \item{Spectrum: By choosing $Z=\mb{i}$, (\ref{general matrix 4 vertex}) is an antisymmetric matrix with purely imaginary matrix elements. Its eigenvalues $\lambda$ are thus real and come in pairs $\lambda=\pm |\lambda|$. Moreover its eigenvectors $\Psi^\lambda=(\Psi^\lambda_1,\ldots,\Psi^\lambda_D)^T$ are orthogonal and there exists a unitary matrix
  \be\label{Diag Matrix of Q}
   U_D=\left(\begin{array}{cccc}
               \Psi^{\lambda_1}_1 & \Psi^{\lambda_2}_1&\cdots&\Psi^{\lambda_D}_1\\
	       \Psi^{\lambda_1}_2 & \Psi^{\lambda_2}_2&\cdots&\Psi^{\lambda_D}_2\\
	       \vdots&\vdots&\cdots&\vdots\\
	       \Psi^{\lambda_1}_D & \Psi^{\lambda_2}_D&\cdots&\Psi^{\lambda_D}_D\\
	       
           \end{array}\right)
  \ee
  such that $U_D\Lambda_D = \widehat{Q}_D U_D$,~~ $\Lambda_N=U_D^{-1}~\widehat{Q}_D~U_D=U^\dagger_D~ \widehat{Q}_D~ U_D$,~~ $\Lambda_N=\mbox{diag}(\lambda_1,\ldots,\lambda_D)$}
  \item{0-eigenvalues: It has been found in \cite{Volume Paper} that $0$ as a single eigenvalue is contained only in the spectrum of odd dimensional matrices $\widehat{Q}_D$.}
\end{enumerate}
\subsection{Simplicity of the Spectrum of the Matrix $\widehat{Q}_D$} 

By inspection an eigenvector $\Psi^\lambda=(\Psi^\lambda_1,\ldots,\Psi^\lambda_D)^T$ for the eigenvalue $\lambda$ of (\ref{general matrix 4 vertex}) with $(\widehat{Q}_D-\lambda \mb{1})\Psi^\lambda=0$ will fulfill:
\be\label{Recursion Relations}\begin{array}{rrclclc}
    \mbox{(I)} & -\lambda \Psi^\lambda_1 -q_1 \Psi^\lambda_2 &=&0&& \\ 
    \mbox{(II)}& q_{k-1}\Psi^\lambda_{k-1}-\lambda\Psi^\lambda_k-q_k\Psi^\lambda_{k+1}&=&0 &&\fbox{$1 < k < D$}\\
    \mbox{(III)} & q_{N-1}\Psi^\lambda_{D-1}-\lambda \Psi^\lambda_D&=&0&& \\ 
\end{array}\ee

Now suppose $\lambda\ne 0$.\footnote{The case $\lambda= 0$ is already discussed in \cite{Volume Paper}.}
 Setting $\Psi^\lambda_1=0$ leads, together with (I) and the fact that $q_k \ne 0$ for $k=1,\ldots N-1$, to $\Psi^\lambda_2\stackrel{!}{=}0$, which then in turn implies together with (II) that $\Psi^\lambda_k\stackrel{!}{=}0$  $\forall k$. Therefore in order to get a non-vanishing eigenvector $\Psi^\lambda$, $\lambda\ne 0$ we must have $\Psi^\lambda_1\stackrel{!}{\ne}0$.

Now suppose there exists an eigenvalue $\lambda\ne 0$ of $\widehat{Q}_D$ that has multiplicity $>1$ : 
Let $\Psi^\lambda,\tilde{\Psi}^\lambda$ be two eigenvectors for $\lambda$.  
Then we can find two constants $\alpha,\tilde{\alpha}\in\mb{C}$, $\alpha,\tilde{\alpha}\ne0$, such that we can construct an eigenvector $\Phi^\lambda= \alpha\Psi^\lambda + \tilde{\alpha}\tilde{\Psi}^\lambda$ with the property $\Phi^\lambda_1=0$.
As we have seen this implies $\Phi^\lambda_k\stackrel{!}{=}0$  $\forall k$, and thus $0= \alpha\Psi^\lambda + \tilde{\alpha}\tilde{\Psi}^\lambda$, that is $\Psi^\lambda,\tilde{\Psi}^\lambda$ are linearly dependent and therefore there exists only one eigenvector $\Psi^\lambda$ for each eigenvalue $\lambda\ne 0$ of $\widehat{Q}_D$. We have thus proved the following
\begin{Theorem}{Spectral Simplicity of $\widehat{Q}_D$} \label{Spectral Simplicity Thm}\\
  (i) The spectrum $\mbox{spec}(\widehat{Q}_D)\ne 0$ is simple (consists of $D$ distinct real numbers)\\
  (ii) If $\Psi^\lambda=(\Psi^\lambda_1,\ldots,\Psi^\lambda_D)^T$ is an eigenvector of $\widehat{Q}_D$ ($\widehat{Q}_D\Psi^\lambda=\lambda\Psi^\lambda$, $\lambda\ne 0$) then $\Psi^\lambda_1\ne 0, \Psi^\lambda_D\ne 0$ 
  
\end{Theorem} 

\subsection{Eigenvectors of the Matrix $\widehat{Q}_D$} 
Starting from (I) in (\ref{Recursion Relations}) by setting $\Psi_1^\lambda=x=\mathrm{const}$ one can explicitly construct the components $\Psi^\lambda_k$. One gets, in agreement with  \cite{Carbone Carfora Marzouli} (using the integer number $L\ge 1$ with $L \le \frac{D}{2}$ for even $D$ and $L \le \frac{D-1}{2}$ for odd $D$):
\be\label{Eigenvectors 4 vertex}
\begin{array}{ll}
\cmt{15.5}{
\begin{footnotesize}
\[\begin{array}{lll}
   \Psi^\lambda_{2L}\!\!\!\!&=\displaystyle\frac{-x\lambda}{\prod\limits_{k=1}^{2L-1}q_k}
   \left[\lambda^{2(L-1)}+\sum_{M=1}^{L-1}\lambda^{2(L-1-M)} \hspace{-2mm}
            \sum_{k_M=2M-1}^{2L-2}\:
	    \sum_{k_{M-1}=2(M-1)-1}^{k_M-2}\hspace{-2mm}
	    \ldots
	    \sum_{k_l=2(M-l)-1}^{k_{l-1}-2} \hspace{-2mm}
	    \ldots  
	    \sum_{k_2=3}^{k_3-2}\:
	    \sum_{k_1=1}^{k_2-2} 
	    q_{k_M}^2 q_{k_{M-1}}^2\ldots q_{k_l}^2\ldots q_{k_2}^2 q_{k_1}^2\right]
   \\	    
   \Psi^\lambda_{2L+1}\!\!\!\!&=\displaystyle\frac{x}{\prod\limits_{k=1}^{2L} q_k}
   \left[\lambda^{2L~~~}+\sum_{M=1}^{L}\lambda^{2(L-M)~~~} \hspace{-2mm}
            \sum_{k_M=2M-1}^{2L-1}\:
	    \sum_{k_{M-1}=2(M-1)-1}^{k_M-2}\hspace{-2mm}
	    \ldots
	    \sum_{k_l=2(M-l)-1}^{k_{l-1}-2} \hspace{-2mm}
	    \ldots  
	    \sum_{k_2=3}^{k_3-2}\:
	    \sum_{k_1=1}^{k_2-2} 
	    q_{k_M}^2 q_{k_{M-1}}^2\ldots q_{k_l}^2\ldots q_{k_2}^2 q_{k_1}^2\right]
  \end{array}\] 
\end{footnotesize}
}&~~~~~~~~
\end{array}
\ee
where $L$ is a positive integer $\leq \frac{N}{2}$.
When $M=1$ the upper bound of the $k_1$ sum is that of the $k_M$ sum.
One may explicitly check that these states fulfill (II) in (\ref{Recursion Relations}). 
In order to fulfill (III) we find the conditions
\be\label{Abbruchbedingung}\begin{array}{ll}
   \multicolumn{1}{l}{\fbox{$D=2L$, even}}
   \\
   \displaystyle 0\stackrel{!}{=}\frac{-x}{\prod\limits_{k=1}^{2L-1}q_k}
   \left[\lambda^{2L}+\sum_{M=1}^{L}\lambda^{2(L-M)} \hspace{-2mm}
            \sum_{k_M=2M-1}^{2L-1}~~ 
	    \sum_{k_{M-1}=2(M-1)-1}^{k_M-2}\hspace{-2mm}
	    \ldots\ldots  
	    \sum_{k_2=3}^{k_3-2}~~
	    \sum_{k_1=1}^{k_2-2} 
	    q_{k_M}^2q_{k_{M-1}}^2\ldots\ldots q_{k_2}^2 q_{k_1}^2\right]
   \\
   \\
      \multicolumn{1}{l}{\fbox{$D=2L+1$, odd}}
   \\
   \displaystyle 0\stackrel{!}{=}\frac{x\lambda}{\prod\limits_{k=1}^{2L} q_k}
   \left[\lambda^{2L}+\sum_{M=1}^{L}\lambda^{2(L-M)} \hspace{-2mm}
            \sum_{k_M=2M-1}^{2L}~~ 
	    \sum_{k_{M-1}=2(M-1)-1}^{k_M-2}\hspace{-2mm}
	    \ldots\ldots  
	    \sum_{k_2=3}^{k_3-2}~~
	    \sum_{k_1=1}^{k_2-2} 
	    q_{k_M}^2q_{k_{M-1}}^2\ldots\ldots q_{k_2}^2 q_{k_1}^2\right]
   \\
\end{array}\ee
Thus, in order for (\ref{Abbruchbedingung}) to be satisfied, $\lambda$ has to be 
a root of the characteristic polynomial 
\be \label{char poly}
\begin{array}{lclllll}
   \pi_{2L}(\lambda)&=&\displaystyle\lambda^{2L}+\sum_{M=1}^L b_M^{(k_0)}\lambda^{2(L-M)} &~~~~~~&D=2L~~\mbox{even}&~~~~~&k_0=2L+1=D+1  
   \\
   \pi_{2L+1}(\lambda)&=&\displaystyle\lambda\Big[\lambda^{2L}+\sum_{M=1}^L b_M^{(k_0)}\lambda^{2(L-M)}\Big] &~~~~~~&D=2L+1~~\mbox{odd}&~~~~~&k_0=2L+2=D+1~~~~~~~~    
\end{array}
\ee
with coefficients
\be\label{char poly coeff}
   b_M^{(k_0)}=\sum_{k_M=2M-1}^{k_0-2}\:
	    \sum_{k_{M-1}=2(M-1)-1}^{k_M-2}\hspace{-2mm}
	    \ldots
	    \sum_{k_l=2(M-l)-1}^{k_{l-1}-2} \hspace{-2mm}
	    \ldots  
	    \sum_{k_2=3}^{k_3-2}
	    \sum_{k_1=1}^{k_2-2} 
	    q_{k_M}^2q_{k_{M-1}}^2\ldots q_{k_l}^2\ldots q_{k_2}^2 q_{k_1}^2
\ee 
Since the bracketed 
terms in the characteristic polynomials (\ref{char poly}) contain only even powers of $\lambda$ we may replace $\lambda^{2k}=\Lambda^k$ with $\Lambda=|\lambda|^2$ in order to arrive at a reduced purely real notation for them:
\ba
   0&=&\lambda^{2L}+\sum_{M=1}^L b_M^{(k_0)}\lambda^{2(L-M)} 
   \nonumber
   \\
   &=&\Lambda^L + \sum_{M=1}^L (-1)^{M}~\big|b^{(k_0)}_M\big|~\Lambda^{L-M}
\ea
where we have pulled out a prefactor $(-1)^M$ coming from the purely imaginary nature of the matrix elements (\ref{general matrix element 4 vertex}).

\subsubsection{Traces of $({\widehat{Q}_D}^2)^{n}$}

Knowing the coefficients (\ref{char poly coeff}) of the characteristic polynomial, together with Newton's equation valid for an arbitrary matrix $A$ with $\tr[ A^n]=s_n$
\be
   s_n=-\Big[n\cdot b^{(k_0)}_n +\sum\limits_{l=1}^{n-1}b^{(k_0)}_{n-l} ~~s_l\Big]     ~~~~~~~~~~~~~n>1
\ee
we are now enabled to successively calculate the traces of all powers of $({\widehat{Q}_D}^2)^n$ starting from   
 $s_1=-2\sum\limits_{k=1}^{D-1}q_k^2$.
This is of special relevance because it allows us  
to reconstruct the spectral density function $\rho(\lambda)$ of eigenvalues $\lambda$ of the hermitian operator $\widehat{Q}_D$ in question (see e.g.\ \cite{Schwabl}, chapter 2).

 \begin{samepage}
 $\rho(\lambda)$ fulfills the properties of a probability distribution\\[2mm]
\begin{tabular}{cll}
(i) & $\rho(\lambda)\ge 0~~~\forall~\lambda$ & positivity\\
(ii) & $\displaystyle\int\limits_{-\infty}^{+\infty} \rho(\lambda)=1$ & normalization\\
(iii) & $\rho(\lambda)=\sum\limits_{k=1}^D \delta(\lambda-\lambda_k)$ & for a discrete spectrum, as is the case for $\widehat{Q}_D$
\end{tabular}
 \end{samepage}
\\[1mm]
Then
 \be\label{spectral reconstruction 1}
    \tr[(\widehat{Q}_D)^n]=\sum\limits_{k=1}^n(\lambda_k)^n= m_n=\int\limits_{-\infty}^{+\infty}\lambda^n~\rho(\lambda)~d\lambda
 \ee
 is called the $\mbox{n}^{\mbox{th}}$ moment of the distribution $\rho(\lambda)$. The Fourier transform of $\rho(\lambda)$
 \be\label{spectral reconstruction 2}
    \chi(\tau)=\int\limits_{-\infty}^{+\infty} \mb{e}^{-\mb{i} \lambda \tau}~\rho(\lambda)~d\lambda
 \ee
is called the characteristic function. By inversion of the Fourier transform $\rho(\lambda)$ can be reconstructed from $\chi(\tau)$: 
\be\label{spectral reconstruction 3}
   \rho(\lambda)=\frac{1}{2\pi}\int\limits_{-\infty}^{+\infty} \mb{e}^{\mb{i} \lambda \tau}~\chi(\tau)~d\tau
\ee
Now  by expanding the exponent in (\ref{spectral reconstruction 2}) and inserting 
(\ref{spectral reconstruction 1}), the Fourier transform $\chi(\tau)$ of the eigenvalue distribution $\rho(\lambda)$ can be expressed in terms of the moments (\ref{spectral reconstruction 1}) as
\ba\label{spectral reconstruction 4}
   \chi(\tau)
   &=&\sum_{n=0}^{\infty}\frac{(-\mb{i})^n}{n!}~\tau^n
              \left[\int\limits_{-\infty}^{+\infty} \lambda^n~\rho(\lambda)~d\lambda\right]
   =\sum_{n=0}^{\infty}\frac{(-\mb{i})^n}{n!}~\tau^n~m_n
\ea
Inserting this back into (\ref{spectral reconstruction 3}) one finally obtains $\rho(\lambda)$ in terms of the moments. However this formal way of obtaining $\rho(\lambda)$ will not be directly accessible in most cases, since the series (\ref{spectral reconstruction 4}) is not convergent,
and one needs a closed expression for $\chi(\tau)$, as otherwise inserting the formal power series (\ref{spectral reconstruction 4}) into (\ref{spectral reconstruction 3}) will result in ill-defined divergent expressions. One then has to put in by hand an assumption on the kind of distribution one is working with (see e.g.\ \cite{Dyson}). 
A note on the results obtained in \cite{Meissner} is given in the companion paper \cite{NumVolSpecLetter}.
There are also numerical procedures to reconstruct a statistical distribution from a finite number of its moments using spline interpolation \cite{John et al}.   

\subsection{Upper and Lower Bounds on the Eigenvalues of $\widehat{Q}_D$} 
\subsubsection{Upper Bound}
 
 One can find upper bounds for the eigenvalues by applying the theorem of $Ger\check{s}gorin$ (see
e.g.\ \cite{Marcus Minc}, \cite{Gantmacher}):

\begin{Theorem}{$Ger\check{s}gorin$} \label{Gersgorin discs}
   
Every eigenvalue  
$\lambda$ of an ($n\times n$)-matrix $Q$ lies at least in one of the discs\\
   \[
     |q_{ii}-\lambda| \le \displaystyle\sum_{{j=1}\atop{j\ne i}}^n |q_{ij}| \hspace{1cm} i=1,\ldots,n
   \] 
\end{Theorem}
That is every eigenvalue lies in a disc centered at the diagonal element $q_{ii}$ with radius the sum of 
moduli of the off-diagonal-elements $q_{ij},~i\ne j$ of the $i^{\mathrm{th}}$ row or column.
In the case of the gauge invariant 4-vertex this theorem simplifies, due to the banded matrix structure of (\ref{general matrix 4 vertex}) and the fact that $q_{ii}=0$, to
\be
  |\lambda|\le \sum\limits_{j \ne i}|q_{ij}| = |q_{i~i-1}| + |q_{i~i+1}|
\ee  

A general upper bound was found already in \cite{Volume Paper}  by directly employing theorem \ref{Gersgorin discs}.

\subsubsection{\label{Lower Bound}Lower Bound}
Finding a lower bound on the modulus of the smallest non-zero eigenvalue turns out to be a much harder task in \cite{Volume Paper}. The idea is to invert the matrix $\widehat{Q}_D$ and find an upper bound on the eigenvalues of the inverse $\widehat{Q}_D^{-1}$, which then in turn would serve as a lower bound on the spectrum of $\widehat{Q}_D$. The problem is that in the case of odd dimensional $\widehat{Q}_D$ we cannot directly invert the matrix but have to project out the null space first. That results in a messy matrix which has lost some of the banded structure of the original $\widehat{Q}_D$ given in (\ref{general matrix 4 vertex}).

We will show now that this procedure is not necessary. In order to see this consider the $(D,D)$ component of the resolvent  of $\widehat{Q}_D$:
\be
   \left[\frac{1}{\widehat{Q}_D-\lambda \mb{1}_D} \right]_{DD}  
   = \frac{\widetilde{\left[\widehat{Q}_D-\lambda \mb{1}_D \right]}_{DD}}{\det\left[\widehat{Q}_D-\lambda\mb{1}_D \right]}
   =\frac{\det[\widehat{Q}_{D-1}-\lambda \mb{1}_{D-1}]}{\det\left[\widehat{Q}_D-\lambda\mb{1}_D \right]}
   =\frac{\pi_{D-1}(\lambda)}{\pi_D(\lambda)}
\ee
where we have used the general definition of the matrix element of the inverse matrix 
$[\widehat{Q}^{-1}]_{jk}=(-1)^{j+k}\frac{\widetilde{\widehat{Q}}_{kj}}{\det \widehat{Q}}$, $\widetilde{\widehat{Q}}_{kj}$ is obtained by deleting row $k$ and column $j$ from $\widehat{Q}$ and taking the determinant of the remaining submatrix. Moreover $\pi_D(\lambda)$ denotes the characteristic polynomial of $\widehat{Q}_D$, $\pi_{D-1}(\lambda)$ is the characteristic polynomial of the $D-1$ dimensional submatrix $\widehat{Q}_{D-1}$ one obtains by deleting row $D$ and column $D$ of $\widehat{Q}_D$.
In what follows we will drop the index $D$ from $\widehat{Q}_D$, $U_D$ etc.\ for clarity.
\\

From (\ref{Diag Matrix of Q}) we know that $\widehat{Q}=U\Lambda U^\dagger$, $\Lambda=\mbox{diag}(\lambda_1,\ldots,\lambda_D)$, and $U^\dagger U=UU^\dagger=\mb{1}$, $U$ being a unitary matrix. Therefore we know that
\ba
   \widehat{Q}_{lm}&=&\sum_{r,s}U_{lr}~\Lambda_{rs}~U^\dagger_{sm}
   \nonumber
   =\sum_{r,s}U_{lr}~\big[\lambda_r\cdot\delta_{rs}]~U^\dagger_{sm}
   \nonumber
   =\sum_{r}\lambda_r U_{lr}~U^\dagger_{rm}
\ea
and, since $U$ diagonalizes $(\widehat{Q}-\lambda\mb{1})^{-1}$ as well: 
\ba
   \big[(\widehat{Q}-\lambda\mb{1})^{-1}\big]_{lm}
   &=&\sum_{r,s}U_{lr}~\big[(\Lambda-\lambda\mb{1})^{-1} \big]_{rs}~U^\dagger_{sm}
   =\sum_{r,s}U_{lr}~\big[(\lambda_r-\lambda)^{-1}\delta_{rs} \big]~U^\dagger_{sm}
   =\sum_{r}\frac{1}{\lambda_r-\lambda} U_{lr}~U^\dagger_{rm}  
\ea
Thus we find using (\ref{Diag Matrix of Q}): 
\ba
   \big[(\widehat{Q}-\lambda\mb{1})^{-1}\big]_{DD}=\frac{\pi_{D-1}(\lambda)}{\pi_D(\lambda)}
   &=&\sum_{r}\frac{U_{Dr}~U^\dagger_{rD}}{\lambda_r-\lambda}
   =\sum_{r}\frac{\Psi^{\lambda_r}_D\overline{\Psi^{\lambda_r}_D}}{\lambda_r-\lambda}
   =\sum_{r}\frac{\big|\Psi^{\lambda_r}_D\big|^2}{\lambda_r-\lambda}   
\ea
where the overline denotes complex conjugation. 

\begin{figure}[!hbt]
\begin{center}
      \psfrag{lkm1}{$\lambda_{k-1}$}
      \psfrag{lk}{$\lambda_{k}$}
      \psfrag{lkp1}{$\lambda_{k+1}$}
      \psfrag{Vla}{$V(\lambda)$}
      \psfrag{la}{$\lambda$}
      \includegraphics[height=6cm]{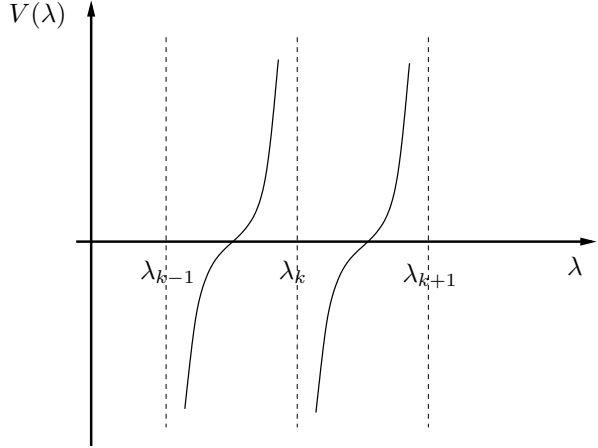}
      \caption{The function $V(\lambda)$.}
      \label{resolvent pic}
\end{center}
\end{figure}

\begin{figure}[!hbt]
\begin{center}
  \begin{minipage}[t]{6cm} 
  \center
      \psfrag{Nodd}{\fbox{D odd}}
      \psfrag{0}{$0$}
      \psfrag{ldots}{$\ldots$}
      \psfrag{vdots}{$\vdots$}
      \psfrag{QN}{$\widehat{Q}_D$}
      \psfrag{QNm1}{$\widehat{Q}_{D-1}$}
      \psfrag{QNm2}{$\widehat{Q}_{D-2}$}
      \includegraphics[height=3.8cm]{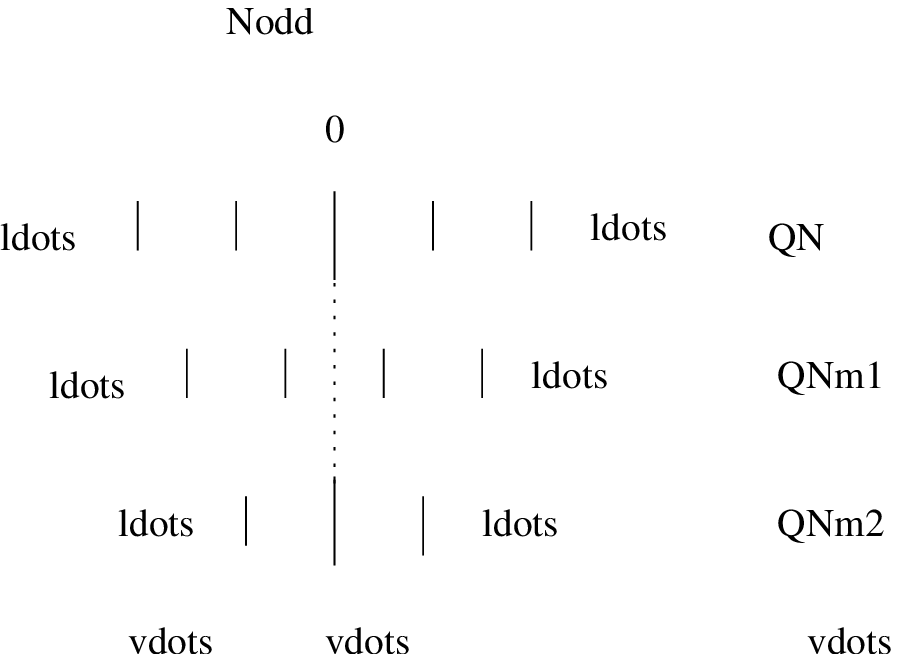}
      \caption{Interlacing of the spectra for odd dimension $D$.}
      \label{odd spec pic}
   \end{minipage}
   ~~~ 
  \begin{minipage}[t]{6cm} 
  \center
      \psfrag{Neven}{\fbox{D even}}
      \psfrag{0}{$0$}
      \psfrag{ldots}{$\ldots$}
      \psfrag{vdots}{$\vdots$}
      \psfrag{QN}{$\widehat{Q}_D$}
      \psfrag{QNm1}{$\widehat{Q}_{D-1}$}
      \psfrag{QNm2}{$\widehat{Q}_{D-2}$}
      \includegraphics[height=3.8cm]{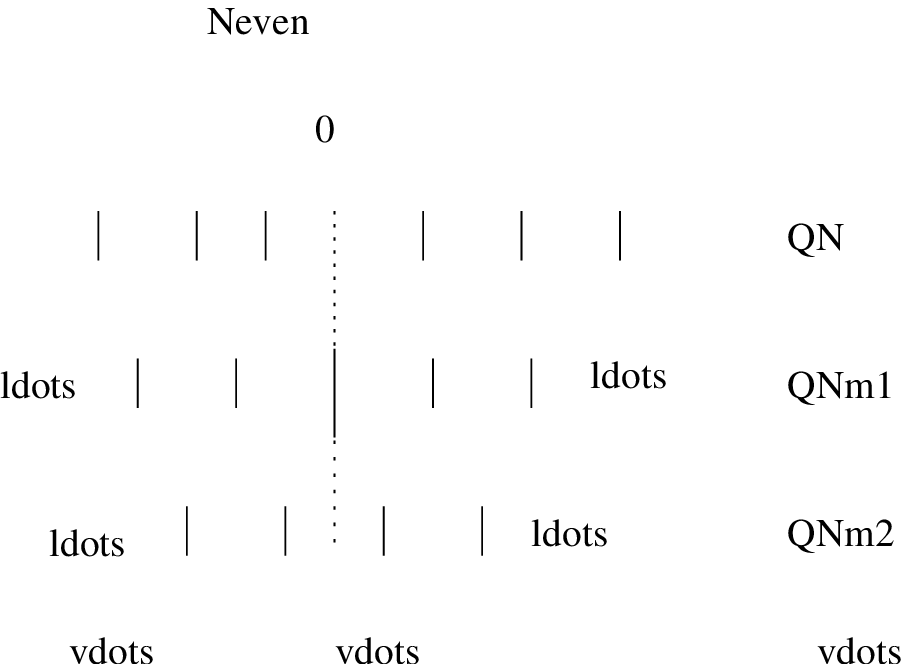}
      \caption{Interlacing of the spectra for even dimension $D$.}
      \label{even spec pic}
   \end{minipage} 
\end{center}
\end{figure}

Now from theorem \ref{Spectral Simplicity Thm} we know that we must have $\Psi^{\lambda_r}_D\ne 0 $ $\forall r=1\ldots D$.
Therefore we know that 
\be
   V(\lambda)=\frac{\pi_{D-1}(\lambda)}{\pi_D(\lambda)}
   =\sum_{r}\frac{\big|\Psi^{\lambda_r}_D\big|^2}{\lambda_r-\lambda}   
\ee 
is a real valued function and its derivative
\be
    \frac{d}{d\lambda} V(\lambda)
   =\sum_{r}\frac{\big|(\Psi^{\lambda_r})_D\big|^2}{(\lambda_r-\lambda)^2} >0~~\forall~\lambda   
\ee 
That is, $V(\lambda)$ is monotonic increasing $\forall~\lambda\in\mb{R}$, has poles at every $\lambda=\lambda_r$, $r=1\ldots D$, and is finite $\forall~\lambda\ne\lambda_r$. So we conclude that $V(\lambda)$ has the form illustrated in figure \ref{resolvent pic}. The function $V(\lambda)$ has precisely one root in  
the interval $[\lambda_k,\lambda_{k+1}]$.  
This means that the roots of 
$\pi_{D-1}(\lambda)$, $\pi_D(\lambda)$ interlace: in between two eigenvalues $\lambda_k,\lambda_{k+1}$ of $\widehat{Q}_D$ there is precisely one eigenvalue of its submatrix $\widehat{Q}_{D-1}$. 
\setlength{\arraycolsep}{-1mm}
From this we conclude that the structure for the spectra of $\widehat{Q}_D$ and its submatrices $\widehat{Q}_{D-1},\widehat{Q}_{D-2},\ldots$ is as illustrated in figures \ref{odd spec pic} and \ref{even spec pic}.
The $D$ odd case in figure \ref{odd spec pic} shows that a lower bound for the smallest non-zero eigenvalues of $\widehat{Q}_D$ is given by the smallest eigenvalue of the even dimensional submatrix $\widehat{Q}_{D-1}$.
However, as mentioned above, one can show \cite{Volume Paper} that even dimensional matrices of type of $\widehat{Q}_D$ have no zero eigenvalue in their spectrum and thus it seems feasible to give a lower bound for their eigenvalues by finding an upper bound for the eigenvalues of its inverse matrix $\widehat{Q}_D^{-1}$. 

Set $\displaystyle L^{(M,N)}=\frac{1}{q_{2M-1}}\prod\limits_{l=M}^{N-1}\frac{q_{2l}}{q_{2l+1}}$, for  $M<N$  and $\displaystyle L^{(M,M)}=\frac{1}{q_{2M-1}}$, such that e.g.\  
$L^{(1,1)}=\frac{1}{q_1}$, $L^{(1,2)}=\frac{q_2}{q_1q_3}$ or $L^{(2,2)}=\frac{1}{q_3}$, $L^{(2,3)}=\frac{q_4}{q_3q_5}$, $L^{(2,4)}=\frac{q_4q_6}{q_3q_5q_7}$. Then if $\widehat{Q}_D$ is as given in 
(\ref{general matrix 4 vertex}) and $D$ is even we have for its inverse (using $n=\frac{D}{2}$):
\be\label{inverse matrix even dim}
  \widehat{Q}_D^{-1}= \left(\begin{array}{cccccccccccccccc}
 0& L^{(1,1)} &0 &  L^{(1,2)} & 0  &  L^{(1,3)}&&\cdots&  0&L^{(1,n-2)} &  0&  L^{(1,n-1)}     & 0&L^{(1,n)}   \\ 
 -L^{(1,1)}   &   0  & 0 &  0   &   0  &  0   &&\cdots&  0     &  0     &  0     &  0   &  0   \\ 
 0     & 0  &  0   &L^{(2,2)}  &   0  &  L^{(2,3)}   &&\cdots&  0     &  L^{(2,n-2)}     &  0     &  L^{(2,n-1)}   &  0&-L^{(2,n)}   \\ 
 -L^{(1,2)}     & 0    & -L^{(2,2)}  &  0   & 0 &  0   &&\cdots &  0     &  0     &  0     &  0   &  0   \\
 0     &0     &  0   & 0  &   0  & L^{(3,3)} &&\cdots &  0     &  L^{(3,n-2)}     &  0     &  L^{(3,n-1)}   &  0&L^{(3,n)}   \\
 -L^{(1,3)}     &0     &  -L^{(2,3)}   &  0   &  -L^{(3,3)} &  0   &     & \cdots      &0  &0  &0  &0&0\\ 
 \vdots&\vdots&\vdots&\vdots&\vdots&\vdots&&&   \vdots    &\vdots  &\vdots  &\vdots  &\vdots&\vdots\\ 
 \vdots&\vdots&\vdots&\vdots&\vdots&\vdots&&&   \vdots    &\vdots  &\vdots  &\vdots  &\vdots&\vdots\\ 
   0  &   0  &   0  &   0  &   0  &0&\cdots&&0&  L^{(n-2,n-2)}     &0&L^{(n-2,n-1)}&  0   &L^{(n-2,n)}   \\ 
  
    -L^{(1,n-2)}  &   0  &   -L^{(2,n-2)}  &   0  &   -L^{(3,n-2)}  &0&\cdots&&-L^{(n-2,n-2)}&  0     &0&   0    &  0   &  0   \\ 
    0  &   0  &   0  &   0  &   0  &0&\cdots&&   0   &0 &   0    &L^{(n-1,n-1)}&  0   &  L^{(n-1,n)}   \\ 
    -L^{(1,n-1)}  &   0  &   -L^{(2,n-1)}  &   0  &   -L^{(3,n-1)}  &0&\cdots&&   -L^{(n-2,n-1)}   &  0     & -L^{(n-1,n-1)}&   0    & 0 &0 \\ 
    0  &   0  &   0  &   0  &   0  &0&\cdots&&   0   &  0     &    0   &0 &   0 &L^{(n,n)}\\
    -L^{(1,n)}  &   0  &   -L^{(2,n)}  &   0  &   -L^{(3,n)}  &0&\cdots&&  -L^{(n-2,n)}   &  0     &    -L^{(n-1,n)}   &  0   &  -L^{(n,n)}  &0 
\end{array} 
   \right)\\~\\~
\ee

We are now left with the task of finding an upper bound for the row/column sums of $\widehat{Q}_D^{-1}$ in order to   apply theorem \ref{Gersgorin discs}.
For this we have to use the explicit form of the matrix elements (\ref{general matrix element 4 vertex}).
Let us introduce the following shorthands:
\ba\label{shorthands x(2l)}
   R_k^{(m,n)}&~=~&j_m+j_n-a_2^{(min)}-k+1
   \NN
   U_k^{(m,n)}&=&(j_m+j_n+a_2^{(min)}+k+1)(-j_m+j_n+a_2^{(min)}+k)(j_m-j_n+a_2^{(min)}+k)
\ea
which enable us to write for \fbox{$M<N$} 
\ba\label{upper bound L^(MN)}
   \Big|L^{(M,N)} \Big|^2
   &~=~&\Big|\frac{1}{q_{2M-1}}\prod\limits_{l=M}^{N-1}\frac{q_{2l}}{q_{2l+1}}\Big|^2
   \nonumber
   \\
   &~=~&\big[2(a_2^{(min)}+2M-1)-1\big]\big[2(a_2^{(min)}+2M-1)+1\big]\prod_{l=M}^{N-1}
   \frac{\big[2(a_2^{(min)}+2l+1)-1\big]\big[2(a_2^{(min)}+2l+1)+1\big]}
   {\big[2(a_2^{(min)}+2l)-1\big]\big[2(a_2^{(min)}+2l)+1\big]}
   \nonumber
   \\
   &&\times~\frac{1}{U_{2M-1}^{(1,2)}~U_{2M-1}^{(3,4)}}\prod_{l=M}^{N-1}
            \frac{U_{2l}^{(1,2)}~U_{2l}^{(3,4)}}{U_{2l+1}^{(1,2)}~U_{2l+1}^{(3,4)}} 
     ~\times~\frac{1}{R_{2M-1}^{(1,2)}~R_{2M-1}^{(3,4)}}\prod_{l=M}^{N-1}
             \frac{R_{2l}^{(1,2)}~R_{2l}^{(3,4)}}{R_{2l+1}^{(1,2)}~R_{2l+1}^{(3,4)}} 
   \nonumber
   \\[5mm]
   &~=~&\frac{\big(2a_2^{(min)}+4M-3 \big)\big(2a_2^{(min)}+4N-1 \big)}
      {U_{2M-1}^{(1,2)}U_{2M-1}^{(3,4)} R_{2N-1}^{(1,2)} R_{2N-1}^{(3,4)}}~~ 
      \prod_{l=M}^{N-1}
           \underbrace{\frac{R_{2l}^{(1,2)}R_{2l}^{(3,4)}U_{2l}^{(1,2)}U_{2l}^{(3,4)}}
                {R_{2l-1}^{(1,2)}R_{2l-1}^{(3,4)}U_{2l+1}^{(1,2)}U_{2l+1}^{(3,4)}}}_{\displaystyle
                =:x(2l)} 
\ea
Note that by construction (\ref{upper bound L^(MN)}) also holds for the case \fbox{$M=N$} just by leaving out the product of the $x(2l)$, as can be seen from the fact that 
\be\label{upper bound L^(MM)}
 \Big|L^{(M,M)}\Big|^2=\big|q_{2M-1}\big|^{-2}
 =\frac{\big(2a_2^{(min)}+4M-3 \big)\big(2a_2^{(min)}+4M-1 \big)}
      {U_{2M-1}^{(1,2)}U_{2M-1}^{(3,4)} R_{2M-1}^{(1,2)} R_{2M-1}^{(3,4)}}
\ee
In the last line of (\ref{upper bound L^(MN)}) we have moreover introduced
\be
  x(2l):=\frac{R_{2l}^{(1,2)}R_{2l}^{(3,4)}U_{2l}^{(1,2)}U_{2l}^{(3,4)}}
                {R_{2l-1}^{(1,2)}R_{2l-1}^{(3,4)}U_{2l+1}^{(1,2)}U_{2l+1}^{(3,4)}}
\ee

Now due to positivity of all terms contained in $R_k^{(m,n)},U_k^{(m,n)}$ we always have \fbox{$x\le 1$} in (\ref{upper bound L^(MN)}).
Let us take a closer look at the product terms contained in (\ref{upper bound L^(MN)}) for the case that \fbox{$M<N$}.\newline Using \fbox{$x^{(m,n)}=j_m+j_n-a_2^{(min)}+1$} we may write

\ba\label{evaluate Rmn}
   \prod_{l=M}^{N-1}\frac{R_{2l}^{(m,n)}}{R_{2l-1}^{(m,n)}}
&~~=~~&
   \prod_{l=M}^{N-1}\frac{x^{(m,n)}-2l}{x^{(m,n)}-2l+1}
\NN
&~~=~~&
   \prod_{l=-M}^{-N+1}\frac{x^{(m,n)}+2l}{x^{(m,n)}+2l+1}
         ~~~~~~~~~~~~~~~~~~~~~~~~~~~~\mbox{introduce: $r=l+N$}
\NN
&=&
   \prod_{r=1}^{N-M}\frac{x^{(m,n)}-2N+2r}{x^{(m,n)}-2N+2r+1}
\NN
&=&\frac{\Gamma(1+\frac{x^{(m,n)}}{2}-M)}{\Gamma(1+\frac{x^{(m,n)}}{2}-N)}\cdot
   \frac{\Gamma(\frac{3}{2}+\frac{x^{(m,n)}}{2}-N)}{\Gamma(\frac{3}{2}+\frac{x^{(m,n)}}{2}-M)}
\ea
Using \fbox{$x_1^{(m,n)}=j_m+j_n+a_2^{(min)}+1$},
     \fbox{$x_2^{(m,n)}=-j_m+j_n+a_2^{(min)}$},
     \fbox{$x_3^{(m,n)}=j_m-j_n+a_2^{(min)}$}
we can write
\ba\label{evaluate Umn}
      \prod_{l=M}^{N-1}\frac{U_{2l}^{(m,n)}}{U_{2l+1}^{(m,n)}}
   &~~=~~&
      \prod_{l=M}^{N-1}\left[\prod_{\mu=1}^3\frac{x_{\mu}+2l}{x_{\mu}+2l+1}\right]
            ~~~~~~~~~~~~~~~~~~~~~~~~~~\mbox{introduce: $s=l-M+1$}
   \NN 
   &=&
      \prod_{\mu=1}^3\left[\prod_{s=1}^{N-M}\frac{x_{\mu}+2M+2(s-1)}{x_{\mu}+2M+2(s-1)+1}\right] 
   \NN
   &=&
     \prod_{\mu=1}^3\left[\prod_{s=0}^{N-M-1}\frac{x_{\mu}+2M+2s}{x_{\mu}+2M+2s+1}\right]
   \NN
   &=&
     \prod_{\mu=1}^3\left[\left(\prod_{s=1}^{N-M}\frac{x_{\mu}+2M+2s}{x_{\mu}+2M+2s+1}\right)
            \cdot\frac{x_{\mu}+2N+1}{x_{\mu}+2N}
            \cdot\frac{x_{\mu}+2M}{x_{\mu}+2M+1}\right]
   \NN
   &=&
     \prod_{\mu=1}^3\left[
         \frac{\Gamma(1+\frac{x_{\mu}}{2}+N)}{\Gamma(1+\frac{x_{\mu}}{2}+M)}
         \cdot
         \frac{\Gamma(\frac{3}{2}+\frac{x_{\mu}}{2}+M)}{\Gamma(\frac{3}{2}+\frac{x_{\mu}}{2}+N)}
         ~~\cdot~~
         \frac{x_{\mu}+2N+1}{x_{\mu}+2N}
         \cdot\frac{x_{\mu}+2M}{x_{\mu}+2M+1}\right]   
   \NN
   &\le&
     \prod_{\mu=1}^3\left[
         \frac{\Gamma(1+\frac{x_{\mu}}{2}+N)}{\Gamma(1+\frac{x_{\mu}}{2}+M)}
         \cdot
         \frac{\Gamma(\frac{3}{2}+\frac{x_{\mu}}{2}+M)}{\Gamma(\frac{3}{2}+\frac{x_{\mu}}{2}+N)}
         \right]   
\ea
since $\frac{x_{\mu}+2N+1}{x_{\mu}+2N} 
\frac{x_{\mu}+2M}{x_{\mu}+2M+1}\le 1$ because $M<N$ and the function $f(y)=\frac{y}{y+1}$ is monotonically increasing.
Using the expansion of the $\Gamma$-function \cite{Abramowitz} one may check that
\be
   \mb{e}^{-x} x^{x-\frac{1}{2}} \sqrt{2\pi}~< ~\Gamma(x)~<~ 2\cdot\mb{e}^{-x} x^{x-\frac{1}{2}} \sqrt{2\pi}
   ~~~~~~~~~~~~~~~~~~~~~~~~~~~\forall~x\in\mb{R}~~~~x>1
\ee
moreover we have
\be
   \left(1+\frac{1}{x} \right)^x~<~\mb{e}~ <~  2\cdot\left(1+\frac{1}{x} \right)^x
   ~~~~~~~~~~~~~~~~~~~~~~~~~~~~~~~~~~~~~~~~\forall~x\in\mb{R}~~~~x>1
\ee
 and we can estimate
\ba
   \frac{\Gamma(x)}{\Gamma(x+\frac{1}{2})}\le\frac{2\sqrt{2}}{\sqrt{x}}
   &~~~~~~~~\mbox{and}~~~~~~~~&
   \frac{\Gamma(x+\frac{1}{2})}{\Gamma(x)}\le 2\sqrt{x}
\ea
and finally obtain from  (\ref{evaluate Rmn}) 
\ba\label{evaluate Rmn 2}
   \prod_{l=M}^{N-1}\frac{R_{2l}^{(m,n)}}{R_{2l-1}^{(m,n)}}
&~~=~~&
  \frac{\Gamma(1+\frac{x^{(m,n)}}{2}-M)}{\Gamma(1+\frac{x^{(m,n)}}{2}-N)}\cdot
   \frac{\Gamma(\frac{3}{2}+\frac{x^{(m,n)}}{2}-N)}{\Gamma(\frac{3}{2}+\frac{x^{(m,n)}}{2}-M)}
~~\le~~
   4\sqrt{2}\left[\frac{1+\frac{x^{(m,n)}}{2}-N}{1+\frac{x^{(m,n)}}{2}-M}\right]^{\frac{1}{2}} 
\ea
For (\ref{evaluate Umn}) we find
\ba\label{evaluate Umn 2}
      \prod_{l=M}^{N-1}\frac{U_{2l}^{(m,n)}}{U_{2l+1}^{(m,n)}}
  ~~\le~~
     \prod_{\mu=1}^3\left[
         \frac{\Gamma(1+\frac{x_{\mu}^{(m,n)}}{2}+N)}{\Gamma(1+\frac{x_{\mu}^{(m,n)}}{2}+M)}
         \cdot
         \frac{\Gamma(\frac{3}{2}+\frac{x_{\mu}^{(m,n)}}{2}+M)}{\Gamma(\frac{3}{2}+\frac{x_{\mu}^{(m,n)}}{2}+N)}
         \right]   
  &~~\le~~&
     (4\sqrt{2})^3\prod_{\mu=1}^3
      \left[\frac{1+\frac{x^{(m,n)}_{\mu}}{2}+M}{1+\frac{x^{(m,n)}_{\mu}}{2}+N}\right]^{\frac{1}{2}}  
\ea
Now recall that 
\be\label{ordered spins 4}
  j_1\le j_2 \le j_3 \le j_4
\ee
where $j_4$ equals the maximal spin $\jmax$.
Note the ranges of the intermediate recoupling spin $a_2=a_2^{(min)}+k~$ according to (\ref{range of a2}). Now without loss of generality assume \fbox{$a_2^{(min)}=j_2-j_1$}.\footnote{Note that we might equivalently choose  $a_2^{(min)}=j_4-j_3$ here, as (\ref{upper bound L^(MN)}) is symmetric with respect to interchanging $(j_1,j_2)\leftrightarrow (j_3,j_4)$. }  Then the dimension of the matrix $\widehat{Q}_D$ is given by
\be\label{dimension of Q}
   \dim \widehat{Q}_D=a_2^{(max)}-a_2^{(min)}+1=2j_1 +1
\ee
which implies $M\le N\le j_1+\frac{1}{2}$.
So finally by using the definitions for $x^{(m,n)}$, $x_{\mu}^{(m,n)}$ above 
with \fbox{$a_2^{(min)}=j_2-j_1$} :
\barr{lclclcl}
   x^{(1,2)}  &~~=~~& 2j_1+1 &~~\hspace{1.5cm}~~& x^{(3,4)}  &~~=~~& ~~j_3+j_4-j_2+j_1+1\\\\
   x_1^{(1,2)}&  =  & 2j_2+1 &                  & x_1^{(3,4)}&  =  & ~~j_3+j_4+j_2-j_1+1\\
   x_2^{(1,2)}&  =  & 2(j_2-j_1) &              & x_2^{(3,4)}&  =  & -j_3+j_4+j_2-j_1  \\
   x_3^{(1,2)}&  =  & 0      &                  & x_3^{(3,4)}&  =  & ~~j_3-j_4+j_2-j_1  \\
\earr
we can estimate using (\ref{evaluate Rmn 2})
\ba\label{evaluate Rmn 3}
    \prod_{l=M}^{N-1}\frac{R_{2l}^{(1,2)}}{R_{2l-1}^{(1,2)}}
                    \frac{R_{2l}^{(3,4)}}{R_{2l-1}^{(3,4)}}
  &~~\le~~&
   32\cdot\left[\frac{(3+2j_1-2N)}{(3+2j_1-2M)}
                \frac{(3+j_3+j_4-j_2+j_1-2N)}{(3+j_3+j_4-j_2+j_1-2M)}\right]^{\frac{1}{2}}       
\ea
moreover with (\ref{evaluate Umn 2}):
\ba\label{evaluate Umn 3}
      \prod_{l=M}^{N-1}\frac{U_{2l}^{(1,2)}}{U_{2l+1}^{(1,2)}}
                       \frac{U_{2l}^{(3,4)}}{U_{2l+1}^{(3,4)}}
  &~~\le~~&
     (32)^3 \left[\frac{(3+2j_2+2M)}{(3+2j_2+2N)}\right]^{\frac{1}{2}}
            \left[\frac{(3+j_3+j_4+j_2-j_1+2M)}{(3+j_3+j_4+j_2-j_1+2N)}\right]^{\frac{1}{2}} 
  \NN
  &&~~\times~ 
            \left[\frac{(1+j_2-j_1+M)}{(1+j_2-j_1+N)}\right]^{\frac{1}{2}}
            \left[\frac{(2-j_3+j_4+j_2-j_1+2M)}{(2-j_3+j_4+j_2-j_1+2N)}\right]^{\frac{1}{2}}
  \NN
  &&~~\times~ 
            \left[\frac{(1+M)}{(1+N)}\right]^{\frac{1}{2}}
            \left[\frac{(2+j_3-j_4+j_2-j_1+2M)}{(2+j_3-j_4+j_2-j_1+2N)}\right]^{\frac{1}{2}}   
\ea
Combining (\ref{evaluate Rmn 3}) and (\ref{evaluate Umn 3}) we finally get an upper bound for the last product on the right hand side of (\ref{upper bound L^(MN)}) by taking into account that $0<M<N$:
\ba\label{explicit final upper bound x(2l)}
     \prod_{l=M}^{N-1}x(2l)
   &~~=~~&
     \prod_{l=M}^{N-1}\frac{R_{2l}^{(1,2)}}{R_{2l-1}^{(1,2)}}
                      \frac{R_{2l}^{(3,4)}}{R_{2l-1}^{(3,4)}}
                      \frac{U_{2l}^{(1,2)}}{U_{2l+1}^{(1,2)}}
                      \frac{U_{2l}^{(3,4)}}{U_{2l+1}^{(3,4)}}
   \NN
   &\le&
   (32)^4   \left[\frac{(3+2j_1-2N)}{(3+2j_1-2M)}\right]^{\frac{1}{2}}
            \left[\frac{(3+j_3+j_4-j_2+j_1-2N)}{(3+j_3+j_4-j_2+j_1-2M)}\right]^{\frac{1}{2}}
  \NN 
  &&~~\times~
            \left[\frac{(3+2j_2+2M)}{(3+2j_2+2N)}\right]^{\frac{1}{2}}
            \left[\frac{(3+j_3+j_4+j_2-j_1+2M)}{(3+j_3+j_4+j_2-j_1+2N)}\right]^{\frac{1}{2}} 
            \left[\frac{(1+j_2-j_1+M)}{(1+j_2-j_1+N)}\right]^{\frac{1}{2}}
  \NN
  &&~~\times~ 
            \left[\frac{(2-j_3+j_4+j_2-j_1+2M)}{(2-j_3+j_4+j_2-j_1+2N)}\right]^{\frac{1}{2}}
            \left[\frac{(1+M)}{(1+N)}\right]^{\frac{1}{2}}
            \left[\frac{(2+j_3-j_4+j_2-j_1+2M)}{(2+j_3-j_4+j_2-j_1+2N)}\right]^{\frac{1}{2}}
\ea 
By construction we have already seen that 
\ba\label{final upper bound x(2l)}
     \prod_{l=M}^{N-1}x(2l)
   &~~\le~~&1
\ea
however the reason for the explicit calculation of (\ref{explicit final upper bound x(2l)}) will become clear in a moment. 
Using (\ref{final upper bound x(2l)}) we  
give a first upper bound on $|L^{(M,N)} |^2$ in (\ref{upper bound L^(MN)})
\ba\label{upper bound inv ME}
   \Big|L^{(M,N)} \Big|^2
   &~\le~&\frac{\big(2a_2^{(min)}+4M-3 \big)\big(2a_2^{(min)}+4N-1 \big)}
      {U_{2M-1}^{(1,2)}U_{2M-1}^{(3,4)} R_{2N-1}^{(1,2)} R_{2N-1}^{(3,4)}}
   \nonumber
   \\
   &=&\frac{\overbrace{\big(2a_2^{(min)}+4M -3 \big)\big(2a_2^{(min)}+4N-1 \big)}^{\displaystyle T_3}}
      {\underbrace{(j_1+j_2+a_2^{(min)}+2M)(-j_1+j_2+a_2^{(min)}+2M-1)(j_1-j_2+a_2^{(min)}+2M-1)}
      _{\displaystyle T_1}}
   \nonumber
   \\
   &&\times~
  \frac{1}{\underbrace{(j_3+j_4+a_2^{(min)}+2M)(-j_3+j_4+a_2^{(min)}+2M-1)(j_3-j_4+a_2^{(min)}+2M-1)}_{\displaystyle T_2}}
  \nonumber
  \\
  &&\times~\frac{1}{\underbrace{(j_1+j_2-a_2^{(min)}-2N+2)(j_3+j_4-a_2^{(min)}-2N+2)}_{\displaystyle T_4}}        
\ea

In order to get an estimate for the smallest non-zero eigenvalue we will now perform a power counting in $j_4=\jmax$ for all the different factors $T_1\ldots T_4$ in (\ref{upper bound inv ME}), that is while respecting (\ref{ordered spins 4}) we have to minimize the obvious inverse power of $j_4=\jmax$ in  (\ref{upper bound inv ME}). To do so we make the following ansatz:
\barr{lllcl}\label{ansatz for spins}
   j_1=\alpha_1~\jmax+\beta_1 &~~~~~~~~~~~&M&~~=~~&\alpha_M~\jmax+\beta_M\\
   j_2=\alpha_2~\jmax+\beta_2 &&N&~~=~~&\alpha_N~\jmax+\beta_N\\
   j_3=\alpha_3~\jmax+\beta_3\\
   j_4=~\jmax
\earr
which is justified if in the factor terms  $T_1,T_2,T_4$ contained in (\ref{upper bound inv ME}) the power of $\jmax$ should be lowered (raised in $T_3$) while simultaneously respecting (\ref{ordered spins 4}) as $\jmax\rightarrow \infty$  \footnote{Certainly the ansatz (\ref{ansatz for spins}) qualifies only to estimate the largest inverse power of the maximal spin in the limit $\jmax\rightarrow \infty$. We could in principle assume the spins to be arbitrary polynomials of $\jmax$. However we want to look for the smallest inverse power of $\jmax$, that is we want the spins to be configured such that any higher inverse power of $\jmax$ is compensated in (\ref{upper bound inv ME}). In the limit $\jmax\rightarrow \infty$ this can only be achieved if the spins scale with the same power of $\jmax$.}.
Moreover by inspection we find that in (\ref{explicit final upper bound x(2l)}) only $M,N$ matter (and $M<N$), since all other factors in the numerator and denominator are identical and it is therefore sufficient to look at (\ref{explicit final upper bound x(2l)}) alone.

Upon inserting  (\ref{ansatz for spins}) into (\ref{upper bound inv ME}) and again assuming \fbox{$a_2^{(min)}=j_2-j_1$} one obtains

\ba\label{upper bound inv ME 2}
   \Big|L^{(M,N)} \Big|^2
  &~\le~&\frac{2\cdot 2\big[(\alpha_2-\alpha_1+2\alpha_M)\cdot j_4 +\beta_2-\beta_1+2\beta_M-\frac{3}{2} \big]
               \big[(\alpha_2-\alpha_1+2\alpha_N)\cdot j_4 +\beta_2-\beta_1+2\beta_N-\frac{1}{2} \big]}
      {2\big[c_1\cdot j_4 +\beta_2 +\beta_M\big]
       2\big[c_2\cdot j_4 +\beta_2-\beta_1 +\beta_M-\frac{1}{2}\big]
       2\big[c_3\cdot j_4 +\beta_M-\frac{1}{2}\big]}
      \nonumber
   \\
   &&\times~
  \frac{1}{
  \big[c_4\!\cdot\! j_4 \!+\!\beta_3\!+\!\beta_2\!-\!\beta_1\!+\!2\beta_M\big]
  \big[c_5\!\cdot\! j_4 \!-\!\beta_3\!+\!\beta_2\!-\!\beta_1\!+\!2\beta_M\!-\!1\big]
  \big[c_6\!\cdot\! j_4 \!+\!\beta_3\!+\!\beta_2\!-\!\beta_1\!+\!2\beta_M\!-\!1\big]}
  \nonumber
  \\
  &&\times~\frac{1}{
  2\big[c_7\!\cdot\! j_4 \!+\!\beta_1\!-\!\beta_N\!+\!1\big]
  \big[c_8\!\cdot\! j_4 \!+\!\beta_3 \!-\!\beta_2 \!+\!\beta_1 \!-\!2\beta_N \!+\!2\big]}        
\ea
where we have introduced the shorthands  
\be\label{c system}
   \begin{array}{lclclclclclc}
      c_1&~~=~~&\alpha_2+\alpha_M &~\hspace{10mm}~&c_4&~~=~~&~~1+\alpha_3+\alpha_2-\alpha_1+2\alpha_M &~\hspace{10mm}~&c_7&~~=~~&\alpha_1-\alpha_N\\
      c_2&=&\alpha_2-\alpha_1+\alpha_M &&c_5&=&~~1-\alpha_3+\alpha_2-\alpha_1+2\alpha_M &&c_8&=&1+\alpha_3-\alpha_2+\alpha_1-2\alpha_N\\
      c_3&=&\alpha_M&& c_6&=&-1+\alpha_3+\alpha_2-\alpha_1+2\alpha_M
   
   \end{array}
\ee
where $c_k\ge0$ for all $k=1\ldots8$ due to positivity of the according factors in (\ref{upper bound inv ME 2}). 
The set of shorthands (\ref{c system}) can be seen as a system of 8 equalities for the 5 variables $\alpha_1,\alpha_2,\alpha_3,\alpha_M,\alpha_N$. Thus we can eliminate these 5 variables by using 5 equations which results in three remaining consistency equations among the constants $c_1,\ldots,c_8$.
\ba\label{c consistency}
   2c_1 &~~=~~& 2+c_6+2c_7-c_8
   \NN
   2c_2 &=& -2 c_3 +c_5 +c_6
   \NN
   c_4 &=& 2+c_6 ~~~~~~~~~~~~~~~~~~~~~~\mbox{and}~~c_k\ge0~~~~\forall~~k=1\ldots8
\ea
The question is now how one can choose the maximal number of $c$'s equal to 0 consistently. By inspection of (\ref{c consistency}) we can simultaneously set at most 6 of the $c$'s to zero if:

\begin{tabular}{lc|l}  
  \cmt{6}{\[ \begin{array}{llll}
      \fbox{Case 1}~~~~~&\multicolumn{3}{l}{c_1=1 ~~~c_4=2}\\[1mm]
      &\mbox{and}\\[1mm]
      &\multicolumn{3}{l}{c_2=c_3=c_5=c_6=c_7=c_8=0}\\[5mm]
      &\multicolumn{3}{l}{\mbox{Then the set (\ref{c system}) obeys a solution:}}\\[1mm]
      &\alpha_1~~~~&=~~&1\\
      &\alpha_2 &=&1\\
      &\alpha_3 &=&1\\
      &\alpha_M &=&0\\
      &\alpha_N &=&1\\
   \end{array} \]} &~~~~~~~~~~~~~~~~~&~~~~~~~
    \cmt{6}{\[ \begin{array}{llll}
      \fbox{Case 2}~~~~~&\multicolumn{3}{l}{c_8=2 ~~~c_4=2}\\[1mm]
      &\mbox{and}\\[1mm]
      &\multicolumn{3}{l}{c_1=c_2=c_3=c_5=c_6=c_7=0}\\[5mm]
      &\multicolumn{3}{l}{\mbox{Then the set (\ref{c system}) obeys a solution:}}\\[1mm]
      &\alpha_1~~~~&=~~&0\\
      &\alpha_2 &=&0\\
      &\alpha_3 &=&1\\
      &\alpha_M &=&0\\
      &\alpha_N &=&0\\
   \end{array} \]}
\end{tabular}

\paragraph[Case 1]{\fbox{Case 1}\\}
Here (\ref{upper bound inv ME 2}) reads as
\ba\label{upper bound inv ME Case 1}
   \Big|L^{(M,N)} \Big|^2
   &~\le~&\frac{2\big( \beta_2-\beta_1+2\beta_M -\frac{3}{2}\big)~ 2\big(2j_4+\beta_2-\beta_1+2\beta_N-\frac{1}{2}\big)}
      {2~(j_4+\beta_2+\beta_M)~2~(\beta_2-\beta_1+\beta_M-\frac{1}{2})~2~(\beta_M-\frac{1}{2})(2j_4+\beta_3+\beta_2-\beta_1+2\beta_M)
       (-\beta_3+\beta_2-\beta_1+2\beta_M-1)}
      \nonumber
   \\
   &&\times~
  \frac{1}{(\beta_3+\beta_2-\beta_1+2\beta_M-1)~2~(\beta_1-\beta_N+1)(\beta_3-\beta_2+\beta_1-2\beta_N+2)}
  \NN\NN 
  &\sim& ~\frac{1}{j_4}      
\ea
However (\ref{explicit final upper bound x(2l)}) gives in this case
\ba\label{upper bound x(2l) Case 1}
     \prod_{l=M}^{N-1}x(2l)
   &~~\le~~&
   (32)^4  \left[\frac{(3+2\beta_1-2\beta_N)}{(3+2j_4+2\beta_1-2\beta_M)}\right]^{\frac{1}{2}}
           \left[\frac{(3+\beta_3-\beta_2+\beta_1-2\beta_N)}{(3+2j_4+\beta_3-\beta_2+\beta_1-2\beta_M)}\right]^{\frac{1}{2}}   
   \NN
   &&~~\times~
            \left[\frac{(3+2j_4+2\beta_2+2\beta_M)}{(3+2j_4+2\beta_2+2j_4+2\beta_N)}\right]^{\frac{1}{2}}
            \left[\frac{(3+j_4+\beta_3+j_4+\beta_2-\beta_1+2\beta_M)}{(3+j_4+\beta_3+j_4+\beta_2-\beta_1+2j_4+2\beta_N)}\right]^{\frac{1}{2}} 
  \NN
  &&~~\times~ 
            \left[\frac{(1+\beta_2-\beta_1+\beta_M)}{(1+\beta_2-\beta_1+j_4+\beta_N)}\right]^{\frac{1}{2}}
            \left[\frac{(2-\beta_3+\beta_2-\beta_1+2\beta_M)}{(2-\beta_3+\beta_2-\beta_1+2j_4+2\beta_N)}\right]^{\frac{1}{2}}
  \NN
  &&~~\times~ 
            \left[\frac{(1+\beta_M)}{(1+j_4+\beta_N)}\right]^{\frac{1}{2}}
            \left[\frac{(2+\beta_3-\beta_4+\beta_2-\beta_1+2\beta_M)}{(2+\beta_3-\beta_4+\beta_2-\beta_1+2j_4+2\beta_N)}\right]^{\frac{1}{2}}
  \NN\NN 
  &\sim& ~\frac{1}{(j_4)^3}
\ea 
We therefore find that in \fbox{Case 1} $\big|L^{(M,N)} \big|$ is bounded from above by a quantity of the order $\frac{1}{(\jmax)^2}$. Moreover according to (\ref{dimension of Q}) the dimension of the matrix $\widehat{Q}$ is given by $\dim \widehat{Q}=2j_1+1=2j_4+2\beta_1+1$ which is proportional to $\jmax$. Therefore upon summing up all terms in the row/column sums in order to apply theorem \ref{Gersgorin discs}, we could get an additional factor of $\jmax$. However, taking this sum implies increasing $M$ and taking $M=\beta_M \sim j_4$. This contradicts our minimization of the inverse power of $\jmax$ above immediately. Despite this we still may assume $j_1\le j_2\le j_3\le j_4$ and if we want the dimension $D$ of $\widehat{Q}_D$ to scale with the maximal spin then (\ref{range of a2}) forces us to take $j_1\sim j_2\sim j_3 \sim j_4=\jmax$. But then in (\ref{c system}) all constants but $c_7,c_8$ are non zero because for all $\alpha$'s we have $0\le \alpha \le 1$. 
Inserting this into 
(\ref{upper bound inv ME Case 1}) and (\ref{upper bound x(2l) Case 1}) gives
\be
  \Big|L^{(M,N)} \Big|^2\sim\frac{1}{(j_4)^4}
  ~\hspace{1cm}\mbox{and}\hspace{1cm}
  \prod_{l=M}^{N-1}x(2l)\sim~ \frac{1}{j_4}~~~~\mbox{unless $\beta_M=j_4$}
\ee
and therefore taking the row sum will not change the inverse power of $\jmax$, because $\lim\limits_{M\rightarrow N} |L^{(M,N)}|\sim (\jmax)^{-\frac{5}{2}}$ and thus the maximal row sum will be of an order $(\jmax)^{-r}$ where $r>1$.

Note that the case $\fbox{M=N}$ contradicts the assumptions of \fbox{case 1}, since for $M=N$ it must hold that $\alpha_M=\alpha_N$. This case will be included in the following discussion.

\paragraph[Case 2]{\fbox{Case 2}\\}
Using (\ref{ansatz for spins}) we have
\barr{lllllcl}\label{ansatz for spins 2 b}
   j_1=\beta_1 &~~~~~&j_3=\jmax+\beta_3&~~~~~~~~~~~~&M&~~=~~&\beta_M\\
   j_2=\beta_2 &&j_4=~\jmax      &&N&~~=~~&\beta_N
\earr
Here (\ref{upper bound inv ME 2}) reads as
\ba\label{upper bound inv ME Case 2}
   \Big|L^{(M,N)} \Big|^2
   &~\le~&\frac{2\big( \beta_2-\beta_1+2\beta_M -\frac{3}{2}\big)~ 2\big(\beta_2-\beta_1+2\beta_N-\frac{1}{2}\big)}
      {2~(\beta_2+\beta_M)~2~(\beta_2-\beta_1+\beta_M-\frac{1}{2})~2~(\beta_M-\frac{1}{2})(2j_4+\beta_3+\beta_2-\beta_1+2\beta_M)
       (-\beta_3+\beta_2-\beta_1+2\beta_M-1)}
      \nonumber
   \\
   &&\times~
  \frac{1}{(\beta_3+\beta_2-\beta_1+2\beta_M-1)~2~(\beta_1-\beta_N+2)(2j_4+\beta_3-\beta_2+\beta_1-2\beta_N+2)}
  \NN\NN
  &\sim& ~\frac{1}{(j_4)^2}      
\ea
By observation, since $M,N$ do not scale with the largest spin, (\ref{explicit final upper bound x(2l)}) is only sensitive to $M,N$ and thus becomes a constant in the limit $j_4\rightarrow \infty$. Since $j_1\le j_2$ we must have $\beta_1\le \beta_2$. Moreover $j_3\le j_4$ and thus $\beta_3\stackrel{!}{=}0$. Also recall that $1\le M\le N\le j_1+ \frac{1}{2}$ which implies, together with (\ref{ansatz for spins 2 b}), that $1\le \beta_M\le \beta_N\le \beta_1+\frac{1}{2}$. Additionally we observe that the row/column sum contains at most $N=\beta_N$ terms.
Hence
\ba\label{upper bound inv ME Case 2.1}
   \Big|L^{(M,N)} \Big|^2
   &~\le~&\frac{\big( \beta_2-\beta_1+2\beta_M -\frac{3}{2}\big)~ \big(\beta_2-\beta_1+2\beta_N-\frac{1}{2}\big)}
      {4~(\beta_2+\beta_M)~(\beta_2-\beta_1+\beta_M-\frac{1}{2})~(\beta_M-\frac{1}{2})(2j_4
       +\beta_2-\beta_1+2\beta_M)
       (\beta_2-\beta_1+2\beta_M-1)}
      \nonumber
   \\
   &&\times~
  \frac{1}{(\beta_2-\beta_1+2\beta_M-1)~(\beta_1-\beta_N+2)(2j_4-\beta_2+\beta_1-2\beta_N+2)}    
\ea
Now by inspection of (\ref{upper bound inv ME Case 2.1}), we find that $ \Big|L^{(M,N)} \Big|^2$
scales as $(\beta_2)^{-4},(\beta_1)^{-4},(\beta_M)^{-5},(\beta_N)^{-1}$. 
The row sum of the inverse matrix $\widehat{Q}_D^{-1}$ contains $\beta_N$ terms. 
However $\beta_N\le\beta_1+\frac{1}{2}$ and hence $\beta_N$ cannot be increased without increasing $\beta_1$ and thus $\beta_2$. This overcompensates the possible $\beta_N$ terms of type (\ref{upper bound inv ME Case 2.1}) in a row sum.
Therefore we will set $\beta_2=\beta_1=:\beta_{12}$ and $\beta_M=1$. Then
\ba\label{upper bound inv ME Case 2.2}
   \Big|L^{(1,N)} \Big|^2
   &~\le~&\frac{\big( \frac{1}{2}\big)~ \big(2\beta_N-\frac{1}{2}\big)}
      {4~(\beta_{12}+1)~(\frac{1}{2})~(\frac{1}{2})(2j_4
       +2)
       (1)}
      \times~
  \frac{1}{(1)~(\beta_{12}-\beta_N+2)(2j_4-2\beta_N+2)}
  \NN
  \NN
  &\le&\frac{ \beta_N}
      {4(\beta_{12}+1)~(j_4+1)(\beta_{12}-\beta_N+2)(j_4-\beta_N+1)}       
  \NN
  &\le&\frac{ \beta_N}
      {4\beta_{12}~j_4(\beta_{12}-\beta_N+2)(j_4-\beta_N+1)} 
\ea
If we demand (\ref{upper bound inv ME Case 2.2}) to be valid already for
$j_4=\jmax=\frac{1}{2}$, then we have to fix $\beta_N=1$ due to positivity of
$\Big|L^{(M,N)} \Big|$. In order to still obtain an upper limit for
$\Big|L^{(M,N)} \Big|$, this implies we minimize $\beta_{12}$ by setting it to
$\frac{1}{2}$. Then
\ba\label{upper bound inv ME Case 2.3}
   \Big|L^{(1,1)} \Big|
   &~\le~&\frac{ 1}
      {\sqrt{3}~j_4} < \frac{ 1}{j_4}
\ea
Recall that always $x(2l)<1$, but from (\ref{explicit final upper bound x(2l)}) it can be seen that for  \fbox{Case 2} the quantity $x(2l)$ does not scale with inverse powers of $\jmax$, as $\jmax$ is increased.  Therefore we can obtain a row sum of an order $(\jmax)^{-1}$. 

Note that giving an upper bound on $L^{(M,N)}$ leads us to set $(M,N)=(1,1)$, which gives the special case of (\ref{upper bound L^(MM)}).
Now the finite upper bound (\ref{upper bound inv ME Case 2.3}) for the spectrum of the inverse matrix $\widehat{Q}_D^{-1}$ provides us with a 
 lower bound for the smallest non-zero eigenvalue of $\widehat{Q}_D$ in leading order of the maximal spin $\jmax$.
This lower bound is given, for the spins $j_1= j_2 =
\frac{1}{2},j_3=j_4=\jmax$, as
\be\label{gauge inv 4-vertex: lower bound non zero eval}
   \lambda_{\widehat{Q}}^{(min)}\ge \frac{1}{\Big|L^{(1,1)} \Big|}
   =   \jmax   
\ee
This may be compared to the numerical result $\lambda_{\widehat{Q}}^{(min)}= 2 \cdot\sqrt{ \jmax(\jmax+1)}$ obtained in \cite{Volume Paper} contributed by the spin configuration $j_1=j_2=\frac{1}{2}$ and $j_3=j_4=\jmax$.
Our estimate above explains this combination of spins.
Recall that $\mineval=\sqrt{|\lambda_{\widehat{Q}}^{(min)}|}$. Therefore we conclude that:
~\\
{\it The smallest non-zero eigenvalue $\lambda_{\hat{V},\mathrm{4-vertex}}^{(min)}$ of the volume operator $\hat{V}$ acting on a gauge invariant four valent vertex $v$ with spin configuration $j_1\le j_2 \le j_3 \le j_4=\jmax$ is bounded from below by\\
 \be
   \lambda_{\hat{V},\mathrm{4-vertex}}^{(min)}(\jmax)\ge \ell_P^3\sqrt{|Z|\cdot|\sigma(123)|\cdot j_{max}}
 \ee
 where $Z$ is the regularization constant in (\ref{Volume Definition}),
$\sigma(123)=0,\pm 2,\pm 4$,  
and $\ell_P$  
is the Planck length.
}

\section{Summary}

In this paper and its companion \cite{NumVolSpecLetter} we have presented  a comprehensive analysis of the spectral properties of the volume operator $\hat{V}$ in Loop Quantum Gravity (LQG), which may serve as a starting point  
for a computer based analysis of the action of constraint operators in the full theory. We have shown how the volume operator $\hat{V}$ can be handled analytically  in case of a 4-valent vertex,  and numerically  
in the general case. 
In this context we have analyzed how the spatial diffeomorphism invariant properties of the graphs underlying the states of the kinematical Hilbert space, as encoded in the sign factors $\epsilon(IJK)$, become relevant in a practical computation. 
This aspect of the theory, although known, has not been analyzed 
in detail before the present work. 

A further summary and outlook is given in section 6 of the companion paper \cite{NumVolSpecLetter}.

\section{Acknowledgments}
We would like to thank Thomas Thiemann for encouraging us to start this project and for suggestions and discussion on the issue of the sign factors.
His continuous, reliable support for this project and his helpful suggestions on the manuscript are gratefully acknowledged. 
We thank Steve White for pointing out to us the relative merit of singular
value decomposition, because of its numerical stability.  This resulted in a
significant improvement in the quality of our numerical results.
Moreover we thank Jonathan Thornburg for providing  
perspective into the problem of handling numerical error.

We would like to thank Luciano Rezzolla and the numerical relativity group
 at the Albert Einstein Institute Potsdam for support and use of computational resources at the AEI, in particular the Peyote cluster.
In addition we thank the Perimeter Institute for Theoretical Physics, where parts of this project were completed, for hospitality.

JB thanks the Gottlieb-Daimler and Carl-Benz foundation, the German Academic Exchange Office as well as the Albert Einstein Institute for financial support. Furthermore JB has been supported in part by the Emmy-Noether-Programm of the Deutsche Forschungsgemeinschaft under grant FL 622/1-1. 
The work of DR was supported in part through the Marie Curie Research and
Training Network ENRAGE (MRTN-CT-2004-005616).

We thank Simone Speziale for pointing out \cite{Carbone Carfora Marzouli} to us.

We are grateful for valuable suggestions on improving the manuscript by
anonymous referees of Classical and Quantum Gravity.

\begin{appendix}

\section*{Appendices}

This paper and its companion \cite{NumVolSpecLetter} make heavy use of
special properties of the spin network functions, in particular the fact that the
action of the flux operators on spin network basis states can be mapped
to the problem of evaluating angular momentum operators on angular momentum
eigenstates, the latter of which is familiar
from ordinary quantum mechanics.  
This not only results in a great simplification, but also provides a convenient way to work in the $SU(2)$-gauge-invariant regime by using powerful techniques from the recoupling theory of angular momenta.

We have added these appendices in order to provide the reader with details on the conventions we use, for example in the construction of recoupling schemes. 
Moreover the reader not familiar with these techniques  
is given a brief summary, in order to be able to redo the computations presented here. 

The appendices are organized as follows: In appendix \ref{App C} we give basic properties of matrix representations of $SU(2)$, whose matrix elements are used for the definition of spin network functions. Appendix \ref{Angular Momentum Theory} then summarizes the theory of angular momentum from quantum mechanics, and discusses an extended notion of recoupling of an arbitrary number of angular momenta in terms of recoupling schemes, as we use them. Appendix \ref{6j} completes this presentation with the definition of so called $6j$-symbols, along with an elaboration 
of their basic properties. This provides an explicit notion of recoupling schemes in terms of polynomials of  quantum numbers.
Finally \ref{App C},\ref{Angular Momentum Theory},\ref{6j}  are connected in appendix \ref{Explanation SNF -> Spin States}, in which the correspondence between spin network states and angular momentum states is reviewed in detail.

\section{\label{App C}Representations of $SU(2)$}
Irreducible matrix representations of $SU(2)$ can be constructed in $(2j+1)$-dimensional linear vector spaces, where $j\ge 0$ is a 
half integer number; $j=0$ denotes the trivial representation.  
Note that we write  for the defining representation of a group element $h\in SU(2)$ $\PI{h}{j=\frac{1}{2}}{mn}=h_{AB}$ and for general representation matrices  $\PI{h}{j}{mn}$: the rows $m$ are labelled from top to bottom by $m=j\ldots -j$ and the columns $n$ from left to right by $n=j\ldots -j$.

In the following an overline denotes complex conjugation, while $\cdot^T$ indicates
the transpose.

\subsection[General Conventions --- Defining Representation $j=\frac{1}{2}$]{General Conventions --- Defining Representation \fbox{$j=\frac{1}{2}$}}

\subsubsection{Generators}
\arraycolsep=0.2em
As generators of $SU(2)$ we use the $\tau$-matrices given by $\tau_k:=-\mb{i}\sigma_k$, with $\sigma_k$ being the Pauli-matrices:
\be
\begin{array}{ccccc}
      \sigma_1=\left(\begin{array}{cc}
                     0 & 1 \\
		     1 & 0
                 \end{array}\right)
    &&\sigma_2=\left(\begin{array}{cc}
                     0 & -\mb{i}~ \\
		     \mb{i} & 0
                 \end{array}\right)		 
    &&\sigma_3=\left(\begin{array}{cc}
                     1 & 0 \\
		      0 & -1
                 \end{array}\right)	~~~~~~~\mbox{for which}
~~~~~\big[\sigma_i,\sigma_j \big]=2\mb{i}\,\epsilon^{ijk}\sigma_k	 
        \\\\
  
      \tau_1=\left(\begin{array}{cc}
                     ~0 & -\mb{i} \\
		     -\mb{i}~ & 0
                 \end{array}\right)
    &&\tau_2=\left(\begin{array}{cc}
                     0 & -1~ \\
		     1 & 0
                 \end{array}\right)		 
    &&\tau_3=\left(\begin{array}{cc}
                     -\mb{i}~ & 0~ \\
		      0 & \mb{i}
                 \end{array}\right)	~~~~~~~\mbox{for which}~~~~~\big[\tau_i,\tau_j \big]=2\,\epsilon^{ijk}\tau_k	 
  \end{array}
\label{defining rep definitions}
\ee

  Additionally we use
  \be\label{Definition epsilon}\begin{array}{cc}\epsilon=-\tau_2=\left(\begin{array}{cc}
                                 	0 & 1\\
				       -1~ & 0   
                                \end{array} \right) 
				~~~~~\mbox{with the obvious properties}~~~~~
				\epsilon^{-1}=\epsilon^T=-\epsilon=\left(\begin{array}{cc}
                                 	0 & -1\\
				        1 &  0   
                                \end{array} \right)
		   \end{array}\ee

\subsubsection{$SU(2)$ Representations}

In the defining representation of $SU(2)$, for a group element $h\in SU(2)$,
we have
  \be\label{simple parametrization of SU(2) group element}
        h=\left(\begin{array}{cc}
                    ~a & b \\
		    -\overline{b}~ & \overline{a}
		\end{array}\right)~~~~~~~~~~~~~  \mbox{for which}~~\det{h}=|a|^2+|b|^2=1 
   \ee
where $a,b\in \mb{C}$.  Moreover
   \be   
      h^{-1}=\epsilon~h^T~\epsilon^{-1} = \left(\begin{array}{cc}
                                                  \overline{a} & -b \\
		                                  \overline{b}~ & ~a
		                                  \end{array}\right)
		                                  ~~~~~~~~~~~~~~~~~~~~~~~~~~~~~~~~~~~
      \\\\	
   \ee
   and we have the additional properties that
   \be\label{SU(2) symmetries}
      \overline{h}=\big[h^{-1}\big]^T
                   =\big[\epsilon~h^T~\epsilon^T\big]^T 
		   =\epsilon~h~\epsilon^T
		   =\epsilon~h~\epsilon^{-1}~~~~~~~
   \ee

As mentioned above, we use the following convention for the matrix elements of $\PI{h}{j=\frac{1}{2}}{AB}$, with $A$ being the row and $B$ the column index
\[\begin{array}{ccccccccccc}
    h_{1\,1}&=&h_{\frac{1}{2}~\frac{1}{2}}& =& a
    &~~~~~~~&
    h_{1\,2}&=&h_{\frac{1}{2}~-\frac{1}{2}}&=& b\\
    h_{2\,1}&=&h_{-\frac{1}{2}~\frac{1}{2}}&=& -\bar{b}
    &~~~~~~~&
    h_{2\,2}&=&h_{-\frac{1}{2}~-\frac{1}{2}}&=&\bar{a}
\end{array}\] \\[-14mm]\be \label{Konvention tau} \ee   
   
   For the $\tau_k$'s we additionally have
   \be\label{tau identity}
     -\overline{\tau_k}^T = \tau_k =-(\tau_k)^{-1}
   \ee

\subsection{General Conventions for $(2j+1)$-dimensional $SU(2)$-Representation Matrices}
Here we follow \cite{Sexl Urbantke,Consistency Check I,Consistency Check II}.

\subsubsection{General Formula for $SU(2)$ Matrix Element}
The $(2j+1)$-dimensional representation matrix $\PI{h}{j}{}$ of  $h\in SU(2)$, given in terms of the parameters of the defining representation (\ref{simple parametrization of SU(2) group element}), 
can be written as
     \be\label{SL(2,C) matrixelement derivation 4}
\fbox{$\displaystyle \PI{h}{j}{mm'}=\sum_{\ell}~(-1)^\ell
     \frac{\sqrt{(j+m)!~(j-m)!~(j+m')!~(j-m')!}}{(j-m-\ell)!~(j+m'-\ell)!~(m-m'+\ell)!~\ell!}
     ~~a^{j+m'-\ell}~(\overline{a})^{j-m-\ell}~b^{m-m'+\ell}~(\overline{b})^{\ell}$}
\ee
where $\ell$ takes all integer values such that none of the factorials in the denominator gets a negative argument.
By construction every representation of $SU(2)$ consists of special unitary matrices $\PI{g}{j}{}$
(such that
$\det \PI{g}{j}{} =1$ and $\PI{g}{j}{}\PI{g}{j}{}^\dagger = \PI{g}{j}{}^\dagger \PI{g}{j}{}=\PI{\mb{1}}{j}{}$ (with $\PI{g}{j}{}^\dagger =\overline{\PI{g}{j}{}}^T$)). Using (\ref{SL(2,C) matrixelement derivation 4}) we obtain a generalization of (\ref{SU(2) symmetries}):
\ba\label{SU(2) unitarity properties}
   \PI{\epsilon}{j}{}\PI{h}{j}{}^T\PI{\epsilon^{-1}}{j}{}
   &=&\PI{\epsilon^{-1}}{j}{}\PI{h}{j}{}^T\PI{\epsilon}{j}{}
   = \PI{h^{-1}}{j}{}=\PI{h}{j}{}^\dagger = \PIbar{h}{j}{}^T
\ea

\subsubsection{Generators and $\epsilon$-Metric}
Upon applying the representation matrix element formula  (\ref{SL(2,C) matrixelement derivation 4}) and the ansatz, 
\be
  \PI{\tau_k}{j}{mn}=\frac{d}{dt}\big(\PI{\mb{e}^{t\tau_k}\mb{1}}{j}{mn} \big)\Big|_{t=0}
\ee 
where in the defining two dimensional representation the exponential can be explicitly evaluated,
one obtains,\footnote{ See \cite{Consistency Check I,Consistency Check II} for details.} using $A(j,m)=\big[j(j+1)-m(m-1)\big]^\frac{1}{2}$ with the obvious property that $A(j,-\!m)=A(j,m\!+\!1)$,
\ba\label{SU(2) unitarity properties II}
   \PI{\tau_1}{j}{mn}&=& -\mb{i}~A(j,m)~ \delta_{m\:n+1}~~-\mb{i}~A(j,m+1)~\delta_{m\:n-1}
   \NN
   \PI{\tau_2}{j}{mn}&=&~-A(j,m)~ \delta_{m\:n+1}~~~+~A(j,m+1)~\delta_{m\:n-1}
   \NN
   \PI{\tau_3}{j}{mn}&=&~-2~\mb{i}~m~ \delta_{m\:n}
\ea
Moreover  one finds by plugging (\ref{Definition epsilon}) into (\ref{SL(2,C) matrixelement derivation 4}) that 
\be\label{representation of epsilon}
  \PI{\epsilon}{j}{mn}=(-1)^{j-m}~\delta_{m\:-\!n} ~~~~\mbox{and}~~~~ \PI{\epsilon^{-1}}{j}{mn}=(-1)^{j+m}~\delta_{m\:-\!n}
\ee

\section{\label{Angular Momentum Theory}Angular Momentum Theory}

\subsection{Basic Definitions}
In this section we will summarize the results given in \cite{Edmonds}. We have an angular momentum orthonormal basis $u(j,m;n)=\Sket{j}{m}{;n}$ of a general $(2j+1)$ dimensional representation of $SU(2)$.
The index $n$ stands for additional quantum numbers, not affected by the action of the angular momentum operators $J^k$ fulfilling the commutation relations
\be
   \big[J^i,J^j\big]=\mb{i}~\epsilon^{ijk}~J^k
\ee

We can formulate ladder operators given as
$ J^{+}=J^1+\mb{i}J^2 $ and $ J^{-}=J^1-\mb{i}J^2$
where, using the shorthand $A(j,m)=\sqrt{j(j+1)-m(m-1)}$ with the obvious property that 
$A(j,\!-m)=A(j,m\!+\!1)$, we have
\ba \label{usual angualr momentum ops}
   J^{-}\Sket{j}{m}{;n}&=&A(j,m)\Sket{j}{m\!-\!1}{;n}
   \NN
   J^{+}\Sket{j}{m}{;n}&=&A(j,m\!+\!1)\Sket{j}{m\!+\!1}{;n}
\ea
The $\Sket{j}{m}{;n}$ simultaneously diagonalize the two operators: the squared total angular momentum\footnote{Note that we denote single components of the angular momentum by $J^i$, $i=1,2,3$, whereas the total angular momentum is denoted by $(J)$.} $(J)^2$ and the 
magnetic quantum number $J^3$ \cite{Edmonds}:
\ba\label{operators in single HS 1}
   (J)^2\Sket{j}{m}{;n}  &=&\big( (J^1)^2\!+\!(J^2)^2\!+\!(J^3)^2\big)\Sket{j}{m}{;n}
   =\Big(\frac{1}{2}\big(J^-J^+\!+\! J^+J^-\big)+(J^3)^2\Big)\Sket{j}{m}{;n}
   \NN[-2mm]
   &=&j(j+1)~\Sket{j}{m}{;n}
   \\
   J^3\Sket{j}{m}{;n}  &=& m \Sket{j}{m}{;n}
   \nonumber
\ea
That is, $\Sket{j}{m}{;n}$ is a maximal set of simultaneous eigenvectors of $(J)^2$ and $J^3$.

One then finds the following commutation relations
\be\label{CR J}
   [J^3,(J)^2]=[J^+,(J)^2]=[J^-,(J)^2]=0~~~~~
    [J^3,J^{+}]=J^{+}~~~~~
    [J^3,J^{-}]=-J^{-}~~~~~
    [J^{+},J^{-}]=2J^3 
\ee
such that we obtain for the $(2j+1)$-dimensional matrix representation with arbitrary weight $j$
\barr{lllllllll}\label{angular momentum operators}
    (J^3)_{m'm}&=&m~\delta_{m'\:m}
    \\[2mm]
    (J^+)_{m'm}&=&A(j,m+1)~\delta_{m'\:m\!+\!1}
    &&
    (J^1)_{m'm}=& 
\!\!\!\frac{1}{2}(J^++J^-)_{m'm}
    &=&
    & 
\!\!\!\frac{1}{2}A(j,m+1)~\delta_{m'\:m\!+\!1} + \frac{1}{2}A(j,m)~\delta_{m'\:m\!-\!1} 
    \\[2mm]
    (J^-)_{m'm}&=&A(j,m)~\delta_{m'\:m\!-\!1}
    &&
    (J^2)_{m'm}=-&
\!\!\!\frac{\mb{i}}{2}(J^+-J^-)_{m'm}
    &=&
    -&
\!\!\!\frac{\mb{i}}{2}A(j,m+1)~\delta_{m'\:m\!+\!1} +\frac{\mb{i}}{2}A(j,m)~\delta_{m'\:m\!-\!1} 
\earr

\subsection{Fundamental Recoupling}

Now we can easily understand what happens if we couple
several angular momenta. For that we first repeat the well known theorem of Clebsch \& Gordan on
tensorized representations of $SU(2)$:
\begin{Theorem}{Clebsch \& Gordan} \label{Clebsh Gordan theorem}\\
   Having two irreducible representations $\pi_{j_1}$, $\pi_{j_2}$ of $SU(2)$
   with weights $j_1$ and $j_2$,
   their tensor product space splits into a direct sum of irreducible representations $\pi_{j_{12}}$ with 
   $|j_1-j_2|\le j_{12}\le j_1+j_2$ such that
   
   \[ \pi_{j_1} \otimes \pi_{j_2}=\pi_{j_1+j_2}\oplus \pi_{j_1+j_2-1} \oplus \ldots 
                                \oplus \pi_{|j_1-j_2+1|}\oplus \pi_{|j_1-j_2|}  \]
\end{Theorem}
~\\
Equivalently we can write for the resulting representation space 
$\mathcal{H}_{(D)}=\mathcal{H}_{(D_1)} \otimes \mathcal{H}_{(D_2)}$ 
(where $D_1=2j_1+1$, $D_2=2j_2+1$, $D=D_1\cdot D_2$ denote the dimensions of the Hilbert spaces):
\be \label{Hilbert space decay}
  \mathcal{H}_{(D)} =\mathcal{H}_{(D_1)} \otimes \mathcal{H}_{(D_2)}
   =\bigoplus\limits_{j_{12}=|j_1-j_2|}^{j_1+j_2}\mathcal{H}_{(2j_{12} +1)} 
\ee
In other words, 
if we couple two angular momenta $j_1,j_2$, we can get resulting angular momenta $j_{12}$
varying in the range $|j_1-j_2| \le j_{12} \le j_1+j_2$.
The tensor product space of two representations of $SU(2)$ decomposes into a direct sum of
representation spaces,  
with one space for every possible value of recoupling  $j_{12}$ with the
according dimension $2j_{12}+1$.

\subsection{Recoupling of $n$ Angular Momenta --- $3nj$-Symbols\label{introduce recoupling scheme}}

As mentioned earlier the successive coupling of three angular momenta to a resulting $j$
can be generalized. For this purpose let us first comment on the generalization principle before we go into detailed
definitions. 

Theorem \ref{Clebsh Gordan theorem} can be applied to an $n$-fold tensor product of
representations $\pi_{j_1}\otimes \pi_{j_2} \otimes \ldots \otimes \pi_{j_n}$ by reducing out step by step
every pair of representations. 
This procedure has to be carried out until all tensor products are reduced out.
One then ends up with a direct sum of representations, each of which has a weight corresponding 
to an allowed value of the total angular momentum to which the $n$ single angular momenta 
$j_1,j_2,\ldots,j_n$ can couple.
However, there is an arbitrariness in how one couples the $n$ angular momenta together, that is the order 
in which $\pi_{j_1}\otimes \pi_{j_2} \otimes \ldots \otimes \pi_{j_n}$ is reduced out (by applying 
\ref{Clebsh Gordan theorem}) matters.
\\

Consider a system of $n$ angular momenta.
First we fix a labelling of these momenta, such that we have $j_1,j_2,\ldots,j_n$. 
Again the first choice would be a tensor basis
$\ket{\vec{j}~\vec{m}}$ of all single angular momentum states $\ket{j_k~m_k}$, $k=1 \ldots n$ defined by:
\be \label{tensor basis n angular momenta}
  \ket{\vec{j}~\vec{m}}=\ket{(j_1,j_2,\ldots,j_n)~(m_1,m_2,\ldots,m_n)}:=
                   \bigotimes\limits_{k=1}^n \ket{j_k~m_k}
\ee
with the maximal set of $2n$ commuting operators $(J_I)^2,J^3_I$, $(I=1,\ldots,n)$.

Now we proceed in order to find a basis in which the total angular
momentum $(J_{tot})^2=(J)^2=(J_1+J_2+\ldots+J_n)^2$ is diagonal (quantum number $j$)
together with the total magnetic 
quantum number $J_{tot}^3=J^3=J_1^3+J_2^3+\ldots+J_n^3$ (quantum number $M$).
As $(J)^2$ and $J^3$ are only two operators, we need $2(n-1)$ more
mutually commuting operators 
to have again a maximal set.
We choose therefore the $n$ operators $(J_I)^2$, $I=1,\ldots,n$ of total single angular momentum (quantum
numbers $(j_1,\ldots,j_n):=\vec{j}$). Then we
are left with the task of finding an additional $n-2$ commuting operators.
For this purpose we define:
\begin{Definition}{Recoupling Scheme} \label{Def Recoupling scheme}\\
   A recoupling scheme $\ket{\vec{g}(IJ)~\vec{j}~j~m}$ is an orthonormal basis which diagonalizes, besides $(J)^2$,
   $J^3$, and $(J_I)^2$ 
($I=1,\ldots,n$) the squares of the additional $n-2$ operators
   $G_2,G_3,\ldots ,G_{n-1}$ defined as: \footnote{Note that formally $G_n:=G_{n-1}+J_n=J_{total}$.}
\begin{eqnarray*}
G_1 & := & J_I,~G_2:=G_1+J_J,~G_3:=G_2+J_1,~G_4:=G_3+J_2,~\ldots,\\
G_I & := & G_{I-1}+J_{I-2},~ G_{I+1}:=G_I+J_{I-1},~G_{I+2}:=G_{I+1}+J_{I+1},~G_{I+3}:=G_{I+2}+J_{I+2}, ~\ldots,\\
G_J & := & G_{J-1}+J_{J-1},~ G_{J+1}:=G_J+J_{J+1},~G_{J+2}:=G_{J+1}+J_{J+2},~\ldots,\\
G_{n-1} & := & G_{n-2}+J_{n-1}
\end{eqnarray*}
The vector
\begin{eqnarray*}
\vec{g}(IJ):=\big( & g_2(j_I,j_J),~g_3(g_2,j_1),~\ldots,~g_{I+1}(g_I,j_{I-1}),~
    g_{I+2}(g_{I+1},j_{I+1}),~\ldots,~ g_J(g_{j-1},j_{j-1}),~  
    g_{j+1}(g_J,j_{j+1}),~\ldots,\\
& g_{n-1}(g_{n-2},j_{n-1})\big)
\end{eqnarray*}
carries as quantum numbers the $n-2$ eigenvalues of the operators
$(G_2)^2,~\ldots ,~(G_{n-1})^2$.
\end{Definition}
So we recouple first the angular momenta labelled by $I,J$ where $I<J$ and secondly all the other angular
momenta successively (all labels are with respect to the fixed label set), by taking into account the
allowed values for each recoupling according to theorem \ref{Clebsh Gordan theorem}. 

Let us define
furthermore the so called $standard~recoupling~scheme$ or $standard~basis$:
\begin{Definition}{Standard Basis} \label{Def Standard basis}\\
   A recoupling scheme based on the pair $(I,J)=(1,2)$ with
   \[ G_K=\sum\limits_{L=1}^K J_L \hspace{10cm}\]
   is called the standard basis.
\end{Definition}

Using definition \ref{Def Recoupling scheme} with the commutation relations (\ref{CR J}) and the fact 
that single 
angular momentum operators acting on different single angular momentum Hilbert spaces
commute\footnote{That is $\big[ J_I^i,J_J^j \big]=0$ whenever $I \ne J$.}, one can easily check
that for every recoupling scheme
\begin{itemize}
   \item[(i)]the $G_I$'s fulfill the angular momentum algebra \ref{CR J}
   \item[(ii)] $(J)^2,~(J_I)^2,~(G_I)^2,~J^3$ $\forall I=1\ldots n$ commute with each other 
\end{itemize}
Note that it is sufficient to prove these two points in the standard basis $\vec{g}(12)$, 
because every other basis $\vec{g}(IJ)$ is related to it by simply relabelling the $n$ angular momenta.\\

We have thus succeeded in giving an alternative description of a system of $n$ angular momenta by all 
possible occurring intermediate recoupling stages $G_I$, instead of using the individual magnetic
quantum numbers.
Every recoupling scheme $\ket{\vec{g}(IJ)~\vec{j}~j~m}$, labeled by the index pair $(IJ)$, gives a distinct orthonormal basis.
We are then in need of a transformation
connecting the different bases, i.e.\ we wish to express one basis, belonging to the pair $(IJ)$, in terms of
another
basis belonging to the pair $(KL)$. This leads to the following definition.
\begin{Definition}{$3nj$-Symbol} \label{Def 3nj-symbol}\\
   The generalized expansion coefficients of a recoupling scheme in terms of the standard recoupling
   scheme are called $3nj$-symbols:
   \[ \ket{~\vec{g}(IJ)~\vec{j}~j~m} = \sum\limits_{all~\vec{g}'(12)} 
                      \underbrace{\scaprod{\vec{g}'(12)~\vec{j}~j~m~}{~\vec{g}(IJ)~\vec{j}~j~m}}_{3nj-symbol}~
		      \ket{~\vec{g}'(12)~\vec{j}~j~m} 
   \]
   The summation has to be extended over all possible values of the intermediate recouplings \\ $\vec{g}'(12)=\big(g_2'(j_1,j_2),~g_3'(g_2',j_3),~\ldots,~g_{n-1}'(g_{n-2}',j_{n-1})\big)$, that
   is all values of each component $g_k'$ allowed by theorem \ref{Clebsh Gordan theorem}. 
\end{Definition} 

In calculations we will suppress the quantum numbers $\vec{j},j,m$, since they are identical 
in every expression, and
write the $3nj$-symbols as $\scaprod{\vec{g}(IJ)}{\vec{g}'(12)}$. 
Note additionally the following properties of the $3nj$-symbols:
\begin{itemize}
   \item[(i)] They are unitary and real, due to the  
the fact that they can be expressed in terms of Clebsch-Gordan-coefficients\footnote{Which are unitary and real.}:
               \[\scaprod{\vec{g}(IJ)}{~\vec{g}'(12)}~=~\scaprod{\vec{g}'(12)~}{~\vec{g}(IJ)}\]
   \item[(ii)]They are rotationally invariant, i.e.\ independent of the magnetic quantum numbers $m_k$
              occurring in (\ref{tensor basis n angular momenta}).   
\end{itemize}

\subsection{\label{Properties of Recoupling Schemes}Properties of Recoupling Schemes}
In this section we will briefly review the properties of recoupling schemes as defined in section \ref{introduce recoupling scheme}.
In what follows we will frequently use \fbox{$\tilde{m}_k:=m_1+m_2+\ldots+m_k$}.

\subsubsection{A General (Standard-)Recoupling Scheme}
A general standard recoupling scheme is defined as follows: Fix a labelling $j_1,\ldots,j_N$ of the $N$ spins to recouple. Then one constructs 
\begin{footnotesize}
\[\begin{array}{ccl}
   \lefteqn{\big|\,\vec{a}(12)~J~M~;~\vec{n}\, \big>=}\\\\
   && \multicolumn{1}{l}{\hspace{1mm}=\big|\,a_2(j_1\,j_2)~a_3(a_2\,j_3)\ldots
       a_{K-1}(a_{K-2}\,j_{K-1})~a_K(a_{K-1}\,j_K)~a_{K+1}(a_K\,j_{K+1})\ldots
       a_{N-1}(a_{N-2}\,j_{N-1})~J(a_{N-1}\,j_N)~M~;~n_1\ldots n_N\, \big>}\\
   &&\left.\begin{array}{cl}
             =~~\displaystyle\sum_{m_1+\ldots+m_N=M}&
	       \big<~j_1~m_1~;~j_2~m_2 ~\big|~a_2(j_1\,j_2)~\tilde{m}_2 ~\big>\\[-3mm]
	     & \big<~a_2~\tilde{m}_2~;~j_3~m_3~\big|~a_3(a_2\,j_3)~\tilde{m}_3~\big>\\
	     & ~~\vdots\\
	     &\big<~a_{K-2}~\tilde{m}_{K-2}~;~j_{K-1}~m_{K-1}~\big|
	           ~a_{K-1}(a_{K-2}\,j_{K-1})~\tilde{m}_{K-1}~\big>\\
             &\big<~a_{K-1}~\tilde{m}_{K-1}~;~j_{K}~m_{K}~\big|
	           ~a_{K}(a_{K-1}\,j_{K})~\tilde{m}_{K}~\big>\\
             &\big<~a_{K}~\tilde{m}_{K}~;~j_{K+1}~m_{K+1}~\big|
	           ~a_{K+1}(a_{K}\,j_{K+1})~\tilde{m}_{K+1}~\big>\\		   
	     & ~~\vdots\\
             &\big<~a_{N-2}~\tilde{m}_{N-2}~;~j_{N-1}~m_{N-1}~\big|
	           ~a_{N-1}(a_{N-2}\,j_{N-1})~\tilde{m}_{N-1}~\big>\\
             &\big<~a_{N-1}~\tilde{m}_{N-1}~;~j_{N}~m_{N}~\big|
	           ~J(a_{N-1}\,j_{N})~M~\big>\\		   
	     
           \end{array}\right\} \fbox{N-1 factors} \\\\
    
     &&\hspace{2.7cm}	   	   
       \big|\,j_1\,m_1\,;\,n_1 \,\big> \,\otimes
       \big|\,j_2\,m_2\,;\,n_2 \,\big> \,\otimes\,\ldots\,\otimes
       \big|\,j_{K-1}\,m_{K-1}\,;\,n_{K-1} \,\big> \,\otimes
       \big|\,j_K\,m_K\,;\,n_K \,\big> \,\otimes
       \big|\,j_{K+1}\,m_{K+1}\,;\,n_{K+1} \,\big>\,\otimes \\
     &&\hspace{64.5mm}  \,\otimes\,\ldots\,\otimes
       \big|\,j_{N-1}\,m_{N-1}\,;\,n_{N-1} \,\big> \,\otimes
       \big|\,j_N\,m_N\,;\,n_N \,\big> 
         
\end{array}\] \\[-8mm]\be\label{Definition recoupling scheme} \ee\\[-8mm]
\end{footnotesize}

\subsubsection{Orthogonality Relations Between Recoupling Schemes} 
For the scalar product of two recoupling schemes we have:
\[ 
\begin{array}{cc@{~}|crclc}
 \big<\,a_2'~a_3'~\ldots~a_{N-1}'~J'~M' \,\big|\,a_2~a_3~\ldots~a_{N-1}~J~M \,\big>~=~
 \delta_{MM'}
                        \delta_{a_2a_2'}\delta_{a_3a_3'}\ldots
			\delta_{a_{N-1}a_{N-1}'}\delta_{JJ'}~
			\delta_{j_1'j_1}\ldots \delta_{j_N'j_N}
			\delta_{n_1'n_1}\ldots\delta{n_N'n_N}
\end{array}\]
This result  can be easily understood by recalling the definition of a recoupling scheme \\ 
$\ket{a_2(j_1~j_2)~a_3(a_2~j_3)\ldots a_{N-1}(a_{N-2}~j_{N-1})~J(a_{N-1}~j_N)~M}$
as the simultaneous eigenstate for the operators $(G_2)^2=(J_1+J_2)^2$, $(G_3)^2=(G_2+J_3)^2$, \ldots, $(G_{N-1})^2=(G_{N-2}+J_{N-1})^2$, $J^2=(G_{N-1}+J_N)^2=(J_1+\ldots+J_N)^2$ with eigenvalues $a_2(a_2+1),~ a_3(a_3+1),~\ldots,~ g_{N-1}(g_{N-1}+1),~ J(J+1)$.

\subsubsection{Partial Orthogonality Relations Between Recoupling Schemes}

The same argument can also be applied to cases where we have to calculate the scalar product of two recoupling schemes of different recoupling order. For illustration let us consider two recoupling schemes 
\begin{footnotesize}
\ba
   \big|~\vec{a}~J~M~\big>
   &=&\big|\,a_2(j_1\,j_2)~a_3(a_2\,j_3)\ldots a_{K\!-\!1}(a_{K\!-\!2}\,j_{K\!-\!1})~a_K(a_{K\!-\!1}\,j_K)\ldots 
   a_L(a_{L\!-\!1}\,j_L)~a_{L\!+\!1}(a_{L}\,j_{L\!+\!1}) \ldots a_{N\!-\!1}(a_{N\!-\!2}\,j_{N\!-\!1})~J(a_{N\!-\!1}\,j_N)~M\big>
   \nonumber
   \\
   \nonumber
   \\
   \big|~\vec{g}~J~M~\big>
   &=&\big|\,g_2(j_1\,j_2)~g_3(a_2\,j_3)\ldots g_{K\!-\!1}(g_{K\!-\!2}\,j_{K\!-\!1})~g_K(g_{K\!-\!1}\,j_P)\ldots 
   g_L(g_{L\!-\!1}\,j_R)~ g_{L\!+\!1}(g_{L}\,j_{L\!+\!1})~ \ldots g_{N\!-\!1}(g_{N\!-\!2}\,j_{N\!-\!1})~ J(g_{N\!-\!1}\,j_N)~M\big>
  \nonumber
\ea  
\end{footnotesize}
Here from $2\ldots K-1$ the spins $j_1,j_2 \ldots j_{K-1}$ are coupled in $\vec{a}$ and $\vec{g}$ in the same order. Then $j_K \ldots j_L$ are coupled in the standard way for $\vec{a}$ but in a different order for $\vec{g}$. After that $j_{L+1} \ldots j_{N}$ are successively coupled to each scheme again.
Now it is clear that $\vec{a}$ and $\vec{g}$ simultaneously diagonalize not only $(G_2)^2=(J_1+J_2)^2,~ (G_3)^2=(G_2+J_3)^2,~ \ldots,$ $ (G_{K-1})^2=(G_{K-2}+J_{K-1})^2$ but also $(G_L)^2 \ldots (G_{N-1})^2,~ J^2=(J_1+\ldots+j_N)^2$. Therefore we can write down immediately
\ba\label{Partial orthogonal recoupling schemes}
   \big<~\vec{a}~J~M~\big|~\vec{g}~J~M~\big>
   &=&
   \big< a_K(a_{K\!-\!1}\,j_K)\ldots 
   a_{L\!-\!1}(a_{L\!-\!2}\,j_{L\!-\!1})~ a_L(a_{L\!-\!1}\,j_L)~ \big|~g_K(a_{K\!-\!1}\,j_P)\ldots g_{L\!-\!1}(g_{L\!-\!2}\,j_Q)~a_L(g_{L\!-\!1}\,j_R)~\big>~ 
   \nonumber   \\
   &&\times
   \delta_{a_2g_2}\ldots\delta_{a_{K-1}g_{K-1}}~\times~\delta_{a_Lg_L}\ldots \delta_{a_{N-1}g_{N-1}}~\times
   ~\delta_{JJ'} ~\times~\delta_{MM'}       
\ea  
For a more detailed derivation of (\ref{Partial orthogonal recoupling schemes}) see lemmas 5.1 and 5.2 of \cite{TT:vopelm}.

\section{Properties of the $6j$-Symbols \label{6j}}
In this appendix we will give an overview of the $6j$-symbols, because they are the basic structure we will
use in our recoupling calculations, as every coupling of $n$ angular momenta can be expressed in terms of  
them. 
For further details we refer the reader to \cite{Varshalovich},\cite{Edmonds}.
\renewcommand{\arraystretch}{1.6}
\arraycolsep=0.2em
\subsection{Definition}
   The $6j$-symbol is defined in \cite{Edmonds}, p 92, as:
   \ba 
   \left\{ \begin{array}{ccc} \label{definition 6j symbol}
   j_1 & j_2 & j_{12}\\
   j_3 & j   & j_{23}
   \end{array} \right\} 
   &:=&[(2j_{12}+1)(2j_{23}+1)]^{-\frac{1}{2}}(-1)^{j_1+j_2+j_3+j}
   \scaprod{j_{12}(j_1,j_2),j(j_{12},j_3)}{j_{23}(j_2,j_3),j(j_1,j_{23})}
   \nonumber\\
   & = &[(2j_{12}+1)(2j_{23}+1)]^{-\frac{1}{2}}(-1)^{j_1+j_2+j_3+j}
   \NN
   &&\times
   \sum_{m_1~m_2}\bigg[\scaprod{j_1~m_1;~j_2~m_2}{j_1~j_2~j_{12}~m_1+m_2} 
   \scaprod{j_{12}~m_1+m_2;~j_3~m-m_1-m_2}{j_{12}~j_3~j~m} \nonumber\\
   &&\times \scaprod{j_2~m_2;~j_3~m-m_1-m_2}{j_2~j_3~j_{23}~m-m_1} 
    \scaprod{j_1~m_1;~j_{23}~m-m_1}{j_1~j_{23}~j~m} \bigg]
   \ea
   The factors in the summation are  
Clebsch-Gordon coefficients.

\subsection{Explicit Evaluation of the $6j$-Symbols}

   A general formula for the numerical value of the $6j$-symbols has been derived by Racah
   \cite{Edmonds}, p.99:
   \be \label{A1}
      \left\{ \begin{array}{ccc}
      j_1 & j_2 & j_{12}\\
      j_3 & j   & j_{23}
      \end{array} \right\}
      =\Delta(j_1,j_2,j_{12})\Delta(j_1,j,j_{23})\Delta(j_3,j_2,j_{23})
      \Delta(j_3,j,,j_{12})w
      \left\{ \begin{array}{ccc}
      j_1 & j_2 & j_{12}\\
      j_3 & j   & j_{23}
      \end{array} \right\}
   \ee
where   
   \be     
      \Delta(a,b,c)=\sqrt{\frac{(a+b-c)!(a-b+c)!(-a+b+c)!}{(a+b+c+1)!}}
      \nonumber
    \ee
and    
    \ba 
      w
      \left\{ \begin{array}{ccc}
      j_1 & j_2 & j_{12}\\
      j_3 & j   & j_{23}
      \end{array} \right\}
      &=&\sum_n (-1)^n (n+1)! 
       [(n-j_1-j_2-j_{12})!(n-j_1-j-j_{23})!(n-j_3-j_2-j_{23})!
      (n-j_3-j-j_{12})!]^{-1}\times
      \nonumber\\
      &&~~~~\times [(j_1+j_2+j_3+j-n)!(j_2+j_{12}+j+j_{23}-n)!
      (j_{12}+j_1+j_{23}+j_3-n)!]^{-1} 
      \ea
      The sum is extended over all positive integer values of $n$ such that no factorial in
      the denominator has a negative argument. That is:
\begin{displaymath}
      \max[j_1+j_2+j_{12},j_1+j+j_{23},j_3+j_2+j_{23},j_3+j+j_{12}] \le n \le
       \min[j_1+j_2+j_3+j,j_2+j_{12}+j+j_{23},j_{12}+j_1+j_{23}+j_3]
\end{displaymath}

      \begin{description} \label{integer conditions}
	\item[Remark]From (\ref{A1}) we are provided with some additional requirements the
	arguments of the $6j$-symbols must fulfill: Certain sums or differences of them
	must be integers to be proper ($\equiv$integer) arguments for the factorials:\\
	\\
        From $\Delta(a,b,c)$ one gets:
	  \begin{itemize}
	     \item $a$, $b$, $c$ must fulfill the triangle inequalities:
		$(a+b-c)\ge 0$,~ 
		$(a-b+c)\ge 0$,~ 
		$(-a+b+c) \ge 0$,
	     \item $(\pm a \pm b \pm c)$ must be an integer
	  \end{itemize}
	 From the $w$-coefficient one gets:
	  \begin{itemize}
	     \item  $j_1+j_2+j_3+j,~j_2+j_{12}+j+j_{23},~j_{12}+j_1+j_{23}+j_3$ ~are 
	      integers. 
	  \end{itemize} 
      \end{description}
      The following (trivial but important) relations are frequently used in calculations involving 
      $6j$-symbols: 
      \ba \label{integer exponents}
          (-1)^{z} & = & (-1)^{-z} ~~~\forall z \in \mathbb{Z} \nonumber\\
	  (-1)^{2z} & = & 1 ~~~~~~~~~~~\forall z \in \mathbb{Z} \nonumber\\
          (-1)^{3k} & = & (-1)^{-k}~~~\forall k=\frac{z}{2} ~~~\mathrm{with} ~z \in \mathbb{Z}
      \ea
\subsection{\label{Symmetry Properties}Symmetry Properties} 
    The $6j$-symbols are invariant under
    \begin{itemize}
      \item any permutation of the columns:
      \ba \label{symmetry1}
      \left\{ \begin{array}{ccc}
      j_1 & j_2 & j_3\\
      j_4 & j_5 & j_6
      \end{array} \right\} 
      =
      \left\{ \begin{array}{ccc}
      j_2 & j_3 & j_1\\
      j_5 & j_6 & j_4
      \end{array} \right\} 
      =
      \left\{ \begin{array}{ccc}
      j_3 & j_1 & j_2\\
      j_6 & j_4 & j_5
      \end{array} \right\} 
      =
      \left\{ \begin{array}{ccc}
      j_2 & j_1 & j_3\\
      j_5 & j_4 & j_6
      \end{array} \right\} 
      =
      \left\{ \begin{array}{ccc}
      j_1 & j_3 & j_2\\
      j_4 & j_6 & j_5
      \end{array} \right\} 
      =
      \left\{ \begin{array}{ccc}
      j_3 & j_2 & j_1\\
      j_6 & j_5 & j_4
      \end{array} \right\} 
      \ea
    \item simultaneous interchange of the upper and lower arguments of two columns, e.g. 
      \begin{equation} \label{symmetry2}
        \left\{ \begin{array}{ccc}
	j_1 & j_2 & j_3\\
	j_4 & j_5 & j_6
	\end{array} \right\} 
	=
	\left\{ \begin{array}{ccc}
	j_1 & j_5 & j_6\\
	j_4 & j_2 & j_3
	\end{array} \right\} 	  
      \end{equation}
    \end{itemize}

 \subsection{Orthogonality and Sum Rules}
    \begin{description}
       \item[Orthogonality Relations]
	  \ba \label{orthogonality relation}
	     \sum_{j_{23}}(2j_{12}+1)(2j_{12}^{'}+1)	
	        \left\{ \begin{array}{ccc}
		j_1 & j_2 & j_{12}\\
		j_3 & j & j_{23}
		\end{array} \right\}
  		\left\{ \begin{array}{ccc}
		j_1 & j_2 & j_{12}^{'} \\
		j_3 & j & j_{23}
		\end{array} \right\} 
		& = &
		\delta_{j_{12} j_{12}^{'}}
 	  \ea
       \item[Composition Relation]
	  \ba
	     \sum_{j_{23}}(-1)^{j_{23}+j_{31}+j_{12}}(2j_{23}+1)	
	        \left\{ \begin{array}{ccc}
		j_1 & j_2 & j_{12}\\
		j_3 & j & j_{23}
		\end{array} \right\}
  		\left\{ \begin{array}{ccc}
		j_2 & j_3 & j_{23} \\
		j_1 & j & j_{31}
		\end{array} \right\} 
		& = &
  		\left\{ \begin{array}{ccc}
		j_3 & j_1 & j_{31} \\
		j_2 & j & j_{12}
		\end{array} \right\} 
	  \ea
       \item[Sum Rule of Elliot and Biedenharn]
	   \ba \label{EBI}
	       \left\{ \begin{array}{ccc}
		j_1 & j_2 & j_{12}\\
		j_3 & j_{123} & j_{23}
		\end{array} \right\}
  		\left\{ \begin{array}{ccc}
		j_{23} & j_1 & j_{123} \\
		j_4 & j & j_{14}
		\end{array} \right\} 
		&=&
	        (-1)^{j_1+j_2+j_3+j_4+j_{12}+j_{23}+j_{14}+j_{123}+j}
	        \\
	        &&\times \sum_{j_{124}}
	        (-1)^{j_{124}}~(2j_{124}+1)	
	        \left\{ \begin{array}{ccc}
		j_3 & j_2 & j_{23}\\
		j_{14} & j & j_{124}
		\end{array} \right\}
  		\left\{ \begin{array}{ccc}
		j_2 & j_1 & j_{12} \\
		j_4 & j_{124} & j_{14}
		\end{array} \right\} 
		\left\{ \begin{array}{ccc}
		j_3 & j_{12} & j_{123} \\
		j_4 & j & j_{124}
		\end{array} \right\} \nonumber
	      \ea   
    \end{description}

\section{\label{Explanation SNF -> Spin States}Spin Networks and Representation Theory}
Concrete calculations performed in Loop Quantum Gravity heavily rest on the fact that the representation theory of $SU(2)$ is closely related to the theory of angular momenta familiar from quantum mechanics. Employing this correspondence we are able to use the powerful techniques provided by recoupling theory of angular momenta in Loop Quantum Gravity. However, as we will describe in this section, there are certain subtleties which have to be worked out carefully in order to fully establish the indicated correspondence. 

 \subsection{Spin Network Functions}

\subsubsection{Defining a Basis of Angular Momentum Eigenstates}

Due to the $Peter~\&~Weyl$ theorem we have \cite{TT:big script} :  
\be\label{Peter & Weyl}
   \int_{SU(2)}d\mu_H(h) \PIbar{h}{j'}{m'n'} \PI{h}{j}{mn}
   =\frac{1}{2j+1}\delta_{j'j}\delta_{m'm}\delta_{n'n}
\ee
which states that the representation matrix element functions $\PI{h}{j}{mn}$ are orthogonal with respect to the Haar measure $d\mu_H(h)$ on $SU(2)$. Theorem (\ref{Peter & Weyl}) enables us to introduce an orthonormal basis\footnote{We
will use Dirac's bracket notation here.} on the total Hilbert space
\be
  \mathcal{H}=\bigoplus_j \mathcal{H}^{(j)}
\ee
by 
\be\label{preliminary basis definition}
   \big<h\big|jmn\big>:=\sqrt{2j+1}\PI{h}{j}{mn}
   ~~~~~\mbox{and}~~~~~
   \big<jmn\big|h\big>:=\sqrt{2j+1}~~\PIbar{h}{j}{mn}=\sqrt{2j+1}\PI{h}{j}{nm}^{-1}
\ee 
where by construction
\ba\label{orthonormality of the preliminary basis} 
   \big<jmn\big|j'm'n'\big>&:=&\int_{SU(2)}d\mu_H(h)\big<jmn\big|h\big>\big<h\big|j'm'n'\big>=
\sqrt{(2j'+1)(2j+1)}\int_{SU(2)}d\mu_H(h) \PIbar{h}{j}{mn} \PI{h}{j'}{m'n'} 
 \NN
 &=& \delta_{j'j}\delta_{m'm}\delta{n'n}
\ea
Therefore the meaning of the representation matrices is twofold: On the one hand the rescaled representation matrix element functions $\sqrt{2j+1}\PI{\cdot}{j}{mn}=:\Sket{j}{m}{n}$ provide an orthonormal basis, whereas the representation matrix $\PI{h}{j}{mn}$ of a specific group  element $h\in SU(2)$ should act as a unitary transformation matrix on the basis states  as 
\be\label{wanted unitary trafo of rep matrix on basis state}
   \PI{h}{j}{}~\Sket{j}{m}{n}=\sum_{m'} \PI{h}{j}{m'm}~\Sket{j}{m'}{n}
\ee

\subsection{Right Invariant Vector Fields}

\subsubsection{Action on Group Valued Functions}

The action of right invariant vector fields $\big(X_k f\big)(g)=\frac{d}{dt}f\big(\mb{e}^{t\tau_k}g\big)\big|_{t=0}$ on group valued functions $f(g)$, $g\in SU(2)$ is evaluated as:
\ba\label{action of right invariant vector fields on general functions}
   \big(X_kf\big)(g)
   &=&\frac{d}{dt} f(\mb{e}^{t\tau_k}g)\Big|_{t=0}
   =\sum_{A~B}\frac{\partial f(\mb{e}^{t\tau_k}g)}{\partial(\mb{e}^{t\tau_k}g)_{AB}}       
      \frac{d}{dt}(\mb{e}^{t\tau_k}g)_{AB}\Big|_{t=0}
  =\sum_{A\,B\,C}\frac{\partial f(\mb{e}^{t\tau_k}g)}{\partial(\mb{e}^{t\tau_k}g)_{AB}}       
      ~~\frac{d}{dt}(\mb{e}^{t\tau_k})_{AC}~~(g)_{CB}\Big|_{t=0}
   \NN
   &=&\tr\big[\tau_k~g~\frac{\partial}{\partial g}\big]~f(g)   
\ea                                   
where we have used the defining $(j=\frac{1}{2}$) representation of $SU(2)$ in order to define the derivation of the group element. In the last line we then express the sum over the defining representation indices as a trace. Note, however, that using the defining representation here is a matter of choice, we are free to choose any other representation and accordingly the trace in (\ref{action of right invariant vector fields on general functions}) is then defined with respect to the chosen representation.
From (\ref{action of right invariant vector fields on general functions}) the commutation relations of the $X_k$ are found to be
\be\label{commutator ri vf}
  \Big(\big[X_i,X_j\big]f\Big)(g)=-2\epsilon^{ijk}\big(X_kf\big)(g)  
\ee

The action of the right invariant vector fields on  representation matrix element functions can be evaluated as:
\ba\label{action of right invariant vector fields 1}
   \big(X_k\PI{\cdot}{j}{mn}\big)(g)
   &=&\frac{d}{dt} \PI{\mb{e}^{t\tau_k}g}{j}{mn}\Big|_{t=0}
   =\sum_r\frac{d}{dt} \PI{\mb{e}^{t\tau_k}\mb{1}}{j}{mr}\Big|_{t=0}~\PI{g}{j}{rn}
   =\sum_r \PI{\tau_k}{j}{mr}\PI{g}{j}{rn}
   \NN
   &\equiv&\sum_r~[\tau_k]^{(j)}_{m~r}\PI{g}{j}{rn}
\ea                                   
where we use the representation property\footnote{$\PI{g_1g_2}{j}{}=\PI{g_1}{j}{}\PI{g_2}{j}{}$} and the definition of the $\tau_k$'s as a basis for the tangent space of the group at the identity element $\mb{1}_{SU(2)}$.
 Note that we have changed the notation in the last line to indicate that the resulting $(2j+1)$ dimensional matrix $[\tau_k]^{(j)}_{r~m}$ is an element of the Lie algebra $su(2)$ rather than a group element. 
 Moreover as mentioned above we choose here the representation of weight $j$ in order to evaluate the trace. 

However (\ref{action of right invariant vector fields 1} )  yields a possible obstacle: while we would like to write the action of any transformation on the basis states (\ref{preliminary basis definition}) to be according to (\ref{wanted unitary trafo of rep matrix on basis state}), the action of the right invariant vector fields (\ref{action of right invariant vector fields 1}) is defined as a matrix multiplication, that is it transforms the basis states (\ref{preliminary basis definition}) as vectors rather than as basis states.
Therefore we have to take a closer look at the group multiplication properties.

\subsubsection{Group Multiplication --- Introducing a New Basis}

For the representation of a general group multiplication we find
\ba\label{evaluation of group multiplication 1}
  \PI{gh}{j}{mn}
    &=& \sum_{m''}\PI{g}{j}{mm''}\PI{h}{j}{m''n}=\frac{1}{\sqrt{2j+1}}\sum_{m''}\PI{g}{j}{m''m}^T
    \big<h\Sket{j}{m''}{n}
    \NN 
  &=&\frac{1}{\sqrt{2j+1}} \sum_{m''}\PI{\epsilon^{-1}~g^{-1}~\epsilon}{j}{m''m}
     \big<h\Sket{j}{m''}{n}
  \NN 
  &=&\frac{1}{\sqrt{2j+1}} \sum_{r~s~m''}
         \PI{\epsilon}{j}{sm}~\PI{g^{-1}}{j}{rs}~\PI{\epsilon^{-1}}{j}{m''r}
     \big<h\Sket{j}{m''}{n}
\ea

By observation, (\ref{evaluation of group multiplication 1}) suggests a solution to the problem above: If we slightly modify our preliminary basis (\ref{preliminary basis definition}) by absorbing the according $\epsilon$ components, we can introduce a  new basis $\Sket{j}{m}{;n}$ as 
\ba\label{final basis definition}
   \Sket{j}{m}{;n} &:=& \sum_r \PI{\epsilon^{-1}}{j}{rm}\Sket{j}{r}{n}
                   =(-1)^{j-m}\Sket{j}{-\!m}{n}
                   =(-1)^{j-m}\sqrt{2j+1}\PI{\cdot}{j}{-\!m\,n}
                   \NN\NN
   \Sbra{j}{m}{;n} &:=& \sum_r \PI{\epsilon}{j}{mr}\Sbra{j}{r}{n} 
                    =(-1)^{j-m}\Sbra{j}{-\!m}{n}
                   =(-1)^{j-m}\sqrt{2j+1}~~\PIbar{\cdot}{j}{-\!m\,n}
\ea
which is orthonormal as well by the unitarity of $\PI{\epsilon}{j}{mn}$\footnote{
Also this can easily be seen from (\ref{orthonormality of the preliminary basis}) and the fact that $(-1)^{2(j-m)}=1$.}.
Note that there is an ambiguity in the definition (\ref{final basis definition}) coming from the fact that due to (\ref{SU(2) unitarity properties}) 
$\PI{\epsilon^{-1}~g^{-1}~\epsilon}{}{}=\PI{\epsilon~g^{-1}~\epsilon^{-1}}{}{}$, and therefore we can exchange $\epsilon^{-1}\leftrightarrow\epsilon$ consistently in (\ref{final basis definition}).  

Now continuing from (\ref{evaluation of group multiplication 1}), and using 
(\ref{final basis definition}), group multiplication can be rewritten as
\ba\label{evaluation of group multiplication 2}
  \PI{gh}{j}{mn}
  &=& \frac{1}{\sqrt{2j+1}} \sum_{r~s}
         \PI{\epsilon^{-1}}{j}{ms}~\PI{g^{-1}}{j}{rs}~\big<h\Sket{j}{r}{;n}
  = \frac{1}{\sqrt{2j+1}} \sum_{r~s}
         (-1)^{j+m}\delta_{m~-\!s}~\PI{g^{-1}}{j}{rs}~\big<h\Sket{j}{r}{;n}
  \NN       
  &=& \frac{1}{\sqrt{2j+1}} \sum_{r}
         (-1)^{j+m}~\PI{g^{-1}}{j}{r\,-\!m}~\big<h\Sket{j}{r}{;n}                 
\ea
We have now formally achieved a form analogous to the transformation property (\ref{wanted unitary trafo of rep matrix on basis state}) on the redefined basis (\ref{final basis definition}).

\subsubsection{Rewriting the Action of the Right Invariant Vector Fields}

Let us evaluate the action of the modified right invariant vector fields $X_k$ on the representation matrix element functions $\PI{\cdot}{j}{mn}$. 
By using (\ref{tau identity}) as $-\overline{\tau}_k^T=\tau_k=-\tau_k^{-1}$  we can evaluate the action of the $X_k$ on the states $\Sket{j}{m}{;n}$ as follows
\ba\label{x_k as true basis transformation....} 
  X_k~\big<h\Sket{j}{m}{;n}
  &=&
   \sqrt{2j+1}(-1)^{j-m} \Big(X_k \PI{\cdot}{j}{-\!m~n}\Big)(h)
  =\sqrt{2j+1}(-1)^{j-m}\frac{d}{dt}\Big(\PI{\mb{e}^{t\tau_k}h}{j}{-\!m~n} \Big)~\bigg|_{t=0}
  \NN
  &\stackrel{(\ref{evaluation of group multiplication 2})}{=}&
   \sum_{r} (-1)^{2(j-m)}\frac{d}{dt}\Big(\PI{\mb{e}^{t\tau_k}\mb{1}}{j}{r~m}^{-1} \Big)~\bigg|_{t=0}
     \big<h~\Sket{j}{r}{;n}
   \NN
  &=& - \sum_{r} [\tau_k]^{(j)}_{r~m}~
     \big<h~\Sket{j}{r}{;n}
\ea
where the action of the right invariant vector fields may now be interpreted as a basis transformation. The prefactor $(-1)^{2(j-m)}=1$ because $j-m$ is an integer number. Here we again have changed the notation in the last line to indicate that the resulting $(2j+1)$-dimensional matrix $[\tau_k]^{(j)}_{r~m}$ is an element of the Lie algebra $su(2)$ rather than a group element. However,  
we will abuse the notation a tad by writing  
$[\tau_k]^{(j)}_{r~m}$ as $\PI{\tau_k}{j}{rm}$.

\subsection{\label{relation SNF spin system}Correspondence to Angular Momentum Theory}

Now we introduce the modified vector fields\footnote{Note that we can raise and lower $su(2)$-indices with the Cartan metric on $su(2)$.  Since this is simply $\delta_{ij}=\delta^{ij}$, there is no difference between upper and lower indices. We thus place $su(2)$-indices as is convenient throughout the paper.} $Y^k:=-\frac{\mb{i}}{2}X_k$, and their commutation relations as evaluated on group valued functions, which easily follows from (\ref{commutator ri vf}):
\be\label{commutation of Y_k on final basis states}
   \big[Y^i,Y^j\big]~\big<h~\Sket{j}{m}{;n}=\mb{i}\epsilon^{ijk}Y^k~\big<h~\Sket{j}{m}{;n}
\ee
These are the commutation relations of angular momentum operators. Moreover using 
(\ref{x_k as true basis transformation....}) the $Y_k$ act on the basis states
 (\ref{final basis definition}) as   
\ba\label{action of Y_k ao final basis states} 
  Y^k
  &=& \frac{i}{2} \sum_{r}\PI{\tau_k}{j}{r~m}~
     \big<h~\Sket{j}{r}{;n}
\ea
Thus we can identify the action of $Y^k$ on the new basis states $\Sket{j}{m}{;n}$ with the usual action of the angular momentum operators $Y^k=J^k$ on basis states of an abstract spin system $\big|j~m\big>_{(n)}$\footnote{Here n denotes an additional quantum number which is irrelevant for the action of $J^k$.}:
\be
   Y^k~\Sket{j}{m}{;n}~~\Leftrightarrow~~J^k~\big|j~m\big>_{(n)}
\ee
 and obtain the usual angular momentum and ladder operator algebra and the algebra of ladder operators as can be found in \cite{Edmonds}.
Upon defining the action of the $Y$'s in this way there is a correspondence 
$Y^\pm\leftrightarrow J^\pm$ and therefore one realizes the following commutation relations (which are compared to the commutation relations of the usual angular momentum operators acting on spin states $\ket{j~m}$).
\barr{llllllllllllllll}\label{Y Commutation Relations}
   \big[(Y)^2,Y^k\big]&=&0 & ~~~&\big[(J)^2,J^k]&=&0 
   &~\hspace{2cm}~&\big[Y^3,Y^+\big]&=&Y^+ &~~~&\big[J^3,J^+\big]&=&J^+ \\
   \big[Y^i,Y^j\big]&=&\mb{i}\epsilon^{ijk}Y^k && \big[J^i,J^j\big]&=&\mb{i}\epsilon^{ijk}J^k
   &~\hspace{2cm}~&\big[Y^3,Y^-\big]&=&-Y^- &~~~&\big[J^3,J^-\big]&=&-J^- \\
   &&&&&&
   &~\hspace{2cm}~&\big[Y^+,Y^-\big]&=&2Y^3 &~~~&\big[J^+,J^-\big]&=&2J^3
\earr

\subsubsection{Rewriting Right Invariant Vector Field Expressions }
We can now give the general prescription for translating the action of the modified right invariant vector fields $Y_k$ on holonomies into the action of usual angular momentum operators on abstract spin systems.
\be\label{correspodence new basis, spin system}
\fbox{$\displaystyle      Y^k~\sqrt{2j+1}\PI{\cdot}{j}{mn}
   =  Y^k~(-1)^{j+m}~\Sket{j}{-\!m}{;n}  
   ~~~~\Leftrightarrow~~~~  
      (-1)^{j+m}~ J^k~\big|~j~-\!m~\big>_{(n)}$}
\ee

\subsubsection{Recoupling Schemes }
Due to (\ref{correspodence new basis, spin system}) we can therefore apply recoupling theory to the states $\Sket{j}{m}{;n}$ as follows:
\ba\label{Recoupling Schemes final basis states}
   \big|~\vec{a}~J~M~;~\vec{n}~\vec{j}~\big>
   &:=&\sum_{m_1+m_2+\ldots+m_N=M} 
            \CGC{j_1\,m_1}{j_2\,m_2}{a_2\,\tilde{m}_2}
            ~\times~\ldots~\times
            \CGC{a_{N-1}\,\tilde{m}_{N-1}}{j_N\,m_N}{J\,M}
   \NN[-2mm]
   &&\hspace{5cm}
      \times
      \Sket{j_1}{m_1}{;n_1} \otimes \Sket{j_2}{m_2}{;n_2}\otimes\ldots  \otimes\Sket{j_N}{m_N}{;n_N}
   \NN
\ea

\subsection{Historical Remark }
For the sake of completeness we would like to add a remark on the development of correspondence
(\ref{correspodence new basis, spin system}) here. The necessity of (\ref{wanted unitary trafo of rep matrix on basis state}) leading to 
\ba
  Y^k\big<h~\Sket{j}{m}{;n}
  &=& \frac{\mb{i}}{2} \sum_{r}\PI{\tau_k}{j}{r\,m}~
     \big<h~\Sket{j}{r}{;n}
\ea
was realized for the first time in \cite{Consistency Check I,Consistency Check II}, where the authors evaluated the detailed matrix elements for the $Y_k$. In calculations in the spin network formalism before \cite{Consistency Check I,Consistency Check II}  
one always used the preliminary basis (\ref{preliminary basis definition}) and directly evaluated the  action of the $Y_k$ resulting from the group multiplication (\ref{action of right invariant vector fields 1}) as
\ba\label{old action of Y_k}
  Y^k\big<h~\Sket{j}{m}{n}
  &=& -\frac{\mb{i}}{2} \sum_{r}\PI{\tau_k}{j}{m~r}~
     \big<h~\Sket{j}{r}{;n}
\ea
This was justified because the commutation relations of the $Y^k$ resulting from 
(\ref{old action of Y_k})
\be
   \big[Y^i,Y^j\big]~ \big<h~\Sket{j}{m}{n}=\mb{i}\epsilon^{ijk}Y^k ~\big<h~\Sket{j}{m}{n}
\ee 
are identical to those of (\ref{commutation of Y_k on final basis states}). As a result one directly identified
\be
   Y^k~\Sket{j}{m}{n}~~~~~~~\leftrightarrow~~~~~~~~J^k~\Sket{j}{m}{}_{(n)}
\ee
However, comparing the operators obtained from (\ref{old action of Y_k}) on a representation of weight $j$ we find:
\ba
   \big[Y_{(j)}^1 \big]_{mn}
   &=&-\frac{\mb{i}}{2}\PI{\tau_1}{j}{mn}
      =-\frac{1}{2}~A(j,n+1)~ \delta_{m~n+1}~~-~~\frac{1}{2}~A(j,n)~\delta_{m~n-1}
   \NN
   \big[Y_{(j)}^2 \big]_{mn}
   &=&-\frac{\mb{i}}{2}\PI{\tau_2}{j}{mn}
      =~~\frac{\mb{i}}{2}~A(j,n+1)~ \delta_{m~n+1}~~-~~\frac{\mb{i}}{2}~A(j,n)~\delta_{m~n-1} 
   \NN
   \big[Y_{(j)}^3 \big]_{mn}
   &=&-\frac{\mb{i}}{2}\PI{\tau_3}{j}{mn} =~-~m~ \delta_{m~n}  
\ea

One can already see an overall minus sign compared to (\ref{angular momentum operators}).
From this one can formulate ladder operators
\ba
  Y_{(j)}^+&:=&\big[Y_{(j)}^1-\mb{i} Y_{(j)}^2 \big]=-A(j,m+1)~ \delta_{m\,n\!-\!1}
  \NN
  Y_{(j)}^-&:=&\big[Y_{(j)}^1+\mb{i} Y_{(j)}^2 \big]=-A(j,m)~ \delta_{m\,n\!+\!1}
\ea
but notices the difference in their definition as compared to (\ref{usual angualr momentum ops}), due to the assumed matrix multiplication instead of the contragredient action. One finds 
\barr{lclclcl}\label{action of Y operators on basis states}
   \displaystyle\sum_r [Y_{(j)}^+]_{mr}\Sket{j}{r}{n}&=& -A(j,m+1)~\Sket{j}{m\!+\!1}{n}
   &~~~~~&
   \displaystyle\sum_r [Y_{(j)}^-]_{mr}\Sket{j}{r}{n}&=& -A(j,m)~\Sket{j}{m\!-\!1}{n}
   \\
   \displaystyle\sum_r [Y_{(j)}^3]_{mr}\Sket{j}{r}{n}&=& -m~\Sket{j}{m}{n}
   &~~~~~&
   \displaystyle\sum_r [(Y_{(j)})^2]_{mr}\Sket{j}{r}{n}&=& j(j+1)~\Sket{j}{m}{n}
\earr
where $(Y)^2=(Y^1)^2+(Y^2)^2+(Y^3)^2=\frac{1}{2}Y^-Y^+ +\frac{1}{2}Y^+Y^- + (Y^3)^2$.
Upon defining the action of the $Y$'s in this way there is a correspondence 
$Y^\pm\leftrightarrow J^\mp$ and therefore one realizes the following commutation relations
(as compared to the commutation relations of the usual angular momentum operators acting on spin states $\ket{j~m}$).
\barr{llllllllllllllll}\label{Y Commutation Relations on preliminary basis}
   \big[(Y)^2,Y^k\big]&=&0 & ~~~&\big[(J)^2,J^k]&=&0 
   &~\hspace{2cm}~&\big[Y^3,Y^+\big]&=&-Y^+ &~~~&\big[J^3,J^+\big]&=&J^+ \\
   \big[Y^i,Y^j\big]&=&\mb{i}\epsilon^{ijk}Y^k && \big[J^i,J^j\big]&=&\mb{i}\epsilon^{ijk}J^k
   &~\hspace{2cm}~&\big[Y^3,Y^-\big]&=&Y^- &~~~&\big[J^3,J^-\big]&=&-J^- \\
   &&&&&&
   &~\hspace{2cm}~&\big[Y^+,Y^-\big]&=&-2Y^3 &~~~&\big[J^+,J^-\big]&=&2J^3
\earr
Although the $Y$'s obey the usual angular momentum commutation relations on the left hand side of (\ref{Y Commutation Relations on preliminary basis}), there is a difference in the signs of the ladder operator commutator algebra,  
providing possible phase convention obstacles when applying techniques of recoupling theory of angular momentum.

\subsubsection{Recoupling Schemes}
In (\ref{correspodence new basis, spin system}) we have found a matching between the basis states $\Sket{j}{m}{;n}$ as defined in (\ref{final basis definition}) to the states $\Sket{j}{m}{}_{(n)}$ of an ordinary angular momentum system. 
Using (\ref{final basis definition}), where
\ba\label{relation preliminary final}
   \Sket{j}{m}{;n} &=&(-1)^{j-m}\Sket{j}{-\!m}{n}
   ~~~~~~\mbox{and}~~~~~~
   \Sbra{j}{m}{;n} =(-1)^{j-m}\Sbra{j}{-\!m}{n}
\ea
we can relate the states $\Sket{j}{-\!m}{n}$ to the abstract spin states $\Sket{j}{m}{}_{(n)}$ and can relate the recoupling scheme (\ref{Recoupling Schemes final basis states})
\ba
   \big|~\vec{a}~J~M~;~\vec{n}~\vec{j}~\big>
   &:=&\sum_{m_1+m_2+\ldots+m_N=M} 
            \CGC{j_1\,m_1}{j_2\,m_2}{a_2\,\tilde{m}_2}
            ~\times~\ldots~\times
            \CGC{a_{N-1}\,\tilde{m}_{N-1}}{j_N\,m_N}{J\,M}
   \NN[-2mm]
   &&\hspace{5cm}
      \times
      \Sket{j_1}{m_1}{;n_1} \otimes \Sket{j_2}{m_2}{;n_2}\otimes\ldots  \otimes\Sket{j_N}{m_N}{;n_N}
   \NN
   &:=&\sum_{m_1+m_2+\ldots+m_N=M} 
            \CGC{j_1\,m_1}{j_2\,m_2}{a_2\,\tilde{m}_2}
            ~\times~\ldots~\times
            \CGC{a_{N-1}\,\tilde{m}_{N-1}}{j_N\,m_N}{J\,M}
 \NN
   &&~~~~~\times (-1)^{j_1+\ldots+j_N}(-1)^{-m_1-\ldots-m_N}
      \times
      \Sket{j_1}{-\!m_1}{n_1}\otimes \Sket{j_2}{-\!m_2}{n_2}\otimes\ldots  \otimes\Sket{j_N}{-\!m_N}{n_N}
   \NN
   &&\fcmt{7}{Introduce $\mu_k=-m_k$, $\tilde{\mu}_k=-\tilde{m}_k$, $\mu=-M$}
   \NN
   &=&\sum_{\mu_1+\mu_2+\ldots+\mu_N=\mu} \hspace{-0.6cm}
            \CGC{j_1\,-\!\mu_1}{j_2\,-\!\mu_2}{a_2\,-\!\tilde{\mu}_2}
            ~\times~\ldots~\times
            \CGC{a_{N-1}\,-\!\tilde{\mu}_{N-1}}{j_N\,-\!\mu_N}{J\,-\!\mu}
 \NN
   &&~~~~~\times (-1)^{j_1+\ldots+j_N}(-1)^{\mu}
      \times
      \Sket{j_1}{\mu_1}{n_1} \otimes\Sket{j_2}{\mu_2}{n_2}\otimes\ldots  \otimes\Sket{j_N}{\mu_N}{n_N}
   \NN
\ea
Using a symmetry property (See e.g.\ \cite{Edmonds}, p.42) of the Clebsch-Gordan coefficients with respect to permutation of their arguments\footnote{The sum $j_1+j_2-a_2$ is an integer number.}
\be
  \CGC{j_1~-\!m_1}{j_2~-\!m_2}{a_2~-\!\tilde{m}_2}
  =(-1)^{-j_1-j_2+a_2} \CGC{j_1~m_1}{j_2~m_2}{a_2~\tilde{m}_2}
\ee
we can continue
\ba\label{Relation Recoupling Schemes final basis states preliminary basis states}
   \big|~\vec{a}~J~M~;~\vec{n}~\vec{j}~\big>
  &=&\sum_{\mu_1+\mu_2+\ldots+\mu_N=\mu} \hspace{-0.6cm}
            \CGC{j_1\,\mu_1}{j_2\,\mu_2}{a_2\,\tilde{\mu}_2}
            ~\times~\ldots~\times
            \CGC{a_{N-1}\,\tilde{\mu}_{N-1}}{j_N\,\mu_N}{J\,\mu}
 \NN
   &&~~~~~\times (-1)^{J}(-1)^{\mu}
      \times
      \Sket{j_1}{\mu_1}{n_1}\otimes \Sket{j_2}{\mu_2}{n_2}\otimes\ldots  \otimes\Sket{j_N}{\mu_N}{n_N}
   \NN
\ea
in complete analogy to (\ref{relation preliminary final}). 
A similar calculation gives for the bra state
\ba
   \big<~\vec{a}~J~M~;~\vec{n}~\vec{j}~\big|
  &=&\sum_{\mu_1+\mu_2+\ldots+\mu_N=\mu} \hspace{-0.6cm}
            \CGC{j_1\,\mu_1}{j_2\,\mu_2}{a_2\,\tilde{\mu}_2}
            ~\times~\ldots~\times
            \CGC{a_{N-1}\,\tilde{\mu}_{N-1}}{j_N\,\mu_N}{J\,\mu}
 \NN
   &&~~~~~\times (-1)^{J}(-1)^{\mu}
      \times
      \Sbra{j_1}{\mu_1}{n_1}\otimes \Sbra{j_2}{\mu_2}{n_2}\otimes\ldots \otimes \Sbra{j_N}{\mu_N}{n_N}
   \NN
\ea

We can thus conclude that for gauge invariant states $J=0,$ $M=\mu=0$ the recoupling schemes built from the preliminary basis $\Sket{j}{m}{n}$ and the final basis $\Sket{j}{m}{;n}$ are identical and thus as long as one works with gauge invariant states the ``historical'' approach coincides with the construction implemented in the previous section. Moreover, since $J+\mu=J-M$ is an integer number, the expectation values of operators diagonal\footnote{By diagonal we mean that these operators require identical $J,M$ upon evaluation in a recoupling scheme basis.} with respect to $J,M$ will also be identical using the conventions (\ref{preliminary basis definition}), (\ref{old action of Y_k}) with respect to our conventions (\ref{final basis definition}),
(\ref{action of Y_k ao final basis states}). This is in particular the case for the volume operator.

\end{appendix}

\end{document}